\pdfoutput=1
\documentclass[twocolumn,english,rmp,aps,longbibliography]{revtex4}
\usepackage[latin9]{inputenc}
\usepackage{amsmath}
\usepackage{amssymb}
\usepackage{graphicx}
\usepackage{esint}

\makeatletter

\providecommand{\tabularnewline}{\\}

\@ifundefined{textcolor}{}
{%
 \definecolor{BLACK}{gray}{0}
 \definecolor{WHITE}{gray}{1}
 \definecolor{RED}{rgb}{1,0,0}
 \definecolor{GREEN}{rgb}{0,1,0}
 \definecolor{BLUE}{rgb}{0,0,1}
 \definecolor{CYAN}{cmyk}{1,0,0,0}
 \definecolor{MAGENTA}{cmyk}{0,1,0,0}
 \definecolor{YELLOW}{cmyk}{0,0,1,0}
 }

\usepackage{amsfonts}
\providecommand{\U}[1]{\protect\rule{.1in}{.1in}}
\usepackage{lmodern}
\usepackage[T1]{fontenc}

\makeatother

\usepackage{babel}
\begin{document}

\title{Cavity Optomechanics}

\author{Markus Aspelmeyer}

\email{markus.aspelmeyer@univie.ac.at}
\email{optomechanicsrmp@gmail.com}

\selectlanguage{english}%

\affiliation{Vienna Center for Quantum Science and Technology (VCQ), Faculty of
Physics, University of Vienna, 1090 Vienna, Austria}

\author{Tobias J. Kippenberg}

\email{tobias.kippenberg@epfl.ch}

\selectlanguage{english}%

\affiliation{Ecole Polytechnique Fédérale de Lausanne (EPFL), 1015 Lausanne, Switzerland }

\author{Florian Marquardt}

\email{Florian.Marquardt@physik.uni-erlangen.de}

\selectlanguage{english}%

\affiliation{University of Erlangen-Nürnberg, Institute for Theoretical Physics,
Staudtstr. 7, 91058 Erlangen, Germany;\\
and Max Planck Institute for the Science of Light, Erlangen, Germany}
\begin{abstract}
We review the field of cavity optomechanics, which explores the interaction
between electromagnetic radiation and nano- or micromechanical motion.
This review covers the basics of optical cavities and mechanical resonators,
their mutual optomechanical interaction mediated by the radiation
pressure force, the large variety of experimental systems which exhibit
this interaction, optical measurements of mechanical motion, dynamical
backaction amplification and cooling, nonlinear dynamics, multimode
optomechanics, and proposals for future cavity quantum optomechanics
experiments. In addition, we describe the perspectives for fundamental
quantum physics and for possible applications of optomechanical devices.
\end{abstract}
\maketitle
\tableofcontents{}

\section{Introduction}

\label{sec:Introduction}

Light carries momentum which gives rise to radiation pressure forces.\ These
forces were already postulated in the 17$^{th}$ century by Kepler,
who noted that the dust tails of comets point away from the sun during
\ a comet transit \cite{1619_Kepler_CometBook}. The first unambiguous
experimental demonstrations of the radiation pressure force predicted
by Maxwell were performed using a light mill configuration \cite{Nichols1901,Lebedew1901}.
A careful analysis of these experiments was required to distinguish
the phenomenon from thermal effects that had dominated earlier observations.
As early as 1909, Einstein derived the statistics of the radiation
pressure force fluctuations acting on a moveable mirror \cite{Einstein1909},
including the frictional effects of the radiation force, and this
analysis allowed him to reveal the dual wave-particle nature of blackbody
radiation. In pioneering experiments, both the linear and angular
momentum transfer of photons to atoms and macroscopic objects were
demonstrated by Frisch \cite{1933_Frisch} and by Beth \cite{Beth1936},
respectively.

\begin{figure}[ptb]
\centering\includegraphics[width=3in]{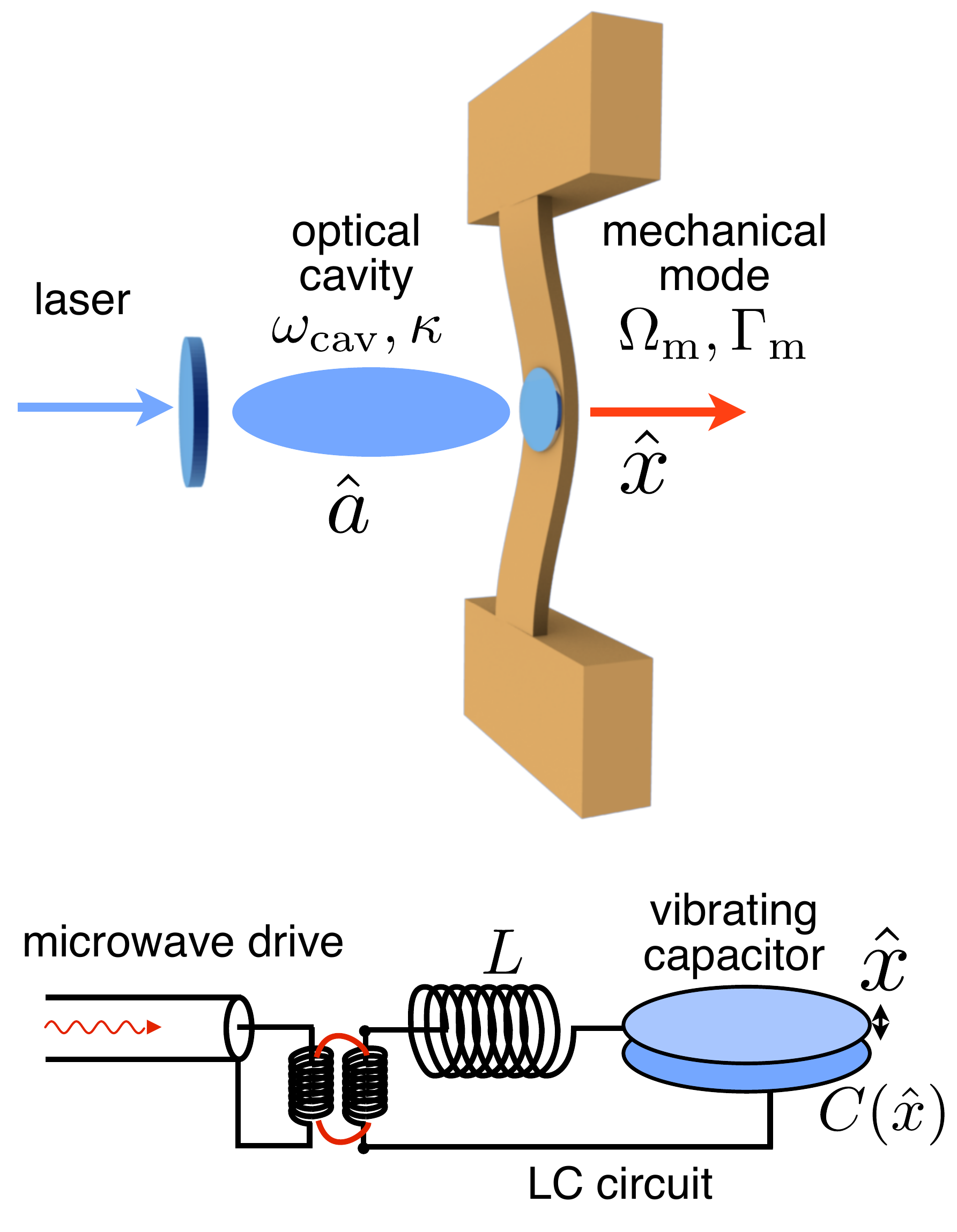}\caption{\label{Fig:GenericSetup}Schematic of a generic optomechanical system,
both in the optical domain (top), with a laser-driven optical cavity
and a vibrating end mirror, as well as in the microwave domain (bottom),
with a vibrating capacitor. Here we have depicted a microwave drive
entering along a transmission line that is inductively coupled to
the LC circuit representing the microwave resonator.}
\end{figure}

In the 1970s Arthur Ashkin demonstrated that focused lasers beams
can be used to trap and control dielectric particles, which also included
feedback cooling \cite{Ashkin1978,Ashkin2006}. The non-conservative
nature of the radiation pressure force and the resulting possibility
to use it for cooling atomic motion was first pointed out by Hänsch
and Schawlow and by Dehmelt and Wineland \cite{Hansch1975,Wineland1975a}.\ Laser
cooling was subsequently realized experimentally in the 1980s and
has become since then an extraordinarily important technique \cite{Stenholm1986}.
It has, for example, allowed cooling of ions to their motional ground
state and it is the underlying resource for ultracold atom experiments.
Many applications have been enabled by laser cooling \cite{Metcalf1999},
including optical atomic clocks, precision measurements of the gravitational
field, and systematic studies of quantum many-body physics in trapped
clouds of atoms \cite{Bloch2008}. 

The role of radiation pressure and its ability to provide cooling
for larger objects was already investigated earlier by Braginsky in
the context of interferometers. Braginsky considered the dynamical
influence of radiation pressure on a harmonically suspended end-mirror
of a cavity. His analysis revealed that the retarded nature of the
force, due to the finite cavity lifetime, provides either damping
or anti-damping of mechanical motion, two effects that he was able
to demonstrate in pioneering experiments using a microwave cavity
\cite{Braginsky1967,Braginsky1970}. In later experiments, these phenomena
were also observed in microwave-coupled $kg$-scale mechanical resonators
\cite{Cuthbertson1996}. Independently, similar physics was explored
theoretically for solid-state vibrations \cite{Dykman1978a}. In the
optical domain, the first cavity optomechanical experiment \cite{Dorsel1983a}
demonstrated bistability of the radiation pressure force acting on
a macroscopic end-mirror.

Braginsky also addressed the fundamental consequences of the quantum
fluctuations of radiation pressure and demonstrated that they impose
a limit on how accurately the position of a free test mass (e.g. a
mirror) can be measured \cite{Braginsky1977,Braginsky1995}. A detailed
analysis by Caves clarified the role of this ponderomotive quantum
noise in interferometers \cite{Caves1980b}. These works established
the standard quantum limit for continuous position detection, which
is essential for gravitational wave detectors such as LIGO or VIRGO. 

During the 1990s, several aspects of quantum cavity optomechanical
systems started to be explored theoretically. These include squeezing
of light \cite{Fabre1994b,Mancini1994} and quantum non-demolition
(QND) detection of the light intensity \cite{Jacobs1994,Pinard1995},
which exploit the effective Kerr nonlinearity generated by the optomechanical
interaction. It was also pointed out that for extremely strong optomechanical
coupling the resulting quantum nonlinearities could give rise to nonclassical
and entangled states of the light field and the mechanics \cite{Mancini1997,Bose1997}.
Furthermore, feedback cooling by radiation pressure was suggested
\cite{Mancini1998}. Around the same time, in a parallel development,
cavity-assisted laser-cooling was proposed as a method to cool the
motion of atoms and molecules that lack closed internal transitions
\cite{Hechenblaikner1998,Vuletic2000}.

On the experimental side, optical feedback cooling based on the radiation
pressure force was first demonstrated in \cite{Cohadon1999a} for
the vibrational modes of a macroscopic end-mirror. This approach was
later taken to much lower temperatures \cite{Kleckner2006a,Poggio2007}.
At the same time, there was a trend to miniaturize the mechanical
element. For example, the thermal motion of a $mm$-scale mirror was
monitored in a cryogenic optical cavity \cite{Tittonen1999}. Producing
high-quality optical Fabry-Perot cavities below that scale, however,
turned out to be very challenging. In spite of this, it was still
possible to observe optomechanical effects of retarded radiation forces
in microscale setups where the forces were of photothermal origin,
effectively replacing the cavity lifetime with a thermal time constant.
Examples include demonstration of the optical spring effect \cite{Vogel2003a},
feedback damping \cite{Mertz1993b}, self-induced oscillations \cite{Zalalutdinov2001b,HoehbergerProceedings2004},
and cavity cooling due to the dynamical backaction of retarded photothermal
light forces \cite{HohbergerMetzger2004}. 

Yet, for future applications in quantum coherent optomechanics it
is highly desirable to be able to exploit the non-dissipative radiation
pressure force. Both the advent of optical microcavities and of advanced
nanofabrication techniques eventually allowed to enter this regime.
In 2005 it was discovered that optical microtoroid resonators with
their high optical finesse at the same time contain mechanical modes
and thus are able to display optomechanical effects, in particular
radiation-pressure induced self-oscillations \cite{Rokhsari2005a,Carmon2005,Kippenberg2005}
(i.e. the effect Braginsky termed {}``parametric instability''%
\footnote{Braginsky called the process of dynamical backaction amplification
and the concomitant self-induced coherent oscillations {}``parametric
oscillatory instability'', as this effect is undesirable in gravitational
wave interferometers which were the basis of his analysis.%
}). In 2006 three different teams demonstrated radiation-pressure cavity
cooling, for suspended micromirrors \cite{Gigan2006,Arcizet2006a}
and for microtoroids \cite{Schliesser2006a}. Since then, cavity optomechanics
has advanced rapidly and optomechanical coupling has been reported
in numerous novel systems. These include membranes \cite{Thompson2008}
and nanorods \cite{Favero2009} inside Fabry-Perot resonators, whispering
gallery microdisks \cite{Wiederhecker2009,Jiang2009} and microspheres
\cite{Ma2007,Park2009,2009_Carmon_StimulatedBrillouinScattering},
photonic crystals \cite{Eichenfield2009a,Eichenfield2009}, and evanescently
coupled nanobeams \cite{Anetsberger2009a}. In addition, cavity optomechanics
has been demonstrated for the mechanical excitations of cold atom
clouds \cite{Murch2008,Brennecke2008}. Optomechanical interactions
are also present in optical waveguides - as first studied and observed
in the context of squeezing, where the confined mechanical modes of
fibers lead to Guided Acoustic Wave scattering \cite{Shelby1985}.
Nowadays there are a number of systems where such optomechanical interactions
are explored in the absence of a cavity, such as waveguides in photonic
circuits or photonic crystal fibres, see e.g. \cite{Li2008,2009_Russell_AcousticPhononsNonlinear}.
These setups lie somewhat outside the scope of the concepts presented
in this review, but we emphasize that they are very promising for
applications due to their large bandwidth.

Optomechanical coupling has also been realized using microfabricated
superconducting resonators, by embedding a nanomechanical beam inside
a superconducting transmission line microwave cavity \cite{Regal2008}
or by incorporating a flexible aluminum membrane into a lumped element
superconducting resonator \cite{2011_Teufel_StrongCouplingMicrowave}.
In these systems the mechanical motion capacitively couples to the
microwave cavity. This approach ties cavity optomechanics to an independent
development that has been garnering momentum since the late 1990s,
which is concerned with measuring and controlling the motion of nano-
and micromechanical oscillators using electrical and other non-optical
coupling techniques. Examples include coupling of mechanical oscillators
to single electron transistors \cite{Cleland2002,LaHaye2004a,Naik2006}
or a quantum point contact \cite{Cleland2002,Flowers-Jacobs2007}.
Besides a wealth of possible applications for such devices in sensitive
detection \cite{Cleland1998,Rugar2004,LaHaye2004}, these methods
provide the possiblility of realizing mechanical quantum devices \cite{Schwab2005,2005_Blencowe_Review,Ekinci2005}
by direct interaction with two-level quantum systems \cite{Cleland2004,Rugar2004,Wilson-Rae2004,LaHaye2009,O'Connell2010,Arcizet2011,Kolkowitz2012}.
For recent comprehensive general reviews of nanomechanical systems
(in particular electro-mechanical devices), we refer the reader to
\cite{2005_Blencowe_Review,2012_PootZant_MechSysQuantumRegime_PhysReports,Greenberg2010}. 

It should be noted that in atomic systems quantum coherent control
of mechanical motion is state of the art since early pioneering experiments
with trapped ions \textendash{} for reviews see \cite{Leibfried2003,Blatt2008,Jost2009}.
In fact, quantum information processing in these systems relies on
using the quantum states of motion to mediate interactions between
distant atomic spins \cite{Cirac1995}. In contrast, the fabricated
nano- and micromechanical structures that form the subject of this
review will extend this level of control to a different realm, of
objects with large masses and of devices with a great flexibility
in design and the possibility to integrate them in on-chip architectures.

There are several different motivations that drive the rapidly growing
interest into cavity optomechanics. On the one side, there is the
highly sensitive optical detection of small forces, displacements,
masses, and accelerations. On the other hand, cavity quantum optomechanics
promises to manipulate and detect mechanical motion in the quantum
regime using light, creating nonclassical states of light and mechanical
motion. These tools will form the basis for applications in quantum
information processing, where optomechanical devices could serve as
coherent light-matter interfaces, for example to interconvert information
stored in solid-state qubits into flying photonic qubits. Another
example is the ability to build hybrid quantum devices that combine
otherwise incompatible degrees of freedoms of different physical systems.
At the same time, it offers a route towards fundamental tests of quantum
mechanics in an hitherto unaccessible parameter regime of size and
mass. 

A number of reviews \cite{Kippenberg2007,Kippenberg2008,Marquardt2009,Favero2009a,Genes2009,Aspelmeyer2010,Schliesser2010,Cole2012,Aspelmeyer2012,Meystre2012}
and brief commentary papers \cite{Karrai2006,Cleland2009,Marquardt2011,Cole2011b}
on cavity optomechanics have been published during the past few years,
and the topic has also been discussed as part of a larger reviews
on nanomechanical systems \cite{2012_PootZant_MechSysQuantumRegime_PhysReports,Greenberg2010}.
Here we aim for a comprehensive treatment that incorporates the most
recent advances and points the way towards future challenges.

This review is organized follows: We first discuss optical cavities,
mechanical resonators, the basic optomechanical interaction between
them and the large range of experimental setups and parameters that
are now available. We then go on to derive the basic consequences
of the interaction (such as optomechanical damping and the optical
spring effect), describe various measurement schemes, and present
the quantum theory of optomechanical cooling. After studying nonlinear
effects in the classical regime, we address multimode setups and the
wide field of proposed applications in the quantum domain, before
concluding with an outlook.

\begin{table*}
\begin{tabular}{|c|l|}
\hline 
Symbol & Meaning\tabularnewline
\hline 
\hline 
$\Omega_{{\rm m}}$ & Mechanical frequency\tabularnewline
\hline 
$\Gamma_{{\rm m}}$ & Mechanical damping rate\tabularnewline
\hline 
$Q_{{\rm m}}$ & Mechanical quality factor, $Q_{{\rm m}}=\Omega_{{\rm m}}/\Gamma_{{\rm m}}$\tabularnewline
\hline 
$\Delta$ & Laser detuning from the cavity resonance, $\Delta=\omega_{L}-\omega_{{\rm cav}}$\tabularnewline
\hline 
$\kappa$ & Overall cavity intensity decay rate, from input coupling and intrinsic
losses, $\kappa=\kappa_{{\rm ex}}+\kappa_{0}$.\tabularnewline
\hline 
$g_{0}$ & Optomechanical single-photon coupling strength, in $\hat{H}_{{\rm int}}=-\hbar g_{0}\hat{a}^{\dagger}\hat{a}(\hat{b}+\hat{b}^{\dagger})$\tabularnewline
\hline 
$g$ & Light-enhanced optomechanical coupling for the linearized regime,
$g=g_{0}\sqrt{\bar{n}_{{\rm cav}}}$\tabularnewline
\hline 
$G$ & Optical frequency shift per displacement, $g_{0}=Gx_{{\rm ZPF}}$\tabularnewline
\hline 
$x_{{\rm ZPF}}$ & Mechanical zero-point fluctuation amplitude, $x_{{\rm ZPF}}=\sqrt{\hbar/2m_{{\rm eff}}\Omega_{{\rm m}}}$\tabularnewline
\hline 
$\hat{a}$ & Photon annihilation operator, with $\hat{a}^{\dagger}\hat{a}$ the
circulating photon number\tabularnewline
\hline 
$\hat{b}$ & Phonon annihilation operator, with $\hat{b}^{\dagger}\hat{b}$ the
phonon number\tabularnewline
\hline 
$\bar{n}$ & Average number of phonons stored in the mechanical resonator, $\bar{n}=\left\langle \hat{b}^{\dagger}\hat{b}\right\rangle $\tabularnewline
\hline 
$\bar{n}_{{\rm th}}$ & Average phonon number in thermal equilibrium, $\bar{n}_{{\rm th}}=(e^{\hbar\Omega_{{\rm m}}/k_{B}T}-1)^{-1}$\tabularnewline
\hline 
$\bar{n}_{{\rm cav}}$ & Photon number circulating inside the cavity, $\bar{n}_{{\rm cav}}=\left\langle \hat{a}^{\dagger}\hat{a}\right\rangle $\tabularnewline
\hline 
$\chi_{aa}(\omega)$ & Optical susceptibility of the cavity, $\chi_{aa}(\omega)=[\kappa/2-i(\omega+\Delta)]^{-1}$\tabularnewline
\hline 
$\chi_{xx}(\omega)$ & Mechanical susceptibility, $\chi_{xx}(\omega)=\left[m_{{\rm eff}}(\Omega_{m}^{2}-\omega^{2})-im_{{\rm eff}}\Gamma_{m}\omega\right]^{-1}$\tabularnewline
\hline 
$S_{xx}(\omega)$ & Quantum noise spectrum, $S_{xx}(\omega)\equiv\int dt\, e^{i\omega t}\left\langle \hat{x}(t)\hat{x}(0)\right\rangle $
(Sec.~\ref{sub:MechanicalResonatorsNoiseSpectra})\tabularnewline
\hline 
$\bar{S}_{xx}^{{\rm ZPF}}(\omega)$ & (Symmetrized) mechanical zero-point fluctuations, $\bar{S}_{xx}^{{\rm ZPF}}(\omega)=\hbar\left|{\rm Im}\chi_{xx}(\omega)\right|$\tabularnewline
\hline 
$ $ & $\bar{S}_{xx}^{{\rm add}}(\omega)\geq\bar{S}_{xx}^{{\rm ZPF}}(\omega)$:
standard quantum limit result for added noise in displacement measurement
(Sec.~\ref{sub:MeasurementsDisplacementSensingAndSQL})\tabularnewline
\hline 
$\Gamma_{{\rm opt}}$ & Optomechanical damping rate (Sec.~\ref{sub:DynamicalBackactionOptomechanicalDampingRate}):
max. $4\bar{n}_{cav}g_{0}^{2}/\kappa$ for $\kappa\ll\Omega_{{\rm m}}$ \tabularnewline
\hline 
$\delta\Omega_{{\rm m}}$ & Optical spring (mechanical frequency shift, Sec.~\ref{sub:DynamicalBackactionOpticalSpring}):
$2\bar{n}_{cav}g_{0}{}^{2}/\Delta$ for $\kappa\ll\Omega_{{\rm m}},\left|\Delta\right|$\tabularnewline
\hline 
$\bar{n}_{{\rm min}}$ & Minimum reachable phonon number in laser-cooling, $\bar{n}_{{\rm min}}=(\kappa/4\Omega_{{\rm m}})^{2}$
for $\kappa\ll\Omega_{{\rm m}}$\tabularnewline
\hline 
\end{tabular}

\caption{List of most important symbols and some formulas used in this review}
\end{table*}

\section{Optical cavities and mechanical resonators}

\label{sec:OpticalCavitiesAndMechanicalResonators}

In this section we recall the basic aspects of optical cavities and
of mechanical resonators, as needed to describe cavity optomechanical
systems. Much more about these topics can be found in standard textbooks
on quantum optics, e.g. \cite{Walls1994}, and on nanomechanical systems
\cite{Cleland2003}.

\subsection{Optical resonators}

Optical resonators can be realized experimentally in a multitude of
forms of which several types will be discussed later in the review.
Here we give a unifying description of the optical properties and
provide the mathematical description of a cavity that is pumped with
a single monochromatic laser source.

\label{sub:OpticalResonators}

\subsubsection{Basic properties}

\label{sub:OpticalResonatorsClassicalProperties}

We first consider the classical response of a simple Fabry-Perot resonator,
which will allow to introduce the relevant parameters to characterize
an optical cavity. A Fabry-Perot resonator or etalon consisting of
two highly reflective mirrors, separated by a distance $L$, contains
a series of resonances which are given by the angular frequency $\omega_{m}\approx m\cdot\pi\frac{c}{L}.$
Here $m$ is the integer mode number. The separation of two longitudinal
resonances is denoted as the free spectral range (FSR) of the cavity:
\begin{equation}
\Delta\omega_{FSR}=\pi\frac{c}{L}
\end{equation}
In the following, we will almost always focus on a single optical
mode, whose frequency we will denote $\omega_{{\rm cav}}$.

Both the finite mirror transparencies and the internal absorption
or scattering out of the cavity lead to a finite photon (intensity)
cavity decay rate%
\footnote{In this review we shall use $\kappa$ for the photon (energy) decay
rate, such that the amplitude decay rate is given by $\kappa/2$.
In some papers the latter is denoted as $\kappa$.%
} $\kappa$. 

A further useful quantity is the optical finesse, $\mathcal{F\,}$,
which gives the average number of round-trips before a photon leaves
the cavity:
\begin{equation}
\mathcal{F}\equiv\frac{\Delta\omega_{FSR}}{\kappa}\label{eq:FinesseDef}
\end{equation}
The optical finesse is a useful parameter as it gives the enhancement
of the circulating power over the power that is coupled into the resonator.
Alternatively, we can introduce the quality factor of the optical
resonator,
\begin{equation}
Q_{opt}=\omega_{{\rm cav}}\tau
\end{equation}
where $\tau=\kappa^{-1}$ is the photon lifetime. Note that the quality
factor is also used to characterize the damping rate of mechanical
resonators (see below). Generally speaking, the cavity decay rate
$\kappa$ can have two contributions, one from losses that are associated
with the (useful) input (and output) coupling and a second contribution
from the internal losses. It is useful to differentiate these two
contributions. For the case of a high-Q cavity, the total cavity loss
rate can be written as the sum of the individual contributions: 

\[
\kappa=\kappa_{ex}+\kappa_{0}.
\]

Here, $\kappa_{ex}$ refers to the loss rate associated with the input
coupling, and $\kappa_{0}$ refers to the remaining loss rate. For
example, in the case of a waveguide coupled to a microtoroidal or
microsphere resonator, $\kappa_{ex}$ is the loss rate associated
with the waveguide-resonator interface and $\kappa_{0}$ describes
the light absorption inside the resonator. For the case of a Fabry-Perot
cavity, $\kappa_{ex}$ is the loss rate at the input cavity mirror
and $\kappa_{0}$ summarizes the loss rate inside the cavity, including
transmission losses at the second cavity mirror as well as all scattering
and absorption losses behind the first mirror. Note that by splitting
the total decay rate into these two contributions, we are assuming
that the photons going into the $\kappa_{0}$ decay channel will not
be recorded. More generally, one could distinguish between more decay
channels (e.g. input mirror, output mirror, absorption).

\subsubsection{Input-output formalism for an optical cavity}

\label{sub:OpticalResonatorsQuantumFluctuations}

A quantum mechanical description of a cavity that is coupled to the
outside electromagnetic environment can be given either via master
equations (if only the internal dynamics is of interest) or via a
framework known as input-output theory, if one also wants to access
the light field being emitted by (or reflected from) the cavity. Input-output
theory allows us to directly model the quantum fluctuations injected
from any coupling port (such as the input mirror) into the cavity.
In addition, it takes into account any coherent laser drive that may
be present. For more details beyond the brief discussion provided
below, see e.g. \cite{Gardiner2004,Clerk2008a}. 

\begin{figure}
\includegraphics[width=1\columnwidth]{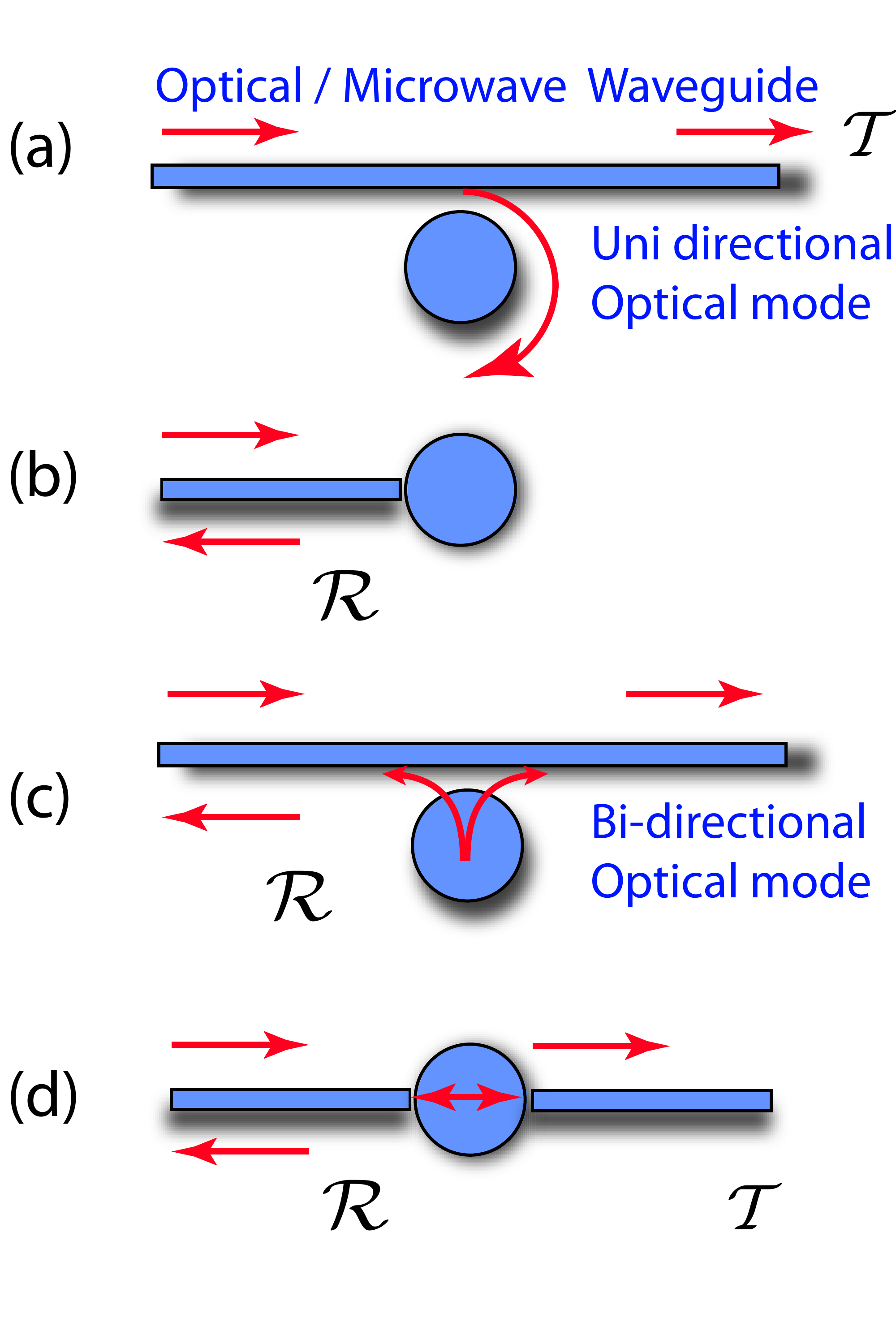}

\caption{Comparison of uni- and bi-directionally coupled cavities and the notion
of reflected and transmitted field amplitude. (a) denotes a waveguide-coupled
unidirectional resonator (e.g. WGM cavity coupled to waveguide). (b)
denotes a single-sided cavity coupled in reflection (or a double-sided
cavity where the transmission signal is disregarded). (c) denotes
a waveguide coupled bi-directional cavity, which can decay in both
forward and backward propagating waveguide modes (i.e. tapered fiber
coupled photonic crystal mode, or a waveguide-coupled quarter-wave
stripline resonators). (d) denotes the coupling to a double-sided
Fabry Perot resonator, in which both transmitted and reflected field
are measured. \label{fig:Comparison-of-cavity-geometries}}
\end{figure}

Input-output theory is formulated on the level of Heisenberg equations
of motion, describing the time-evolution of the field amplitude $\hat{a}$
inside the cavity. One finds that the amplitude $\hat{a}$ experiences
decay at a rate $\kappa/2$. At the same time, its fluctuations are
constantly replenished via the quantum noise entering through the
various ports of the cavity. In the present case, we distinguish between
the channels associated with the input coupling (decay rate $\kappa_{{\rm ex}}$)
and the other loss processes (overall decay rate $\kappa_{0}$, including
loss through the second mirror). The equation of motion reads:

\begin{equation}
\dot{\hat{a}}=-\frac{\kappa}{2}\hat{a}+i\Delta\hat{a}+\sqrt{\kappa_{ex}}\hat{a}_{\mathrm{in}}+\sqrt{\kappa_{0}}\hat{f}_{\mathrm{in}}\label{eq:InputOutputEqOfMotion}
\end{equation}
In the classical case, $\hat{a}$ would be replaced by a properly
normalized complex amplitude of the electric field of the cavity mode
under consideration. Indeed, the classical version of this equation
(and the following ones) can be obtained by simply taking the average,
such that $\hat{a}\mapsto\left\langle \hat{a}\right\rangle $. We
have chosen a frame rotating with the laser frequency $\omega_{L}$,
i.e. $\hat{a}^{{\rm orig}}=e^{-i\omega_{L}t}\hat{a}^{{\rm here}}$
and have introduced the laser detuning $\Delta=\omega_{L}-\omega_{{\rm cav}}$
with respect to the cavity mode (see also Sec.~\ref{sub:OptomechanicalCouplingHamiltonianFormulation}).
Note that a similar equation can also be written down for the mechanical
oscillator in order to describe its dissipation and the associated
noise force, comprising quantum and thermal contributions (see Sec.~\ref{sub:OptomechanicalEquationsOfMotion}).

The input field $\hat{a}_{{\rm in}}(t)$ should be thought of as a
stochastic quantum field. In the simplest case, it represents the
fluctuating vacuum electric field coupling to the cavity at time $t$,
plus a coherent laser drive. However, the same formalism can also
be used to describe squeezed states and other more complex field states.
The field is normalized in such a way that

\[
P=\hbar\omega_{{\rm cav}}\left\langle \hat{a}_{{\rm in}}^{\dagger}\hat{a}_{{\rm in}}\right\rangle 
\]
is the input power launched into the cavity, i. e. $\left\langle \hat{a}_{{\rm in}}^{\dagger}\hat{a}_{{\rm in}}\right\rangle $
is the \emph{rate} of photons arriving at the cavity. The same kind
of description holds for the {}``unwanted'' channel associated with
$\hat{f}_{{\rm in}}$.

According to the input-output theory of open quantum systems, the
field that is reflected from the Fabry Perot resonator (or coupled
back into the coupling waveguide) is given by:
\begin{equation}
\hat{a}_{\mathrm{out}}=\hat{a}_{in}-\sqrt{\kappa_{\mathrm{ex}}}\hat{a}\label{eq:InputOutputReflectedField}
\end{equation}
Note that this input-output relation describes correctly the field
reflected from the input mirror of a Fabry-Perot resonator. The above
equation describes also the transmitted pump field of an evanescently
coupled uni-directional waveguide resonator system, such as a whispering
gallery mode resonator coupled to a waveguide \cite{Cai2000a}. In
this case the above expression would yield the transmitted pump field.

We still have to consider the case of a two-sided cavity, e.g. a two-sided
Fabry Perot cavity. Other examples in this review include a waveguide
coupled to superconducting stripline cavities or fiber-taper coupled
photonic crystal defect cavities. In these cases there are both transmitted
and reflected fields. In all of these cases there are two options
for the description. If the field transmitted through the second mirror
is not of interest to the analysis, one may lump the effects of that
mirror into the decay rate $\kappa_{0}$, which now represents both
internal losses and output coupling through the second mirror. If,
however, the field is important, it should be represented by an additional
term of the type $\sqrt{\kappa_{{\rm ex}}^{(2)}}\hat{a}_{{\rm in}}^{(2)}$
in Eq.~(\ref{eq:InputOutputEqOfMotion}). Then an equation analogous
to Eq.~(\ref{eq:InputOutputReflectedField}) will hold for the output
field $\hat{a}_{{\rm out}}^{(2)}$ at that second mirror. 

In the following, we will not be concerned with noise properties,
but focus instead on classical average quantities (for a single-sided
cavity), taking the average of Eqs.~(\ref{eq:InputOutputEqOfMotion})
and (\ref{eq:InputOutputReflectedField}).

We can solve the equation (\ref{eq:InputOutputEqOfMotion}) first
for the steady-state amplitude in the presence of a monochromatic
laser drive whose amplitude is given by $\left\langle \hat{a}_{{\rm in}}\right\rangle $.
Noting that $\langle\hat{f}_{\mathrm{in}}\rangle=0,$ we obtain:

\begin{equation}
\left\langle \hat{a}\right\rangle =\frac{\sqrt{\kappa_{{\rm ex}}}\left\langle \hat{a}_{{\rm in}}\right\rangle }{\frac{\kappa}{2}-i\Delta}\,.\label{eq:AvgAmplitude}
\end{equation}
The expression linking the input field to the intracavity field will
be referred to as the optical susceptibility, 

\[
\chi_{aa}(\omega)=\frac{1}{-i(\omega+\Delta)+\kappa/2}
\]
Thus, the steady-state cavity population $\bar{n}_{{\rm cav}}=\langle\hat{a}^{\dagger}\hat{a}\rangle$,
i.e the average number of photons circulating inside the cavity, is
given by: 
\begin{equation}
\bar{n}_{cav}=\left|\left\langle \hat{a}\right\rangle \right|^{2}=\frac{\kappa_{{\rm ex}}}{\Delta^{2}+(\kappa/2)^{2}}\frac{P}{\hbar\omega_{L}}
\end{equation}
were $P$ is the input power launched into the cavity. The reflection
or transmission amplitude (for the case of a Fabry-Perot cavity or
a waveguide-coupled resonator, respectively) can be calculated by
inserting Eq.~(\ref{eq:AvgAmplitude}) into Eq.~(\ref{eq:InputOutputReflectedField}).
Using the symbol $\mathcal{R}$ for the reflection amplitude in the
sense of figure \ref{fig:Comparison-of-cavity-geometries} case (b),
we obtain:

\begin{equation}
\mathcal{R}=\frac{\left\langle \hat{a}_{{\rm out}}\right\rangle }{\left\langle \hat{a}_{{\rm in}}\right\rangle }=\frac{(\kappa_{0}-\kappa_{{\rm ex}})/2-i\Delta}{(\kappa_{0}+\kappa_{{\rm ex}})/2-i\Delta}\,.
\end{equation}
The square $\left|\mathcal{R}\right|^{2}$ of this amplitude gives
the probability of reflection from the cavity (for Fabry-Perot) or
transmission in the case of a uni-directional waveguide resonator
system. From this expression, several regimes can be differentiated.
If the external coupling $\kappa_{{\rm ex}}$ dominates the cavity
losses ($\kappa_{{\rm ex}}\approx\kappa\gg\kappa_{0})$, the cavity
is called {}``overcoupled''. In that case $\left|\mathcal{R}\right|^{2}\approx1$
and the pump photons emerge from the cavity without having been absorbed
or lost via the second mirror (a property that is important as discussed
below in the context of quantum limited detection). The case where
$\kappa_{0}=\kappa_{ex}$ refers to the situation of {}``critical
coupling''. In this case, $\mathcal{R}(\Delta=0)=0$ on resonance.
This implies the input power is either fully dissipated within the
resonator or fully transmitted through the second mirror (in the case
of a Fabry-Perot cavity with $\kappa_{0}$ denoting the decay through
the second mirror). The situation $\kappa_{ex}$ $\ll\kappa_{0}$
is referred to as {}``undercoupling'' and is associated with cavity
losses dominated by intrinsic losses. For many experiments this coupling
condition is not advantageous, as it leads to an effective loss of
information.

The physical meaning of reflection (or transmission) depends sensitively
on the experimental realization under consideration. One can distinguish
four scenarios, which are outlined in the figure \ref{fig:Comparison-of-cavity-geometries}.

\subsection{Mechanical resonators}

\label{sub:MechanicalResonators}

\subsubsection{Mechanical normal modes}

The vibrational modes of any object can be calculated by solving the
equations of the linear theory of elasticity under the appropriate
boundary conditions that are determined by the geometry%
\footnote{A powerful simulation approach in this context are finite element
(FEM) simulations.%
} \cite{Cleland2003}. This eigenvalue problem yields a set of normal
modes and corresponding eigenfrequencies $\Omega_{(n)}$. The mechanical
displacement patterns associated with mechanical motion are given
by the strain field $\vec{u}_{n}(\vec{r}),$ where $n$ designates
the normal mode. 

For the purposes of this review, we will mostly focus on a single
normal mode of vibration of frequency $\Omega_{{\rm m}}$ (where '${\rm m}$'
stands for 'mechanical'), assuming that the mode spectrum is sufficiently
sparse such that there is no spectral overlap with other mechanical
modes. The loss of mechanical energy is described by the (energy)
damping rate $\Gamma_{m},$ which is related to the mechanical quality
factor%
\footnote{In the context of mechanical dissipation often the loss tangent $\delta\Phi$is
quoted, its relation to the quality factor being $Q_{m}=\frac{1}{\delta\Phi}$.%
} by $Q_{m}=\Omega_{m}/\Gamma_{m}.$ If one is interested in the equation
of motion for the global amplitude $x(t)$ of the motion, one can
utilize a suitably normalized (see below) dimensionless mode function
$\vec{u}(\vec{r},t)$, such that the displacement field would be $\vec{u}(\vec{r},t)=x(t)\cdot\vec{u}(\vec{r})$.
Then the temporal evolution of $x(t)$ can be described by the canonical
simple equation of motion of a harmonic oscillator of effective mass
$m_{{\rm eff}}$: 
\begin{equation}
m_{\mathrm{eff}}\frac{dx^{2}(t)}{dt^{2}}+m_{\mathrm{eff}}\Gamma_{m}\frac{dx(t)}{dt}+m_{\mathrm{eff}}\Omega_{m}^{2}x(t)=F_{ext}(t)\label{eq:OscillatorEqMotion}
\end{equation}
Here $F_{ext}(t)$ denotes the sum of all forces that are acting on
the mechanical oscillator. In the absence of any external forces,
it is given by the thermal Langevin force (see Sec.~\ref{sub:MechanicalResonatorsNoiseSpectra}).
In the above equation the (energy) damping rate $\Gamma_{m}$ has
been assumed to be frequency independent. Deviations of this model
are treated for example in \cite{Saulson1990}.

A brief remark about the effective mass $m_{{\rm eff}}$ is necessary
at this point \cite{Cleland2003,Pinard1999}. The normalization that
has been chosen for the mode function $\vec{u}(\vec{r})$ affects
the normalization of $x(t)$. However, it will always be true that
the potential energy is given by $m_{{\rm eff}}\Omega_{{\rm m}}^{2}\left\langle x^{2}(t)\right\rangle /2$.
This value can then be compared to the expression for the potential
energy that arises from a calculation according to the theory of elasticity.
Demanding them to be equal yields the correct value for the effective
mass $m_{{\rm eff}}$ (which therefore is seen to depend on the normalization
that was chosen for the mode function). Of course, for the simple
case of a center-of-mass oscillation of a solid object, a natural
definition of $x(t)$ is the center of mass displacement in which
case the effective mass will be on the order of the total mass of
the object. A treatment of effective mass in optomechanical experiments
is found in \cite{Pinard1999}.

Eq.~(\ref{eq:OscillatorEqMotion}) can be solved easily, which is
best done in frequency space. We introduce the Fourier transform via
$x(\omega)=\int_{-\infty}^{+\infty}dt\, e^{i\omega t}x(t)$. Then
$\delta x(\omega)=\chi_{xx}(\omega)F_{{\rm ext}}(\omega)$ defines
the susceptibility $\chi_{xx}$, connecting the external force to
the response $\delta x$ of the coordinate:

\begin{equation}
\chi_{xx}(\omega)=\left[m_{{\rm eff}}(\Omega_{m}^{2}-\omega^{2})-im_{{\rm eff}}\Gamma_{m}\omega\right]^{-1}.\label{eq:MechSusc-1}
\end{equation}
The low frequency response is given by $\chi_{xx}(0)=(m_{eff}\Omega_{m}^{2})^{-1}=1/k$
where $k$ is the spring constant%
\footnote{To describe the response of a high Q oscillator near resonance $\omega\approx\Omega_{{\rm m}}$one
can approximate $\chi_{xx}$ by a Lorentzian, i.e. using $\Omega_{m}^{2}-\omega^{2}=(\Omega_{m}-\omega)(\Omega_{m}+\omega)\approx2(\Omega_{m}-\omega)\Omega_{m}$
yields $\chi_{xx}(\omega)=\left(m_{{\rm eff}}\Omega_{m}\left[2(\Omega_{m}-\omega)-i\Gamma_{m}\right]\right)^{-2}.$%
}. 

The quantum mechanical treatment of the mechanical harmonic oscillator
leads to the Hamiltonian
\[
\hat{H}=\hbar\Omega_{m}\hat{b}^{\dagger}\hat{b}+\frac{1}{2}\hbar\Omega_{m}
\]
Here the phonon creation ($\hat{b}^{\dagger}$) and annihilation ($\hat{b}$)
operators have been introduced, with

\[
\hat{x}=x_{{\rm ZPF}}(\hat{b}+\hat{b}^{\dagger}),\,\hat{p}=-im_{{\rm eff}}\Omega_{{\rm m}}x_{{\rm ZPF}}(\hat{b}-\hat{b}^{\dagger}),
\]
where 

\[
x_{{\rm ZPF}}=\sqrt{\frac{\hbar}{2m_{{\rm eff}}\Omega_{{\rm m}}}}
\]
is the zero-point fluctuation amplitude of the mechanical oscillator,
i.e. the spread of the coordinate in the ground-state: $\left\langle 0\left|\hat{x}^{2}\right|0\right\rangle =x_{{\rm ZPF}}^{2}$,
and where $|0\rangle$ denotes the mechanical vacuum state. The position
and momentum satisfy the commutator relation $[\hat{x}_{ZPF},\hat{p}_{ZPF}]=i\hbar$.
The quantity $\hat{b}^{\dagger}\hat{b}$ is the phonon number operator,
whose average is denoted by $\bar{n}=\langle\hat{b}^{\dagger}\hat{b}\rangle$.
In the following, we will typically not display explicitly the contribution
$\frac{1}{2}\hbar\Omega_{m}$ of the zero-point energy to the energy
of the oscillator.

We briefly discuss the effect of dissipation. If the mechanical oscillator
is coupled to a high temperature bath, the average phonon number will
evolve according to the expression:

\[
\frac{d}{dt}\langle n\rangle=-\Gamma_{m}(\langle n\rangle-\bar{n}_{th})
\]
For an oscillator which is initially in the ground state, $\langle n\rangle(t=0)=0$
this implies a simple time dependence of the occupation according
to $\langle n\rangle(t)=\bar{n}_{th}(1-e^{-t/\Gamma_{m}})$, where
$\bar{n}_{th}$ is the average phonon number of the environment. Consequently,
the rate at which the mechanical oscillator heats out of the ground
state is given by:

\[
\frac{d}{dt}\langle n(t=0)\rangle=\bar{n}_{th}\cdot\Gamma_{m}\approx\frac{k_{B}T_{bath}}{\hbar Q_{m}}
\]
The latter is often referred to as the \emph{thermal decoherence}
rate, and given by the inverse time it takes for one quantum to enter
from the environment. In the above expression the high temperature
limit has been taken, i.e. $\bar{n}_{th}\approx k_{B}T_{bath}/\hbar\Omega_{m}.$
This expression shows that to attain low decoherence a high mechanical
Q factor and a low temperature bath are important. The change of population
of a certain Fock state can be described within the framework of the
Master equation approach. This approach allows to calculate the decoherence
rate of other quantum states such as a Fock state $|n\rangle$. The
latter is given by (see e.g.\cite{Gardiner2004}):

\[
(n+1)\bar{n}_{th}\Gamma_{m}+n(\bar{n}_{th}+1)\Gamma_{m}
\]
revealing that higher Fock states exhibit a progressively higher rate
of decoherence.

\label{sub:MechanicalResonatorsNormalModes}

\label{sec:II.B.1}

\subsubsection{Mechanical dissipation}

\label{sub:MechanicalResonatorsDissipation}

The loss of mechanical excitations, i.e. phonons, is quantified by
the energy dissipation rate $\Gamma_{m}=\Omega_{{\rm m}}/Q_{{\rm m}}$.
The origins of mechanical dissipation have been intensively studied
over the last decades and comprehensive reviews are found for example
in \cite{Cleland2003,Ekinci2005}. The most relevant loss mechanisms
include: 
\begin{itemize}
\item viscous damping, which is caused by interactions with the surrounding
gas atoms or by compression of thin fluidic layers \cite{Vignola2006,Karabacak2007,Verbridge2008};
\item clamping losses, which are due to the radiation of elastic waves into
the substrate through the supports of the oscillator \cite{Nguyen2000,Cross2001,Mattila2002,Park2004,Photiadis2004,Clark2005,Bindel2005,Judge2007,Wilson-Rae2008,Anetsberger2008,Eichenfield2009,Cole2011,Jockel2011};
\item fundamental anharmonic effects such as thermoelastic damping (TED)
and phonon-phonon interactions \cite{Zener1938,Lifshitz2000,Kiselev2008,Duwel2006};
\item materials-induced losses, which are caused by the relaxation of intrinsic
or extrinsic defect states in the bulk or surface of the resonator
\cite{Yasumura2000,Mohanty2002,Southworth2009,Venkatesan2010,Unterreithmeier2010}.
Such losses have been successfully described by a phenomenological
model involving two level defect states, which are coupled to the
strain via the deformation potential \cite{Anderson1972,Phillips1987,Tielburger1992,Hunklinger1973,Seoanez2008,Remus2009}.
In the context of nano- and micromechanical oscillators the two level
fluctuator damping has been revisited \cite{Seoanez2008,Remus2009}.
\end{itemize}
The various dissipation processes contribute independently to the
overall mechanical losses and hence add up incoherently. The resulting
mechanical quality factor $Q_{total}$ is given by $\frac{1}{Q_{total}}=\sum\frac{1}{Q_{i}}$,
where $i$ labels the different loss mechanisms.

Another helpful quantity is the so-called {}``$Q\cdot f$'' product,
which plays an important role in the phase noise performance of oscillators.
In the context of optomechanics, it quantifies the decoupling of the
mechanical resonator from a thermal environment. Specifically, $\frac{\Omega_{m}}{\bar{n}_{th}\cdot\Gamma_{m}}=Q_{m}\cdot f_{m}\times(\frac{h}{k_{B}T})$,
denotes the number of coherent oscillations in the presence of thermal
decoherence, and evidently scales with $Q\cdot f$. 

\begin{figure}
\includegraphics[width=1\columnwidth]{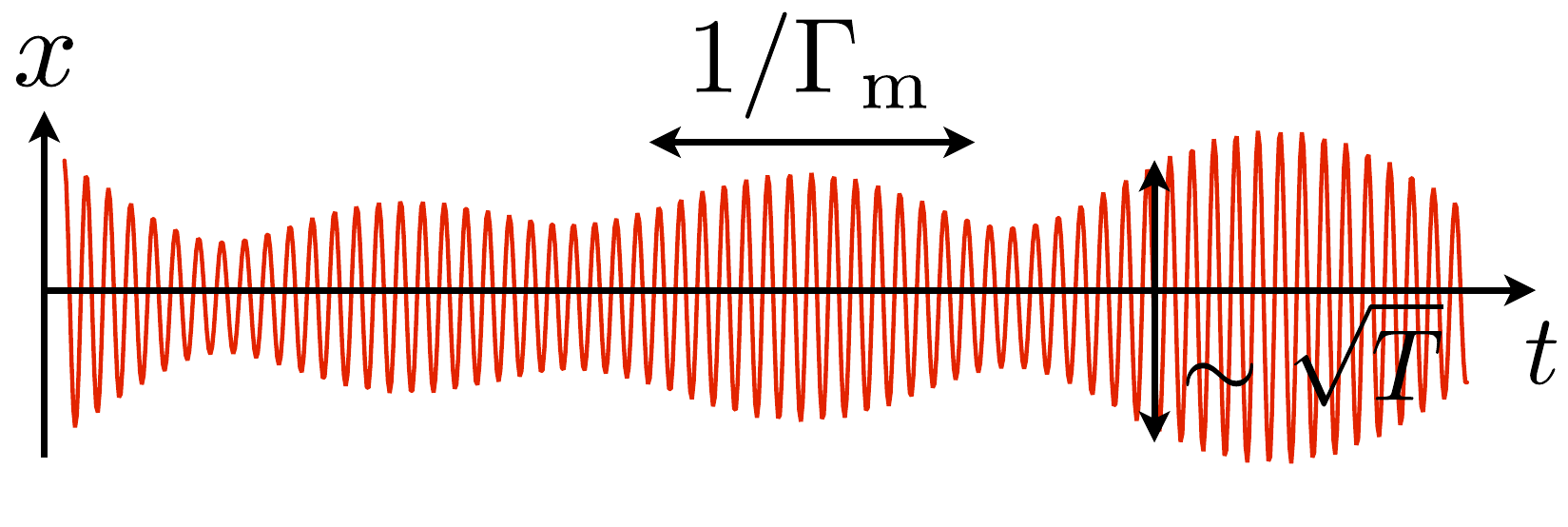}

\caption{\label{fig:Brownian-motion-Osc}Brownian motion (thermal fluctuations)
of a nanomechanical resonator in the time-domain (schematic), with
amplitude and phase fluctuating on a time scale set by the damping
time $\Gamma_{{\rm m}}^{-1}$.}
\end{figure}

\begin{figure}
\includegraphics[width=0.8\columnwidth]{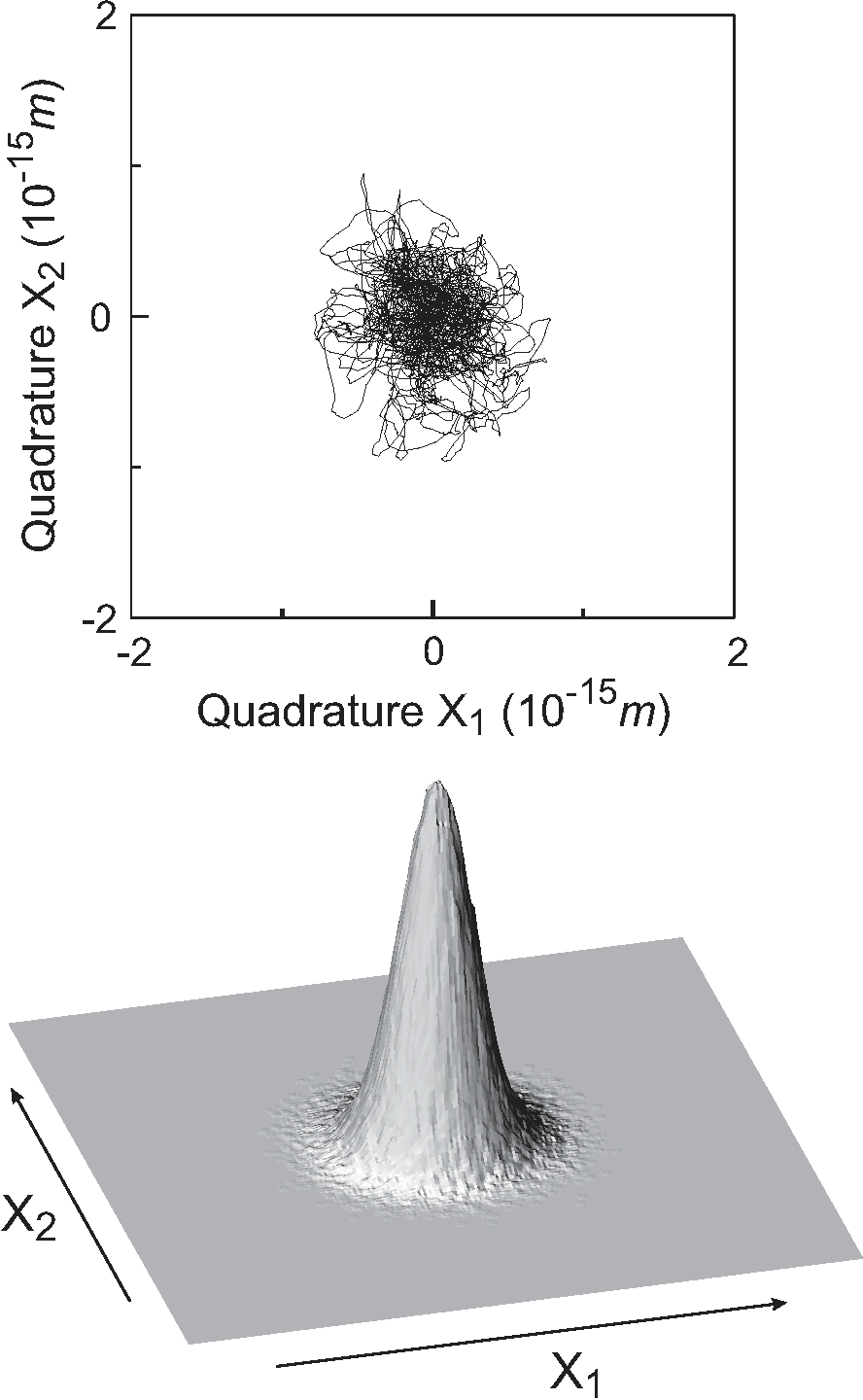}

\caption{\label{fig:BrownianMotion}Brownian motion of a mechanical resonator
obtained in an optomechanical setup. The two {}``quadratures'' $x(t)=X_{1}(t)cos(\Omega_{m}t)+X_{2}(t)sin(\Omega_{m}t)$,
are displayed in a frame rotating at the angular mechanical resonance
frequency $\Omega_{m}$, such that the unperturbed undamped motion
would correspond to a stationary single point sitting at the origin
of the phase space. The fluctuations are a consequence of the thermal
Brownian motion (figure courtesy of A. Heidmann, see also \cite{Hadjar1999}).}
\end{figure}

\subsubsection{Susceptibility, noise spectra and fluctuation dissipation theorem}

\label{sub:MechanicalResonatorsNoiseSpectra}

If one measures the motion of a single harmonic oscillator in thermal
equilibrium, one will observe a trajectory $x(t)$ oscillating at
the eigenfrequency $\Omega_{{\rm m}}$. However, due to the influence
of both mechanical damping and the fluctuating thermal Langevin force,
these oscillations will have a randomly time-varying amplitude and
phase. Both amplitude and phase change on the time scale given by
the damping time $\Gamma_{{\rm m}}^{-1}$. Such real-time measurements
have been performed in optomechanical systems \cite{Hadjar1999} (see
Fig.~\ref{fig:BrownianMotion}).

\begin{figure}
\includegraphics[width=1\columnwidth]{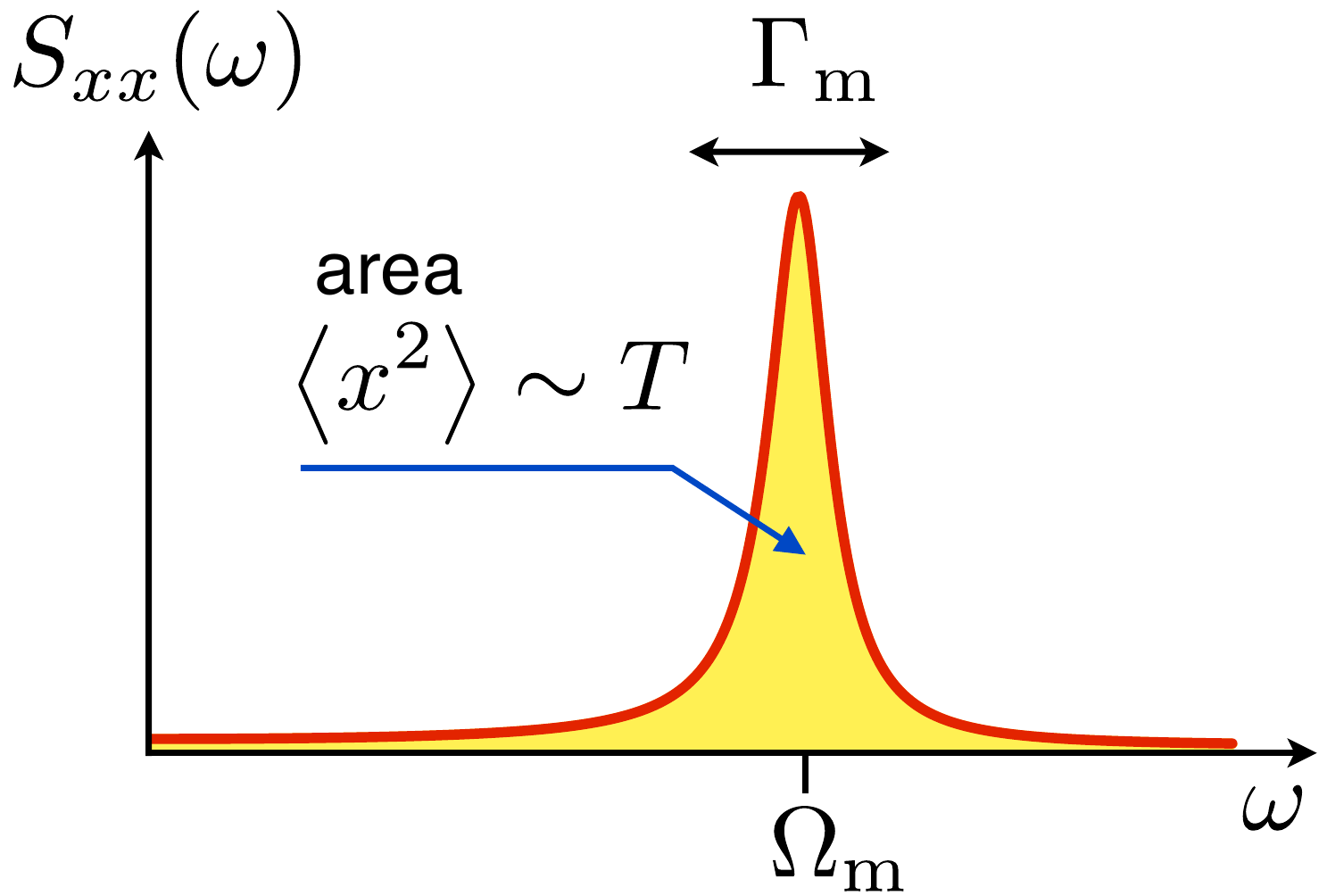}

\caption{\label{fig:Noise-spectrumLorentzianTheory}Noise spectrum of a damped
harmonic oscillator in thermal equilibrium (symmetric in $\omega\mapsto-\omega$
in the classical limit $k_{B}T\gg\hbar\omega$).}
\end{figure}

In experiments, the mechanical motion is often not analyzed in real-time
but instead as a noise spectrum in frequency space. This allows to
easily separate the contributions from different normal modes. We
briefly recapitulate the relevant concepts. Given one particular realization
of the trajectory $x(t)$ obtained during a measurement time $\tau$,
we define the gated Fourier transform over a finite time interval
$\tau$: 
\begin{equation}
\tilde{x}(\omega)=\frac{1}{\sqrt{\tau}}\int_{0}^{\tau}x(t)e^{i\omega t}dt\,.
\end{equation}
Averaging over independent experimental runs, we obtain the spectral
density $\left\langle |\tilde{x}(\omega)|^{2}\right\rangle $. In
the limit $\tau\rightarrow\infty$, the Wiener-Khinchin theorem connects
this to the Fourier transform $S_{xx}(\omega)$ of the autocorrelation
function, also called the noise power spectral density: 
\begin{equation}
\left\langle |\tilde{x}(\omega)|^{2}\right\rangle =S_{xx}(\omega)\,.\label{eq:Wiener-Khinchin}
\end{equation}
Here we have defined:

\begin{equation}
S_{xx}(\omega)\equiv\int_{-\infty}^{+\infty}\left\langle x(t)x(0)\right\rangle e^{i\omega t}\, dt\,.\label{eq:Def-NoiseSpectrumFromCorrelator}
\end{equation}
The only assumption which has been made is that $x(t)$ is a stationary
random process. From Eqs.~(\ref{eq:Wiener-Khinchin},\ref{eq:Def-NoiseSpectrumFromCorrelator}),
we immediately obtain the important result that the area under the
experimentally measured mechanical noise spectrum yields the variance
of the mechanical displacement, $\left\langle x^{2}\right\rangle $:

\begin{equation}
\int_{-\infty}^{+\infty}S_{xx}(\omega)\frac{d\omega}{2\pi}=\left\langle x^{2}\right\rangle \,.\label{eq:AreaNoiseSpectrum}
\end{equation}
Furthermore, in thermal equilibrium, the fluctuation-dissipation theorem
(FDT) relates the noise to the dissipative part of the linear response,

\begin{equation}
S_{xx}(\omega)=2\frac{k_{B}T}{\omega}{\rm Im}\chi_{xx}(\omega)\,,
\end{equation}
where $\chi_{xx}(\omega)$ denotes the mechanical susceptibility introduced
above and we have treated the high-temperature (classical) case. For
weak damping ($\Gamma_{m}\ll\Omega_{m}$), this gives rise to Lorentzian
peaks of width $\Gamma_{{\rm m}}$ in the noise spectrum, located
at $\omega=\pm\Omega_{m}$ (see Fig.~\ref{fig:Noise-spectrumLorentzianTheory}).
Integration of $S_{xx}(\omega)$ according to Eq.~(\ref{eq:AreaNoiseSpectrum})
yields the variance, which for weak damping is set by the equipartition
theorem: $\left\langle x^{2}\right\rangle =k_{B}T/m_{{\rm eff}}\Omega_{{\rm m}}^{2}$.

In the quantum regime, the natural generalization of Eq.~(\ref{eq:Def-NoiseSpectrumFromCorrelator})
contains the product of Heisenberg time-evolved operators, $\left\langle \hat{x}(t)\hat{x}(0)\right\rangle $,
which do not commute. As a consequence, the spectrum $S_{xx}(\omega)$
is asymmetric in frequency. The quantum FDT 

\begin{equation}
S_{xx}(\omega)=\frac{2\hbar}{1-e^{-\hbar\omega/k_{B}T}}{\rm Im}\chi_{xx}(\omega)\,\label{eq:FDT}
\end{equation}
implies that $S_{xx}(\omega)=0$ for $\omega<0$ at $T=0$. Our discussion
of dynamical backaction cooling will mention that this means the $T=0$
bath is not able to supply energy, as there are no thermal excitations.
In this review we will also consider the symmetrized noise spectrum,
$\bar{S}_{xx}(\omega)=\left\{ S_{xx}(\omega)+S_{xx}(-\omega)\right\} /2$.
For more on noise spectra, we refer to \cite{Clerk2008a}.

\section{Principles of optomechanical coupling}

\label{sec:PrinciplesOfOptomechanicalCoupling}

\subsection{The radiation pressure force and optomechanical coupling}

\label{sub:TheRadiationPressureForce}

In our discussion the fundamental mechanism that couples the properties
of the cavity radiation field to the mechanical motion is the momentum
transfer of photons, i.e. radiation pressure. The simplest form of
radiation pressure coupling is the momentum transfer due to reflection
that occurs in a Fabry Perot cavity. A single photon transfers the
momentum $|\Delta p|=2h/\lambda$ ($\lambda$: photon wavelength).
As a consequence the radiation pressure force is given by 
\[
\langle\hat{F}\rangle=2\hbar k\frac{\langle\hat{a}^{\dagger}\hat{a}\rangle}{\tau_{c}}=\hbar\frac{\omega}{L}\langle\hat{a}^{\dagger}\hat{a}\rangle
\]
Here $\tau_{c}=2L/c$ denotes the cavity round trip time. Therefore,
$\hbar\frac{\omega}{L}$ describes the radiation pressure force caused
by one intracavity photon. The parameter $G=\omega/L$ which appears
in this expression describes also the change of cavity resonance frequency
with position, i.e. the frequency pull parameter. In the next section,
which introduces a Hamiltonian description of the interaction between
a movable mirror and optical cavity, this relation will be derived
in its full generality.

More generally, the optomechanical coupling can arise for example
by direct momentum transfer via reflection (Fabry-Perot type cavities
with a moveable end-mirror, microtoroids), by coupling via a dispersive
shift of the cavity frequency (membrane in the middle, levitated nano-objects
trapped inside the cavity) or by optical near-field effects (e.g.
nano-objects in the evanescent field of a resonator or a waveguide
just above a substrate). 

Various radiation pressure forces have been investigated in the pioneering
work of Ashkin, who first demonstrated that small dielectric particles
can be trapped in laser light \cite{Ashkin2006}. The relevant forces
are generally referred to as gradient (or dipole) forces, as the force
arises from the gradient of the laser field intensity. The particle
is attracted to the center of the Gaussian trapping laser beam. If
$\vec{E}(\vec{r})$ denotes the laser electric field distribution,
the time-averaged dielectric energy of the particle in the field is
given by $U=-\frac{1}{2}\vec{p}\cdot\vec{E}=-\frac{1}{2}\alpha|\vec{E}(\vec{r})|^{2}$
(with $\alpha$ the polarizability) which correspondingly yields a
force $\vec{F}=-\vec{\nabla}U.$ In addition to the gradient force,
scattering forces occur for a traveling wave. These forces scale with
$\left|\vec{k}\right|$, i.e. the wavenumber of the electromagnetic
radiation, in contrast to the gradient forces. In addition there is
also a contribution from the strain-optical effect, i.e. the strain-dependent
polarizability. The strain-optical coupling is the dominant coupling
mechanism in guided acoustic wave scattering \cite{Shelby1985,Locke1998}.
Independent of the physical interpretation of the force, however,
the optomechanical interaction in an optomechanical system can always
be derived by considering the cavity resonance frequency shift as
a function of displacement (i.e. the {}``dispersive'' shift). This
will be the basis for our Hamiltonian description adopted in the next
section.

It is important to note at this point that a significant difference
between the trapping of particles in free space and micromechanical
systems is the fact that the latter are also subject to radiation
forces based on thermal effects. Absorption of light can heat a structure
and deform it, which corresponds to the action of a force (e.g. in
an asymmetric, bimorph structure, including materials of different
thermal expansion). These photothermal forces can in many ways lead
to effects similar to retarded radiation pressure forces, with the
thermal relaxation time of the structure replacing the cavity photon
lifetime. However, since such forces are based on absorption of light,
they cannot form the basis for future fully coherent quantum optomechanical
setups, since at least the coherence of the light field is thereby
irretrievably lost.

\subsection{Hamiltonian formulation}

\label{sub:OptomechanicalCouplingHamiltonianFormulation}

The starting point of all our subsequent discussions will be the Hamiltonian
describing the coupled system of a radiation mode interacting with
a vibrational mode (Fig.~\ref{Fig:GenericSetup}). For brevity we
will refer to the radiation field as {}``optical'', even though
the important case of microwave setups is included here as well.

We will focus here on the simplest possible model system in cavity
optomechanics, which has been used to successfully describe most of
the experiments to date. In this model, we restrict our attention
to one of the many optical modes, i.e. the one closest to resonance
with the driving laser. Moreover, we also describe only one of the
many mechanical normal modes. This is mostly arbitrary, as the displacement
frequency spectrum will show peaks at any of the mechanical resonances.
Still, as long as the dynamics is linear with independently evolving
normal modes, the model will provide a valid approximation. In some
cases, like sideband-resolved cooling, it may be possible to experimentally
select a particular mechanical mode by adjusting the laser detuning,
whereas in other cases, like nonlinear dynamics, an extended description
involving several mechanical modes may become crucial. 

The uncoupled optical ($\omega_{{\rm cav}}$) and mechanical ($\Omega_{m}$)
mode are represented by two harmonic oscillators, which is typically
an excellent approximation at the displacements generated in the experiments:

\begin{equation}
\hat{H}_{0}=\hbar\omega_{{\rm cav}}\hat{a}^{\dagger}\hat{a}+\hbar\Omega_{m}\hat{b}^{\dagger}\hat{b}
\end{equation}
In the case of a cavity with a movable end mirror the coupling of
optical and mechanical mode is parametric, i.e. the cavity resonance
frequency is modulated by the mechanical amplitude%
\footnote{Note that such a setup is also considered for discussions of the dynamical
Casimir effect, where cavity photons are created by the mechanical
modulation of the boundaries. In the optomechanical scenarios considered
here, however, the mechanical frequencies are too small for this effect
to play a role.%
}:

\[
\omega_{{\rm cav}}(x)\approx\omega_{{\rm cav}}+x\partial\omega_{{\rm cav}}/\partial x+\ldots
\]
For most experimental realizations discussed in this review, it suffices
to keep the linear term, where we define the optical frequency shift
per displacement as $G=-\partial\omega_{{\rm cav}}/\partial x$ (but
see Sec.~\ref{sub:MeasurementsQNDFockState} for another example).
A more detailed derivation of the optomechanical Hamiltonian can be
found in an early paper \cite{Law1995}. 

We mention in passing that other coupling mechanisms have been discussed.
For example, the transparency of a moving Bragg mirror, and hence
$\kappa$, can depend on its velocity \cite{2008_Karrai_DopplerOptomechanics}.
More generally, the displacement may couple to the cavity decay rate,
yielding $\kappa=\kappa(x)$. This case (sometimes termed {}``dissipative
coupling''), which is of practical relevance in some setups \cite{2009_Tang_MicroDiskCoupledWaveguide},
can lead to novel physical effects, e.g. in cooling \cite{Elste2009}.

For a simple cavity of length $L$, we have $G=\omega_{{\rm cav}}/L$.
The sign reflects the fact that we take $x>0$ to indicate an increase
in cavity length, leading to a decrease in $\omega_{{\rm cav}}(x)$
if $G>0$. In general, expanding to leading order in the displacement,
we have:

\begin{equation}
\hbar\omega_{{\rm cav}}(x)\hat{a}^{\dagger}\hat{a}\approx\hbar(\omega_{{\rm cav}}-G\hat{x})\hat{a}^{\dagger}\hat{a}\,.
\end{equation}
Here $\hat{x}=x_{{\rm ZPF}}(\hat{b}+\hat{b}^{\dagger})$, as defined
before. Thus, the interaction part of the Hamiltonian can be written

\begin{equation}
\hat{H}_{{\rm int}}=-\hbar g_{0}\hat{a}^{\dagger}\hat{a}(\hat{b}+\hat{b}^{\dagger})\,,\label{eq:H_optomech_int}
\end{equation}
where

\begin{equation}
g_{0}=Gx_{{\rm ZPF}}
\end{equation}
is the vacuum optomechanical coupling strength, expressed as a frequency.
It quantifies the interaction between a single phonon and a single
photon. We stress that, generally speaking, $g_{0}$ is more fundamental
than $G$, since $G$ is affected by the definition of the displacement
that is to some extent arbitrary for more complicated mechanical normal
modes (see the discussion in Sec.~\ref{sec:II.B.1}). Therefore,
in the following we will almost always refer to $g_{0}$. Further
below, we will also mention $g$, which is an often-used measure for
the effective optomechanical coupling in the linearized regime. It
will be enhanced compared to $g_{0}$ by the amplitude of the photon
field. The Hamiltonian reveals that the interaction of a movable mirror
with the radiation field is fundamentally a \emph{nonlinear} process,
involving three operators (three wave mixing).

The radiation pressure force is simply the derivative of $\hat{H}_{{\rm int}}$
with respect to displacement: 
\begin{equation}
\hat{F}=-\frac{d\hat{H}_{{\rm int}}}{d\hat{x}}=\hbar G\hat{a}^{\dagger}\hat{a}=\hbar\frac{g_{0}}{x_{{\rm ZPF}}}\hat{a}^{\dagger}\hat{a}
\end{equation}

The full Hamiltonian $\hat{H}$ will also include terms that describe
dissipation (photon decay and mechanical friction), fluctuations (influx
of thermal phonons), and driving by an external laser. These effects
are formulated most efficiently using the equations of motion and
the input-output formalism (see Sec.~\ref{sub:OpticalResonatorsQuantumFluctuations},
and also the next section). Here, we just remark that it is convenient
to change the description of the optical mode by switching to a frame
rotating at the laser frequency $\omega_{L}$. Applying the unitary
transformation $\hat{U}=\exp(i\omega_{L}\hat{a}^{\dagger}\hat{a}t)$
makes the driving terms time-independent%
\footnote{$\hat{U}(\hat{a}^{\dagger}e^{-i\omega_{L}t}+\hat{a}e^{+i\omega_{L}t})\hat{U}^{\dagger}=\hat{a}^{\dagger}+\hat{a}$%
}, and generates a new Hamiltonian $\hat{H}=\hat{U}\hat{H}_{{\rm old}}\hat{U}^{\dagger}+i\hbar\partial\hat{U}/\partial t$
of the form

\begin{equation}
\hat{H}=-\hbar\Delta\hat{a}^{\dagger}\hat{a}+\hbar\Omega_{m}\hat{b}^{\dagger}\hat{b}-\hbar g_{0}\hat{a}^{\dagger}\hat{a}(\hat{b}+\hat{b}^{\dagger})+\ldots\,,\label{eq:StandardHamiltonianRotatingFrame}
\end{equation}
where

\begin{equation}
\Delta=\omega_{L}-\omega_{{\rm cav}}
\end{equation}
is the laser detuning introduced already in Sec.~\ref{sub:OpticalResonatorsQuantumFluctuations},
and where we have omitted ($\ldots$) driving, decay, and fluctuation
terms, which will be discussed below. Eq.~(\ref{eq:StandardHamiltonianRotatingFrame})
is the frequently used starting point in cavity optomechanics.

We now introduce the so-called {}``linearized'' approximate description
of cavity optomechanics. To this end, we split the cavity field into
an average coherent amplitude $\left\langle \hat{a}\right\rangle =\bar{\alpha}$
and a fluctuating term:

\begin{equation}
\hat{a}=\bar{\alpha}+\delta\hat{a}\label{eq:FieldShift}
\end{equation}
Then, the interaction part of the Hamiltonian 

\begin{equation}
\hat{H}_{{\rm int}}=-\hbar g_{0}(\bar{\alpha}+\delta\hat{a})^{\dagger}(\bar{\alpha}+\delta\hat{a})(\hat{b}+\hat{b}^{\dagger})
\end{equation}
may be expanded in powers of $\bar{\alpha}$. The first term, $-\hbar g_{0}\left|\bar{\alpha}\right|^{2}(\hat{b}+\hat{b}^{\dagger})$,
indicates the presence of an average radiation pressure force $\bar{F}=\hbar G\left|\bar{\alpha}\right|^{2}$.
It may be omitted after implementing an appropriate shift of the displacement's
origin by $\delta\bar{x}=\bar{F}/m_{{\rm eff}}\Omega_{m}^{2}$. The
second term, of order $|\bar{\alpha}|^{1}$, is the one we keep:

\begin{equation}
-\hbar g_{0}(\bar{\alpha}^{*}\delta\hat{a}+\bar{\alpha}\delta\hat{a}^{\dagger})(\hat{b}+\hat{b}^{\dagger})
\end{equation}
The third term, $-\hbar g_{0}\delta\hat{a}^{\dagger}\delta\hat{a}$,
is omitted as being smaller by a factor $\left|\bar{\alpha}\right|$.
Without loss of generality, we will now assume $\bar{\alpha}=\sqrt{\bar{n}_{{\rm cav}}}$
real-valued. Thus, the Hamiltonian in the rotating frame reads

\begin{equation}
\hat{H}\approx-\hbar\Delta\delta\hat{a}^{\dagger}\delta\hat{a}+\hbar\Omega_{m}\hat{b}^{\dagger}\hat{b}+\hat{H}_{{\rm int}}^{({\rm lin})}+\ldots\,,\label{eq:FullHAfterLinearization}
\end{equation}
where the quadratic interaction part

\begin{equation}
\hat{H}_{{\rm int}}^{(lin)}=-\hbar g_{0}\sqrt{\bar{n}_{{\rm cav}}}(\delta\hat{a}^{\dagger}+\delta\hat{a})(\hat{b}+\hat{b}^{\dagger})\label{eq:LinearizedInteraction}
\end{equation}
\begin{figure}
\includegraphics[width=1\columnwidth]{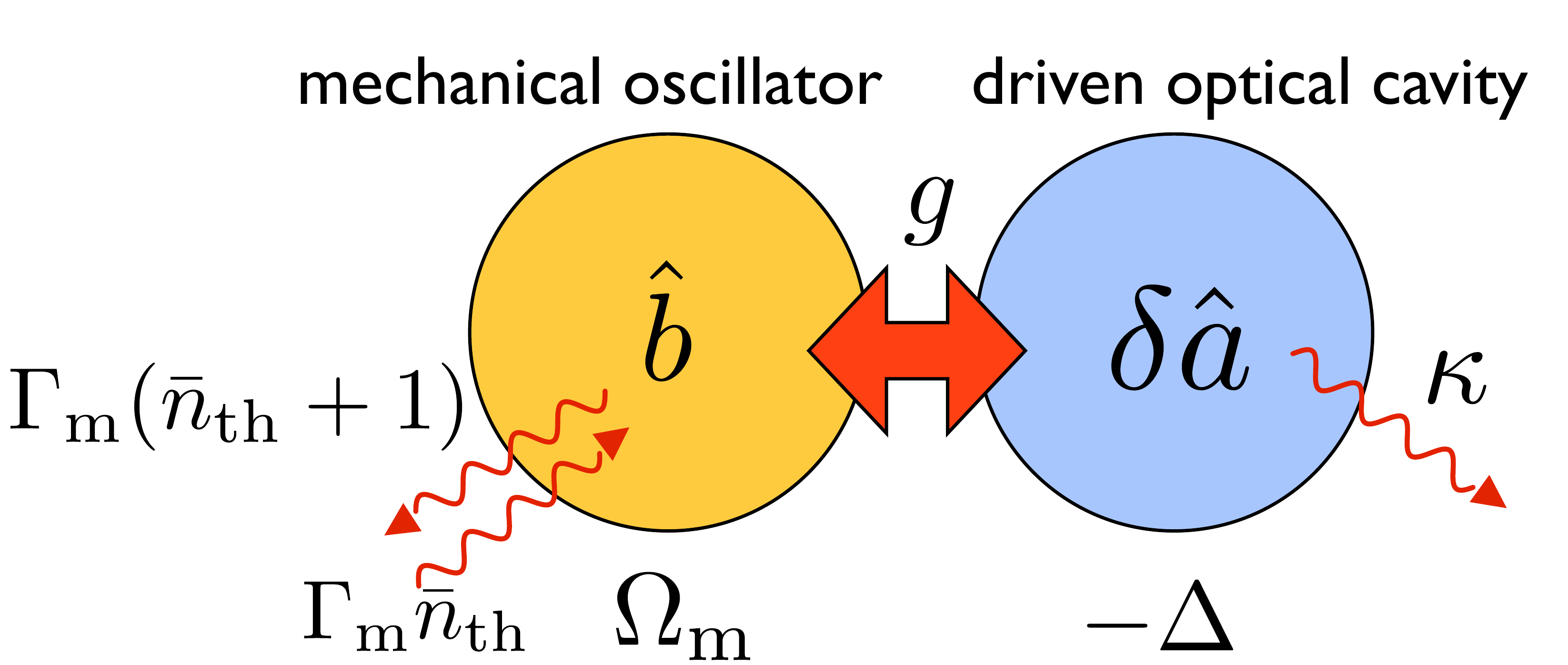}

\caption{Schematic: Optomechanical (linearized) interaction between a driven
optical mode and a mechanical resonator.}
\end{figure}
is referred to as {}``linearized'', since the resulting coupled
equations of motion will be linear in this approximation. Note that
the remaining terms in Eq.~(\ref{eq:FullHAfterLinearization}) no
longer contain the laser driving, as that has already been taken care
of by the shift implemented in Eq.~(\ref{eq:FieldShift}). In the
literature up to now, the combination

\begin{equation}
g=g_{0}\sqrt{\bar{n}_{{\rm cav}}}
\end{equation}
is often referred to as {}``the optomechanical coupling strength''.
Obviously, it depends on the laser intensity and is thus less fundamental
than the single-photon coupling $g_{0}$ (obtained for $\bar{n}_{{\rm cav}}=1$).
In the linearized regime described here, the optomechanical system
can be viewed in analogy to a linear amplifier \cite{2012_StamperKurn_AmplifierModel}
that receives optical and mechanical input fields.

The linearized description can be good even if the average photon
number circulating inside the cavity is not large. This is because
the mechanical system may not be able to resolve the individual photons
if the decay rate $\kappa$ is sufficiently large. The detailed conditions
for the linearized approximation to be valid may depend on the questions
that are asked. We will return to this question in the section on
nonlinear quantum optomechanics (Sec.~\ref{sub:QuantumOptomechanicsNonlinear}). 

We briefly note that $g>\kappa$ is one neccessary condition for the
so-called {}``strong coupling'' regime of cavity optomechanics,
where the mechanical oscillator and the driven optical mode hybridize
(Sec.~\ref{sub:CoolingStrongCoupling}). A much more challenging
condition is to have $g_{0}>\kappa$, i.e. the single-photon optomechanical
coupling rate exceeding the cavity decay rate. In the latter regime,
nonlinear quantum effects will become observable (see Sec.~\ref{sub:QuantumOptomechanicsNonlinear}).

Depending on the detuning, three different regimes can be distinguished
with respect to the interaction (\ref{eq:LinearizedInteraction}),
especially in the sideband-resolved regime ($\kappa\ll\Omega_{m}$,
which we assume in the remainder of this section). For $\Delta\approx-\Omega_{m}$,
we have two harmonic oscillators of (nearly) equal frequency that
can interchange quanta: the mechanical oscillator and the driven cavity
mode. Within the rotating-wave approximation (RWA) we thus can write
the interaction as

\begin{equation}
-\hbar g(\delta\hat{a}^{\dagger}\hat{b}+\delta\hat{a}\hat{b}^{\dagger})\,.\label{eq:BeamSplitterInteraction}
\end{equation}
This is the case relevant for cooling (transferring all thermal phonons
into the cold photon mode; Sec.~\ref{sec:VII.A}) and for quantum
state transfer between light and mechanics (Sec.~\ref{sec:QuantumOptomechanics}).
In the quantum-optical domain, it is referred to as a {}``beam-splitter''
interaction.

For $\Delta\approx+\Omega_{m}$, the dominant terms in RWA

\begin{equation}
-\hbar g(\delta\hat{a}^{\dagger}\hat{b}^{\dagger}+\delta\hat{a}\hat{b})\,\label{eq:TwoModeSqueezingInteraction}
\end{equation}
represent a {}``two-mode squeezing'' interaction that lies at the
heart of parametric amplification \cite{Clerk2008a}. In the absence
of dissipation, this would lead to an exponential growth of the energies
stored both in the vibrational mode and the driven optical mode, with
strong quantum correlations between the two. Thus, it may be used
for efficiently entangling both modes (Sec.~\ref{sec:QuantumOptomechanics}).
Focussing on the mechanical mode alone, the growth of energy can be
interpreted as {}``anti-damping'' or amplification (Sec.~\ref{sub:DynamicalBackactionOptomechanicalDampingRate}).
If the intrinsic dissipation is low enough, this behaviour may trigger
a dynamical instability that leads to self-induced mechanical oscillations.
The resulting features will be discussed in Sec.~\ref{sec:VIII}.

Finally, when $\Delta=0$, the interaction

\begin{equation}
-\hbar g(\delta\hat{a}^{\dagger}+\delta\hat{a})(\hat{b}+\hat{b}^{\dagger})\,
\end{equation}
means that the mechanical position $\hat{x}\propto\hat{b}+\hat{b}^{\dagger}$
leads to a phase shift of the light field, which is the situation
encountered in optomechanical displacement detection (Sec.~\ref{sec:QuantumOpticalMeasurementsOfMechanicalMotion}).
In addition, this interaction Hamiltonian can be viewed as implementing
QND detection of the optical amplitude quadrature $\delta\hat{a}+\delta\hat{a}^{\dagger}$,
since that operator commutes with the full Hamiltonian in this case.

\subsection{Optomechanical equations of motion}

\label{sub:OptomechanicalEquationsOfMotion}

The mechanical motion induces a shift of the optical resonance frequency,
which in turn results in a change of circulating light intensity and,
therefore, of the radiation pressure force acting on the motion. This
kind of feedback loop is known as optomechanical {}``backaction''.
The finite cavity decay rate $\kappa$ introduces some retardation
between the motion and the resulting changes of the force, hence the
term {}``dynamical'' backaction.

A convenient starting point for the analytical treatment of the radiation-pressure
dynamical backaction phenomena (Sec.~\ref{sub:DynamicalBackactionBasics}
and Sec.~\ref{sec:OptomechanicalCooling}) is the input-output formalism.
This formalism (briefly introduced in Sec.~\ref{sub:OpticalResonatorsQuantumFluctuations})
provides us with equations of motion for the cavity field amplitude
$\hat{a}$ and, analogously, for the mechanical amplitude $\hat{b}$.
These equations have the form of Quantum Langevin equations%
\footnote{In the standard approximation adopted here, these equations are Markoffian,
i.e. without memory and with delta-correlated noise.%
}, since both the light amplitude and the mechanical motion are driven
by noise terms that comprise the vacuum noise and any thermal noise
entering the system: 
\begin{align}
\dot{\hat{a}} & =-\frac{\kappa}{2}\hat{a}+i(\Delta+G\hat{x})\hat{a}+\sqrt{\kappa_{ex}}\hat{a}_{{\rm in}}(t)+\sqrt{\kappa_{0}}\hat{f}_{{\rm in}}(t)\label{eq:NonlinEqMotionA}\\
\dot{\hat{b}} & =\left(-i\Omega_{m}-\frac{\Gamma_{m}}{2}\right)\hat{b}+ig_{0}\hat{a}^{\dagger}\hat{a}+\sqrt{\Gamma_{m}}\hat{b}_{in}(t)\label{eq:NonlinEqMotionB}
\end{align}
Please see Sec.~\ref{sub:OpticalResonatorsQuantumFluctuations} for
remarks on the input-output treatment and the optical decay rates
$\kappa,\,\kappa_{{\rm ex}},\,\kappa_{0}$. With regard to the damping
term $-\Gamma_{{\rm m}}/2$ for the mechanical dissipation, we note
that this treatment is correct as long as $\Omega_{{\rm m}}\gg\Gamma_{{\rm m}}$.
Otherwise the equations would have to be formulated on the level of
the displacement $\hat{x}$, with a damping force $-m_{{\rm eff}}\Gamma_{{\rm m}}\dot{\hat{x}}$.

The noise correlators associated with the input fluctuations are given
by:

\begin{align}
\langle\hat{a}_{in}(t)\hat{a}_{in}^{\dagger}(t^{\prime})\rangle & =\delta(t-t^{\prime})\\
\langle\hat{a}_{in}^{\dagger}(t)\hat{a}_{in}(t^{\prime})\rangle & =0\\
\langle\hat{b}_{in}(t)\hat{b}_{in}^{\dagger}(t^{\prime})\rangle & =(\bar{n}_{th}+1)\delta(t-t^{\prime})\\
\langle\hat{b}_{in}^{\dagger}(t)\hat{b}_{in}(t^{\prime})\rangle & =\bar{n}_{th}\delta(t-t^{\prime})
\end{align}
Here we have assumed that the optical field has zero thermal occupation
($k_{B}T/\hbar\omega_{{\rm cav}}\approx0$), which is an approximation
that is valid for optical fields at room temperature, although it
may fail for the case of microwave fields, unless the setup is cooled
to sufficiently low temperatures. In contrast, the mechanical degree
of freedom is typically coupled to a hot environment, with an average
number of quanta given by $\bar{n}_{{\rm th}}\approx k_{B}T/\hbar\Omega_{{\rm m}}\gg1$.
Together with these correlators, the quantum Langevin equations describe
the evolution of the optical cavity field and the mechanical oscillator,
including all fluctuation effects. 

It is equally useful to give the classical, averaged version of these
equations that will be valid for sufficiently large photon and phonon
numbers, in the semiclassical limit. Then, we can write down the equations
for the complex light amplitude $\alpha(t)=\left\langle \hat{a}(t)\right\rangle $
and the oscillator position $x(t)=\left\langle \hat{x}(t)\right\rangle $:

\begin{align}
\dot{\alpha} & =-\frac{\kappa}{2}\alpha+i(\Delta+Gx)\alpha+\sqrt{\kappa_{ex}}\alpha_{{\rm in}}\label{eq:ClassEqMotion-alpha}\\
m_{{\rm eff}}\ddot{x} & =-m_{{\rm eff}}\Omega_{m}^{2}x-m_{{\rm eff}}\Gamma_{{\rm m}}\dot{x}+\hbar G\left|\alpha\right|^{2}\label{eq:ClassEqMotion-x}
\end{align}
Here we have neglected all fluctuations, although these could be added
to describe thermal and even, in a semiclassical approximation, quantum
noise forces. The term $\alpha_{{\rm in}}$ represents the laser drive.
Note that we have also chosen to write the mechanical equation of
motion in terms of the displacement, where $x=2x_{{\rm ZPF}}{\rm Re}\left(\left\langle \hat{b}\right\rangle \right)$.
This becomes equivalent to the equation given above only for weak
damping, $\Gamma_{{\rm m}}\ll\Omega_{{\rm m}}$. These fully nonlinear
coupled differential equations will be the basis for our discussion
of nonlinear phenomena, in particular the optomechanical parametric
instability (also called {}``self-induced oscillations'' or {}``mechanical
lasing'', Sec.~\ref{sec:NonlinearDynamics}).

The equations of motion Eq.~(\ref{eq:NonlinEqMotionA}),~(\ref{eq:NonlinEqMotionB})
(and likewise their classical versions) are inherently nonlinear as
they contain the product of the mechanical oscillator amplitude and
the cavity field (first line) or the radiation pressure force $\propto\hat{a}^{\dagger}\hat{a}$
that is quadratic in photon operators (second line). While they can
still be solved numerically in the classical case, for the quantum
regime they are of purely formal use and in practice cannot be solved
exactly, neither analytically nor numerically. However, in many situations
that we will encounter it is permissible to linearize this set of
equations around some steady-state solution: $\hat{a}=\alpha+\delta\dot{\hat{a}}$.
Using $\alpha=\sqrt{\bar{n}_{{\rm cav}}}$ and keeping only the linear
terms, we find the following set of coupled linear equations of motion:

\begin{align}
\delta\dot{\hat{a}} & =\left(i\Delta-\frac{\kappa}{2}\right)\delta\hat{a}+ig\left(\hat{b}+\hat{b}^{\dagger}\right)+\label{eq:EqsMotionLinQuantum-a}\\
 & \sqrt{\kappa_{ex}}\delta\hat{a}_{in}(t)+\sqrt{\kappa_{0}}\hat{f}_{{\rm in}}(t)\\
\dot{\hat{b}} & =\left(-i\Omega_{m}-\frac{\Gamma_{m}}{2}\right)\hat{b}+ig\left(\delta\hat{a}+\delta\hat{a}^{\dagger}\right)+\sqrt{\Gamma_{m}}\hat{b}_{in}(t).\label{eq:EqsMotionLinQuantum-b}
\end{align}
These correspond to what one would have obtained alternatively by
employing the {}``linearized'' coupling Hamiltonian of Eq.~(\ref{eq:LinearizedInteraction})
and then applying input-output theory. Here we have (as is common
practice) redefined the origin of the mechanical oscillations to take
into account the constant displacement $\hbar G|\alpha|^{2}/m_{{\rm eff}}\Omega_{{\rm m}}^{2}$
that is induced by the average radiation pressure force. It is evident
that now the mutual coupling terms between the optical and mechanical
degrees of freedom are linear in the field operators, and that the
strength is set by the field-enhanced coupling rate $g=g_{0}\sqrt{\bar{n}_{{\rm cav}}}$
. 

As shown in the later sections, these linearized equations are able
to fully describe several phenomena, including optomechanical cooling,
amplification, and parametric normal mode splitting (i.e. strong,
coherent coupling). They can be solved analytically, which is best
performed in the frequency domain (see Sec.~\ref{sub:DynamicalBackactionBasics}).

For completeness, we display the linearized quantum equations in frequency
space:

\begin{align}
-i\omega\delta\hat{a}[\omega] & =\left(i\Delta-\frac{\kappa}{2}\right)\delta\hat{a}[\omega]+ig\left(\hat{b}[\omega]+\left(\hat{b}^{\dagger}\right)[\omega]\right)\nonumber \\
 & +\sqrt{\kappa_{ex}}\hat{a}_{in}[\omega]\\
-i\omega\hat{b}[\omega] & =\left(-i\Omega_{m}-\frac{\Gamma_{m}}{2}\right)\hat{b}[\omega]+ig\left(\delta\hat{a}[\omega]+(\delta\hat{a}^{\dagger})[\omega]\right)\nonumber \\
 & +\sqrt{\Gamma_{m}}\hat{b}_{in}[\omega].
\end{align}
Here $\hat{b}[\omega]=\int_{-\infty}^{+\infty}dt\, e^{i\omega t}\hat{b}(t)$
is the Fourier transform of $\hat{b}$. Note the important relation
$\left(\hat{b}^{\dagger}\right)[\omega]=\left(\hat{b}[-\omega]\right)^{\dagger}$,
which has to be taken care of while solving the equations.

It is equally useful to consider the linearized version of the classical
equations of motion for the light amplitude and the displacement,
Eqs.~(\ref{eq:ClassEqMotion-alpha}) and (\ref{eq:ClassEqMotion-x}):

\begin{align}
\delta\dot{\alpha} & =\left(i\Delta-\frac{\kappa}{2}\right)\delta\alpha+iG\bar{\alpha}x\\
m_{{\rm eff}}\ddot{x} & =-m_{{\rm eff}}\Omega_{{\rm {\rm m}}}^{2}x-m_{{\rm eff}}\Gamma\dot{x}+\hbar G(\bar{\alpha}^{*}\delta\alpha+\bar{\alpha}\delta\alpha^{*})
\end{align}
Finally, we display them in frequency space, in the form that we will
employ in Sec.~\ref{sub:DynamicalBackactionBasics}.

\begin{align}
-i\omega\delta\alpha[\omega] & =\left(i\Delta-\frac{\kappa}{2}\right)\delta\alpha[\omega]+iG\bar{\alpha}x[\omega]\label{eq:ClassicalEqsLinearizedOmega-alpha}\\
-m_{{\rm eff}}\omega^{2}x[\omega] & =-m_{{\rm eff}}\Omega_{{\rm m}}^{2}x[\omega]+i\omega m_{{\rm eff}}\Gamma_{{\rm m}}x[\omega]\label{eq:ClassicalEqsLinearizedOmega-x}\\
 & +\hbar G(\bar{\alpha}^{*}\delta\alpha[\omega]+\bar{\alpha}(\delta\alpha^{*})[\omega])\nonumber 
\end{align}
Again, note $(\delta\alpha^{*})(\omega)=\delta\alpha(-\omega)^{*}$.

\section{Experimental realizations and optomechanical parameters}

\label{sec:ExperimentalRealizationAndParameters}

The increasing availability of high-quality optomechanical devices,
i.e. high-Q mechanical resonators that are efficiently coupled to
high-Q optical cavities, has been driving a plethora of experiments
during the last years that are successfully demonstrating the working
principles of cavity optomechanics. We now discuss some of the most
frequently used architectures. 

\begin{figure*}
\includegraphics[width=1\textwidth]{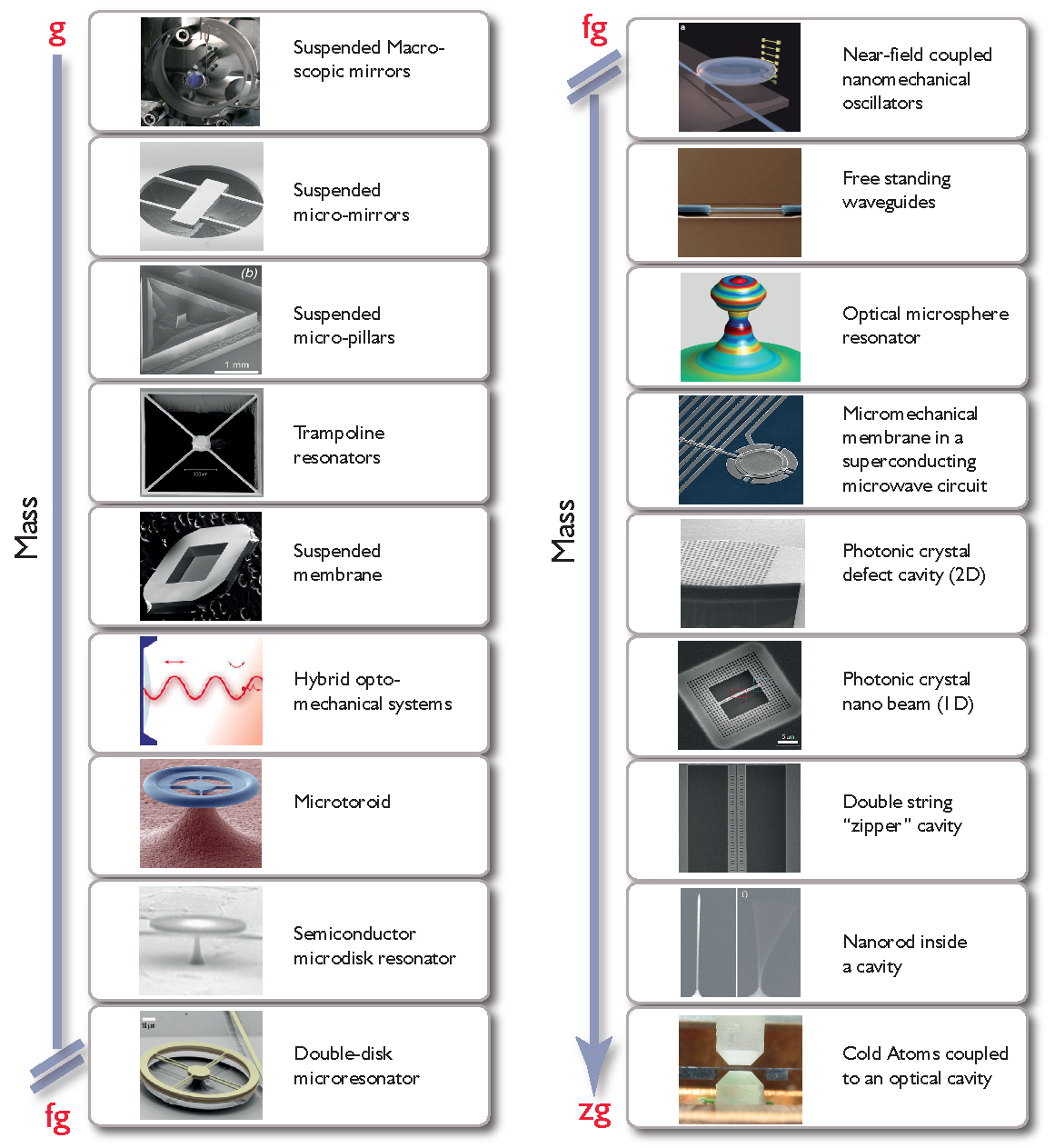}

\caption{A gallery illustrating the variety of optomechanical devices, arranged
according to mass. Pictures courtesy (from top left, down): N. Mavalvala,
M. Aspelmeyer, A. Heidmann, D. Bouwmeester, J. Harris, P. Treutlein,
T. J. Kippenberg, I. Favero, M. Lipson, T. J. Kippenberg/E. Weig/J.
Kotthaus, H. Tang, K. Vahala/T. Carmon, J.~Teufel/K.~Lehnert, I.
Robert, O. Painter, O. Painter, I.~Favero/E.~Weig/K.~Karrai, D.
Stamper-Kurn}
\end{figure*}

\subsection{Optomechanical parameters}

\label{sub:OptomechanicalCouplingParameters}

The following table summarizes the relevant optomechanical parameters
for some typical current experimental implementations. These are:
the mechanical resonator frequency $\Omega_{m}$ and mass $m$; the
fundamental mechanical (phonon) and optical (photon) dissipation rates
$\Gamma_{m}=\Omega_{{\rm m}}/Q_{{\rm m}}$ and $\kappa$, respectively;
the {}``$Q\cdot f$'' product, which is a direct measure for the
degree of decoupling from the thermal environment (specifically, $Q_{{\rm m}}\cdot f=Q_{m}\cdot\Omega_{{\rm m}}/2\pi>k_{B}T/\hbar$
is the condition for neglecting thermal decoherence over one mechanical
period); the sideband suppression factor $\kappa/\Omega_{{\rm m}}$
that determines the ability to realize ground-state cooling (see Sec.\ref{sec:OptomechanicalCooling});
and finally the bare optomechanical coupling rate $g_{0}$, which
corresponds to the cavity frequency shift upon excitation of a single
phonon.

\begin{table*}
\begin{tabular*}{1\textwidth}{@{\extracolsep{\fill}}|l|c|c|c|c|c|c|c|}
\hline 
Reference & $\Omega_{m}/2\pi${[}Hz{]} & $m$ {[}kg{]} & $\Gamma_{m}/2\pi${[}Hz{]} & $Q\cdot f$ {[}Hz{]} & $\kappa/2\pi${[}Hz{]} & $\frac{\kappa}{\Omega_{m}}$ & $g_{0}/2\pi${[}Hz{]}\tabularnewline
\hline 
\hline 
\cite{Murch2008} & $4.2\cdot10^{4}$ & $1\cdot10^{-22}$ & $1\cdot10^{3}$ & $1.7\cdot10^{6}$ & $6.6\cdot10^{5}$ & 15.7 & $6\cdot10^{5}$\tabularnewline
\hline 
\cite{Chan2011c} & $3.9\cdot10^{9}$ & $3.1\cdot10^{-16}$ & $3.9\cdot10^{4}$ & $3.9\cdot10^{14}$ & $5\cdot10^{8}$ & 0.13 & $9\cdot10^{5}$\tabularnewline
\hline 
\cite{Teufel2011} & $1.1\cdot10^{7}$ & $4.8\cdot10^{-14}$ & $32$ & $3.5\cdot10^{12}$ & $2\cdot10^{5}$ & 0.02 & $2\cdot10^{2}$\tabularnewline
\hline 
\cite{Verhagena} & $7.8\cdot10^{7}$ & $1.9\cdot10^{-12}$ & $3.4\cdot10^{3}$ & $1.8\cdot10^{12}$ & $7.1\cdot10^{6}$ & 0.09 & $3.4\cdot10^{3}$\tabularnewline
\hline 
\cite{Thompson2008} & $1.3\cdot10^{5}$ & $4\cdot10^{-11}$ & $0.12$ & $1.5\cdot10^{11}$ & $5\cdot10^{5}$ & $3.7$ & $5\cdot10^{1}$\tabularnewline
\hline 
\cite{Kleckner2011} & $9.7\cdot10^{3}$ & $1.1\cdot10^{-10}$ & $1.3\cdot10^{-2}$ & $9\cdot10^{9}$ & $4.7\cdot10^{5}$ & 55 & $2.2\cdot10^{1}$\tabularnewline
\hline 
\cite{Groblacher2009a} & $9.5\cdot10^{5}$ & $1.4\cdot10^{-10}$ & $1.4\cdot10^{2}$ & $6.3\cdot10^{9}$ & $2\cdot10^{5}$ & $0.22$ & $3.9$\tabularnewline
\hline 
\cite{Arcizet2006a} & $8.14\cdot10^{5}$ & $1.9\cdot10^{-7}$ & $81$ & $8.1\cdot10^{9}$ & $1\cdot10^{6}$ & $1.3$ & $1.2$\tabularnewline
\hline 
\cite{Cuthbertson1996} & $318$ & 1.85 & $2.5\cdot10^{-6}$ & $4.1\cdot10^{10}$ & $275$ & 0.9 & $1.2\cdot10^{-3}$\tabularnewline
\hline 
\end{tabular*}

\caption{Experimental parameters for a representative sampling of published
cavity optomechanics experiments}
\end{table*}

Some parameter combinations are of particular relevance for optomechanical
tasks. The following figures provide an overview of the state of the
art in current experiments. The data is compiled from published experiments.
These are labelled as follows:

1 \cite{Cuthbertson1996}, 2 \cite{Massel2011a}, 3 \cite{Regal2008},
4~\cite{Rocheleau2010}, 5~\cite{Teufel2011}, 6~\cite{Chan2011c},
7~\cite{Gavartin2011}, 8~\cite{Thompson2008,Sankey2010}, 9~\cite{Wilson2009},
10~\cite{Jiang2009}, 11~\cite{Lin2009}, 12~\cite{Wiederhecker2009},
13~\cite{Eichenfield2009a}, 14~\cite{Ding2011}, 15~\cite{Park2009},
16~\cite{Schliesser2008}, 17~\cite{Verhagena}, 18~\cite{Schliesser2009a},
19~\cite{Arcizet2006}, 20~\cite{Favero2007}, 21~\cite{Gigan2006},
22~\cite{Groblacher2009}, 23~\cite{Kleckner2006}, 24~\cite{Mow-Lowry2008},
25~\cite{Kleckner2011}, 26~\cite{Groblacher2009a}, 27~\cite{Schleier-smith2011},
28~\cite{Murch2008}, 29~\cite{Brennecke2008}, 30~\cite{Goryachev2012},
31~\cite{Verlot2009}. Different symbols indicate the different optomechanical
implementations: suspended mirrors ($\blacktriangleleft$), optical
microresonators ($\blacklozenge$), photonic crystal cavities ($\blacktriangle$),
suspended nanoobjects ($\blacktriangledown$), microwave resonators
($\bullet$), cold atoms ($\blacktriangleright$).

\begin{figure}
\includegraphics[width=1\columnwidth]{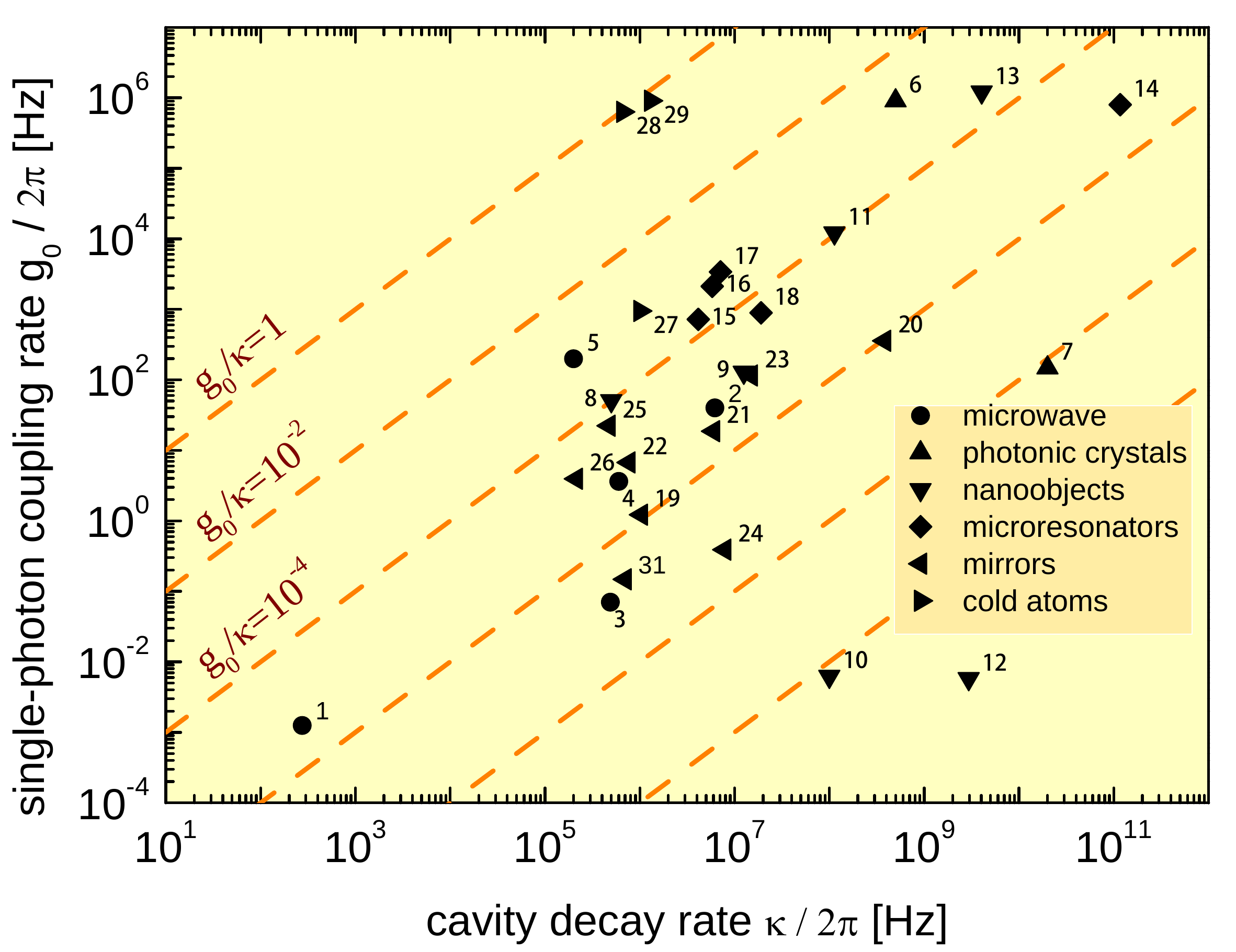}

\caption{\label{fig:g0_vs_kappa}The single-photon optomechanical coupling
strength $g_{0}$, vs. cavity decay rate $\kappa$, for published
experiments (see main text for references). A high value of $g_{0}/\kappa$
would be favourable for \emph{nonlinear} quantum optomechanical experiments,
working with single photons and phonons (Sec.~\ref{sub:QuantumOptomechanicsNonlinear}).}
\end{figure}

\begin{figure}
\includegraphics[width=1\columnwidth]{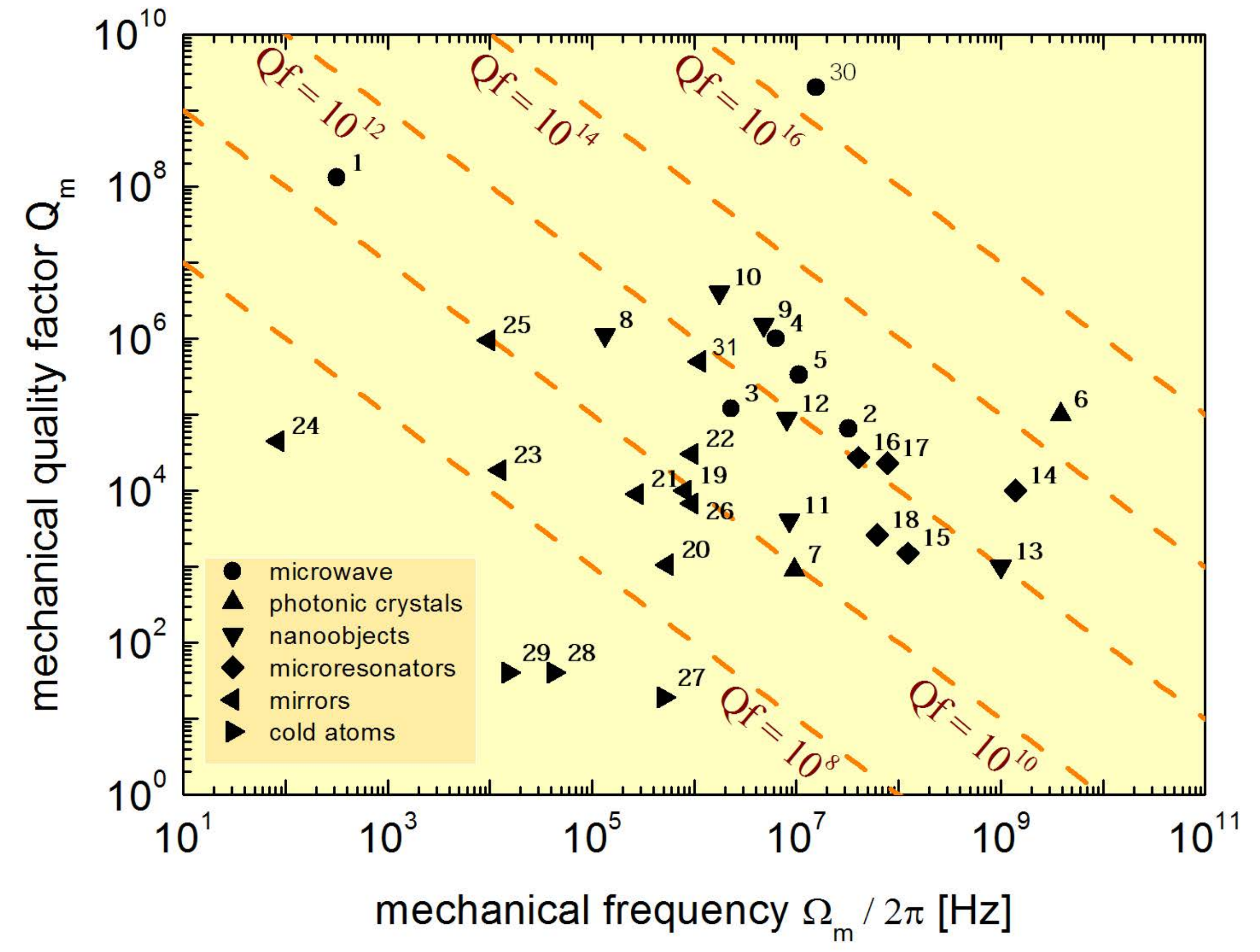}

\caption{\label{fig:Q_vs_Frequency}Mechanical quality factor $Q_{{\rm m}}=\Omega_{{\rm m}}/\Gamma_{{\rm m}}$
vs mechanical frequency $\Omega_{m}=2\pi f_{{\rm m}}$ for published
experiments (see main text for references). The dashed lines represent
$Q_{m}\cdot f_{m}=const.$ for various $Q\cdot f$ values. Note that
full coherence over one mechanical period $1/f_{m}$ is obtained for
$Q_{m}\cdot f_{m}=k_{B}T/\hbar$, i.e. $Q_{m}\cdot f_{m}\gg6\cdot10^{12}$
is a minimum requirement for room-temperature quantum optomechanics.}
\end{figure}

\begin{figure}
\includegraphics[width=1\columnwidth]{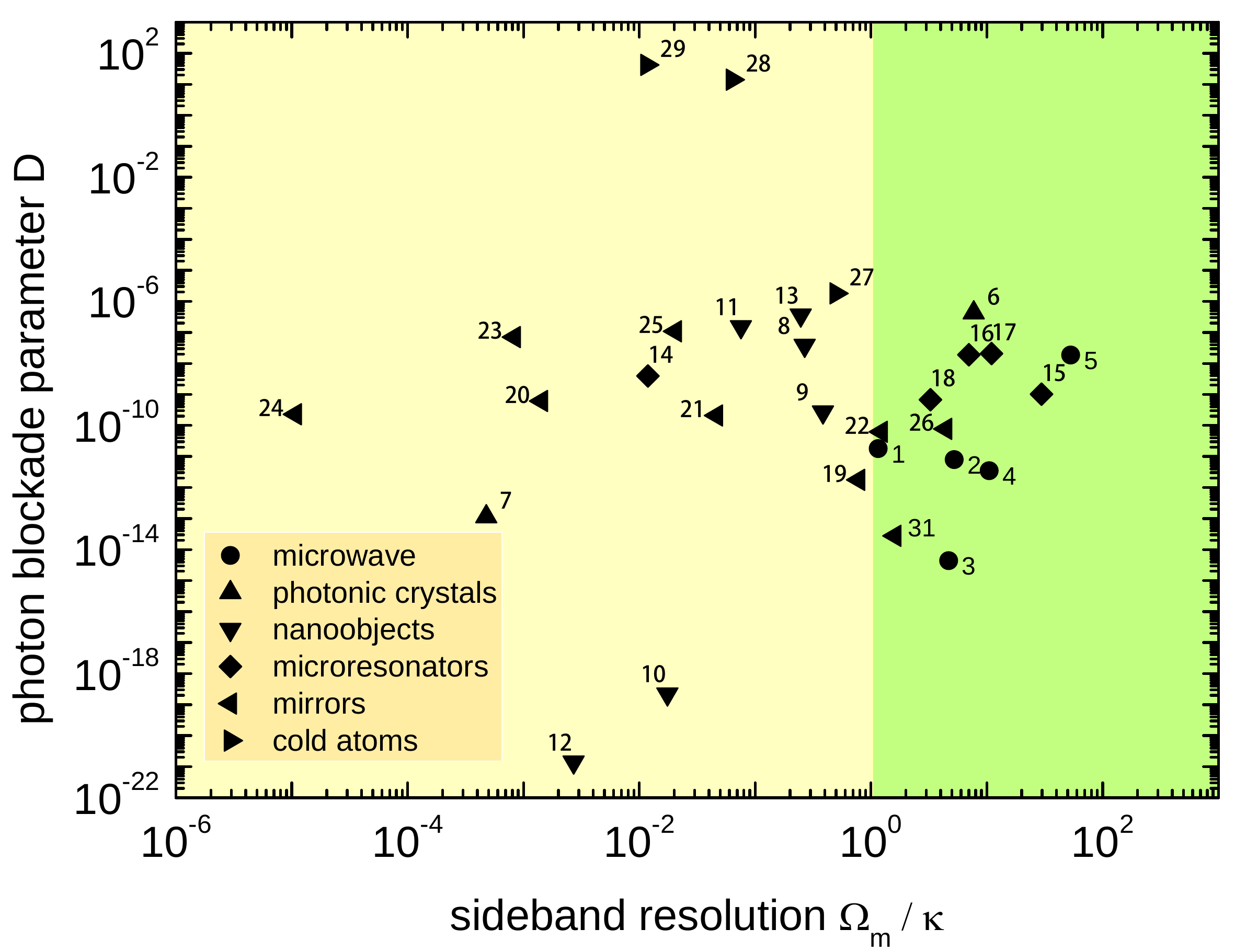}

\caption{\label{fig:PhotonBlockadeParameter_vs_SidebandResolution}Single-photon
blockade parameter $D=g_{0}^{2}/(\Omega_{m}\cdot\kappa)$ vs sideband
resolution $\Omega_{m}/\kappa$, for published experiments (see main
text for references). A single photon induces a cavity frequency shift
$\Delta\omega_{c}=D\cdot\kappa$, which results in a blockade effect
for a subsequent photon for $D>1$, as discussed in Sec.~\ref{sub:QuantumOptomechanicsNonlinear}.
The maximum two-photon suppression scales with $(\kappa/\Omega_{m})^{2}$
and therefore sideband resolution (green shaded area) is an additional
requirement for successful single-photon blockade. }
\end{figure}

\begin{figure}
\includegraphics[width=1\columnwidth]{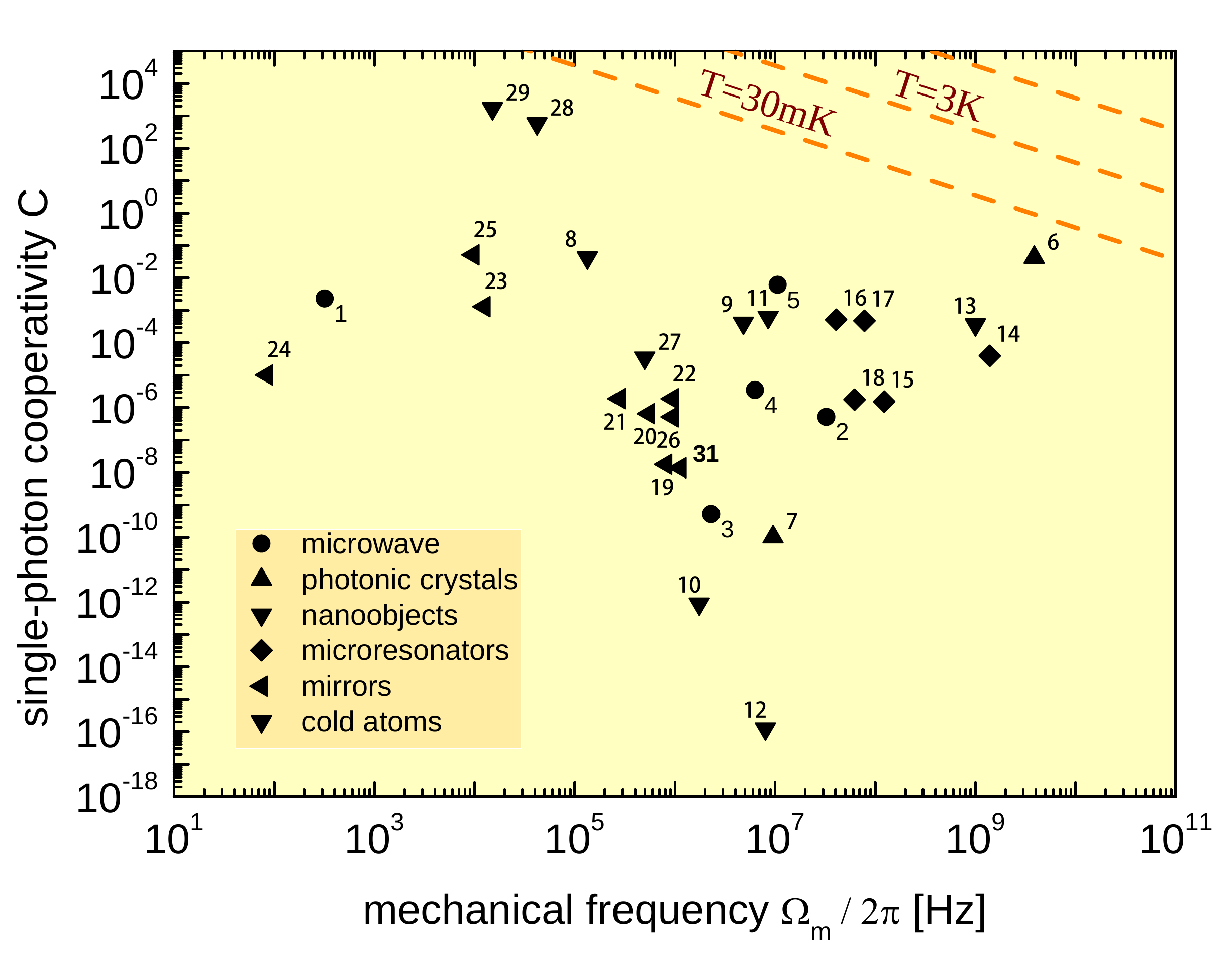}

\caption{\label{fig:Cooperativity_vs_Frequency}The single-photon cooperativity,
$C=g_{0}^{2}/(\kappa\Gamma_{{\rm m}})$, vs. mechanical frequency.
This quantity is important for aspects such as the strength of optomechanically
induced transparency (Sec.~\ref{sub:CoolingOptomechanicallyInducedTransparency}).
Moreover, if $g_{0}^{2}\bar{n}_{{\rm cav}}/(\kappa\Gamma_{{\rm m}}\bar{n}_{{\rm th}})$
is on the order of unity, the state transfer between light and mechanics
is faster than the mechanical decoherence rate. Contour lines indicate
at which temperatures this would be true even for single photons ($\bar{n}_{{\rm cav}}=1$).}
\end{figure}

\subsection{Suspended mirrors}

\label{sub:ExperimentsSuspendedMirrors}

An obvious way to realize optomechanical interactions in a cavity
is to suspend one of the cavity's mirrors. The mechanical motion changes
directly the cavity length and hence the frequency response of such
a {}``rubber cavity''%
\footnote{for the origin of this terminology see the Acknowledgements section
of \cite{Bose1997}.%
} 

The first experimental implementations of this type were a Fabry-Perot
cavity with moving mirrors and date back to the early attempts of
laser interferometeric detection of gravitational waves \cite{Abramovici1992}.
While there the purpose of suspending the macroscopic cavity mirrors
is to achieve acoustic isolation, optomechanical effects, in particular
quantum mechanical radiation pressure fluctuations, eventually pose
the fundamental limit for its interferometric sensitivity \cite{Caves1980b,Unruh1983}
(see Section \ref{sub:MeasurementsDisplacementSensingAndSQL} for
a detailed discussion). At the same time this configuration allows
to exploit cavity optomechanics for the center of mass motion of truly
macroscopic test masses. Experiments of that type have to date resulted
in the demonstration of the optical bistability \cite{Dorsel1983a}
(Sec.~\ref{sub:StaticPhenomenaBistability}), of the optical spring
effect \cite{Sheard2004,Corbitt2007a} (Sec.~\ref{sub:DynamicalBackactionOpticalSpring})
and of optical cooling \cite{Corbitt2007,Mow-Lowry2008} (Secs.~\ref{sub:DynamicalBackactionBasics},
\ref{sec:OptomechanicalCooling}) with suspended mirrors on the gram
scale and even of feedback cooling of suspended mirrors on the kilogram
scale at the LIGO facility \cite{Abbott2009}. A practical challenge
of these experiments is their operation at very low mechanical frequencies
($\Omega_{m}/2\pi<$1 kHz), which requires sophisticated isolation
against acoustic noise. For such experiments this is achieved by suspending
the macroscopic mirrors over several stages. To minimize mechanical
losses it has recently been suggested to levitate the macrosopic mirror
in an optical trap \cite{Singh2010}. Finally, this kind of setup
also allows to monitor and to optomechanically control internal mechanical
modes of macroscopic mirrors \cite{Cohadon1999a,Hadjar1999}, which
are a dominant source for unwanted cavity phase noise e.g. in gravitational
wave detectors \cite{Harry2002,Harry2011}, in cavity QED \cite{Buck2003}
or in frequency stabilization of atomic optical clocks \cite{Numata2004}. 

Another possibility is to use highly reflecting micromechanical devices
as a Fabry-Perot end mirror. These systems include coated cantilevers
\cite{Tittonen1999,HohbergerMetzger2004,Arcizet2006a,Arcizet2008}
and micropillars \cite{Verlot2011}, micrometer-sized mirror pads
on top of cantilevers \cite{Kleckner2006,Favero2007,Groblacher2009,Kleckner2011},
or micromechanically suspended optical coatings \cite{Bohm2006,Cole2008,Cole2010}
and photonic crystal slabs \cite{Antoni2011,Kemiktarak2012}. Efficient
optomechanical coupling in this configuration requires the size of
the mechanical structure to be much larger than the wavelength of
the light: typical cavity lengths range from $10^{-5}{\rm m}$ to
$10^{-2}{\rm m}$, with an optical finesse up to $10^{5}$, which
is generally limited by losses due to the finite cavity mirror sizes.
It has been pointed out that additional interference effects may be
able to overcome this limit significantly \cite{Kleckner2010}. Compared
to the macroscopic mirrors discussed above, these micromechanical
devices allow access to higher mechanical frequencies (up to some
tens of MHz) and, in principle, to higher mechanical quality factors.
In particular, the possibility of exact geometric control via microfabrication
techniques allows to minimize mechanical losses due to clamping \cite{Wilson-Rae2008,Anetsberger2008,Cole2011}. 

The accessible mass and frequency ranges in combination with the restrictions
on cavity length ($L>\lambda$) and achievable cavity finesse set
some practical limitations for this geometry. In particular, sideband-resolution
(small $\kappa$) and large optomechanical coupling $g_{0}$ (small
$L$, leading to a large $\kappa$) impose conflicting conditions
and need to be traded against each other. On the other hand, along
with the macroscopic microwave transducers (see Sec.\ref{sub:ExperimentsMicrowaveResonators}),
this realization provides optomechanical control over by far the largest
range of mass and frequency.

\subsection{Optical microresonators}

\label{sub:ExperimentsOpticalMicroresonators}

A situation similar to the Fabry-Perot case occurs in optical microresonators,
where light is guided in whispering gallery modes along the rim of
a circular resonator \cite{Vahala2003a}. There is a large number
of different mechanical normal modes of vibration of these structures.
The resulting distortions of the structure directly modify the optical
path length of the resonator, shifting its optical resonance frequency
and hence generating optomechanical coupling. The small size of microresonators
allows to achieve large coupling rates $g_{0}$ \cite{Ding2011,Verhagena}
and to access mechanical frequencies from several MHz up to some GHz.
In essence, three different architectures can be distinguished: (i)
Microdisk resonators, which are the standard resonator structure in
planar photonic circuits and can be fabricated with high precision.
Recent experiments have demonstrated large optomechanical coupling
rates up to $g_{0}\approx2\pi\times8\cdot10^{5}$ Hz \cite{Ding2011}.
A fundamental limit in their performance is given by radiation losses
at the sidewalls. Another limitation is due to internal materials
losses, which could be improved by using single-crystalline materials.
A first demonstration in this direction, specifically optomechanical
coupling to internal modes of a single-crystalline ${\rm CaF}_{2}$
resonator, has recently been reported \cite{Hofer2009}. (ii) Microsphere
resonators, which allow a larger optical quality \cite{Park2009,2009_Carmon_StimulatedBrillouinScattering};
there, the mechanical quality is mainly limited by internal materials
losses, in particular for the often used silica microspheres. (iii)
Microtoroidal resonators, which are obtained from microdisk resonators
by a thermal reflow process that melts the sidewalls into a toroidal
topology. The generated smooth surface together with the microfabrication
control provides a combination of high optical \cite{Armani2003}
and high mechanical quality \cite{Anetsberger2008}. This has resulted
in the first demonstration of radiation-pressure driven optomechanical
parametric amplification \cite{Kippenberg2005,Carmon2005,Rokhsari2005a}
as well as of sideband-resolved operation \cite{Schliesser2008}.
Recently, hybrid toroid devices have been developed that combine optomechanical
and electromechanical actuation \cite{Lee2010}.

The practical benefits of these geometries are the availability of
large optical qualities in combination with the resolved sideband
regime $\kappa<\Omega_{{\rm m}}$, essentially owed to the fact that
the mechanical frequencies range from $10\,{\rm MHz}$ to several
${\rm GHz}$. Possible limitations arise from the necessity to propagate
light inside a solid-state medium, which increases the sensitivity
to optical absorption and thermorefractive cavity noise.

\subsection{Waveguides and photonic crystal cavities}

\label{sub:ExperimentsWaveguidesPhotonicCrystals}

On-chip waveguides and photonic crystal cavities offer a different
implementation architecture. 

Photonic crystals are formed by a periodic modulation of the index
of refraction of some material (typically silicon), which leads to
the formation of optical bands, in analogy to the electronic bands
for electron waves propagating in a crystal lattice. Light cannot
propagate in the band-gaps. Thus, when artificial defects are introduced
into the periodic pattern, localized electromagnetic field modes \cite{vahala2004optical}
can form that do not decay into the continuum inside the structure.
These structures are called photonic crystal cavities. To obtain an
optomechanical device, in-plane photonic crystal cavities are underetched
to form nanomechanical beams. The mechanical motion results in modulations
of the cavity boundaries and stresses in the material, both of which
contribute to the optomechanical coupling between the cavity photons
and the mechanical modes of the structure. This has been demonstrated
both for 1D \cite{Eichenfield2009a,Eichenfield2009} and for 2D \cite{2010_Painter_2DPhotonicCrystalCavity,Gavartin2011}
photonic crystal cavities . The small cavity dimensions in combination
with the small mass of the localized mechanical mode result in an
optomechanical coupling strength that is much larger than in regular
Fabry-Perot approaches, with current experiments achieving $g_{0}/2\pi\approx{\rm MHz}$.
The available mechanical frequencies can range from several tens of
${\rm MHz}$ up to several ${\rm GHz}$, which significantly reduces
the environment thermal occupation $\bar{n}_{th}=k_{B}T/\hbar\Omega_{{\rm m}}$.
The idea of creating bandgaps by inducing periodic boundary conditions
can be extended to the modes of the mechanical beam. Introducing a
surrounding periodic structure matched to the phonon wavelength ({}``phonon
shield'') results in a 1D photonic crystal cavity with co-localized
photonic and phononic modes with a significantly increased mechanical
quality $Q_{m}$\cite{Chan2011c}; see \cite{2009_Painter_OMC_Design,Safavi-Naeini2010}
for more on the design of 1D and 2D optomechanical crystals \cite{Safavi-Naeini2010}.
It is also possible to integrate two-level quantum systems inside
the photonic crystal nanobeam, e.g. by fabricating photonic crystal
cavity nanobeams out of diamond \cite{Riedrich-Moller2011} or out
of GaAs \cite{Rundquist2011}, which can include artificial qubits
formed by for example nitrogen vacany (NV) centers or by quantum dots,
respectively. Another possibility is to have hybrid devices with both
optical and electrical actuation \cite{Winger2011}.

Currently, because of the large available coupling rates $g_{0}/\kappa$,
this approach may allow to enter the regime of nonlinear photon-phonon
interactions (see Sec.~\ref{sub:QuantumOptomechanicsNonlinear}).
Moreover, the large mechanical frequencies in the ${\rm GHz}$ range
could allow for low-temperature operation in a regime where the average
phonon number drops below one even without additional laser-cooling.
This would be highly beneficial for quantum applications. Finally,
the in-plane architecture is immediately compatible with the architectures
of integrated (silicon) photonics and provides a direct route to larger-scale
optomechanical arrays, which is interesting in the context of classical
and quantum information processing, and for the study of collective
dynamics (Sec.~\ref{sec:MultimodeOptomechanics}). 

It should be noted that optomechanical forces can become strong even
in the absence of a cavity, for structures with waveguides running
close to a substrate or close to each other. This approach (while
somewhat outside the domain of the concepts covered in the present
review) could be very fruitful for applications, since it does away
with the bandwidth restrictions generated by a cavity \cite{2008_Tang_PhotothermalEffectSiliconWaveguides,Li2008,2009_Tang_BroadbandTransduction,2009_Tang_TheoryWaveguideSubstrateCoupling,2009_Tang_TunableInteractionsWaveguides,Bagheri2011}.
In another equally promising development, the ${\rm GHz}$ acoustic
vibrations of photonic crystal fibres are being excited and controlled
via optomechanical interactions \cite{2006_Russell_StimulatedBrillouinScattering_,2008_Russell_CoherentControlAcousticsResonances,2008_Russell_ExcitationGHz,2009_Russell_AcousticPhononsNonlinear,2010_Russell_GigaHertzForwardScatteringPCF,2011_Russell_OptoacousticIsolator,Butsch2012,Butsch2012b}.

\subsection{Suspended and levitated nano-objects}

\label{sub:ExperimentsNanoObjects}

This class of cavity optomechanics implementations uses a rigid optical
cavity that contains a mechanical element either inside the cavity
or in the near field of the cavity. It allows in particular the efficient
optomechanical coupling to sub-wavelength size mechanical objects,
which has been demonstrated for systems such as high-quality mechanical
membranes made of high-stress SiN \cite{Thompson2008,Sankey2010},
stochiometric SiN \cite{Wilson2009} or AlGaAs \cite{Liu2011a}, and
for carbon nanowires \cite{Favero2009}, which have been suspended
inside state-of-the-art Fabry-Perot cavities. The embedded nano-objects
modify the cavity field either via dispersion \cite{Thompson2008}
or via dissipation, as suggested in \cite{Xuereb2011}.

An alternative approach to Fabry-Perot resonators is to exploit near-field
effects close to the surface of optical microresonators, where the
evanescent optical field allows dispersive coupling to other structures.
In essence, the mechanical motion modulates the distance $d$ between
the interfaces. Due to the near-field character the optomechanical
coupling strength scales exponentially with d and hence allows to
generate large values for $g_{0}$. This has been used to demonstrate
optomechanical coupling between a toroidal microcavity and a nearby
SiN nanomechanical resonator \cite{Anetsberger2009a}. Another related
possibility is to couple two mechanically vibrating microdisk resonators
\cite{Lin2009,Jiang2009,Wiederhecker2009} or two photonic crystal
cavities \cite{Roh2010,Eichenfield2009a} via their optical near field. 

In order to further suppress mechanical clamping losses, it has been
suggested to levitate the mechanical objects either by an additional
optical dipole trap or in the standing wave trap formed by the cavity
field \cite{Chang2010,Romero-Isart2010,Barker2010}. This implementation
allows a direct extension to matter-wave interferometry \cite{Romero-Isart2011}
and may enable fundamental tests of quantum theory in a new macroscopic
parameter regime (see also Sec.~\ref{sub:QuantumHybridSystems}).
The necessary parameter regime for such tests is experimentally challenging
\cite{Romero-Isart2011a} and may even require a space environment
\cite{Kaltenbaek2012}. Levitation of micrometer-size \cite{1977_Ashkin_Feedback,Li2011a}
and sub-micrometer size \cite{Gieseler2012} silica spheres has already
been demonstrated in optical dipole traps in high vacuum. An alternative
approach could be to combine optical trapping with a low-frequency
mechanical suspension \cite{Corbitt2007,Ni2012}, which has been suggested
to lead to thermal decoupling of similar quality as purely optical
trapping ($Q\cdot f\approx10^{18}$) \cite{Chang2010a}.

A prominent feature of such setups, with a nano-object inside the
standing light wave of a cavity mode, is quadratic coupling to position.
The optical frequency shift may be no longer linear but rather quadratic
in the mechanical displacement, if the object is placed at a node
or antinode. This could lead to interesting applications, such as
QND detection of single phonons, as explained in Sec.~\ref{sub:MeasurementsQNDFockState}.
These setups have also been suggested to strongly couple two nano-objects,
for example a mechanical membrane to a single atom \cite{Hammerer2009,Wallquist2010}
(see Sec.\ref{sub:QuantumHybridSystems}).

\subsection{Microwave resonators}

\label{sub:ExperimentsMicrowaveResonators}

Analogous to optical cavities, LC circuits form a resonator for electromagnetic
radiation in the microwave regime, i.e. $\omega_{c}/2\pi\sim GHz$.
The motion of a mechanical element capacitively coupled to this microwave
cavity results in a shift of capacitance, and thereby of the LC resonance
frequency ($\partial C/\partial x\propto\partial\omega_{c}/\partial x$).
Thus, one obtains the standard cavity-optomechanical radiation pressure
interaction. The first experiments along this line have been performed
by Braginsky and co-workers \cite{Braginsky1967,Braginsky1970,Braginsky1977},
and later in the context of resonant bar gravitational wave detection
\cite{Blair1995,Cuthbertson1996}; already back then these works have
demonstrated both cold damping and optomechanical backaction effects
such as cooling and parametric amplification. Later, in the context
of ion-trap physics, cooling of a micromechanical resonator via an
LC circuit was shown \cite{Brown2007}. With the advent of microfabricated
superconducting circuits it has become possible to enter the size
and frequency regime of nanomechanical devices coupled to microwave
cavities \cite{Regal2008}. Typical available mechanical frequencies
range from some ${\rm MHz}$ to some tens of ${\rm MHz}$. In order
to resemble a low-entropy reservoir of the radiation field, which
is of particular importance for quantum optomechanics (see Sec. \ref{sec:QuantumOptomechanics}),
the microwave photons need to be kept at cryogenic temperatures. For
${\rm GHz}$ photons, environment temperatures in the mK regime are
sufficient, which necessitates operation inside a dilution refrigerator.
Although the momentum transfer of microwave photons is several orders
of magnitude smaller compared to photons at optical frequencies, the
bare optomechanical coupling rates $g_{0}$ can be made comparable
to (or larger than) implementations in the optical domain \cite{Rocheleau2010,Teufel2011,Pirkkalainen2012}.
The essential idea is to have a very small coupling gap and to optimize
the fraction of the total capacitance that responds to the mechanical
motion (see also Fig.~ \ref{fig:g0_vs_kappa}). 

A current practical challenge for the microwave schemes is the sparse
availability of quantum optics techniques such as the preparation
and detection of Fock states or of squeezed states of the radiation
field. However, several recent proof-of-concept experiments have demonstrated
their availability in principle \cite{Hofheinz2009,Mallet2011,Eichler2011}. 

As a sidenote, capacitive coupling has also been used to couple nanomechanical
objects directly to two-level quantum systems, e.g. to a superconducting
Cooper-Pair box \cite{LaHaye2009} or to a superconducting phase qubit
\cite{O'Connell2010}. Note finally that the coupling need not be
capacitive. Recently, it was shown that a microwave resonator can
also be coupled via dielectric gradient forces to the vibrations of
a nanobeam \cite{2011_Faust_PlugAndPlayNanomechanics}. This makes
available a larger range of materials, which could be beneficial for
applications.

\subsection{Ultracold atoms}

\label{sub:ExperimentsUltracoldAtoms}

The ideas of cavity optomechanics have also been implemented by using
clouds of up to $10^{6}$ atoms. Their collective motional dynamics
can resemble a single mechanical mode that, for the case of ultracold
atoms, is already pre-cooled to its quantum ground state of motion.
In one case, the collective motion of a cloud of ultracold Rb atoms
inside a Fabry Perot cavity was used to observe signatures of shot-noise
radiation pressure fluctuations \cite{Murch2008}. The dispersive
coupling of the collective motion of the cloud to an optical cavity
field results in a position-dependent frequency shift and therefore
to quantum optomechanical interactions. 

Suppose the single-photon dispersive energy shift of a single atom
sitting at an antinode of the standing light wave pattern is $\delta E=-\hbar(g_{0}^{{\rm at}})^{2}/\Delta_{{\rm at}}$,
with $g_{0}^{{\rm at}}$ the atom-cavity vacuum Rabi frequency, and
$\Delta_{{\rm at}}$ the detuning between atom and cavity resonance.
Then the coupling Hamiltonian between the cavity mode and an atom
cloud of $N$ atoms trapped near position $\bar{x}$ would be $N\delta E\hat{a}^{\dagger}\hat{a}\sin^{2}(k(\bar{x}+\hat{x}))$.
Expanding to lowest order in $\hat{x}$, this yields a bare optomechanical
coupling rate

\[
g_{0}=\frac{\left(g_{0}^{{\rm at}}\right)^{2}}{\Delta_{{\rm at}}}\left(kx_{{\rm ZPF}}^{{\rm atom}}\right)\sin(2k\bar{x})\sqrt{N}\,,
\]
where $x_{{\rm ZPF}}^{{\rm atom}}=\sqrt{\hbar/(2m_{{\rm atom}}\Omega_{{\rm m}})}$
denotes the zero-point fluctuations of a single-atom in the trapping
potential, that is $\sqrt{N}$ times larger than $x_{{\rm ZPF}}^{CM}$
of the center-of-mass motion of the cloud. Here we have assumed that
the extent of the cloud is small with respect to the wavelength. Sometimes
the cloud is actually distributed over several lattice sites. Note
that the trapping potential could be provided by another optical lattice
or magnetically. Incidentally, we note that the same kind of derivation
applies for trapped dielectric particles (Sec.~\ref{sub:ExperimentsNanoObjects}).
We also mention that if the atoms are trapped right at a node or antinode,
the leading optomechanical coupling is to $\hat{x}^{2}$ instead of
$\hat{x}$, which leads to different physics (e.g. as in Sec.~\ref{sub:MeasurementsQNDFockState}). 

In another experiment, cavity optomechanics was used to cool the motion
of a thermal cloud of Cs atoms trapped inside an optical cavity \cite{Schleier-smith2011}.
Finally, density fluctuations in a Bose-Einstein condensate of $10^{6}$
atoms have been used as the mechanical mode inside a Fabry-Perot cavity
\cite{Brennecke2008}. In both ultracold cases, due to the strong
dispersive atomic coupling and the small mass (leading to a large
zero-point motional amplitude), operation was close to the single-photon
strong coupling regime, $g_{0}/\kappa\sim1$. 

More recently, a setup has been demonstrated that couples the motion
of a vibrating mirror to the motion of atoms trapped in a standing
light wave being reflected from that mirror \cite{2011_Treutlein_ColdAtomsAndMembrane},
without an optical cavity.

\section{Basic consequences of the optomechanical interaction}

\label{sec:BasicConsequencesOfOptomechanics}

\subsection{Static phenomena: Optical potential and bistability}

\label{sub:StaticPhenomenaBistability}

\begin{figure}[ptb]
 \centering\includegraphics[width=3in]{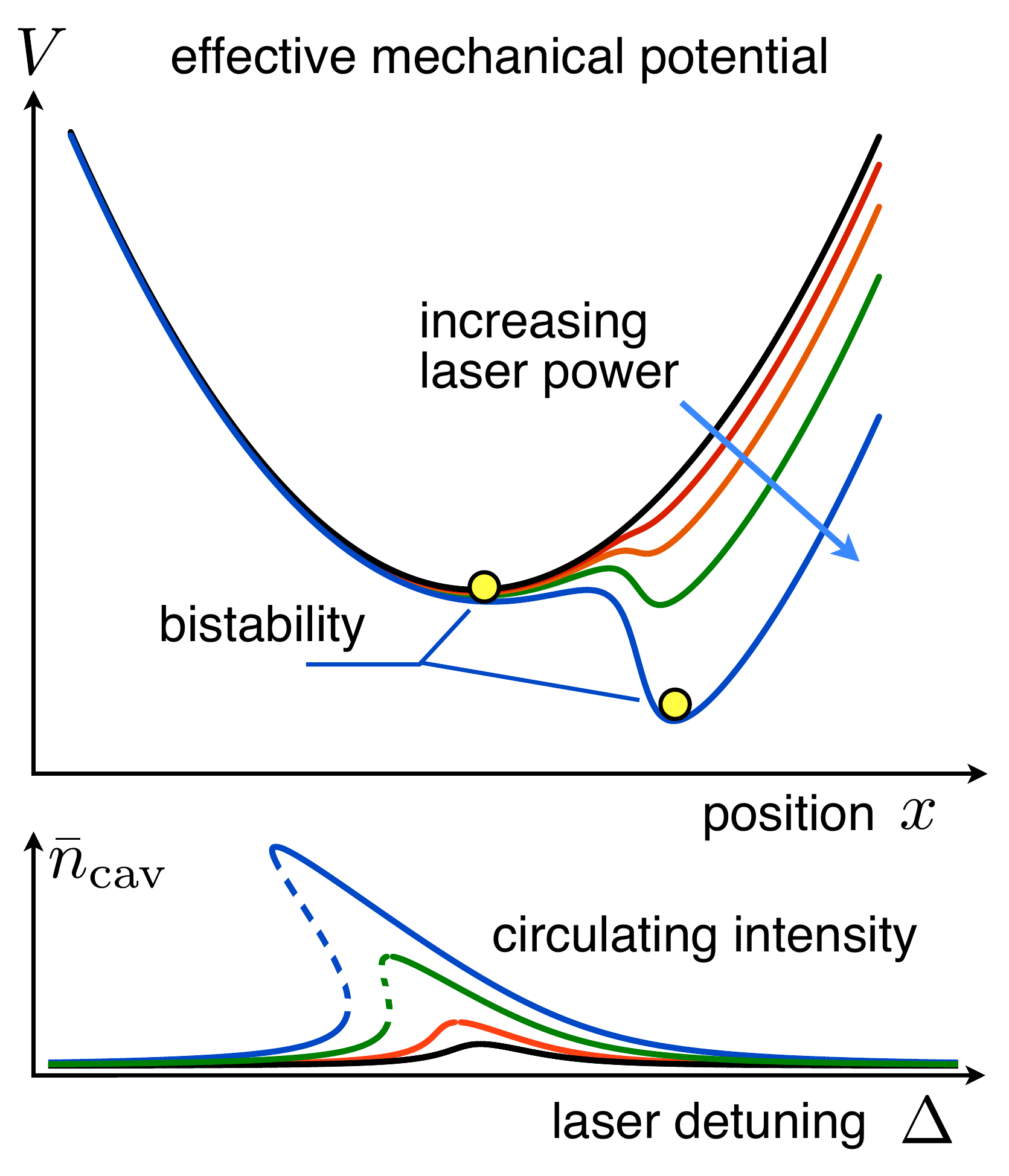}\caption{\label{fig:Bistability}Optomechanical static bistability occurs when
the laser intensity is sufficiently high to generate two stable local
minima in the effective mechanical potential (top). This results in
bistable behaviour and hysteresis when recording the circulating photon
number $\bar{n}_{{\rm cav}}$ or the transmission as a function of
laser detuning (bottom).}
\end{figure}

We first deal with the simplest case, when the light force reacts
instantaneously to the mechanical motion. This will be relevant for
$\kappa\gg\Omega_{m}$. Then, the radiation pressure force $F(x)=\hbar G\bar{n}_{{\rm cav}}(x)$
depends on the displacement $x$ via $\bar{n}_{{\rm cav}}(x)$, the
photon number circulating inside the optical mode. Such a 1D conservative
force can be derived from a potential (Fig.~\ref{fig:Bistability}):

\begin{equation}
F(x)=-\frac{\partial V_{{\rm rad}}(x)}{\partial x}\,.
\end{equation}
For the case of a single, high-finesse optical resonance we have $\bar{n}_{{\rm cav}}(x)=\bar{n}_{{\rm cav}}^{{\rm max}}/[1+(2(Gx+\Delta)/\kappa)^{2}]$,
where $\bar{n}_{{\rm cav}}^{{\rm max}}$ is the maximum number of
circulating photons, obtained at resonance (proportional to the incoming
laser intensity). As a result,

\begin{equation}
V_{{\rm rad}}(x)=-\frac{1}{2}\hbar\kappa\bar{n}_{{\rm cav}}^{{\rm max}}\arctan[2(Gx+\Delta)/\kappa]\,.
\end{equation}
Note that for the case of photothermal forces, the discussion still
applies, only with a different prefactor in $F(x)\propto\bar{n}_{{\rm cav}}(x)$.
The overall potential for the mechanical motion also includes the
intrinsic harmonic restoring potential:

\begin{equation}
V(x)=\frac{m_{{\rm eff}}\Omega_{m}^{2}}{2}x^{2}+V_{{\rm rad}}(x)\,.
\end{equation}
The first effect of the radiation force is to shift the equilibrium
position to $x_{0}\neq0$, with $V'(x_{0})=0$. In addition, the effective
spring constant is changed to its new value

\begin{equation}
k_{{\rm eff}}=V''(x_{0})=m_{{\rm eff}}\Omega_{m}^{2}+V''_{{\rm rad}}(x_{0})\,,
\end{equation}
where the second contribution is called {}``optical spring''. In
particular for low-frequency mechanical modes, this term can be orders
of magnitude larger than the intrinsic mechanical spring \cite{Corbitt2007a}.
Such an approach essentially amounts to a variant of optical trapping
and can be exploited to diminish the unwanted mechanical dissipation
and heating connected with the intrinsic mechanical spring being attached
to a substrate. In the limit of low light intensity, the resulting
correction for the mechanical frequency (obtained from $V_{{\rm rad}}''(0)=m_{{\rm eff}}\delta(\Omega_{m}^{2})$)
is 

\begin{equation}
\delta\Omega_{m}=8\Delta\left(\frac{g_{0}}{\kappa}\right)^{2}\frac{\bar{n}_{{\rm cav}}^{{\rm max}}}{[1+(2\Delta/\kappa)^{2}]^{2}}\,.
\end{equation}
This corresponds to the limit $\kappa\gg\Omega_{m}$ of the dynamical
case discussed below (note the relation between $\bar{n}_{{\rm cav}}$
and $\bar{n}_{{\rm cav}}^{{\rm max}}$).

At larger light intensities $V(x)$ may develop into a double-well
potential, with two local minima, leading to what we may term static
bistability. Both of these minima correspond to stable equilibrium
positions, determined by the nonlinear equation $F(x_{1/2})=m_{{\rm eff}}\Omega_{m}^{2}x_{1/2}$.
Physically, they represent situations with low/high light intensity
and low/high restoring force. The bistable behaviour will occur at
negative detunings, $\Delta<0$, roughly when $\delta\Omega_{m}\sim-\Omega_{m}$,
such that the original equilibrium position becomes unstable. As negative
detunings are necessary for cooling, the bistability limits the achievable
cooling laser intensities for the case $\kappa>\Omega_{m}$. 

More quantitatively, analysis of the equation $F(x)=m_{{\rm eff}}\Omega^{2}x$
shows that bistability sets in first (at a single value of $\Delta$)
when the maximum correction to the spring constant, $-\partial F(x)/\partial x$,
obtained at $Gx+\Delta=-\kappa/(2\sqrt{3})$, equals the intrinsic
spring constant $m_{{\rm eff}}\Omega_{m}^{2}$. This happens at a
critical detuning of $\Delta=-\sqrt{3}\kappa/2$ and at a critical
laser power determined by

\begin{equation}
\frac{3}{2}\sqrt{3}\frac{g_{0}^{2}}{\Omega_{m}\kappa}\bar{n}_{{\rm cav}}^{{\rm max}}=1\,,
\end{equation}
where we employed the relations $G=g_{0}x_{{\rm ZPF}}$ and $x_{{\rm ZPF}}^{2}=\hbar/(2m_{{\rm eff}}\Omega_{{\rm m}})$.
At higher light intensities, the range of detunings for which bistability
is observed widens. In experiments, bistability is revealed in hysteresis,
e.g. when recording the transmission or phase shift while sweeping
the detuning up and down. The first experiments on optomechanical
bistability, with a macroscopic mirror, were reported and analyzed
already in the 1980s, both in the optical \cite{Dorsel1983a,Meystre1985}
and in the microwave domain \cite{Gozzini1985a}.

For low-finesse systems, nearby optical resonances may also become
relevant for the mechanical motion, leading to a more complicated
effective potential, possibly with several local minima (Fig.~\ref{fig:Bistability}).

\subsection{Dynamical backaction}

\label{sub:DynamicalBackactionBasics}

We now turn to dynamical effects, due to the retarded nature of the
radiation pressure force.

\begin{figure}
\includegraphics[width=1\columnwidth]{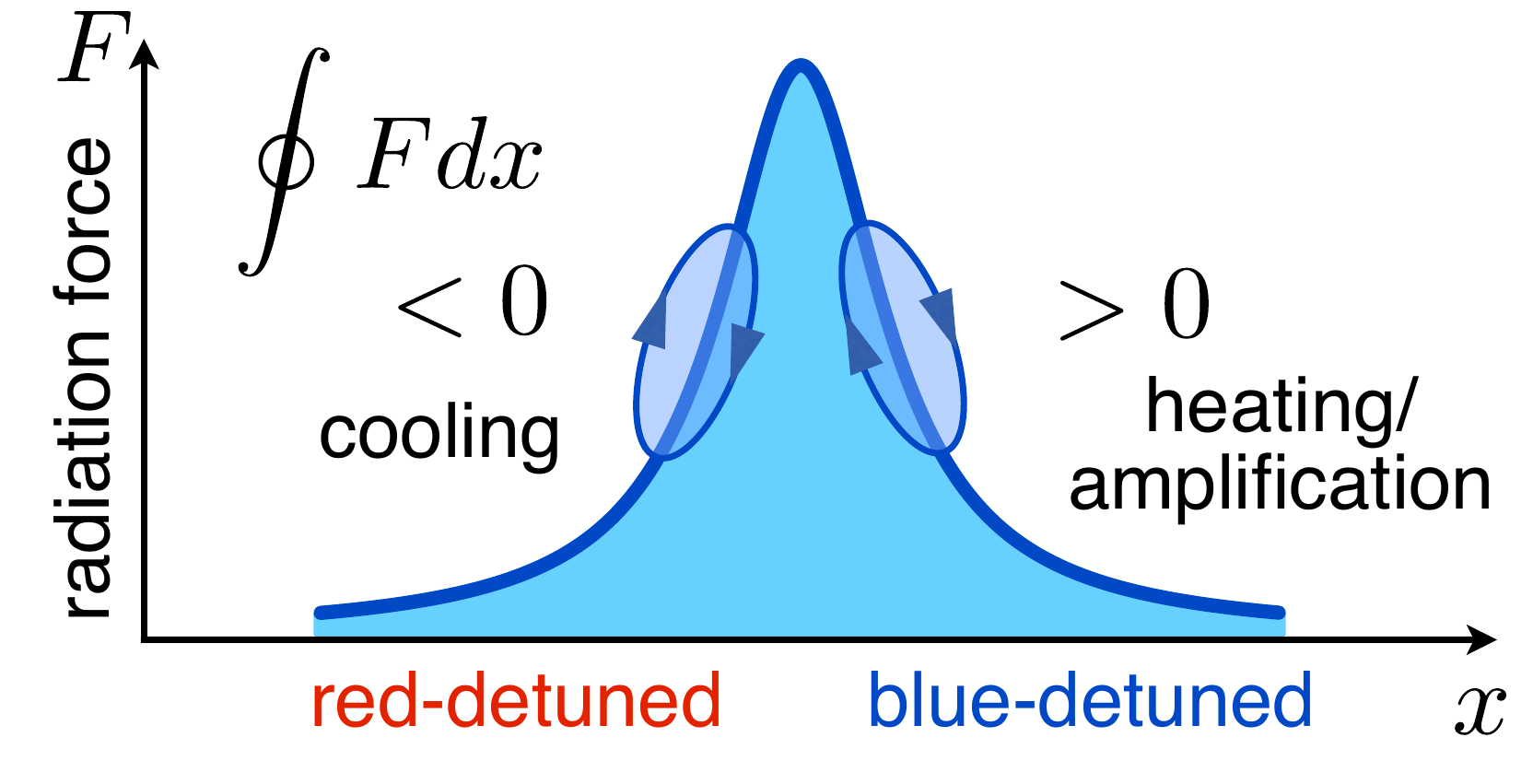}

\caption{\label{fig:DynamicalBackactionWorkDoneDuringCycle}''Thermodynamic''
style schematic diagram depicting the work done by the radiation force
during one cycle of oscillation. The work is given by the finite area
swept in the force-displacement diagram, which is due to the retardation
of the force (finite cavity decay rate). The work is negative or positive,
depending on whether one sits on the red-detuned or blue-detuned side
of the resonance. This then gives rise to damping or amplification,
respectively.}
\end{figure}

To derive the dynamics arising from the optomechanical coupling, one
can solve the linearized coupled equations of motion for the light
and the mechanics, as presented in section \ref{sub:OptomechanicalEquationsOfMotion}.
This is best done in frequency space. We employ the classical linearized
equations Eqs.~(\ref{eq:ClassicalEqsLinearizedOmega-alpha}) and
(\ref{eq:ClassicalEqsLinearizedOmega-x}) as the basis for our following
analysis. This is possible since we will be only interested in the
linear response to an external mechanical force, and the averaged
linearized quantum equations are identical to their classical version
(and do not contain the noise sources anymore).

\begin{figure}
\includegraphics[width=1\columnwidth]{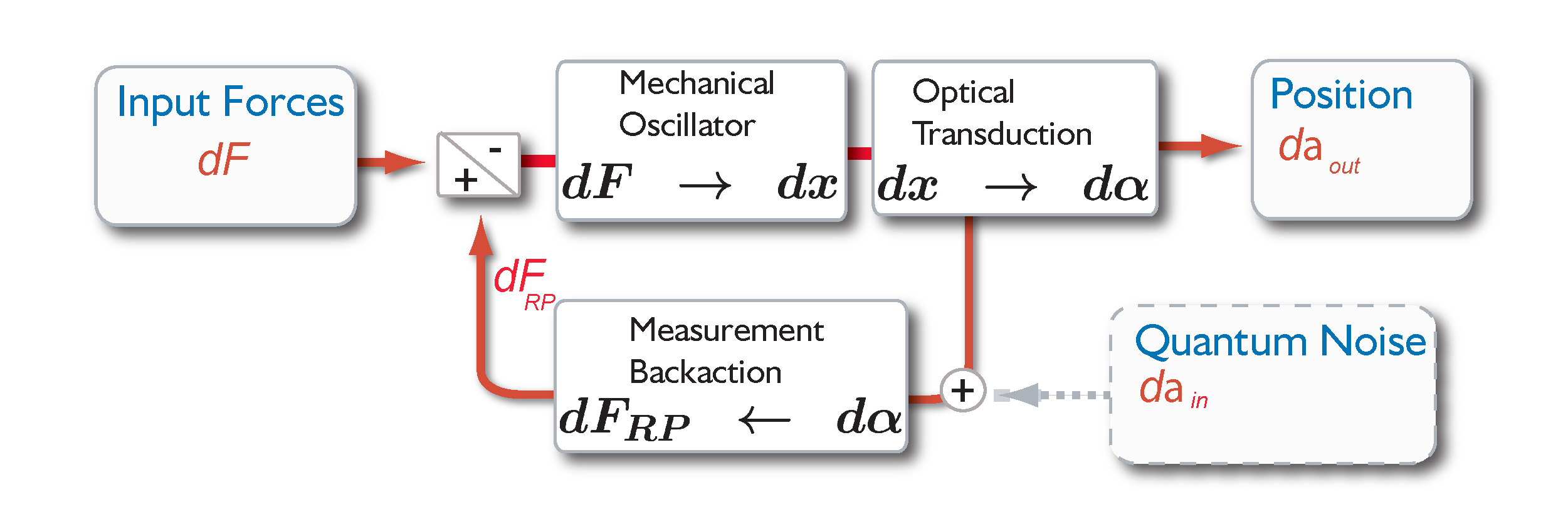}

\caption{\label{fig:Schematic-optomechanical-feedbac}Schematic optomechanical
feedback loop}
\end{figure}

In the absence of optomechanical coupling, the mechanical oscillator
has the susceptibility $\chi_{xx,0}^{-1}(\omega)=m_{{\rm eff}}[(\Omega_{m}^{2}-\omega^{2})-i\Gamma_{{\rm m}}\omega]$
(see Sec.~\ref{sub:MechanicalResonatorsNoiseSpectra}). We will now
assume that a weak test force $F$ acts on the mechanical oscillator
in the presence of the optomechanical interaction. Solving the coupled
set of equations, we can find the mechanical response to that force.
This defines the modified mechanical susceptibility, which can be
expressed in terms of the original susceptibility plus some modification%
\footnote{The notation $\Sigma(\omega)$ reminds us that this modification has
the form of a {}``self-energy'', in the way that it occurs in the
analogous Dyson expression for a Green's function of a particle modified
from the bare Green's function due to interactions.%
}:

\begin{align}
\chi_{xx}^{-1}(\omega) & =\chi_{xx,0}^{-1}(\omega)+\Sigma(\omega)\label{eq:OpticallyModifiedSusceptibility}
\end{align}
The coupled equations (\ref{eq:ClassicalEqsLinearizedOmega-alpha})
and (\ref{eq:ClassicalEqsLinearizedOmega-x}) are solved by expressing
$\delta\alpha[\omega]$ in terms of $x[\omega]$ and inserting the
result into the equation for $x[\omega]$. This yields the modification
of the linear response to an external force:

\begin{equation}
\Sigma(\omega)=2m_{{\rm eff}}\Omega_{{\rm m}}g^{2}\left\{ \frac{1}{(\Delta+\omega)+i\kappa/2}+\frac{1}{(\Delta-\omega)-i\kappa/2}\right\} 
\end{equation}
, where we employed the relation $G=g_{0}/x_{{\rm ZPF}}$ and obtain
$\hbar G^{2}\left|\bar{\alpha}\right|^{2}=2m_{{\rm eff}}\Omega_{{\rm m}}g^{2}$.
For now, we just \emph{define}

\[
\Sigma(\omega)\equiv m_{{\rm eff}}\omega(2\delta\Omega_{m}(\omega)-i\Gamma_{\text{opt}}(\omega))\,,
\]
such that the new terms have the structure that is suggested by the
form of the original susceptibility, leading to:

\begin{equation}
\chi_{xx}^{-1}(\omega)=m_{{\rm eff}}(\Omega_{m}^{2}+2\omega\delta\Omega_{m}(\omega)-\omega^{2}-i\omega[\Gamma_{m}+\Gamma_{\text{opt}}(\omega)])\,
\end{equation}
The real and imaginary parts then yield the frequency-dependent mechanical
frequency shift $\delta\Omega_{{\rm m}}(\omega)$ and optomechanical
damping rate $\Gamma_{{\rm opt}}(\omega)$, whose meaning will be
discussed further in the next section. The explicit expressions are
(by taking real and imaginary parts, $\delta\Omega_{{\rm m}}(\omega)={\rm Re}\Sigma(\omega)/2\omega m_{{\rm eff}}$,
and $\Gamma_{{\rm opt}}(\omega)=-{\rm Im}\Sigma(\omega)/m_{{\rm eff}}\omega$): 

\begin{align*}
\delta\Omega_{{\rm m}}(\omega)= & g^{2}\frac{\Omega_{{\rm m}}}{\omega}\left[\frac{\Delta+\omega}{(\Delta+\omega)^{2}+\kappa^{2}/4}+\frac{\Delta-\omega}{(\Delta-\omega)^{2}+\kappa^{2}/4}\right]\\
\Gamma_{{\rm opt}}(\omega)= & g^{2}\frac{\Omega_{{\rm m}}}{\omega}\left[\frac{\kappa}{(\Delta+\omega)^{2}+\kappa^{2}/4}-\frac{\kappa}{(\Delta-\omega)^{2}+\kappa^{2}/4}\right]
\end{align*}
These expressions provide an exact solution of the linearized problem
that is also valid in the regime of strong coupling, where $g>\kappa/2$
(Sec.~\ref{sub:CoolingStrongCoupling}). Also note that the effect
is linear in the laser drive power, i.e. in the circulating photon
number: $g^{2}=g_{0}^{2}\bar{n}_{{\rm cav}}$. In the next sections
we discuss the resulting physical phenomena, i.e. optical spring effect
and amplification and cooling.

The frequency dependence of $\Sigma(\omega)$ will in general yield
a non-Lorentzian lineshape for the susceptibility that will even turn
into a double-peak structure at strong coupling (Sec.~\ref{sub:CoolingStrongCouplingNormalModeSplitting}).
However, for sufficiently weak laser drive ($g\ll\kappa$), it is
permissible to evaluate $\delta\Omega_{{\rm m}}(\omega)$ and $\Gamma_{{\rm opt}}(\omega)$
at the original, unperturbed oscillation frequency $\omega=\Omega_{{\rm m}}$.
Then, we just have a shifted and broadened mechanical resonance. This
picture also explains why we need the assumption $\kappa\gg\Gamma_{{\rm eff}}$
for this approach to hold. A high-Q mechanical oscillator samples
the optical density of states at $\omega=\pm\Omega_{m}$, with a small
frequency linewidth $\Gamma_{{\rm eff}}$ that can be neglected as
long as $\Gamma_{{\rm eff}}\ll\kappa$. We will now discuss both quantities,
the frequency shift and the damping rate, under this assumption.

\begin{figure}[ptb]
 \centering\includegraphics[width=1\columnwidth]{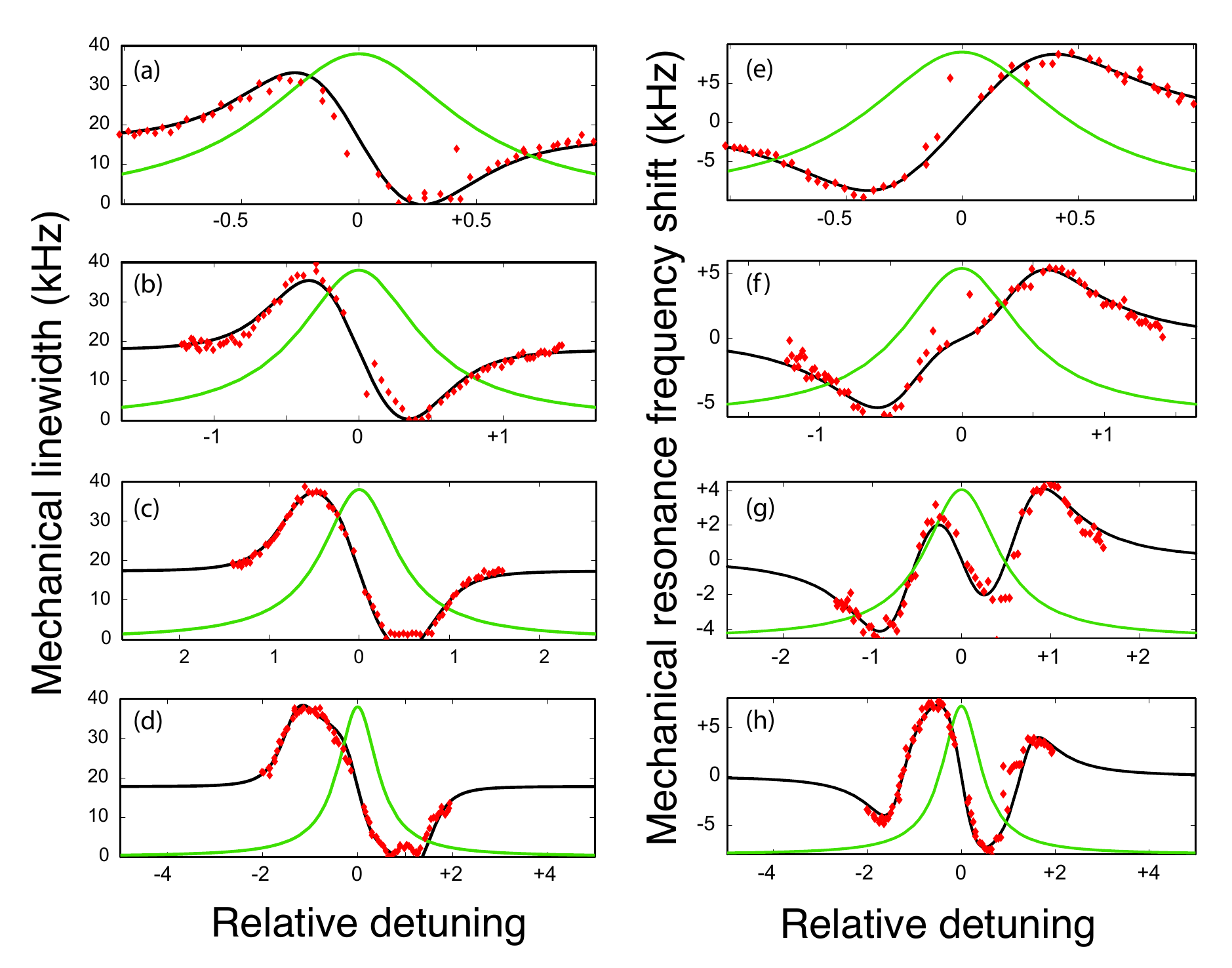}\caption{\label{fig:OptomechanicalDampingRateAndOpticalSpring}Optomechanical
damping rate and optical spring effect vs. detuning and different
sideband parameters, with theory (black lines) and experimental data
(red dots). Left: Full mechanical damping rate $\Gamma_{{\rm eff}}/2\pi$
vs. detuning $\Delta/\kappa$, with decreasing values of $\kappa/\Omega_{{\rm m}}$
from top to bottom ($\kappa/\Omega_{{\rm m}}=3.7,\,2.2,\,1.4,\,0.7$).
The regime where $\Gamma_{{\rm eff}}$ touches zero is the regime
of dynamical instability (mechanical lasing). Right: Optical spring
effect, i.e. light-induced mechanical frequency shift $\delta\Omega_{{\rm m}}/2\pi$.
Data from \cite{Schliesser2010}.}
\end{figure}

\begin{figure}[ptb]
 \centering\includegraphics[width=1\columnwidth]{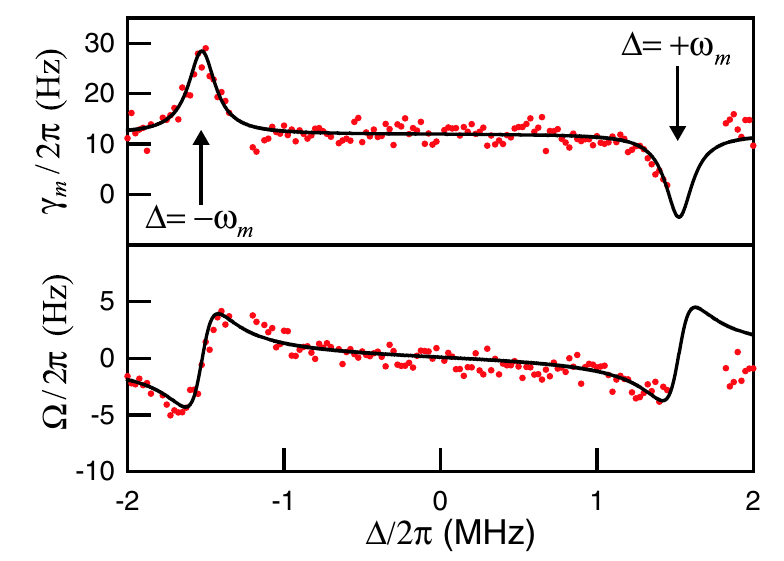}\caption{\label{fig:OptomechanicalDampingRateAndOpticalSpring-1}Optomechanical
cooling rate and frequency shift in the resolved sideband regime,
shown here for $\kappa/\Omega_{m}\ll1$. (Data courtesy J.~Teufel
and K.~Lehnert, from \cite{Teufel2008})}
\end{figure}

\subsubsection{Optical spring effect}

\label{sub:DynamicalBackactionOpticalSpring}

We find, with $\omega=\Omega_{{\rm m}}$, for the frequency shift
of the oscillator induced by the light field: 
\[
\delta\Omega_{m}=g_{0}^{2}\bar{n}_{{\rm cav}}\left(\frac{\Delta-\Omega_{\text{m}}}{\kappa^{2}/4+(\Delta-\Omega_{\text{m}})^{2}}+\frac{\Delta+\Omega_{\text{m}}}{\kappa^{2}/4+(\Delta+\Omega_{\text{m}})^{2}}\right)
\]
In the limit of large cavity decay rate (i.e. the Doppler regime,
$\kappa\gg\Omega_{{\rm m}}$), this formula yields: 
\[
\delta\Omega_{m}(\Delta)|_{\kappa\gg\Omega_{M}}=g^{2}\frac{2\Delta}{\kappa^{2}/4+\Delta^{2}}
\]
This implies that the mechanical oscillator will be spring-softenened
for a red-detuned laser beam ($\Delta<0$), and spring-hardened for
a blue-detuned laser ($\Delta>0$). 

Note that the frequency shift takes a markedly different form when
entering the resolved sideband regime. Here, the optical spring effect
vanishes at certain detunings and the radiation pressure contributes
only to cooling or amplification.

\subsubsection{Optomechanical damping rate}

\label{sub:DynamicalBackactionOptomechanicalDampingRate}

\label{sec:V.B.2}

Using the same approximation as for the optical spring effect, the
optomechanical damping rate is given by the expression 
\begin{align}
\Gamma_{\text{opt}} & =\bar{n}_{cav}g_{0}{}^{2}\left(\frac{\kappa}{\kappa^{2}/4+(\Delta+\Omega_{m})^{2}}-\frac{\kappa}{\kappa^{2}/4+(\Delta-\Omega_{m})^{2}}\right)\label{eq:CoolingRateFullExpression}
\end{align}
This yields the full effective mechanical damping rate: 
\[
\Gamma_{{\rm eff}}=\Gamma_{\text{m}}+\Gamma_{\text{opt}}
\]
Since $\Gamma_{{\rm opt}}$ can be both positive and negative, it
can either increase or decrease the mechanical damping rate, i.e.
cause extra damping or anti-damping. Extra damping leads to cooling
(Sec.~\ref{sec:PrinciplesOfOptomechanicalCoupling}), while anti-damping
can lead to amplification of thermal fluctuations and finally to an
instability if the full damping rate becomes negative, $\Gamma_{{\rm eff}}<0$
(see Sec.~\ref{sec:NonlinearDynamics}). 

This behavior can also be observed experimentally: Figure \ref{fig:OptomechanicalDampingRateAndOpticalSpring}
shows the damping rate and optically induced frequency shift for different
ratios of $\Omega_{m}/\kappa$. 

The physical origin of the optomechanical damping rate can be described
in several ways. 

\begin{figure}
\includegraphics[width=1\columnwidth]{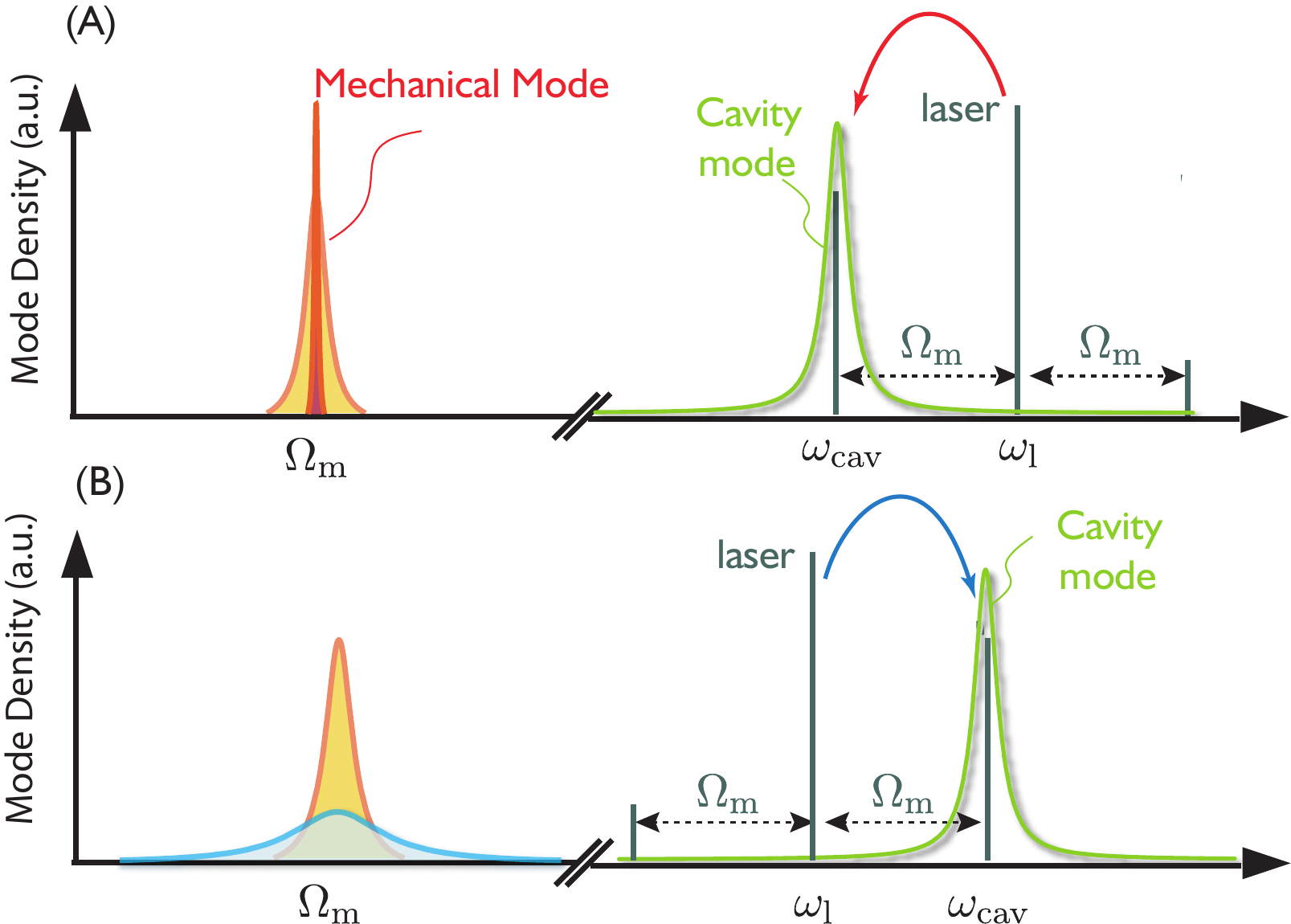}

\caption{Scattering picture of cooling and amplication. (a) Amplification and
heating proceeds by suppressing anti-Stokes scattering and enhancing
Stokes scattering, via the cavity density of states. (b) Cooling proceeds
vice versa, by suppressing the Stokes process and enhancing anti-Stokes
scattering.}
\end{figure}

\textit{Mechanical picture}. Cooling and heating can be understood
by the following mechanical consideration. As the mechanical oscillator
is performing its harmonic motion, it traces a trajectory in the diagram
of radiation pressure force vs displacement, as shown in Fig.~ \ref{fig:DynamicalBackactionWorkDoneDuringCycle}.
In the limit where $\kappa\gg\Omega_{m}$, the intracavity field (and
the associated radiation pressure force) respond instantaneously to
the oscillator motion. If, however, we still take into account the
remaining cavity retardation, this leads to the radiation pressure
force getting out of phase with the mechanical motion. We then can
split the force $F(t)=\bar{F}+\delta F_{I}(t)+\delta F_{Q}(t)$ into
a component $\delta F_{I}(t)$ in-phase with the motion $x(t)$ (responsible
for the optical spring effect) and an out-of-phase quadrature term
$\delta F_{Q}(t)$ (responsible for cooling or heating). In the diagram
this implies that the oscillator motion traces out no longer a line
but an area. The sense in which this area is encompassed gives the
direction of energy flow, i.e cooling or amplification. 

\textit{Scattering picture}. Writing the optomechanical dynamical
backaction in the above fashion allows to gain physical insight into
the origin of cooling and amplification. The expression consists of
two terms which are essentially the cavity buildup factor $\frac{\kappa}{\Delta^{2}+\kappa^{2}/4}$
evaluated at the frequencies $\Delta=\omega+\Omega_{m}$ and $\Delta=\omega-\Omega_{m}$.
These terms describe the strength of the motional sidebands of the
intracavity field, generated due to the cavity frequency oscillating
because of the motion $x(t)=x_{0}\cdot\sin(\Omega_{m}t)$ of the mechanical
oscillator. Perturbation theory analysis of the classical coupled
mode equations reveals that the intracavity field consists of sidebands
\cite{Kippenberg2007}:

\begin{eqnarray}
a(t) & \approx & a_{0}(t)+a_{1}(t)
\end{eqnarray}
where $a_{0}(t)=s\cdot e^{-i\omega_{L}t}\sqrt{\eta\kappa}/(-i\Delta+\kappa/2)$
is the unperturbed field and $a_{1}(t)$ contains the Stokes $a_{s}(t)$
and anti-Stokes $a_{as}(t)$ sidebands: 

\begin{equation}
a_{1}(t)=\frac{g_{0}x_{0}}{2}a_{0}(t)\cdot(\frac{e^{-i\Omega_{m}t}}{-i(\Delta+\Omega_{m})+\kappa/2}+\frac{e^{i\Omega_{m}t}}{-i(\Delta-\Omega_{m})+\kappa/2})
\end{equation}
These two sidebands become asymmetric for nonzero laser detuning due
to the cavity density of states. Applying energy conservation implies
that the mode-density induced sideband asymmetry extracts or adds
power to the mechanical oscillator by shifting the frequency of the
pump photons %
\footnote{The difference power in the motional sidebands has for this reason
to be multiplied with the pre-factor $\frac{\Omega_{m}}{\omega_{L}}$. %
}, i.e. $P_{L}=\frac{\Omega_{m}}{\omega_{L}}\kappa\cdot1/T_{rt}(|a_{s}|^{2}-|a_{as}|^{2})=-\Gamma_{\mathrm{opt}}\cdot m_{eff}\Omega_{m}^{2}x_{0}^{2}$,
which yields a cooling rate identical to the expression derived above
($T_{rt}$ denoting the cavity round trip time). Consequently cooling
and amplification can be viewed to originate from the imbalance of
Stokes and anti-Stokes scattering. 

\begin{figure}[ptb]
 \centering\includegraphics[width=1\columnwidth]{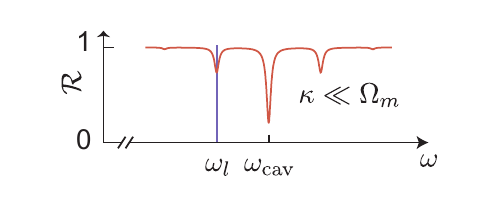}\caption{\label{fig:TransmissionSpectrumOscillatingCavity}Transmission spectrum
of an oscillating cavity in the resolved sideband regime. The mechanical
motion leads to the appearance of a series of sidebands spaced by
multiples of the mechanical frequency, $n\Omega_{m}$, where $n=0,\pm,1,\pm2...$.
In optomechanical experiments only the first pair of sidebands is
relevant. Optomechanical cooling proceeds by pumping the lower sideband,
which leads to cooling of the mechanical oscillator.}
\end{figure}

For the case of resolved sidebands ($\Omega_{m}\gg\kappa$) the cavity
absorption spectrum $1-\mathcal{R}(\Delta)$ develops a series of
sidebands due to the mechanical oscillator's motion, similar to the
absorption spectrum of an oscillating ion \cite{Stenholm1986} (Fig.~\ref{fig:TransmissionSpectrumOscillatingCavity}): 

\begin{equation}
\mathcal{R}(\Delta)\approx1-\kappa^{2}\sum_{n}\frac{J_{n}(\beta)^{2}}{(\Delta+n\Omega_{m})^{2}+\kappa^{2}/4}
\end{equation}
Here $\beta=\frac{G}{\Omega_{m}}x_{0}$ denotes the modulation index.
For the simple case of a weak coherent oscillation (with amplitude
$x_{0}$) one obtains to lowest order two sidebands only, i.e. $J_{\pm1}(\beta)^{2}=\beta^{2}$.
The lower and upper motional sidebands that appear in the spectrum
are, in analogy to trapped-ion physics, related to motional increasing
and motional decreasing scattering processes\cite{Wineland1979a}%
\footnote{From a quantum mechanical picture each photon that is absorbed by
the lower sideband removes one quanta of energy from the mechanical
oscillator ($\hbar\Omega_{m}$). The rate at which photons enter the
cavity via the lower sideband is given by $P\beta^{2}/\hbar\omega_{L}$.
Consequently the rate at which energy is removed from the oscillator
is: $\frac{dE}{dt}=-\frac{P}{\hbar\omega_{L}}\beta^{2}\hbar\Omega_{m}=-\Gamma_{opt}E$,
yielding again the expression $\Gamma_{opt}=(g_{0}^{2}/\Omega_{m}^{2})\frac{P}{\hbar\omega}=4\bar{n_{p}}g_{0}^{2}/\kappa$
in agreement with the classical calculation.%
}.

\textit{Feedback picture.} Finally, one can also understand the cooling
by considering a feedback picture. In this picture the mechanical
oscillator motion modulates the cavity field, the latter gives rise
to a radiation pressure force, which in turn acts back onto the mechanical
oscillator. Cooling arises again from the phase relationship (retardation)
between the mechanical motion and the radiation pressure force. Note
that the feedback is not entirely noiseless: quantum noise adds to
the intracavity radiation pressure force, giving rise to a quantum
limit of cooling treated in section \ref{sub:CoolingQuantumTheory}.

It is instructive to consider several limiting cases of the cooling
rate expression. 

\textit{Resolved-sideband regime}.\textendash{} First, in the limit
where the mechanical frequency is much larger than the cavity decay
rate ($\Omega_{m}\gg\kappa$) the cooling rate exhibits pronounced
maxima and minima near the lower and upper sideband ($\Delta=\pm\Omega_{m}$).
The maximum cooling rate is attained on the lower sideband ($\Delta=-\Omega_{m}$):

\begin{equation}
\Gamma_{\text{opt}}|_{\kappa\ll\Omega_{M}}=4\bar{n}_{cav}\frac{g_{0}^{2}}{\kappa}=\frac{4g^{2}}{\kappa}
\end{equation}

The cooling rate in the resolved sideband regime can also be expressed
in a different way, as detailed in \cite{Schliesser2008}. In the
resolved sideband regime (considering an overcoupled single-sided
cavity for simplicity), the relation between intracavity photon number
and input power is given by $\bar{n}_{cav}=(P/\hbar\omega)\cdot\kappa/\Delta^{2}$,
thus:

\begin{equation}
\Gamma_{\text{opt}}|_{\kappa\ll\Omega_{M}}=\frac{P}{\hbar\omega_{L}}\frac{g_{0}^{2}}{\Omega_{m}^{2}}
\end{equation}
 As one can see, the cooling rate in the sideband limit does not
depend on the cavity linewidth. 

\emph{Doppler regime}. \textendash{} In the unresolved sideband regime
(Doppler case, i.e. $\kappa\gg\Omega_{m}$), the maximum cooling or
amplification rate is reached for a detuning equal to $\kappa/2$:

\begin{eqnarray}
\Gamma_{\text{opt}}(\Delta) & = & g^{2}\Omega_{\text{m}}\frac{-4\Delta\kappa}{(\kappa^{2}/4+\Delta^{2})^{2}}\\
\Gamma_{\text{opt}}(\Delta=-\frac{\kappa}{2}) & = & 8\left(\frac{g}{\kappa}\right)^{2}\Omega_{{\rm m}}
\end{eqnarray}

Note that in this case the mechanical cooling and amplification rate
exhibit a strong dependence on the inverse (cubic) cavity decay rate.
This shows that the cooling rate strongly diminishes in the Doppler
regime.

Note that sub-Doppler cooling in the unresolved sideband regime can
still be achieved by using pulsed optical pumping schemes \cite{Machnes2012,Jacobs2011,2011_Aspelmeyer_PulsedOptomechanics,Vanner2012}.

\section{Quantum optical measurements of mechanical motion}

\label{sec:QuantumOpticalMeasurementsOfMechanicalMotion}

One of the principal advantages of cavity optomechanical systems is
the built-in readout of mechanical motion via the light field transmitted
through (or reflected from) the cavity. In the following, we will
discuss several variants of optical measurement schemes. We first
address the measurement of position, where we will find that quantum
mechanics places fundamental restrictions on the overall precision
in the regime typically employed in experiments (i.e. weak measurements).
Then we discuss alternative schemes, where there are no such limitations.
Some supply a measurement of a selected quadrature of mechanical motion,
i.e. the amplitude of the $\cos(\Omega_{m}t)$ or $\sin(\Omega_{m}t)$
contribution to $x(t)$. Other schemes measure the discrete phonon
number. Both approaches are examples of so-called quantum non-demolition
(QND) measurements. We close the section by pointing out some experimental
issues in phase measurements, and by discussing feedback cooling based
on the possibility of precise read-out.

\subsection{Parametric displacement sensing and the standard quantum limit (SQL)}

\label{sub:MeasurementsDisplacementSensingAndSQL}

Measuring the displacement of a mechanical resonator via the transmission
or reflection phase shift typically involves integrating the data
over a long time, in order to suppress the noise. It is therefore
an example of a weak, continuous measurement. It turns out that such
a measurement applied to the coordinate of a harmonic oscillator cannot
be more precise than what is known as the {}``standard quantum limit''.
In the following we give a brief qualitative and quantitative discussion.
More about this topic can be found in \cite{Caves1980b,Caves1981,Caves1982a,Braginsky1985a,Braginsky1995,Clerk2008a}.

\subsubsection{Introduction and qualitative discussion}

The optical cavity represents an interferometer and thus allows a
direct measurement of the mechanical position via the phase shift
of transmitted or reflected light. In practice this requires either
a homodyne or heterodyne detector, in which the signal is brought
into interference with a local oscillator that serves as a phase reference.
The optomechanical cavity frequency pull $Gx$ is converted into a
phase shift $\theta\propto Gx/\kappa$ imparted on the photons during
their lifetime $1/\kappa$ inside the cavity (assuming slow motion,
$\Omega_{{\rm m}}\ll\kappa$). If one tries to measure this phase
shift using $N$ photons passing through the cavity (and interfering
with a reference beam to read out the phase later), then the fundamental
uncertainty relation between number and phase yields a shot-noise
limited imprecision of $\delta\theta=1/\sqrt{N}$. It seems that this
would allow for an arbitrarily precise readout, provided one uses
a sufficiently large number of photons. Indeed, this would be true
for an instantaneous readout with a very intense flash of light.

However, in many experiments one rather performs a weak measurement:
The noisy signal $x(t)$ determined from the phase measurement is
effectively integrated over many oscillation periods to average away
the noise and get a sufficient signal-to-noise ratio. That this will
pose a problem can be seen from the general quantum-mechanical uncertainty
principle which states that it is impossible to follow the trajectory
$x(t)$ of a particle with arbitrary precision (or, to know both position
and momentum at the same time). It is instructive to see qualitatively
how that limitation is enforced in our case. The fluctuating radiation
pressure force (again, due to the photon shot noise) imprints an unavoidable
jitter on the mechanical motion. Each of the photons imparts a random
kick, and their overall effect on the momentum and position will grow
like $\sqrt{N}$, as in a random walk. That effect is called {}``backaction
noise'' and counteracts the increase of phase readout precision at
large $N$. Thus, there is an optimum at intermediate photon numbers,
where the sum of the two effects is minimal. 

The quantitative analysis outlined below will be phrased in terms
of noise spectra, describing the imprecision and backaction noise
contributions to the overall measurement error. In that context, the
appropriate question to ask is how large the error is given a certain
measurement time (which sets the bandwidth over which the spectra
are to be integrated). In thermal equilibrium, the mechanical oscillator's
phase and amplitude will fluctuate on the scale of the damping time
$1/\Gamma_{{\rm m}}$. Thus, this is the longest reasonable measurement
time at our disposal. The outcome of the analysis will be that one
can determine the trajectory (or rather its two quadratures) up to
a precision given by the oscillator's zero-point fluctuations $x_{{\rm ZPF}}=(\hbar/2m_{{\rm eff}}\Omega_{{\rm m}})^{-1/2}$
during a time $1/\Gamma_{{\rm m}}$. This statement (see Eq.~(\ref{eq:SQLformula}))
is known as the {}``standard quantum limit'' of displacement detection
in a weak measurement. It is independent of whether the oscillator
is in its ground state or at high temperatures.

\begin{figure}
\includegraphics[width=0.8\columnwidth]{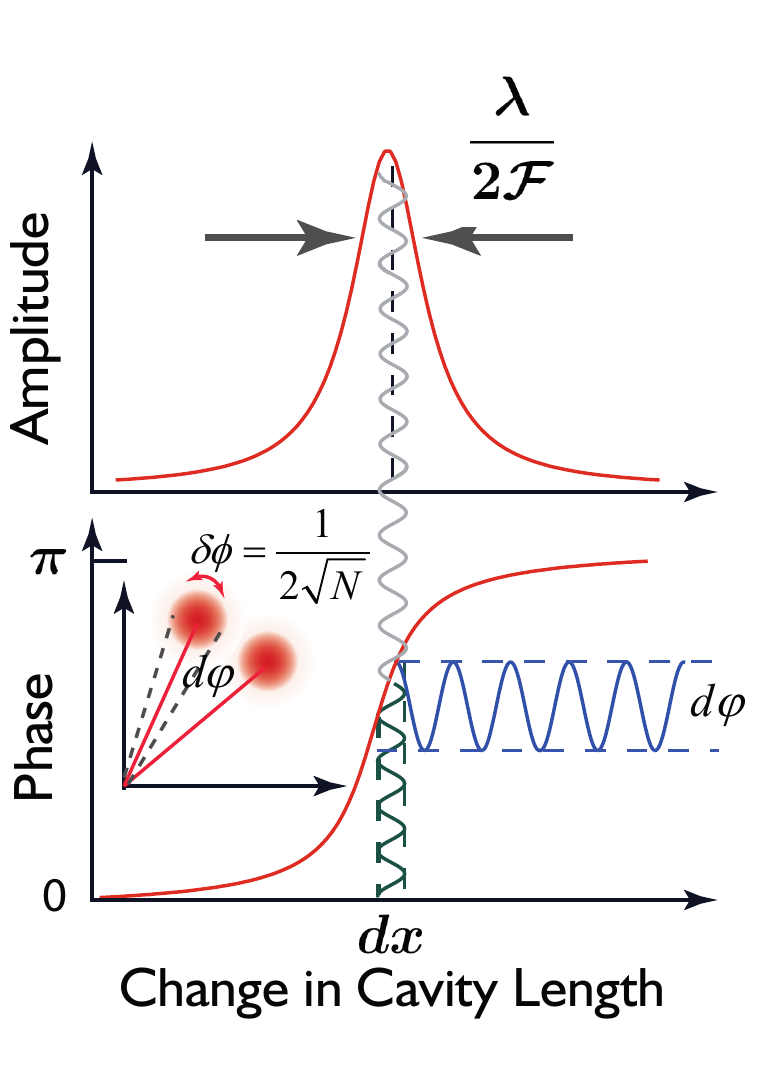}

\caption{\label{fig:PhaseTransduction}Optomechanical systems transduce displacements
into changes of the optically transmitted (or reflected) phase. Upper
panel: amplitude response. Lower panel: phase response.}
\end{figure}

\begin{figure}[ptb]
 \centering\includegraphics[width=1\columnwidth]{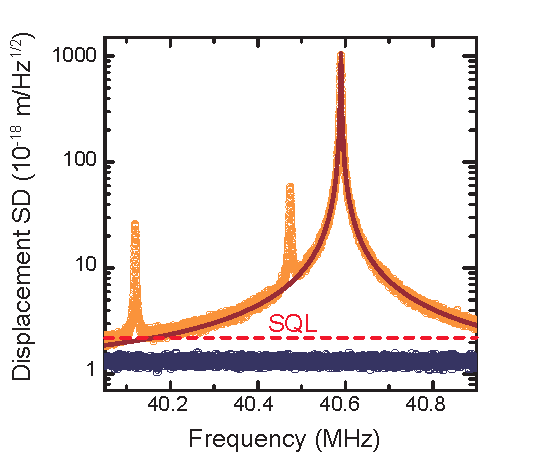}\caption{\label{fig:ExampleMechanicalFrequencySpectrum}Mechanical frequency
spectrum for an example of an optomechanical system (data from \cite{Schliesser2008a}).
The actual imprecision noise floor (dark data points at bottom) is
indicated along with the full noise at the standard quantum limit.
Similar data demonstrating imprecision below that at the SQL exists
for nanomechanical oscillators using a microwave cavity interferometer
\cite{Teufel2009} or nanomechanical oscillators coupled to the near
field of an optical microresonator.}
\end{figure}

\subsubsection{The standard quantum limit}

Let us first state more formally why there must be a standard quantum
limit. The oscillator's trajectory can be decomposed into quadratures:

\begin{equation}
\hat{x}(t)=\hat{X}\cos(\Omega_{{\rm m}}t)+\hat{Y}\sin(\Omega_{{\rm m}}t)\,,
\end{equation}
where $\hat{X}$ and $\hat{Y}$ remain constant during intervals smaller
than the damping time. Since $\hat{X}=\hat{x}(0)$ and $\hat{Y}=\hat{p}(0)/m_{{\rm eff}}\Omega_{{\rm m}}$,
the commutator $\left[\hat{x},\hat{p}\right]=i\hbar$ carries over
to $\hat{X}$ and $\hat{Y}$, yielding $\left[\hat{X},\hat{Y}\right]=i\hbar/m_{{\rm eff}}\Omega_{{\rm m}}$.
Heisenberg's uncertainty relation thus reads

\begin{equation}
\Delta X\cdot\Delta Y\geq\frac{\hbar}{2m_{{\rm eff}}\Omega_{{\rm m}}}=x_{{\rm ZPF}}^{2}\,.
\end{equation}
, and any simultaneous measurement that tries to measure both quadratures
with equal precision is limited to $\Delta X=\Delta Y=x_{{\rm ZPF}}$.%
\footnote{Here $\Delta X=\sqrt{\left\langle (\hat{X}-\langle\hat{X}\rangle)^{2}\right\rangle }$. %
}

Here we only discuss the most important results. For a much more extended
recent discussion of the quantum limits to weak measurements we refer
the reader to \cite{Clerk2008a}. We will consider a single-sided
optical cavity driven at laser detuning $\Delta=0$ (the optimal case),
where the number of photons circulating inside the cavity is $\bar{n}_{{\rm cav}}=4P/(\kappa\hbar\omega_{{\rm cav}})$.
All noise spectral densities will be symmetrized in frequency (indicated
by $\bar{S}$).

The noise in the optical phase readout $\delta\theta\sim1/\sqrt{N}$
induces an imprecision in the $x$ measurement, where $\delta x^{{\rm imp}}\sim\kappa\delta\theta/G=\kappa/(G\sqrt{N})$,
as discussed above. Inserting the photon number $N=t\dot{N}\sim t\kappa\bar{n}_{{\rm cav}}$,
we find $(\delta x^{{\rm imp}})^{2}\sim(\kappa/\bar{n}_{{\rm cav}}G^{2})t^{-1}$.
This can be understood as the integral of a flat noise spectral density
$\bar{S}_{xx}^{{\rm imp}}\sim\kappa/(\bar{n}_{{\rm cav}}G^{2})$ over
the bandwidth $t^{-1}$ set by the measurement time $t$. The complete
expression for the quantum noise limited imprecision noise spectral
density reads:

\begin{equation}
\bar{S}_{\text{xx}}^{{\rm imp}}(\omega)=\frac{\kappa}{16\bar{n}_{cav}G^{2}}\left(1+4\frac{\omega^{2}}{\kappa^{2}}\right),
\end{equation}
where we also kept the growth of the noise at higher values of $\omega/\kappa$.
This is due to the fact that the cavity is a low-pass filter, suppressing
the contribution of motional frequencies $\omega>\kappa$ to the phase
output. On the other hand, the phase noise itself is independent of
$\omega$, so referring it back to the input leads to larger imprecision
at higher $\omega$.

At the same time, the backaction noise force has the following spectral
density (cf. Sec.~\ref{sub:CoolingQuantumTheory}): 
\begin{align}
\bar{S}_{\text{FF}}(\omega) & =\bar{n}_{cav}\frac{4\hbar^{2}G^{2}}{\kappa}\left(1+4\frac{\omega^{2}}{\kappa^{2}}\right)^{-1}\,
\end{align}
In general, the product of imprecision noise and backaction force
noise densities fulfills a fundamental inequality, see e.g. \cite{Clerk2008a},
a variant of the Heisenberg uncertainty relation:
\begin{equation}
\bar{S}_{\text{xx}}^{{\rm imp}}(\omega)\cdot\bar{S}_{\text{FF}}(\omega)\geq\frac{\hbar^{2}}{4}\,.\label{eq:SxxSffProduct}
\end{equation}
In our particular case, we see that the equality {}``$=$'' is realized,
i.e. the cavity displacement detector is as good as allowed by quantum
mechanics. 

The total noise registered at the detector, expressed in terms of
$x$ ({}``referred back to the input''), reads:
\begin{equation}
\bar{S}_{\text{xx}}^{\text{total}}(\omega)=\bar{S}_{\text{xx}}^{\text{th}}(\omega)+\bar{S}_{\text{xx}}^{\text{imp}}(\omega)+\bar{S}_{\text{FF}}(\omega)\left|\chi_{xx}(\omega)\right|{}^{2}\,.
\end{equation}
Here we have added the intrinsic thermal fluctuation spectrum, the
imprecision noise, and the effect of the backaction noise force on
the displacement, calculated via the mechanical susceptibility $\chi_{xx}$
(see Sec.~\ref{sub:MechanicalResonatorsNoiseSpectra}). In doing
so we have assumed a situation where there are no cross-correlations
between the force noise and the imprecision noise. See \cite{Clerk2008a}
for a more complete discussion including the general case. In the
following, we will denote the sum of the imprecision and backaction
noises as the total added noise, $\bar{S}_{xx}^{{\rm add}}$. Inserting
the relation (\ref{eq:SxxSffProduct}), and treating $\bar{S}_{FF}$
as variable (e.g. by tuning $\bar{n}_{{\rm cav}}$), we can minimize
$\bar{S}_{xx}^{{\rm add}}$. The minimum (at any given, fixed frequency)
is reached at $\bar{S}_{FF}=\hbar/(2|\chi_{xx}|)$, and this yields
$\bar{S}_{xx}^{{\rm add}}(\omega)\geq\hbar|\chi_{xx}(\omega)|\geq\hbar\left|{\rm Im}\chi_{xx}(\omega)\right|$.
By using the quantum FDT at $T=0$ (see Eq.~(\ref{eq:FDT})), we
can introduce the spectral density of mechanical zero-point fluctuations,
$\bar{S}_{xx}^{{\rm ZPF}}(\omega)=\hbar\left|{\rm Im}\chi_{xx}(\omega)\right|$.
We arrive at the fundamental inequality

\begin{equation}
\bar{S}_{xx}^{{\rm add}}(\omega)\geq\bar{S}_{xx}^{{\rm ZPF}}(\omega)\,.\label{eq:SQLformula}
\end{equation}
This is the standard quantum limit (SQL) of weak displacement detection.
The measurement adds at least the zero-point noise, on top of the
intrinsic fluctuations. Roughly speaking, the effect on the noise
looks as if the oscillator's energy were increased by $\hbar\Omega_{{\rm m}}/2$,
i.e. half a phonon. However, only the backaction contribution really
corresponds to a physical increase of the oscillator's effective temperature,
and at the SQL this contribution is half of the overall effect, i.e.
$\hbar\Omega_{{\rm m}}/4$, the other half being provided through
the imprecision noise.

If we measure at the mechanical resonance, $\omega=\Omega_{{\rm m}}$,
then the added noise of the cavity displacement detector is $\bar{S}_{xx}^{{\rm add}}(\Omega_{{\rm m}})=\bar{S}_{xx}^{{\rm ZPF}}(\Omega_{{\rm m}})=\hbar/(m_{{\rm eff}}\Omega_{{\rm m}}\Gamma_{{\rm m}})$.
This corresponds to the zero-point fluctuations, in a measurement
time $t\sim\Gamma_{{\rm m}}^{-1}$, just as stated in the introduction.
Obviously, if one is limited by laser power, it is better to have
a high-quality oscillator (small $\Gamma_{{\rm m}}$), which boosts
$\bar{S}_{xx}^{{\rm ZPF}}(\Omega_{{\rm m}})$ and makes it easier
to reach that imprecision level. The power to reach the standard quantum
limit therefore is a natural expression to characterize a transducer.
It is given by the expression:

\begin{equation}
P_{SQL}=\Gamma_{m}\cdot\hbar\omega\frac{\kappa^{2}}{64g_{0}^{2}}\left(1+4\frac{\omega^{2}}{\kappa^{2}}\right)
\end{equation}

\begin{figure}[ptb]
\includegraphics[width=3.2in]{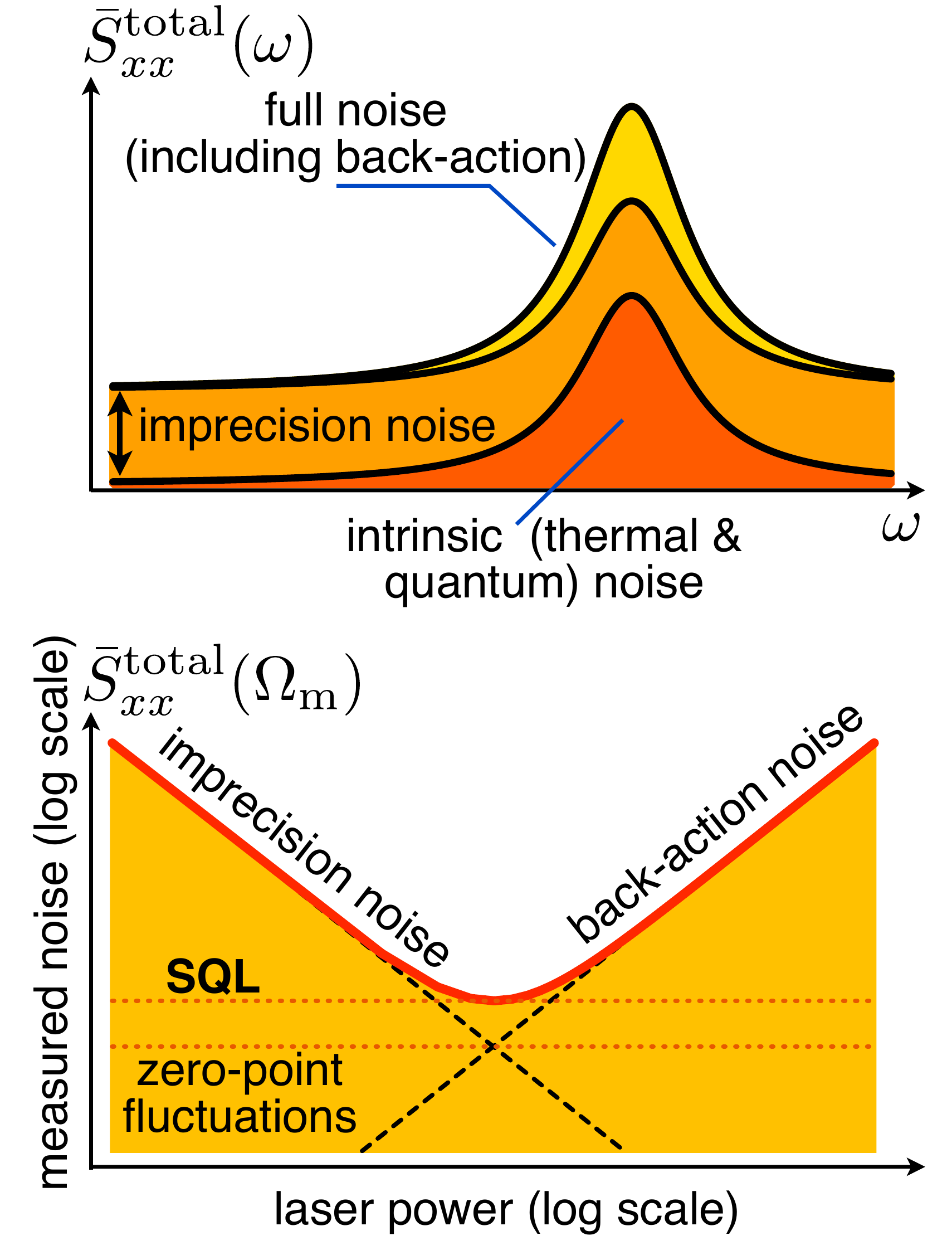}

\caption{\label{fig:SQLSpectrumAndImprecisionPlot}The full measured noise
spectrum contains contributions from the intrinsic fluctuations of
the mechanical oscillator, but also extra noise due to imprecision
in the measurement (typically flat in frequency) and noise due to
the backaction heating of the oscillator. Bottom: The value of the
measured noise spectral density evaluated at the mechanical resonance,
plotted as a function of the power of the measurement beam. At lower
powers, imprecision noise dominates (few photons yield bad phase resolution),
while at higher powers the backaction noise represents the most important
contribution. The standard quantum limit minimal noise is reached
at intermediate powers.}
\end{figure}

Both the imprecision noise and the backaction noise are shown as a
function of laser power (or optomechanical coupling) in Fig.~\ref{fig:SQLSpectrumAndImprecisionPlot}.
When referring to {}``precision beyond the standard quantum limit''
in this context, one wants to emphasize that one can make the imprecision
noise alone lower than the SQL (which implies the backaction noise
is already appreciable). This situation has been achieved in optomechanical
systems for mechanical oscillators of nanoscale \cite{Teufel2009,Anetsberger2009a}
and microscale dimensions \cite{Westphal2012,Schliesser2008a}.

The observed thermal noise at any large temperature can also be used
to obtain the value of the standard quantum limit via the relation

\[
\bar{S}_{xx}^{{\rm th}}(\omega)/\bar{S}_{xx}^{{\rm ZPF}}(\omega)=2\bar{n}_{{\rm th}}(\omega)=2k_{B}T/\hbar\omega
\]
This is a useful expression, since it is independent of the calibration
of the $x$ measurement.

\begin{figure}
\includegraphics[width=1\columnwidth]{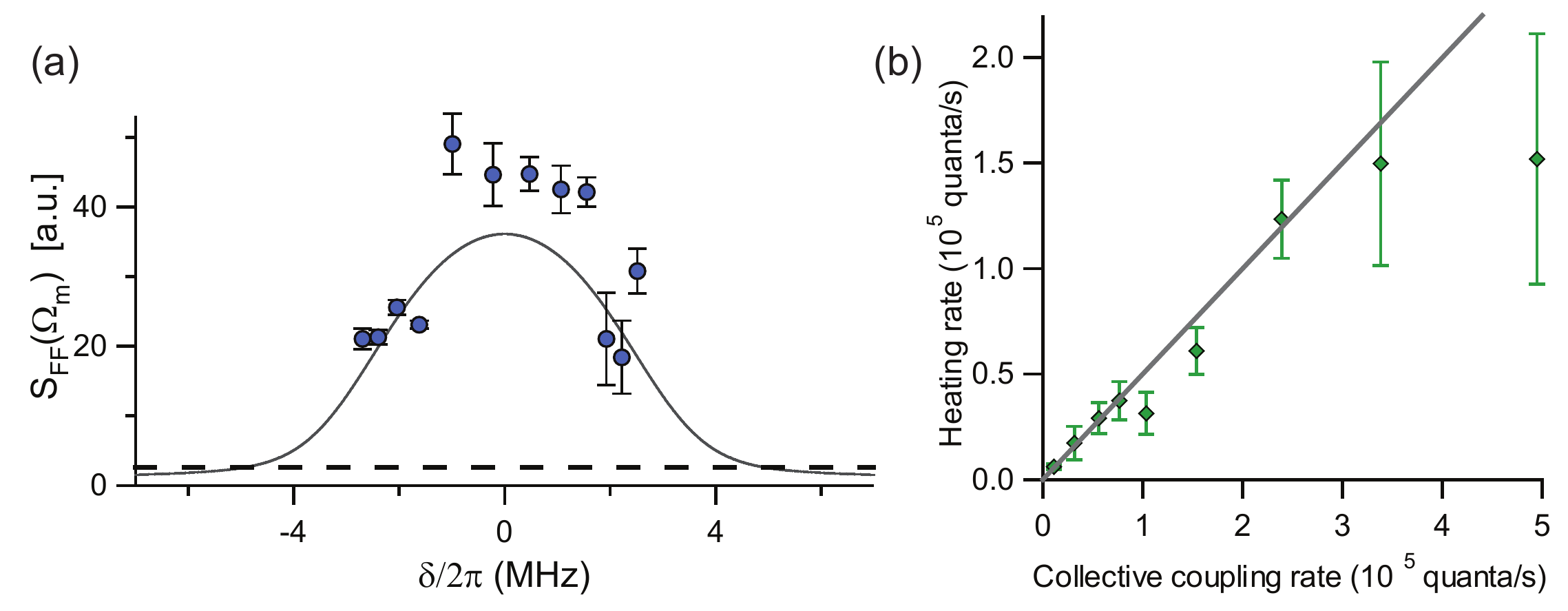}

\caption{\label{Fig:StamperKurnShotNoise}Observation of quantum radiation
pressure force fluctuations through the energy transferred to a near-ground-state
mechanical oscillator. Experiments were performed with an ultracold
atomic gas serving as the mechanical element within a Fabry-Perot
optical cavity. In (a), the energy transferred to the gas was quantified
via the rate at which atoms were ejected from a finite-depth optical
trap. The force fluctuation spectral density at the mechanical oscillation
frequency, $S_{{\rm FF}}(\Omega_{{\rm m}})$, is thereby obtained
at different detunings $\Delta$ between the cavity probe and resonance
frequencies. In (b), from the power difference between the red and
blue motional sidebands observed in the emission of a resonantly driven
optical cavity, one obtains the heat flux into the mechanical system
via the cavity probe. The observed heating, given in units of mechanical
energy quanta per second, matches well to that predicted for intracavity
shot noise from a coherent optical field (gray line). From (a) \cite{Murch2008}
and (b) \cite{2011_StamperKurn_SideBandThermometry} {[}Courtesy of
D. Stamper-Kurn{]}}
\end{figure}

In the context of measurements at the SQL, an important step for optomechanical
experiments is to observe the effects of radiation-pressure shot noise
on the mechanical oscillator. This has now been achieved already in
cold atom setups \cite{Murch2008,2011_StamperKurn_SqueezingViaOptomechanics},
which are conducted routinely at low temperatures, and where the particularly
low effective mass of the atomic cloud leads to a very strong single-photon
coupling rate $g_{0}$. This is of advantage, since the ratio of quantum
backaction to thermal force noise (at $\Delta=0$) is given by: 
\begin{equation}
\bar{S}_{FF}(\Omega_{m})/\bar{S}_{FF}^{th}(\Omega_{m})=g_{0}\Omega_{m}/(2k_{B}T\Gamma_{{\rm m}})\left(P/\omega_{{\rm cav}}\right)(\Omega^{2}+\kappa^{2}/4)^{-1}.
\end{equation}
These atomic cloud experiments have allowed to access the radiation
pressure shot noise spectrum, e.g. via tracking the heating of the
cloud (see Figure \ref{Fig:StamperKurnShotNoise}).

Current solid-state based devices still exhibit both significantly
smaller ratios $g_{0}/\kappa$ and deleterious effects of thermal
noise, which make the observation of radiation pressure shot noise
effects a challenging task. One possible strategy is to measure the
cross-correlations between a strong beam exerting radiation pressure
force fluctuations and another beam measuring the resulting displacement
fluctuations (\cite{Heidmann1997a}; see also \cite{2010_Borkje_RadiationPressureShotNoiseDiscussion}
for a more recent analysis). This idea has been demonstrated for a
model situation with deliberately introduced classical noise instead
of the quantum shot noise of a laser beam \cite{Verlot2009}. In another
experiment, it was demonstrated how the radiation pressure backaction
can be employed for amplifying an interferometric signal, which can
lead to a sensitivity lower than the SQL \cite{2010_Heidmann_BackActionAmplification}.
Recently, first signatures of a direct observation of radiation pressure
shot noise have been reported \cite{Purdy2012}.

\subsection{Optical QND measurements}

\label{sub:MeasurementsQND}

The weak displacement measurements discussed in the last section effectively
try to measure non-commuting observables simultaneously, namely the
two quadrature components of motion. They are therefore limited fundamentally
by the Heisenberg uncertainty principle. However, it is also possible
to perform measurements of a single observable. This observable can
be measured with arbitrary precision, thereby approaching an idealized
projection measurement. Repeating the measurement before the state
had a chance to decay would reproduce the same outcome. This is because
the system's Hamiltonian commutes with the observable (neglecting
decay). Therefore such measurements are called quantum non-demolition
(QND) \cite{Braginsky1980,Braginsky1996,Braginsky1995}. These have
been successfully realized in the quantum optical domain \cite{Lvovsky2009,Haroche2006,1996_Leibfried_WignerDensity}.

\subsubsection{Single quadrature measurements}

\label{sub:MeasurementsQNDSingleQuadrature}

It is possible to optically measure only one quadrature component
of the mechanical motion to arbitrary precision \cite{Clerk2008}.
This is important, since it can be used for a full reconstruction
of the mechanical quantum state, extracting its Wigner density using
quantum state tomography (see below). In addition, it can serve to
measure the correlation of mechanical quadratures with either the
light field quadratures or those of another mechanical object. The
resulting correlators can then be used to test for entanglement. The
fundamentally limited precision of a standard displacement measurement
would not be sufficient for such tests.

One way of achieving this makes use of a simple property of harmonic
motion: Any force applied at time $t$ does not affect the position
a full period later, at $t+2\pi/\Omega_{m}$. Thus, the unavoidable
perturbation of the momentum connected with a position measurement
does not destroy the precision of {}``stroboscopic'' periodic observations
at times $t+n2\pi/\Omega_{m}$. An equivalent but more practical approach
is to do a displacement measurement with a laser beam (at detuning
$\Delta=0$) whose intensity is modulated at frequency $\Omega_{m}$.
This reads out only one quadrature (say, $\hat{X}_{\varphi}$) in
the decomposition

\begin{equation}
\hat{x}(t)=\hat{X}_{\varphi}\cos(\Omega_{m}t+\varphi)+\hat{Y}_{\varphi}\sin(\Omega_{m}t+\varphi)\,,
\end{equation}
whose phase $\varphi$ is determined by the phase of the laser amplitude
modulation. The backaction noise exclusively affects the other quadrature
\cite{Clerk2008}. All this can be derived in the Hamiltonian formulation
(Sec.~\ref{sub:OptomechanicalCouplingHamiltonianFormulation}). Suppose
the modulated drive yields an intracavity amplitude $\alpha=2\alpha_{0}\cos(\Omega_{{\rm m}}t+\varphi)$,
in a frame rotating at the cavity resonance. Then the standard linearization
gives

\begin{equation}
H_{int}=-2\hbar G\alpha_{0}\cos(\Omega_{{\rm m}}t+\varphi)(\delta\hat{a}+\delta\hat{a}^{\dagger})\hat{x}\approx-\hbar G\alpha_{0}(\delta\hat{a}+\delta\hat{a}^{\dagger})\hat{X}_{\varphi}\,,
\end{equation}
where we have omitted rapidly oscillating terms and adopted the rotating
frame for the mechanical resonator in the second step. This is a QND
Hamiltonian for measuring $\hat{X}_{\varphi}$. Deviations from this
idealized picture are small in the resolved sideband limit, i.e. for
$\kappa/\Omega_{m}\ll1$. As a side-effect, the state conditioned
on the measurement result becomes squeezed, since the variance of
$\hat{X}_{\varphi}$ after the measurement approaches zero. First
experimental steps along these lines have been taken in \cite{Hertzberg2009}.

There is an alternative approach for QND detection of single quadratures:
measuring much faster than the oscillation period of the mechanical
resonator. Such an approach requires very short, intense laser pulses
that essentially implement an instantaneous projection measurement
of displacement. Picking another quadrature to measure then simply
involves performing the measurement at another time, when the phase
of the harmonic oscillations has advanced. This approach has been
proposed and analyzed in \cite{1978_Braginsky_PulseMeasurements,Braginsky1995,2011_Aspelmeyer_PulsedOptomechanics}.
Because of the short pulses, this scheme operates in the non-resolved
sideband regime $\kappa\gg\Omega_{m}$. First experiments in this
direction have recently been reported in \cite{Vanner2012}. It has
also been pointed out that properly controlled pulse sequences can
alter the effective optomomechanical interaction, enabling for example
sub-Doppler cooling rates in the non-resolved sideband regime \cite{2011_Retzker_PulsedLaserCooling,Jacobs2011}.

In both of these approaches, quantum state tomography \cite{Lvovsky2009}
then would work by repeatedly preparing the same mechanical state,
measuring the probability densities of the quadratures at a large
number of phases $\varphi$, and applying the inverse Radon transform
to obtain the Wigner density. 

Reconstruction of the Wigner density of quantum states of vibrational
motion has, for example, successfully been achieved for ions \cite{1996_Leibfried_WignerDensity}.

\begin{figure}[ptb]
 \centering\includegraphics[width=3.2in]{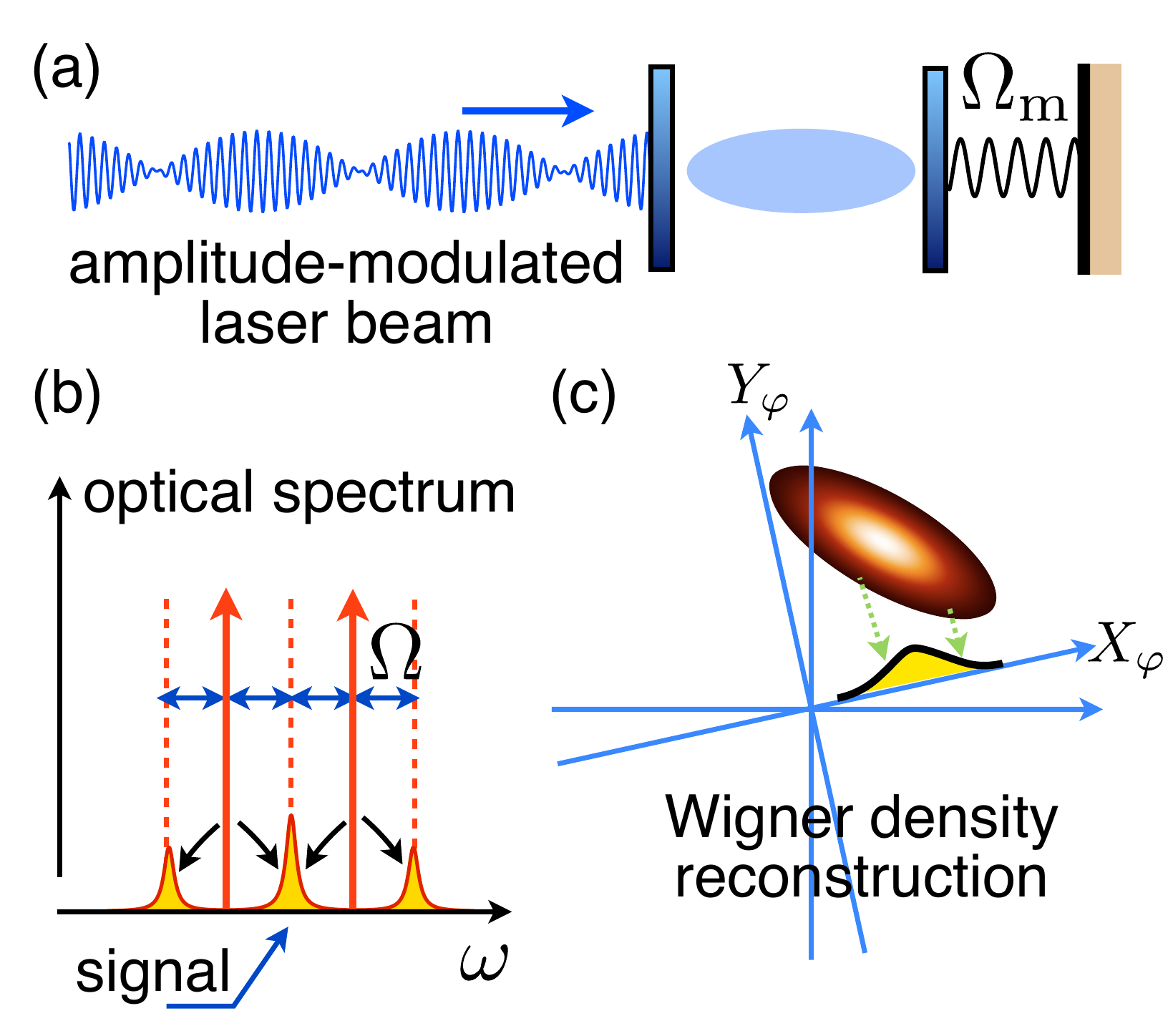}\caption{\label{fig:SingleQuadratureMeasurement}(a) Schematic setup for an
optomechanical single-quadrature measurement, with a laser beam amplitude-modulated
at the mechanical frequency. (b) In optical frequency space, the beam
carries two sidebands, which are then modulated by the motion. (c)
Tomographical reconstruction of the full mechanical Wigner phase space
density is possible based on measurements of the quadratures at different
phases $\varphi$. }
\end{figure}

\subsubsection{Mechanical Fock state detection}

\label{sub:MeasurementsQNDFockState}

Another crucial observable in the mechanical oscillator is the phonon
number $\hat{n}=\hat{b}^{\dagger}\hat{b}$. It is especially important,
since measuring the discrete Fock states $n=0,1,2,\ldots$ is a very
direct proof of the quantum nature of an oscillator. We note that
the first measurements of phonon number in a fabricated, mesoscopic
mechanical oscillator have recently been performed in a nanoelectromechanical
system, exploiting the strong interaction between a piezomechanical
vibration and a superconducting qubit \cite{O'Connell2010}. Here
we discuss a route towards observing quantum jumps between mechanical
Fock states in an optomechanical system \cite{Thompson2008,Jayich2008}.

The idea is that any measurement of $\hat{x}^{2}$ instead of $\hat{x}$
will be closely connected to the oscillator's energy, and thus the
phonon number. This then permits QND detection of the phonon number
in a mechanical resonator (for an early detailed analysis of this
concept in an anharmonic two-mode nanomechanical system, see \cite{Santamore2004}).
In practice, for an optomechanical system this requires changing the
standard setup to another variant where the light field couples to
the square of the displacement. That can be achieved for example by
placing a thin membrane (or any other dielectric object) inside an
optical cavity and positioning it near a node (or antinode) of the
standing light wave that forms one optical mode. In that case, the
light intensity at the object's position, and thereby the object's
effect on the optical resonance frequency, depends quadratically on
the displacement $\hat{x}$. The optomechanical coupling (sometimes
termed {}``dispersive'' in this case) thus reads

\begin{equation}
\hat{H}_{{\rm int}}=\hbar g_{0}^{(2)}(\hat{b}+\hat{b}^{\dagger})^{2}\hat{a}^{\dagger}\hat{a}\,,\label{eq:QuadraticCouplingHamiltonian}
\end{equation}
where $g_{0}^{(2)}=\frac{\partial^{2}\omega_{{\rm cav}}}{\partial x^{2}}x_{{\rm ZPF}}^{2}$.
If the membrane is highly reflecting, an alternative way of deriving
Eq.~(\ref{eq:QuadraticCouplingHamiltonian}) is to focus on two optical
modes (to the left and right of the membrane). Their coupling, provided
by photon tunneling through the membrane, then leads to an avoided
level crossing in the optical spectrum as a function of $x$. At the
degeneracy point, both the upper and lower optical branch frequencies
are extremal \cite{Thompson2008,Sankey2010}, leading to Eq.~(\ref{eq:QuadraticCouplingHamiltonian})
for each of them. Quadratic couplings have recently also been observed
in a cold atom setup \cite{2010_StamperKurn_AtomChipCavityOptomechanics}.

The phase shift observed in reflection from the cavity will measure
$(\hat{b}+\hat{b}^{\dagger})^{2}\propto\hat{x}^{2}$, but averaged
over the photon lifetime $1/\kappa$. If $\kappa\ll\Omega_{m}$, the
time-average of this term reduces to $2\hat{b}^{\dagger}\hat{b}+1$.
Alternatively, one may keep only this contribution in Eq.~(\ref{eq:QuadraticCouplingHamiltonian}),
invoking the rotating wave approximation. Thus, this setup in principle
allows a QND measurement of the phonon number. Cooling the oscillator
to near its ground state and then measuring the phase shift vs. time,
one should be able to observe quantum jumps between mechanical Fock
states, similar to the jumps that have been observed between photon
number states \cite{Guerlin2007}. To this end, the phase shift $\sim g_{0}^{(2)}/\kappa$
induced by a single phonon must be resolved during the lifetime $(\Gamma_{m}\bar{n}_{{\rm th}})^{-1}$
of the ground state, by reflecting a sufficient number of photons
from the cavity. Note that one must suppress any deviations from the
ideal case, such as imperfect placement away from the degeneracy point
\cite{Thompson2008,Jayich2008} or deviations from a perfectly one-sided
cavity. Such deviations lead to extra noise acting on the membrane
that hampers the phonon QND measurement. A careful analysis \cite{2009_Chen_QuantumLimit}
shows that this requires $g_{0}>\kappa_{0}$, where $\kappa_{0}$
includes any absorption in the membrane or cavity, but not the {}``good''
decay channel through the entrance mirror. In practice, for setups
where $\kappa\sim\kappa_{0}$, this is identical to the single-photon
strong coupling regime, to be discussed further in Sec.~\ref{sub:QuantumOptomechanicsNonlinear}.

Detailed quantum jump trajectory simulations of the phonon detection
process have been first presented in \cite{2011_Milburn_PhononNumberQuantumJumps},
for the idealized case of a single optical mode. More recently, studies
of the complete two-mode setup \cite{2012_Ludwig_TwoModeScheme} have
been able to reveal the physics discussed above \cite{2009_Chen_QuantumLimit}.

The detection of quantum jumps between mechanical Fock states is obviously
extremely challenging. However, one might detect quantum signatures
of the mechanical oscillator in another way: When driving it strongly
into a coherent mechanical state of mean phonon number $\bar{n}$,
there will be Poissonian phonon shot noise of size $\sqrt{\bar{n}}$.
A setup like the one described above may then detect this noise, which
reveals the granularity of mechanical quanta \cite{Clerk2010}, much
in the same way that measurements of electrical current noise can
reveal the charge of the charge carriers \cite{Blanter2000}. Furthermore,
higher moments of the phonon shot noise may be observed in this way
as well. These display distinctly nonclassical features \textendash{}
for an in-depth analysis see \cite{2011_Clerk_PhononFullCountingStatistics_PRA}.

\begin{figure}[ptb]
 \centering\includegraphics[width=3.2in]{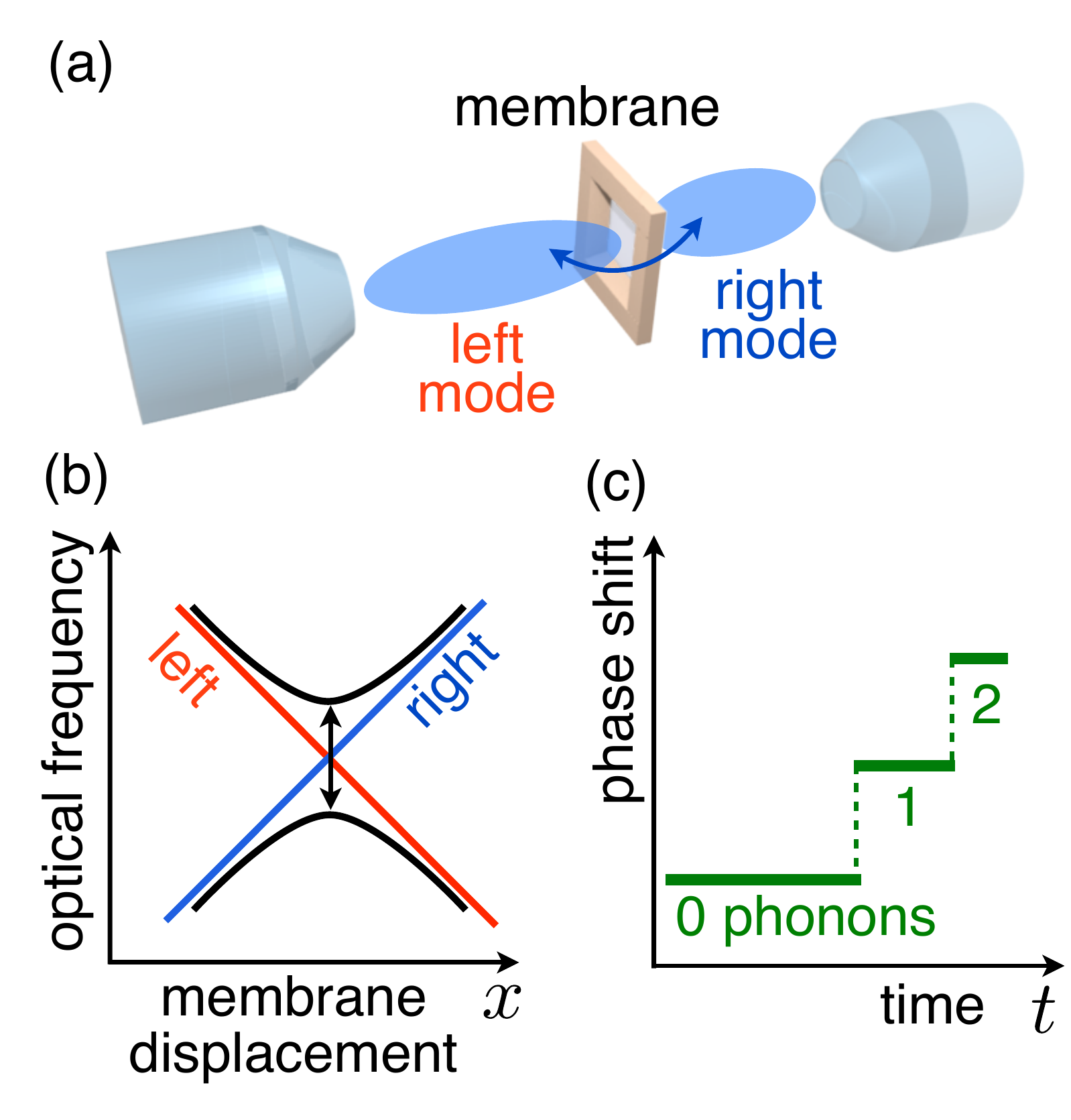}\caption{\label{fig:MembraneInMiddleSetupAndFockStateDetection}(a) The {}``membrane-in-the-middle''
setup used to generate an optomechanical coupling to $\hat{x}^{2}$.
It can be viewed as two optical modes to the left and right of the
membrane, with photon transmission through the membrane coupling those
modes. (b) Resulting optical spectrum, with an avoided crossing between
the left and right optical mode as a function of displacement $x$.
Coupling to $\hat{x}^{2}$ is obtained at the degeneracy point. (c)
Future dispersive setups along these lines (with greatly improved
parameters) may enable detection of quantum jumps between mechanical
Fock states in the phase shift of light reflected from the cavity.
An idealized quantum jump trajectory (without measurement noise) is
depicted here, starting from the mechanical ground state ($0$ phonons).}
\end{figure}

\subsubsection{Optical feedback cooling (cold damping) }

\label{sub:MeasurementsFeedbackCooling}

\begin{figure}
\includegraphics[width=1\columnwidth]{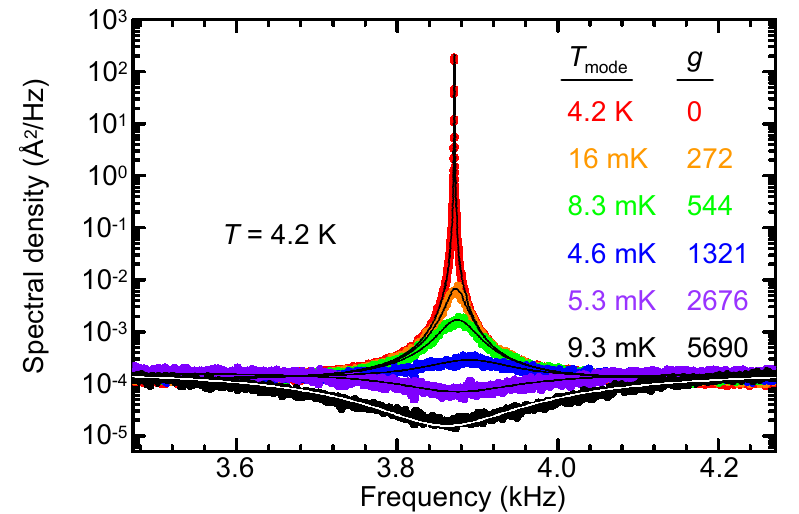}

\caption{\label{fig:Optomechanical-feedback-cooling}Optomechanical feedback
cooling of a cantilever. Note the 'noise squashing' at strong feedback
(reduction of the spectrum below the noise floor due to interference
between the classical noise in the readout and the signal caused by
the mechanical motion) \cite{Poggio2007}.}
\end{figure}

The high sensitivity provided by the cavity readout of mechanical
motion can also be used for directly cooling the mechanical motion
via active feedback. The main idea is to obtain the oscillator position
by a phase-sensitive detection of the cavity output and to use it
to generate a negative feedback on the oscillator, i.e. a force $F=-m_{{\rm eff}}\delta\Gamma\dot{x}$
proportional to the time derivative of the output signal. This increases
the damping rate of the system by $\delta\Gamma$ without increasing
the thermal noise (cold damping). The scheme has been suggested in
\cite{Mancini1998} and has been experimentally realized in several
optomechanical devices \cite{Cohadon1999a,Arcizet2006,Kleckner2006a,Poggio2007}
with radiation pressure as the feedback force. Because the scheme
relies on the precise readout of the instantaneous oscillator position,
the ideal configuration comprises both weak coupling and a fast cavity
decay, i.e. $\kappa\gg\Omega_{m}\gg g$ (adiabatic regime). The quantum
limits of this cooling method have been discussed in \cite{Courty2001a,Vitali2002,Genes2008}
and it has been shown that ground-state cooling by cold damping is
possible. A detailed discussion can be found for example in \cite{Genes2008}.
The maximum amount of cooling is limited by the imprecision of the
read-out. An important aspect here is the phenomenon of {}``noise
squashing'', where the noise on the detector and the noise-driven
motion of the mechanical oscillator become correlated (see Fig.~\ref{fig:Optomechanical-feedback-cooling}).

In the context of cold atoms, feedback cooling of a single neutral
atom has been implemented in a cavity QED setup \cite{2010_Rempe_FeedbackCooling}.

\section{Optomechanical cooling}

\label{sec:OptomechanicalCooling}

Optomechanical quantum control requires the mechanical oscillator
to be in or near its quantum ground state. Unless the mechanical frequency
$\Omega_{m}$ is in the ${\rm GHz}$ range (which is true only for
some recent nanomechanical setups \cite{Eichenfield2009,Chan2011c}),
even dilution refrigerator temperatures of $20{\rm mK}$ are not sufficient
to ensure $k_{B}T\ll\hbar\Omega_{m}$. Thus, additional cooling of
the selected mechanical mode is needed. 

In a previous section (Sec.~\ref{sub:DynamicalBackactionBasics}),
we have already discussed dynamical backaction effects and the resulting
optomechanical damping rate. It is obvious that this can be used for
cooling the mechanical motion. The purpose of the present section
then is to develop the quantum theory of cooling, which in particular
describes the limits for cooling that cannot be obtained from a discussion
of the damping rate alone. We will focus in this section on intrinsic
cavity cooling. For a comparison of this approach to feedback cooling
(discussed above), see \cite{Genes2008}. 

A simple classical theory of an oscillator at initial temperature
$T_{{\rm init}}$ subject to extra damping $\Gamma_{{\rm opt}}$ (proportional
to the laser power) would predict that its temperature is reduced
down to

\begin{equation}
T_{{\rm final}}=T_{{\rm init}}\frac{\Gamma_{{\rm m}}}{\Gamma_{{\rm m}}+\Gamma_{{\rm opt}}}\,.
\end{equation}
However, this classical expression ceases to be valid at sufficiently
low $T_{{\rm final}}$, when the fluctuations of the radiation pressure
force due to photon shot noise set a lower bound to the achievable
temperature. In the following, we discuss the full quantum theory
that permits to calculate the quantum limits to cooling and predicts
that, in many cases, ground state cooling is possible only in the
resolved sideband regime $\kappa\ll\Omega_{{\rm m}}$. 

It should be noted that sideband resolution does not intrinsically
prohibit optomechanical ground state cooling. For example, the displacement
may not only couple to the cavity frequency, $\omega_{{\rm cav}}=\omega_{{\rm cav}}(x)$,
but also to the decay rate, yielding $\kappa=\kappa(x)$. This mechanism
is quite plausible both for nano-objects that scatter light out of
the cavity (depending on where they sit in the standing light wave
pattern), and for vibrating objects that are evanescently coupled
to some tapered fibre, as well as for other geometries. This kind
of behaviour has been observed experimentally \cite{2009_Tang_MicroDiskCoupledWaveguide},
and it has been predicted \cite{Elste2009} that this can yield novel
behaviour for cooling, with the possibility of reaching the quantum
ground state even when $\kappa>\Omega_{{\rm m}}$. Similar results
hold for pulsed cooling schemes \cite{2011_Aspelmeyer_PulsedOptomechanics,Jacobs2011,2011_Retzker_PulsedLaserCooling}
and photothermal cooling.

Another interesting option is to consider a nonlinear medium inside
a cavity, which may enhance the efficiency of cooling \cite{2009_HuangAgarwal_CoolingEnhancementOPA}.

Finally, one could do away with the cavity by exploiting other sharp
changes of radiation forces as a function of wavelength. This has
been proposed e.g. for a Bragg mirror \cite{2008_Karrai_DopplerOptomechanics}
or for microspheres with their narrow internal whispering gallery
resonances \cite{2010_Barker_DopplerCoolingMicroSphere}. 

We will focus on radiation pressure cooling in cavity setups, which
is conceptually the simplest case. Recently it has been argued that
photothermal forces, which have been exploited for optomechanical
cooling early on \cite{HohbergerMetzger2004,2008_Karrai_SelfCoolingPhotothermalLongPaper},
could in principle also lead to the quantum ground state \cite{2008_PinardDantan_PhotothermalCooling,2011_Favero_PhotothermalCoolingTheory}.

\subsection{Quantum theory of radiation pressure cooling }

\label{sub:CoolingQuantumTheory}

\label{sec:VII.A}

In the following, we will work in the weak coupling regime, $g\ll\kappa$,
where a perturbative picture applies. The quantum theory of optomechanical
cooling \cite{Marquardt2007,Wilson-Rae2007,Genes2008} is related
to earlier approaches for trapped ions \cite{Neuhauser1978,Itano1992}
and for cavity-assisted laser cooling of atomic and molecular motion
\cite{Hechenblaikner1998,Vuletic2000}. The idea is best explained
in a Raman-scattering picture. Photons impinging at a frequency red-detuned
from the cavity resonance will, via the optomechanical interaction,
preferentially scatter upwards in energy in order to enter the cavity
resonance, absorbing a phonon from the oscillator in the process.
As a consequence, they will be reflected blue-shifted by $\Omega_{{\rm m}}$,
carrying away a quantum of mechanical energy. These {}``anti-Stokes''
processes happen at a rate $A^{-}$ (to be calculated below). More
precisely, the transition rate from phonon state $n$ to $n-1$ will
include a bosonic factor $n$,

\begin{equation}
\Gamma_{n\rightarrow n-1}=nA^{-}\,.
\end{equation}
The {}``Stokes'' process, where photons return red-shifted and leave
behind an extra phonon, happens at a smaller rate $A^{+}$ (with $\Gamma_{n\rightarrow n+1}=(n+1)A^{+}$)
due to the suppression in the final density of available photon states,
if the laser is red-detuned. Note that we choose to follow the notation
$A^{\pm}$ for these rates, as it is used in the context of atomic
laser cooling.

Given these rates ($A^{+}$ for upward transitions in the mechanical
oscillator, $A^{-}$ for downward transitions), the full optomechanical
damping rate is the net downward rate,

\begin{equation}
\Gamma_{{\rm opt}}=A^{-}-A^{+}\,.
\end{equation}
The average phonon number $\bar{n}=\sum_{n=0}^{\infty}nP_{n}$ (with
the Fock state populations $P_{n}$) changes according to the rates
$\Gamma_{n\rightarrow n\pm1}$, leading to:

\begin{equation}
\dot{\bar{n}}=(\bar{n}+1)(A^{+}+A_{{\rm th}}^{+})-\bar{n}(A^{-}+A_{{\rm th}}^{-}).\label{eq:phonon_number_evolution}
\end{equation}
Here we have introduced the extra transition rates due to the oscillator's
thermal environment, which has a mean phonon number $\bar{n}_{{\rm th}}$:
$A_{{\rm th}}^{+}=\bar{n}_{{\rm th}}\Gamma_{{\rm m}}$ and $A_{{\rm th}}^{-}=(\bar{n}_{{\rm th}}+1)\Gamma_{{\rm m}}$.
In the absence of optomechanical effects, these would establish equilibrium
at $\bar{n}=\bar{n}_{{\rm th}}$. Now, however, the steady-state final
phonon number (requiring $\dot{\bar{n}}=0$ in (\ref{eq:phonon_number_evolution}))
will be

\begin{equation}
\bar{n}_{f}=\frac{A^{+}+\bar{n}_{{\rm th}}\Gamma_{{\rm m}}}{\Gamma_{{\rm opt}}+\Gamma_{{\rm m}}}\,.\label{eq:FinalPhononNumberFirstExpression}
\end{equation}
Even in the optimal case, i.e. in the absence of any coupling to a
mechanical thermal environment ($\Gamma_{{\rm m}}=0$), this leads
to a minimal phonon number of

\begin{equation}
\bar{n}_{{\rm min}}=\frac{A^{+}}{\Gamma_{{\rm opt}}}=\frac{A^{+}}{A^{-}-A^{+}}\,.\label{eq:MinimumPhononNumber}
\end{equation}

\begin{figure}
\includegraphics{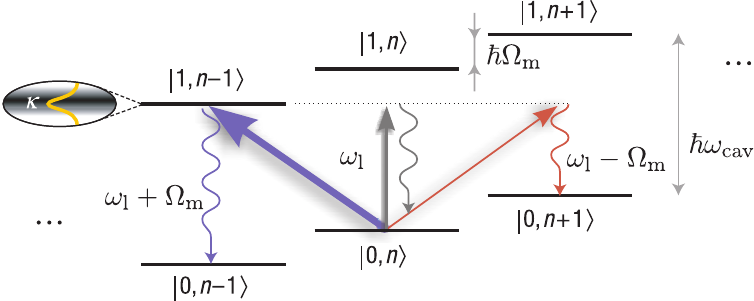}\caption{Principle of cavity optomechanical cooling and corresponding transition
diagram. Three processes can be distinguished. First a photon can
be absorbed by the cavity leaving the motional states unchanged (carrier
transitions). Second, the photon can create a phonon (corresponding
to absorption on the upper motional sideband). Third, the photon leads
to the annihilation of a phonon (corresponding to a photon being absorbed
by the lower motional sideband). Appropriate red-detuning of the laser
can lead to effective suppression of the first two processes and efficient
cooling.}
\end{figure}

The rates $A^{\pm}$ can be calculated using Fermi's Golden Rule,
applied to the interaction of the oscillator with the fluctuating
radiation pressure force: $\hat{H}_{{\rm int}}=-\hat{x}\hat{F}$,
where $\hat{F}=\hbar G\hat{a}^{\dagger}\hat{a}$ according to Eq.~(\ref{eq:H_optomech_int}).
In the weak coupling limit, all the transition rates can be calculated
once the quantum noise spectrum $S_{FF}(\omega)=\int_{-\infty}^{+\infty}dt\, e^{i\omega t}\left\langle \hat{F}(t)\hat{F}(0)\right\rangle $
of the force is known (see e.g. \cite{Clerk2008a} for more on quantum
noise spectra). Re-expressing the result in terms of noise spectra
yields:

\begin{equation}
A^{\pm}=\frac{x_{{\rm ZPF}}^{2}}{\hbar^{2}}S_{FF}(\omega=\mp\Omega_{{\rm m}})=g_{0}^{2}S_{NN}(\omega=\mp\Omega_{{\rm m}})\,,\label{eq:AratesSpectrum}
\end{equation}
where we introduced the photon number noise spectrum $S_{NN}(\omega)=\int_{-\infty}^{+\infty}dt\, e^{i\omega t}\left\langle (\hat{a}^{\dagger}\hat{a})(t)(\hat{a}^{\dagger}\hat{a})(0)\right\rangle $.
Exploiting shifted photon operators, one can show \cite{Marquardt2007}
that the photon number spectrum of a laser-driven cavity is

\begin{equation}
S_{NN}(\omega)=\bar{n}_{{\rm cav}}\frac{\kappa}{\kappa^{2}/4+(\Delta+\omega)^{2}}\,.
\end{equation}
Inserting this into Eq.~(\ref{eq:AratesSpectrum}), one obtains $A^{\pm}$
and from there the optomechanical damping rate $\Gamma_{{\rm opt}}=A^{-}-A^{+}$,
which coincides with the expression obtained using the linearized
equations of motion (Sec.~\ref{sec:V.B.2},\textbf{ }Eq.~(\textbf{\ref{eq:CoolingRateFullExpression}})).
We obtain the final minimum phonon number (\ref{eq:MinimumPhononNumber})
as

\begin{equation}
\bar{n}_{\min}=\left(\frac{A^{-}}{A^{+}}-1\right)^{-1}=\left(\frac{(\kappa/2)^{2}+(\Delta-\Omega_{\text{m}})^{2}}{(\kappa/2)^{2}+(\Delta+\Omega_{\text{m}})^{2}}-1\right)^{-1}\,.
\end{equation}
Experimentally, one can still vary the laser detuning $\Delta$ to
minimize this expression. In the resolved sideband regime $\kappa\ll\Omega_{{\rm m}}$,
this leads to

\begin{equation}
\bar{n}_{\min}=\left(\frac{\kappa}{4\Omega_{{\rm m}}}\right)^{2}<1\,,
\end{equation}
which permits ground state cooling, while in the opposite limit ($\kappa\gg\Omega_{{\rm m}}$)
we find 
\begin{equation}
\bar{n}_{\min}=\frac{\kappa}{4\Omega_{\text{m}}}\gg1\,.
\end{equation}

These two results are almost identical to the case of atomic laser
cooling (in the atomic laser cooling expressions, a scalar prefactor
is different due to the directional dependence of the spontaneous
emission).%
\footnote{The similarity to atomic Doppler cooling is due to the fact that the
Doppler cooling limit is a result of the Heisenberg uncertainty principle.
Each photon that leaves the cavity has an energy uncertainty given
by its photon decay time, i.e. \textbf{$\Delta E=\hbar/\Delta t,$}
where in the cavity cooling case \textbf{$\Delta t$} is the inverse
cavity decay rate \textbf{$\kappa^{-1}.$} The minimum phonon number
can therefore never be smaller than this value, i.e. \textbf{$\Delta E=\hbar\kappa$},
or expressed as an occupancy \textbf{$\bar{n}_{\min}\propto\kappa/\Omega_{{\rm m}}.$}%
}

In the presence of a thermal environment, the final phonon number
(\ref{eq:FinalPhononNumberFirstExpression}) can be written as the
result of coupling to two baths at average occupations $\bar{n}_{{\rm min}}$
and $\bar{n}_{{\rm th}}$ with coupling rates $\Gamma_{{\rm opt}}$
and $\Gamma_{{\rm m}}$, respectively:

\begin{equation}
\bar{n}_{f}=\frac{\Gamma_{{\rm opt}}\bar{n}_{\min}+\Gamma_{{\rm m}}\bar{n}_{th}}{\Gamma_{\text{opt}}+\Gamma_{\text{m}}}\,.\label{eq:FinalPhononNumberWeightedAverage}
\end{equation}

\textit{Finite thermal cavity occupancy. }This formula is modified
when the cooling field has thermal occupation. This is in particular
the case for microwave fields due to their low frequency.\ If the
cavity occupation is given by $\bar{n}_{{\rm cav}}^{{\rm th}}$, then
the final occupation is modified to (in the resolved sideband limit):
\begin{equation}
\bar{n}_{f}=\bar{n}_{{\rm th}}\frac{\Gamma_{{\rm m}}}{\Gamma_{{\rm eff}}}+\bar{n}_{{\rm cav}}^{{\rm th}}+\frac{\kappa^{2}}{16\Omega_{m}^{2}}
\end{equation}
This implies that the final phonon number can never be below the effective
thermal occupation of the drive field \cite{Dobrindt2008}.\ It should
be noted that when the radiation field and the mechanical oscillator
initially have the same bath temperature (as will be the case in equilibrium,
without extra absorption), the equilibration of these two oscillators
of frequency $\omega_{{\rm cav}}$ and $\Omega_{m}$ will lead to
an effective cooling of the lower frequency mechanical oscillator,
as $\bar{n}_{{\rm th}}=\frac{k_{B}T_{bath}}{\hbar\Omega_{m}}\gg\bar{n}_{{\rm cav}}=\frac{k_{B}T_{bath}}{\hbar\omega_{{\rm cav}}}.$

\begin{figure}[ptb]
\centering\includegraphics[width=3in]{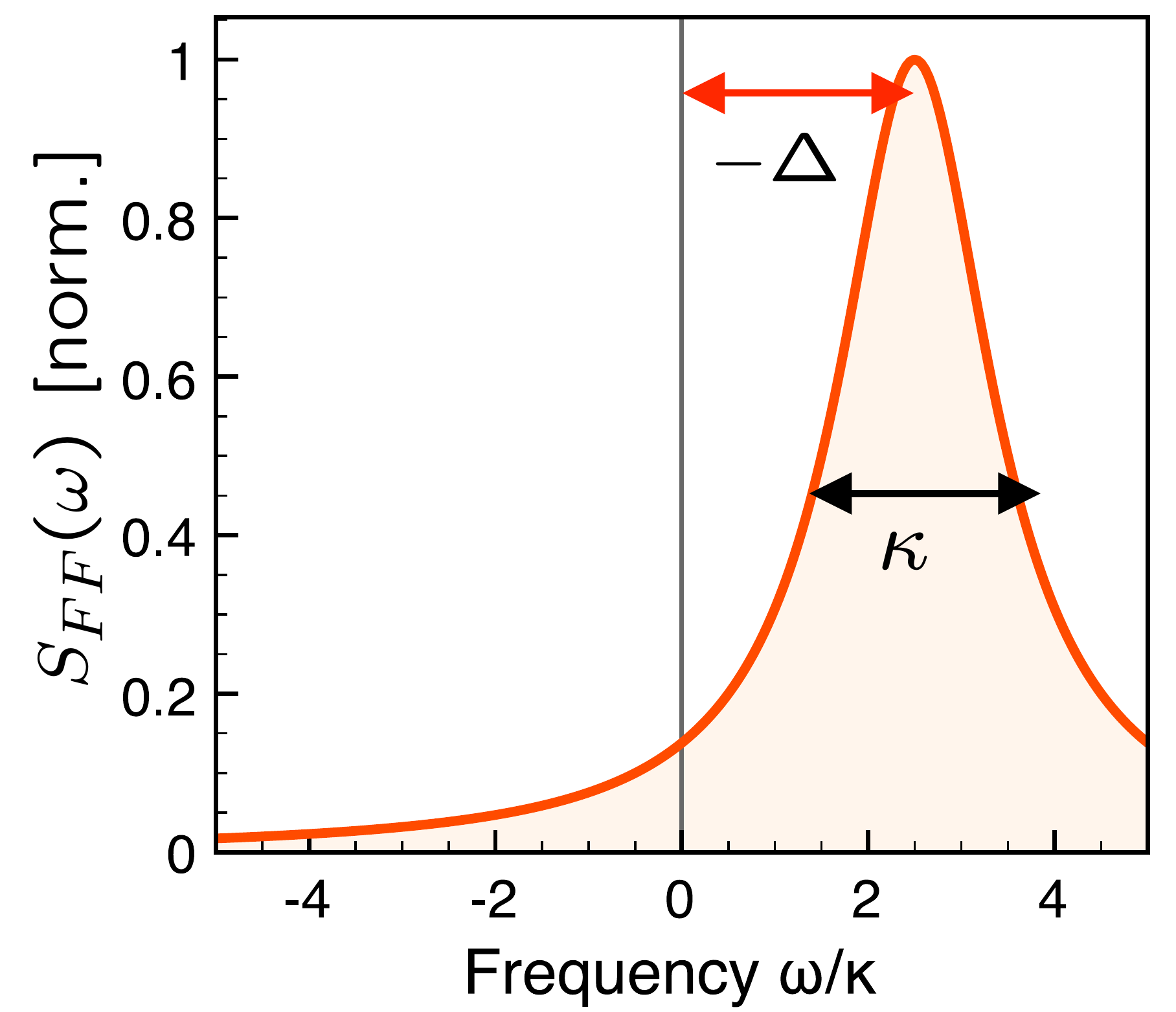}

\caption{\label{fig:CoolingNoiseSpectra}Quantum noise spectrum $S_{FF}(\omega)$
for the radiation pressure force fluctuations acting on the end-mirror
of a laser-driven optical cavity. The spectrum is asymmetric in frequency,
due to the quantum nature of the fluctuations. For red detuning, the
bulk of the spectrum sits at positive frequencies. This implies (see
main text) that the fluctuations induce more downward (cooling) transitions
than upward transitions.}
\end{figure}

\textit{Equations-of-motion approach. }An alternative way to calculate
the final phonon number is to write down the linearized equations
of motion for the oscillator and light field, eliminate the light
field dynamics, and exploit the known spectra of the quantum Langevin
forces driving both the mechanics and the cavity (see Sec.~\ref{sub:OptomechanicalEquationsOfMotion})%
\footnote{Note that there can also be experimental situations where the mechanical
oscillator mediates scattering of photons out of the cavity, such
as in the case of levitated particles in an optical cavity. Denoting
the photon scattering rate caused by the mechanical oscllator as $\gamma_{scat}$
this additional heating causes a modification of the quantum limit
of cooling. In the sideband limit this yields:\textbf{ $\bar{n}_{f}=\frac{\Gamma_{opt}\bar{n}_{\min}+\Gamma_{{\rm m}}\bar{n}_{th}}{\Gamma_{\text{opt}}+\Gamma_{\text{m}}}\,+\frac{\gamma_{scat}}{\kappa}$}.
The analytical form of the scattering rate depends on the geometry
of the mechanical oscillator under consideration. For the case of
a small sphere, trapped within a Fabry Perot cavity, an analytic solution
of $\gamma_{scat}$ can be derived \cite{Chang2010,Romero-Isart2010}.%
}.

We do this by considering the influence of the quantum backaction
force noise driving the mechanical oscillator. Adopting the \emph{classical}
equations of motion for the mechanical oscillator (but keeping the
quantum part of the symmetrized noise spectrum), we find an average
energy of:
\[
\langle E_{m}\rangle=\int_{-\infty}^{+\infty}d\omega\left\vert \chi_{\text{eff}}(\omega)\right\vert ^{2}\bar{S}_{FF}(\omega)m_{eff}\omega^{2}\,,
\]
where $\bar{S}_{FF}$ is the symmetrized version of the spectrum introduced
above and we assumed the laser to be dominant, $\Gamma_{{\rm eff}}\gg\Gamma_{{\rm m}}$,
and also $\Omega_{{\rm m}}\gg\Gamma_{{\rm eff}}$. This leads to the
correct quantum result:
\[
\langle E_{m}\rangle=\frac{\hbar\Omega_{m}}{2}\left(1+\frac{A^{+}}{A^{-}-A^{+}}\right)
\]
This expression reveals that the energy, consisting of the zero point
motion and an additional residual term, is caused by the quantum fluctuations
of the laser field. The fact that the quantum fluctuations of the
laser field also give rise to the zero point energy can be understood
by noting that the dominant bath that the mechanical oscillator is
coupled to is the laser field itself when $\Gamma_{eff}\gg\Gamma_{m}.$
In this sense then, the correct quantum limit for cooling can be formulated
on a semiclassical basis, where it is sufficient to consider only
the quantum fluctuations of the drive field.

\textit{Optical output spectrum. }The output spectrum of the light
that emerges from the optomechanical cavity can be calculated using
the input-output formalism (Sec.~\ref{sub:OptomechanicalEquationsOfMotion}).
In close analogy to the fluorescence spectrum of a laser-cooled trapped
ion, the specrum is given by the expression:
\begin{align}
S(\omega)\,\approx & I_{0}\delta\left(\omega-\omega_{L}\right)\nonumber \\
 & +A^{-}\bar{n}_{f}I_{{\rm side}}(\omega-\Omega_{{\rm m}})\\
 & +A^{+}\left(\bar{n}_{f}+1\right)I_{{\rm side}}(\omega+\Omega_{{\rm m}})\label{eq:SpectrumSidebands}
\end{align}
This spectrum refers to the frequency-resolved photon flux. Here 

\[
I_{{\rm side}}(\omega)=\frac{1}{2\pi}\frac{\Gamma_{{\rm eff}}}{\left(\omega-\omega_{L}\right)^{2}+\frac{\Gamma_{{\rm eff}}^{2}}{4}}
\]
is the Lorentzian shape of the sidebands, where the width of the sidebands
is given by the overall effective mechanical damping rate $\Gamma_{{\rm eff}}$.
The weight of the main peak%
\footnote{It is interesting to consider the total (integrated) power in the
generated sidebands. The weight of the sideband for the case where
the final occupancy is given by $\bar{n}_{f}=1$ is given by $P_{sideband}/P_{carrier}\approx\frac{4g_{0}^{2}}{\Omega_{m}^{2}}$
and is thus independent of the initial occupancy (or temperature).%
} at the laser frequency is (for a single-sided Fabry Perot resonator) 

\[
I_{0}=\frac{\kappa_{ex}}{\kappa}\frac{P}{\hbar\omega_{L}}\left[\frac{\kappa}{\kappa_{ex}}-\frac{\kappa^{2}-\kappa_{ex}\kappa}{\Delta^{2}+\frac{1}{4}\kappa^{2}}\right]\,.
\]

As expected, the spectrum consists (in addition to the carrier) of
blueshifted (anti-Stokes) photons at $\omega=\omega_{L}+\Omega_{{\rm m}}$
(second line of Eq.~(\ref{eq:SpectrumSidebands})) and redshifted
(Stokes) photons at $\omega=\omega_{L}-\Omega_{{\rm m}}$ (third line).
The sideband asymmetry changes as a function of cooling laser power.
Detailed balance causes the initially asymmetric sidebands (as $A^{-}\gg A^{+}$)
to become progressively more symmetric, with $A^{+}(\bar{n}_{f}+1)=A^{-}\bar{n}_{f}$
in the limit where $\bar{n}_{f}$ is entirely determined by the cooling
process ($\Gamma_{{\rm opt}}\bar{n}_{{\rm min}}\gg\Gamma_{{\rm m}}\bar{n}_{{\rm th}}$).
This provides a means to determine the temperature of the mechanical
oscillator via the spectral weight of the two sidebands, a technique
that has been widely used in the trapped ion community and is termed
sideband thermometry \cite{Diedrich1989}. There, the spectral weight
can be directly measuring via optical shelving\cite{Haroche2006}.
A variant of this method has recently been demonstrated in an optomechanical
cooling experiment \cite{2012_Painter_SidebandThermometry}. In this
method the excitation laser is placed on the upper ($\Delta=+\Omega_{m}$
) and subsequently on the lower sideband ($\Delta=-\Omega_{m}$) and
the rate of scattering into the cavity $\dot{N}^{cav}$ is measured.
This yields $\dot{N}^{cav}(\Delta=\Omega_{m})=\frac{\kappa_{ex}}{\kappa}A^{+}(\bar{n}_{f}+1)$
and $\dot{N}^{cav}(\Delta=-\Omega_{m})=\frac{\kappa_{ex}}{\kappa}A^{-}(\bar{n}{}_{f})$
provided the laser is quantum limited. A treatment in the presence
of laser noise, which can yield a contribution to the asymmetry, is
given in Appendix \ref{Appendix:LaserNoiseSidebandThermometry}.

\begin{figure}
\includegraphics[width=1\columnwidth]{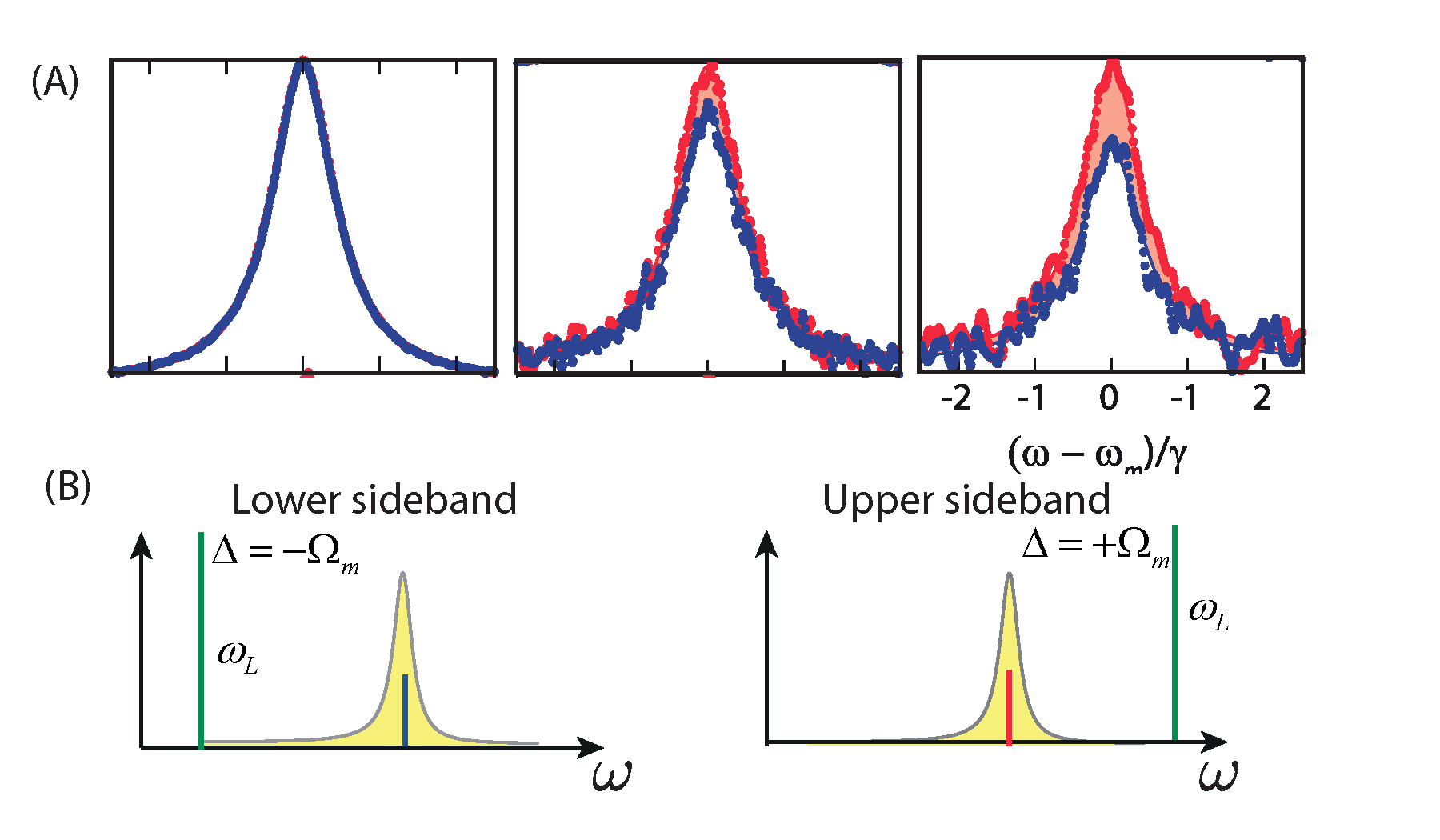}\caption{\label{fig:Sideband-thermometry-,}(a) Sideband thermometry revealing
the growing asymmetry between photons scattered into the cavity resonance
when increasing the cooling laser power (Figure courtesy O.~Painter,
from \cite{2012_Painter_SidebandThermometry}) (b) Scheme of the readout
laser for measuring the photons scattered into the cavity resonance.}
\end{figure}

Experiments and practical limitations

\label{sub:CoolingExperiments}

The first experimental attempts to damp the motion of a mechanical
oscillator with radiation pressure dynamical backaction were carried
out by Braginsky and coworkers \cite{Braginsky1970}. In these experiments
a microwave cavity was employed, and a modification of the damping
rate of the end-mirror pendulum could be observed.\ Microwave cooling
deeply in the resolved-sideband regime was moreover implemented in
the field of gravitational wave detectors, in the form of a high Q
cryogenic sapphire transducer, where it served the role to reduce
the effective noise temperature \cite{Cuthbertson1996}. 

Optical feedback cooling of a micromechanical mirror using the radiation
pressure force was demonstrated in 1999 \cite{Cohadon1999a}. Dynamical
backaction cooling in the optical domain was achieved using photothermal
forces in 2004 \cite{HohbergerMetzger2004}, and using radiation pressure
forces in 2006 \cite{Gigan2006,Arcizet2006a,Schliesser2006a}. 

\begin{figure}
\includegraphics[width=1\columnwidth]{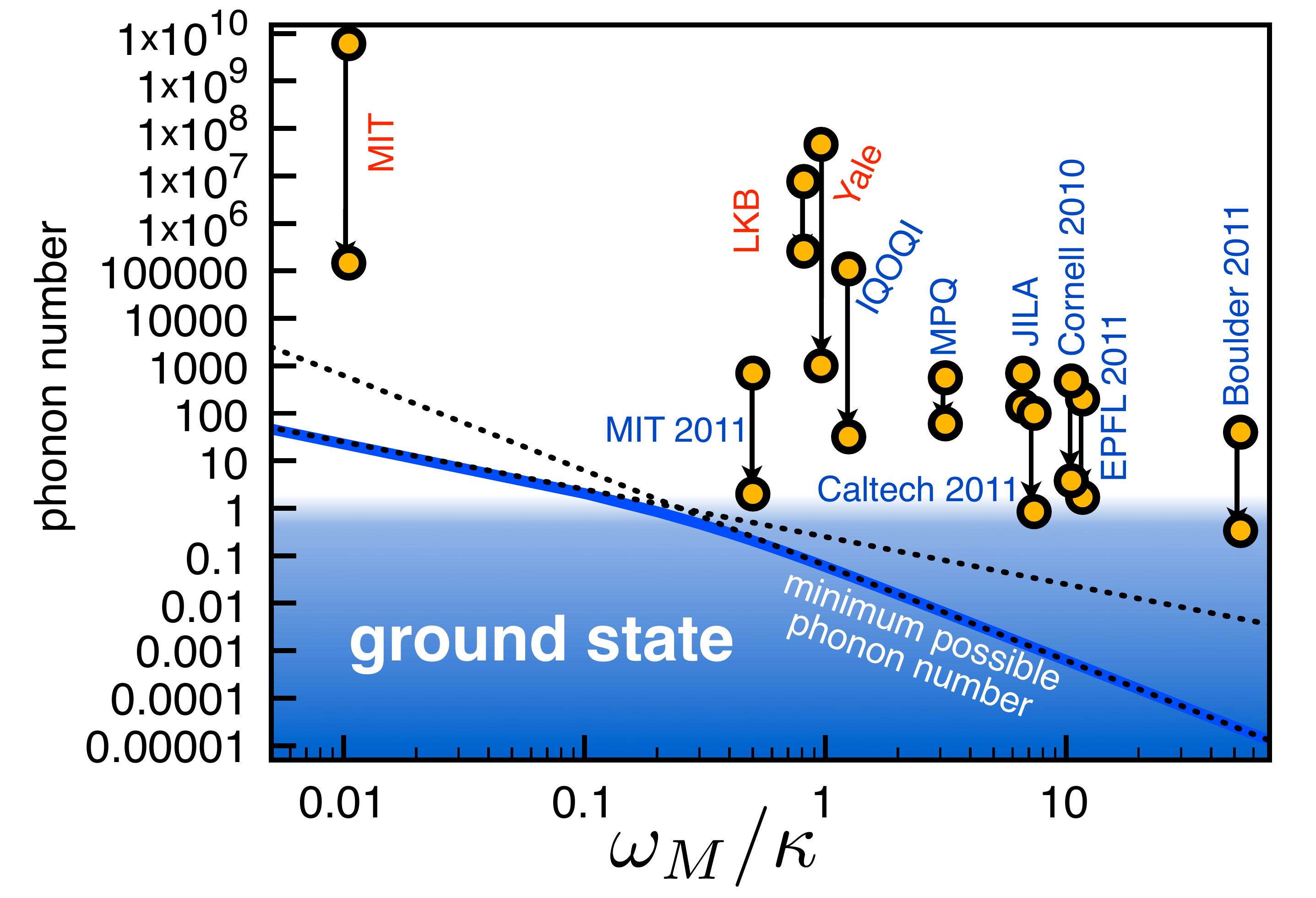}

\caption{\label{fig:CoolingExperimentalResults}Experimental results for optomechanical
laser-cooling. We display the initial and final phonon numbers (log
scale), vs. the sideband resolution parameter $\Omega_{{\rm m}}/\kappa$.
The quantum limit for the minimum achievable phonon number is plotted
as well. MIT: \cite{Corbitt2007a}, LKB: \cite{Arcizet2006a}, Yale:
\cite{Thompson2008}, Vienna: \cite{Groblacher2009}, MPQ: \cite{Schliesser2009a},
JILA: \cite{Teufel2008}, Cornell 2010: \cite{Rocheleau2010}, Caltech
2011: \cite{Chan2011c}, EPFL 2011: \cite{2011_Kippenberg_GroundStateCoolingPRA},
Boulder 2011: \cite{Teufel2011}, MIT 2011: \cite{Schleier-smith2011}}
\end{figure}

These early experiments operated in the Doppler regime ($\kappa>\Omega_{{\rm m}}$).
As outlined in the previous section, an important requirement for
many applications is the resolved sideband regime. This was demonstrated
in the microwave regime for the first time \cite{Cuthbertson1996}
and later in the optical domain \cite{Schliesser2008}. Since then,
a plethora of novel systems (e.g. \cite{Thompson2008}, \cite{Teufel2008},
\cite{Lin2009} and many more) have been realized in the resolved
sideband regime (see Fig.~\ref{fig:CoolingExperimentalResults}).
All these experiments were, however, essentially room temperature
experiments. 

To initialize a mechanical oscillator in the ground state at thermal
equilibrium, the condition $k_{B}T/\hbar\Omega_{{\rm m}}\ll1$ has
to be realized. In general, the probability to find the system in
the ground state is related to the average occupation number $\bar{n}$
by $P_{0}=1/(1+\bar{n})$. Note that in the field of trapped ion experiments,
the motional ground state can nowadays routinely be prepared with
more than $97\%$ probability \cite{Diedrich1989,Leibfried2003}.
Reaching the ground state (often referred to as $\bar{n}<1$) is challenging
for low frequency oscillators, as the thermal freezeout for a $1$
$\operatorname{MHz}$ oscillator would equate to $50$ $\mu K.$ These
temperature are far below those attained with a dilution refridgerator.
Ground state cooling with conventional cryogenics can therefore be
reached only for $\operatorname{GHz}$ oscillators. Indeed, the first
demonstration of quantum control at the single phonon level was demonstrated
with a $6\operatorname{GHz}$ piezoelectric mechanical oscillator
cooled to below $50$ m$\operatorname{K}$ \cite{O'Connell2010}.

It is challenging to reach the quantum ground state of micromechanical
oscillators at lower frequencies. A widely pursued strategy has been
to combine cryogenic precooling with dynamical backaction laser cooling.
The precooling thereby allows to reduce the starting mode temperature.
This technique has over the past years allowed a substantial reduction
of the phonon occupancy. 

While initial cryogenic experiments had demonstrated cooling to a
level of a few dozen quanta in the optical domain \cite{Groblacher2009,Schliesser2009a,Park2009},
further experimental progress allowed to reduce the motional energy
to a level close to the zero point motion in several experiments both
in the microwave domain \cite{Rocheleau2010,Teufel2011} and in the
optical domain \cite{Chan2011c,2011_Kippenberg_GroundStateCoolingPRA},
with ground state probabilities ranging from 0.2 to 0.7. 

Specifically, in the microwave domain, increasing the coupling strength
by improved cavity designs resulted in cooling to around $4$ quanta
\cite{Rocheleau2010} ($P_{0}=0.2$) and to $\bar{n}_{f}=0.38$ ($P_{0}=0.72$)
using a superconducting resonator coupled to micromechanical drum
mode \cite{Teufel2011}, respectively ($\bar{n}_{f}=0.38$). In both
experiments the limitation for the occupancy was set by the fact that
the cavity had a low, yet finite residual thermal occupation. This
also led to the observation of squashing in the mechanical noise spectrum
data of this experiment for some drive powers. All these experiments
have been performed in dilution refrigerators with base temperatures
as low as about 25 mK. 

Also in the optical domain, new geometries allowed to improve the
cooling performance to $\bar{n}_{f}=1.7$ quanta ($P_{0}=0.37$) for
improved microtoroidal resonators \cite{Verhagena} and to 0.8 quanta
($P_{0}=0.54$) for a nanomechanical mode of a photonic crystal beam
\cite{Chan2011c}. \ In \cite{2011_Kippenberg_GroundStateCoolingPRA}
a spoke-supported microresonator was precooled to $\sim200$ quanta
in a He-3 buffer gas cryostat (ca. 750 mK temperature) and optically
pumped by a Ti:Sa laser. In this experiment, the occupancy was limited
by the cavity decay rate $\kappa$ and in addition by reheating of
the sample due to the laser, which in the case of glass reduces the
Q-factor due to two-level systems in the amorphous glass$.$ The experiment
by \cite{Chan2011c} involved a $3.67\operatorname{GHz}$ mechanical
mode and an external cavity diode laser drive. The cooling was limited
by reheating of the sample to its initial temperature of ca. $20\operatorname{K}$
inside a He-4 flowthrough cryostat and the fact that the mechanical
Q factor exhibited a strong temperature dependence, as well as unwanted
effects at higher drive powers. In all cases, higher fidelity ground
state preparation - as achieved in ion trapping - will require further
reduction of the effects of laser heating of the sample, further increases
of the mechanical quality factor or higher optomechanical coupling
rates.

The technique of radiation pressure dynamical backaction cooling has
many similarities to atomic laser cooling \cite{Diedrich1989,Leibfried2003}.
Nevertheless there are substantial differences. For instance, optomechanical
cooling does not proceed by the use of a two-level system, but by
a cavity that is excited with a coherent laser field.\ From an experimental
point of view, optomechanical cooling of a mechanical oscillator to
the ground state is impeded by several challenges. First, in contrast
to atomic laser cooling where the atoms are well isolated, mechanical
systems are usually coupled to a high temperature bath with correspondingly
large motional heating rates. In addition, optomechanical laser cooling
is extraordinarily sensitive to laser phase (i.e. frequency) noise.
In the following two sections the fundamental limits of cooling are
reviewed due to both laser phase noise and thermorefractive cavity
noise.

A selection of cooling experiments is displayed in Fig.~\ref{fig:CoolingExperimentalResults}.
There, we plot the initial and final phonon numbers vs. the sideband
resolution parameter $\Omega_{{\rm m}}/\kappa$ that determines the
minimum achievable phonon number.

\paragraph{Laser phase noise }

In the last section, the laser input noises were considered to be
essentially quantum noises, i.e. the laser beam is considered a perfect
coherent state. However, real laser systems exhibit excess noise,
e.g. due to relaxation oscillations that can be derived from the dynamical
equations of laser theory. In these cases, the laser frequency noise
cannot simply be inferred by the laser linewidth and simplified models
fail to provide an accurate description of the laser's phase and amplitude
noise. Lasers exhibit generally significant excess noise for frequencies
below the relaxation oscillation frequency, which can differ strongly,
from the k$\operatorname{Hz}-\operatorname{MHz}$ range in Nd:YAG
or fiber lasers to several $\operatorname{GHz}$ in the case of diode
lasers \cite{Vahala1982,Wieman1991,Kippenberg2012}. This noise will
lead to radiation pressure fluctuations that heat the mechanical oscillator,
as has been discussed in \cite{Schliesser2008,Diosi2008,Rabl2009},
Moreover, phase noise impacts state transfer and entanglement generation
\cite{Simon2011,Vitali2011} and can impact sideband asymmetry measurements.
For a more detailed discussion we refer the reader to Appendix \ref{Appendix:LaserNoiseSidebandThermometry}.

\paragraph{Cavity frequency noise}

In addition to laser phase noise - which in principle can be mitigated
by properly filtered laser systems - a more fundamental limit to dynamical
backaction laser cooling arises from the fact that at finite tempererature,
the cavity frequency will exhibit thermodynamical fluctuations \cite{Gorodetsky2004}.
These are due to local temperature fluctuations that affect the dielectric
properties of the cavity itself (for microtoroids, microspheres, or
photonic crystals) or of the mirror. For more details, we refer to
Appendix \ref{Appendix:InfluenceOfCavityFrequencyNoise}.

\subsection{Strong coupling regime}

\label{sub:CoolingStrongCoupling}

We now discuss what happens when the laser power $P$ is increased
further. At first, this will just improve cooling, since $\Gamma_{{\rm opt}}\propto P$.
However, as we will see in the next section, qualitatively new features
start to appear when $\Gamma_{{\rm opt}}\sim\kappa$, or equivalently
when $g\sim\kappa$. This regime is referred to as the strong coupling
regime , where the driven optical mode and the mechanical mode hybridize
to form two new modes, with a splitting set by $2g$. Furthermore,
even for lower laser drive powers interesting features in the transmission
spectrum of the cavity appear if it is probed weakly in the presence
of a strong drive. This will be the phenomenon of optomechanically-induced
transparency, discussed in Sec.~\ref{sub:CoolingOptomechanicallyInducedTransparency}.

\subsubsection{Optomechanical normal-mode splitting}

\label{sub:CoolingStrongCouplingNormalModeSplitting}

The strong coupling regime is discussed most easily if we assume the
non-dissipative part of the Hamiltonian to dominate all decay channels,
i.e. $g\gg\kappa,\,\Gamma_{{\rm m}}$. In that case, we can just consider
the following part of the linearized Hamiltonian,

\begin{equation}
\hat{H}=-\hbar\Delta\delta\hat{a}^{\dagger}\delta\hat{a}+\hbar\Omega_{{\rm m}}\hat{b}^{\dagger}\hat{b}-\hbar g(\delta\hat{a}^{\dagger}+\delta\hat{a})(\hat{b}+\hat{b}^{\dagger})\,.
\end{equation}
In the most interesting red-detuned regime, where $\Delta\approx-\Omega_{{\rm m}}$,
we can even start our discussion employing the rotating-wave-approximation
for the coupling, $-\hbar g(\delta\hat{a}^{\dagger}\hat{b}+\delta\hat{a}\hat{b}^{\dagger})$,
which is the beam-splitter Hamiltonian of Eq.~(\ref{eq:BeamSplitterInteraction}).
The Hamiltonian of these two coupled oscillators is then easily diagonalized
by going over to the two eigenmodes. These modes now represent excitations
that are hybrids between the mechanical oscillations ($\hat{b}$)
and the fluctuations of the driven cavity mode ($\delta\hat{a}$)
around the strong coherent amplitude. Their eigenfrequencies are:

\begin{equation}
\omega_{\pm}=\frac{\Omega_{{\rm m}}-\Delta}{2}\pm\sqrt{g^{2}+\left(\frac{\Omega_{{\rm m}}+\Delta}{2}\right)^{2}}\,.\label{eq:StrongCouplingEigenFrequencies}
\end{equation}
In particular, right at resonance $\Delta=-\Omega_{{\rm m}}$, one
observes an avoided crossing, with a splitting of $\omega_{+}-\omega_{-}=2g$
between the two excitation branches. At this point, the eigenmodes
are symmetric and antisymmetric superpositions of light and mechanics,
with new annihilation operators $(\delta\hat{a}\pm\hat{b})/\sqrt{2}$.
Far from resonance, one recovers the two bare frequencies $-\Delta$
and $\Omega_{{\rm m}}$, and the excitations become again of purely
optical and mechanical nature, respectively.

\begin{figure}
\includegraphics[width=1\columnwidth]{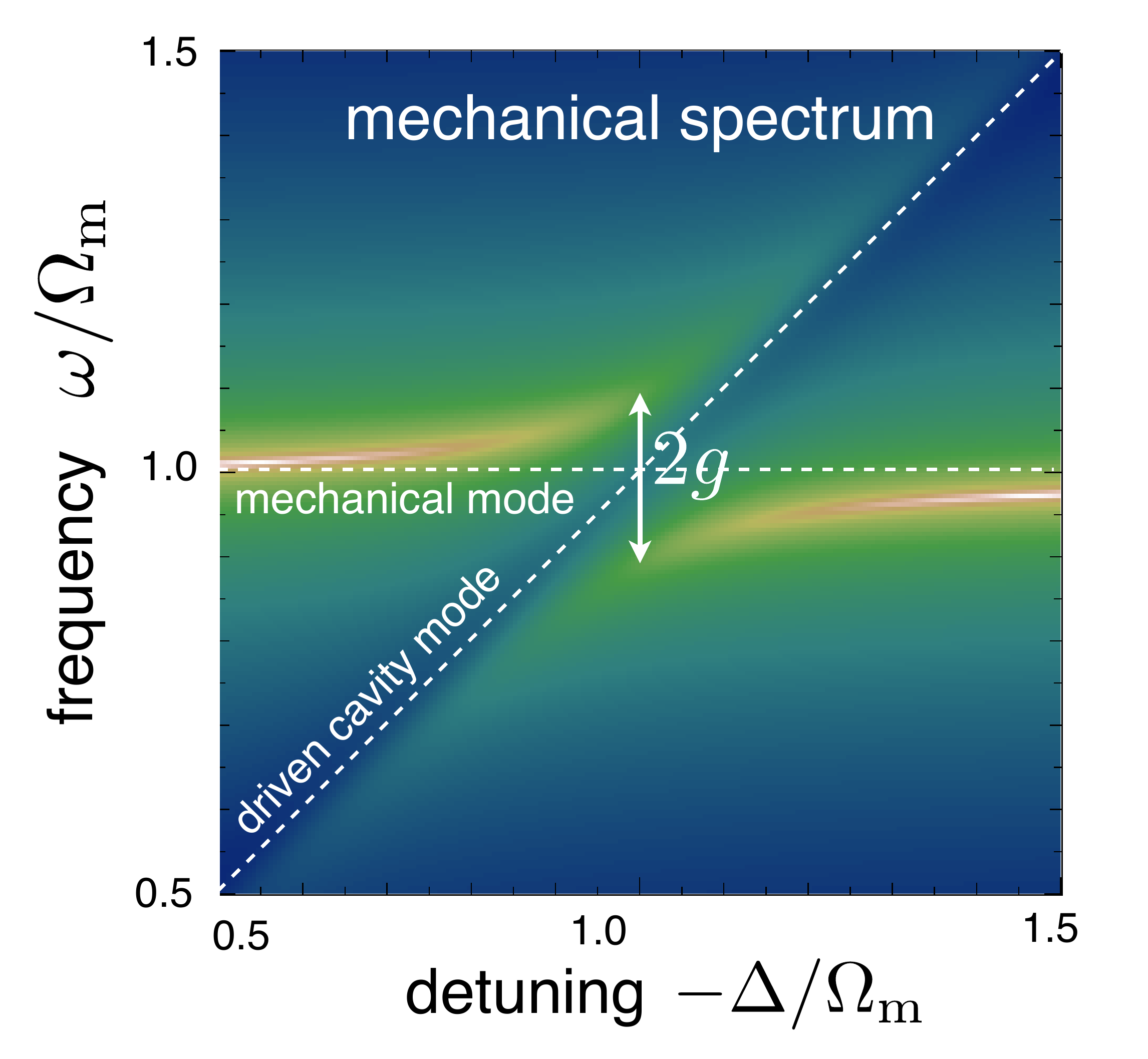}

\caption{\label{fig:StrongCouplingAvoidedLevelCrossing}Mechanical frequency
spectrum (frequency on vertical axis) as a function of laser detuning,
for a strongly coupled optomechanical system. An avoided crossing,
with a splitting of size $2g$, appears when the negative detuning
equals the mechanical resonance frequency. This is due to the hybridization
of the mechanical mode (frequency $\Omega_{{\rm m}}$) and the driven
cavity mode (effective frequency $-\Delta$).}
\end{figure}

This is the picture appropriate for $g\gg\kappa$, where we assume
that $\kappa\gg\Gamma_{{\rm m}}$ is the dominant decay channel. In
the opposite case $g\ll\kappa$ (assumed in the previous sections),
the two peaks at $\omega_{\pm}$ merge and the avoided crossing cannot
be observed. 

In principle, the complete scenario, including the decay channels
and the transition into the strong-coupling regime, is described fully
by the solution of the linearized equations of motion, Eqs.~(\ref{eq:EqsMotionLinQuantum-a})
and (\ref{eq:EqsMotionLinQuantum-b}). As discussed in Sec.~\ref{sub:DynamicalBackactionBasics},
one can solve these coupled equations analytically. In this way, for
example, one arrives at an exact expression for the mechanical susceptibility,
Eq.~(\ref{eq:OpticallyModifiedSusceptibility}). When plotting this
vs. frequency for increasing values of $g$, one observes peaks at
$\omega_{\pm}$ that can be clearly resolved for $g\gg\kappa$. The
evolution of the mechanical spectrum in the strong coupling regime
$g>\kappa/4$ as a function of laser detuning $\Delta$ is displayed
in Fig.~\ref{fig:StrongCouplingAvoidedLevelCrossing}. The same kind
of analysis also applies to the transmission spectrum of the cavity
that can be expressed via the same solution. 

However, instead of referring to these rather lengthy exact expressions,
we can simplify things by considering the regime $\Delta\approx-\Omega_{{\rm m}}$,
which allows us to perform the rotating-wave approximation already
employed above. We can then write down the linearized equations of
motion for the mean values:

\begin{equation}
\left(\begin{array}{c}
\left\langle \delta\dot{\hat{a}}\right\rangle \\
\left\langle \dot{\hat{b}}\right\rangle 
\end{array}\right)=-i\left(\begin{array}{cc}
-\Delta-i\frac{\kappa}{2} & -g\\
-g & \Omega_{{\rm m}}-i\frac{\Gamma_{{\rm m}}}{2}
\end{array}\right)\left(\begin{array}{c}
\left\langle \delta\hat{a}\right\rangle \\
\left\langle \hat{b}\right\rangle 
\end{array}\right)\,.
\end{equation}
Solving for the complex eigenvalues of this non-Hermitian matrix,
we recover the expression for the two branches $\omega_{\pm}$ given
in Eq.~(\ref{eq:StrongCouplingEigenFrequencies}) above, except with
the replacements $\Delta\mapsto\Delta+i\kappa/2$ and $\Omega_{{\rm m}}\mapsto\Omega_{{\rm m}}-i\Gamma_{{\rm m}}/2$.
With $\delta\equiv-\Delta-\Omega_{{\rm m}}$, we have

\begin{align*}
\omega_{\pm} & =\Omega_{{\rm m}}+\frac{\delta}{2}-i\frac{\kappa+\Gamma_{{\rm m}}}{4}\\
 & \pm\sqrt{g^{2}+\left(\frac{\delta+i(\Gamma_{{\rm m}}-\kappa)/2}{2}\right)^{2}}
\end{align*}
In particular, at resonance $\Delta=-\Omega_{{\rm m}}$ ($\delta=0$)
we find $\omega_{\pm}=\Omega_{{\rm m}}-i\frac{\kappa}{4}\pm\sqrt{g^{2}-\left(\frac{\kappa}{4}\right)^{2}}$,
where we assumed $\kappa\gg\Gamma_{{\rm m}}$ to slightly simplify
this formula. Thus, the eigenfrequencies change character at the threshold
$g=\kappa/4$, where the root changes from purely imaginary ($g<\kappa/4$)
to real-valued ($g>\kappa/4$). This corresponds to the transition
into the strong-coupling regime, with two well-resolved peaks. Each
of those peaks is of width (FWHM) $\kappa/2$. This is because both
of these excitations are half optical, and so each of them shares
half of the optical decay rate. Away from the degeneracy point $\Delta=-\Omega_{{\rm m}}$,
the relative contributions of the two decay channels $\Gamma_{{\rm m}}$
and $\kappa$ get re-weighted according to the unequal distribution
of optical and mechanical excitation in the two branches. 

\begin{figure}[ptb]
 \centering\includegraphics[width=3in]{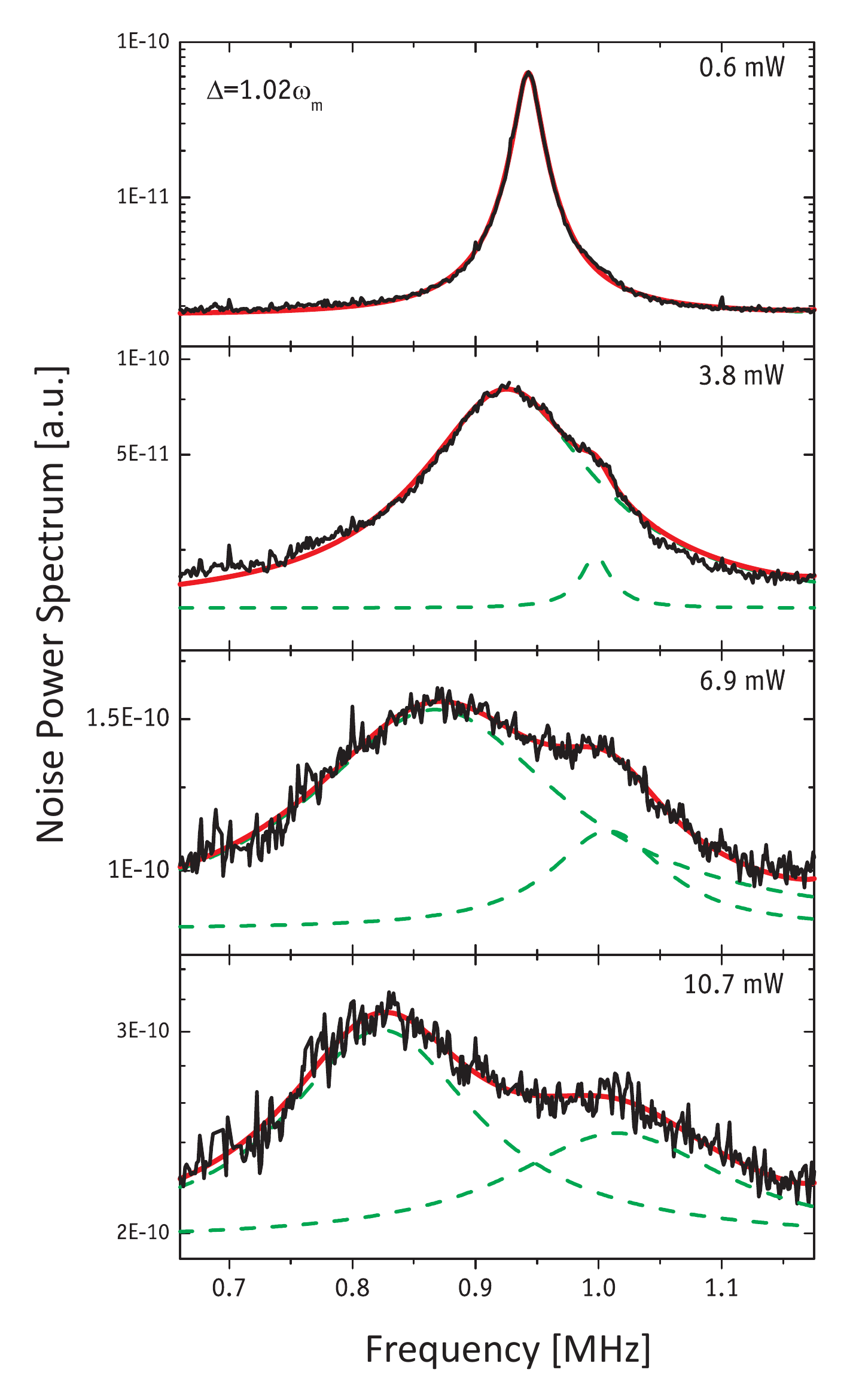}\caption{\label{fig:NormalModeSplittingExperiment}Experimental observation
of strong coupling of a mechanical oscillator to a light field ({}``parametric
normal mode splitting''). From top to bottom, the light intensity
and thereby $g=g_{0}\sqrt{\bar{n}_{{\rm cav}}}$ is increased. Data
from the experiment reported in \cite{Groblacher2009a}. .}
\end{figure}

It turns out that cooling becomes less efficient when one approaches
the strong coupling regime. In fact, the exact solution of the linearized
equations of motion can be employed to derive the appropriate modification
to the formulas for the final occupany in the case of strong optomechanical
coupling:

\begin{equation}
\bar{n}_{f}=\bar{n}_{f}^{(0)}+\bar{n}_{{\rm th}}\frac{\Gamma_{{\rm m}}}{\kappa}+2\bar{n}_{{\rm min}}\frac{\Gamma_{{\rm opt}}}{\kappa}\,.
\end{equation}
Here $\bar{n}_{f}^{(0)}$ is the standard result (\ref{eq:FinalPhononNumberWeightedAverage})
derived above for the weak-coupling regime $g\ll\kappa$, and to simplify
the expression we assumed the resolved-sideband regime $\kappa\ll\Omega_{{\rm m}}$
where $\bar{n}_{{\rm min}}=(\kappa/4\Omega_{{\rm m}})^{2}$ and $\Gamma_{{\rm opt}}=4g^{2}/\kappa$,
as well as strong cooling $\Gamma_{{\rm opt}}\gg\Gamma_{{\rm m}}$,
neglecting terms of still higher orders in $\kappa/\Omega_{{\rm m}}$. 

The peak splitting in the strong coupling regime ($g\gg\kappa,\,\Gamma_{{\rm m}}$)
and the resulting modification to cooling (notably the limitation
arising from the finite cavity decay rate) was predicted in \cite{Marquardt2007}.
It was analyzed extensively in \cite{Dobrindt2008,Wilson-Rae2008a};
see also \cite{Marquardt2008} and the supplementary material of \cite{Groblacher2009a}.
More generally, the full expression \cite{Dobrindt2008,Wilson-Rae2008a}
for $\bar{n}_{f}$ can be given, including any possible thermal occupation
of the cavity field (which has been observed in microwave setups):

\begin{align}
\bar{n}_{f} & =\bar{n}_{{\rm th}}\frac{\Gamma_{m}}{\kappa}\frac{4g^{2}+\kappa^{2}}{4g^{2}+\Gamma_{m}\kappa}+\frac{4g^{2}}{4g^{2}+\Gamma_{m}\kappa}\bar{n}_{{\rm cav}}^{{\rm th}}\nonumber \\
 & +\bar{n}_{{\rm th}}\frac{\Gamma_{m}}{\kappa}\frac{\kappa^{2}}{4\Omega_{m}^{2}}+\left(\bar{n}_{{\rm cav}}^{{\rm th}}+\frac{1}{2}\right)\frac{\kappa^{2}+8g^{2}}{8\Omega_{m}^{2}}\,.\label{eqn:nf-1}
\end{align}

The first experimental observation of strong optomechanical coupling
($g>\kappa,\,\Gamma_{{\rm m}}$) was reported in \cite{Groblacher2009a}.
Subsequently, experiments on other setups have been able to achieve
significantly larger ratios $g/\kappa$, see \cite{2011_Teufel_StrongCouplingMicrowave}.
In addition, the regime of $g>\kappa,\,\Gamma_{{\rm m}}\bar{n}_{{\rm th}}$
(see below) was reached as well \cite{2011_Teufel_StrongCouplingMicrowave,Teufel2011,Verhagena},
where coherent quantum state transfer between light and mechanics
could take place (see Sec.~\ref{sub:QuantumOptomechanicsQuantumProtocols}).

The photon statistics of an optomechanical system in the strong coupling
regime have been found to display interesting antibunching behaviour
and photon correlations \cite{2010_Agarwal_NormalModeSplitting_Antibunching},
and the same analysis discussed four-wave mixing in a setup driven
by a strong pump field and another ({}``Stokes'') field. One may
also consider having a nonlinear medium inside the optical cavity.
In \cite{2009_Agarwal_NormalModeSplitting_ParampCavity}, the effects
of an optical parametric amplifier cavity on the phenomenon of optomechanical
normal mode splitting have been studied.

\begin{figure}[ptb]
 \centering\includegraphics[width=4in]{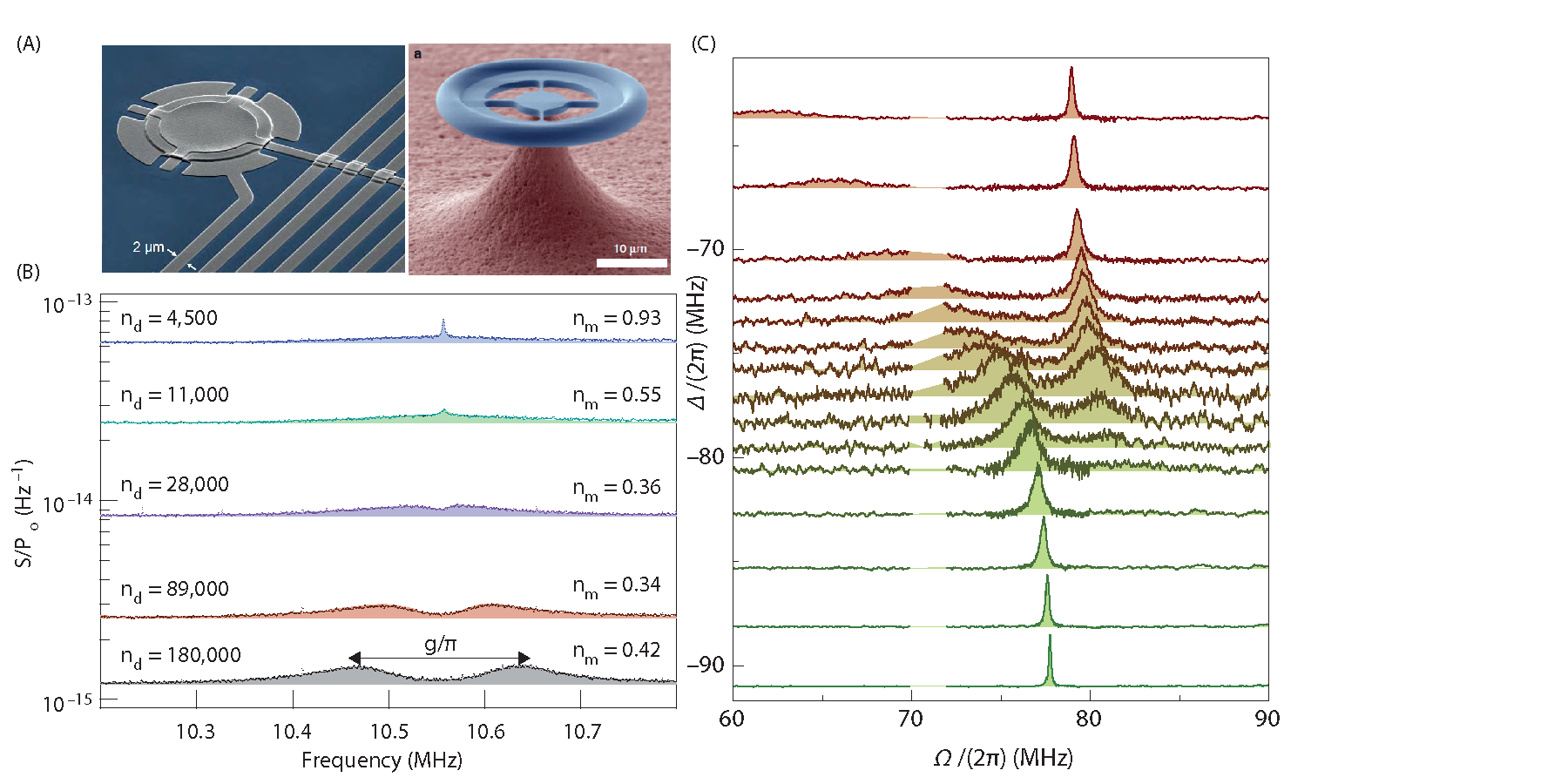}\caption{\label{fig:CoherentCouplingNMS}Experimental observation of the coherent
coupling regime, i.e. $g>\{\kappa,\bar{n}_{m}\Gamma_{m}\}$, between
a mechanical oscillator and a microwave cavity mode as well as an
optical cavity mode. The data shows the splitting in the mechanical
displacement spectrum as a function of increasing drive power (left
panel, for the case of a superconducting drum resonator \cite{2011_Teufel_StrongCouplingMicrowave},
courtesy J.~Teufel) and as a function of the laser detuning around
the lower sideband (right panel, for the case of a spoke supported
microtoroid resonator \cite{Verhagena}, courtesy T. J. Kippenberg).}
\end{figure}

The spectroscopic signatures of strong coupling indicate whether the
coupling exceeds the cavity decay rate. If the coupling rate exceeds
also the thermal decoherence rate ($\gamma=\Gamma_{n}\bar{n}_{m}$)
the interaction between the mechanical oscillator and the light field
becomes quantum coherent, i.e. the timescale of the mutual coupling
is faster than the timescale for one quantum of noise to enter from
the environment:

\[
g>\{\Gamma_{n}\bar{n}_{m},\kappa\}
\]
This parameter regime of coherent coupling is a precondition for many
quantum protocols such as quantum state transfer between the cavity
field and a mechanical mode (see section \ref{sub:QuantumOptomechanicsQuantumProtocols}).
Experimentally, this regime has been reached using a superconducting
membrane coupled to an LC circuit \cite{Teufel2011}, and in the optical
domain using a toroidal spoke-supported microresonator \cite{Verhagena}
(cf. Figure\ref{fig:CoherentCouplingNMS} ).

\subsubsection{Optomechanically induced transparency}

\label{sub:CoolingOptomechanicallyInducedTransparency}

\begin{figure}
\includegraphics[width=1\columnwidth]{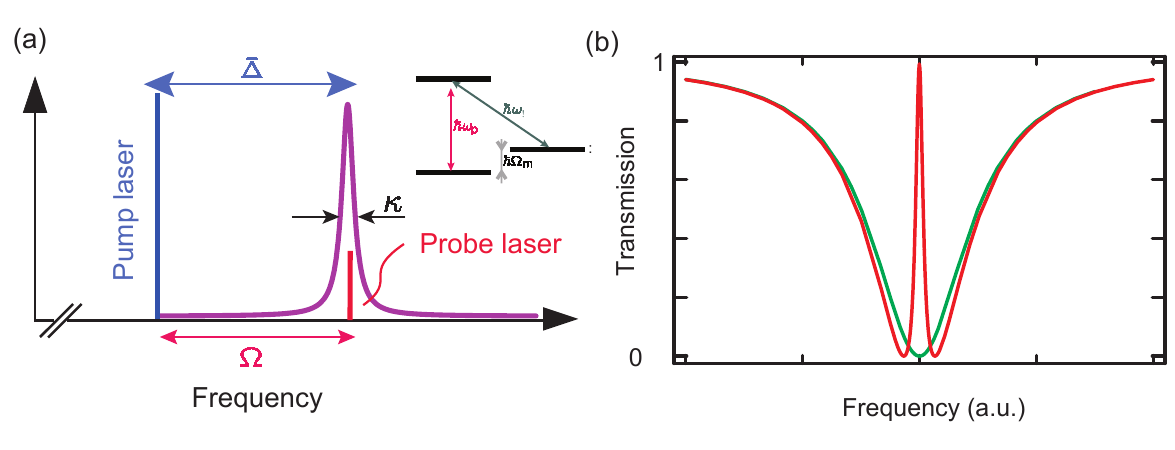}

\caption{(a) Level scheme for optomechanically induced transparency. A strong
control laser on the red sideband drives the optomechanical system,
while a weak probe laser scans across the cavity resonance and takes
the (b) spectrum of the driven system, which displays a sharp transparency
feature (for a side-coupled toroid setup). The linewidth of that feature
is given by the mechanical linewidth (at weak driving), much narrower
than the cavity linewidth. Adapted from \cite{Schliesser2010}.}
\end{figure}

Electromagnetically induced transparency \cite{Fleischhauer2005}
is a phenomenon which occurs in multi-level atoms and manifests itself
as a cancellation of absorption in the presence of an auxiliary laser
field. It arises from electronic interference or, in an equivalent
picture, is due to a dark-state resonance of the excited state. This
phenomenon has been demonstrated for cold atomic ensembles, giving
rise to a host of phenomena, ranging from optical pulse storage to
slowing or advancing of light pulses. An analogous phenomenon also
occurs in optomechanical systems, as predicted theoretically in \cite{2009_Schliesser_Thesis,2010_Agarwal_OMIT}
and analyzed further for optical pulse storage \cite{2011_ChangPainter_SlowingStoppingLightArray}.
Optomechanically induced transparency has been observed in experiments\cite{2010_Weis_ScienceOMIT,2011_Painter_EITOptomechanics}.
Injecting a strong control laser beam into the lower (red-detuned)
sideband of an optomechanical system, the optomechanical interaction
causes the cavity resonance, as seen by a second, weak probe laser
field to be rendered transparent. The simultaneous presence of a strong
control laser ($\bar{s}e^{-i\omega_{c}t}$) and a weak probe laser
($\delta s\cdot e^{-i\omega_{p}t}$) lead to a transmission $|t_{p}|^{2}$
of the weak probe laser given by:

\begin{equation}
|t_{p}|^{2}=\left|1-\eta\kappa\frac{\chi_{aa}(\Omega)}{1+g_{0}^{2}\bar{a}^{2}\chi_{{\rm mech}}(\Omega)\chi_{aa}(\Omega)}\right|^{2}\label{eq:OMITgrandFormula}
\end{equation}
Here $\Delta=\omega_{c}-\omega_{{\rm cav}}$ denotes the detuning
of the strong control field from the cavity resonance $\omega_{{\rm cav}}$,
and $\Omega$ denotes the detuning between the control laser and probe
laser, i.e. $\Omega=\omega_{p}-\omega_{c}$. Moreover, the coupling
efficiency $\eta=\kappa_{{\rm ex}}/\kappa$ has been introduced, and
the mechanical susceptibility%
\footnote{Here it is convenient to adopt a definition different from $\chi_{xx}$
of Sec.~\ref{sub:MechanicalResonatorsNoiseSpectra}.%
} $\chi_{{\rm mech}}^{-1}(\Omega)=-i(\Omega-\Omega_{m})+\Gamma_{m}/2$
as well as the optical susceptibility $\chi_{aa}^{-1}(\Omega)=-i(\Omega+\bar{\Delta})+\kappa/2$.
Note that, with regard to the transmission, we used the terminology
appropriate for a waveguide-coupled unidirectional cavity (e.g. whispering
gallery mode resonator), and we follow the discussion to be found
in the supplementary material of \cite{2010_Weis_ScienceOMIT}. Plotting
the expression $|t_{p}|^{2}$ reveals that when the resonance condition
$\Omega\approx\Omega_{m}$ is met, a transparency window arises. When
the coupling laser is placed on the lower sideband ($\Delta=-\Omega_{{\rm m}}$),
the expression for the transmission of the probe in the vicinity of
$\Omega=\Omega_{{\rm m}}$ (which corresponds to $\omega_{p}\approx\omega_{{\rm cav}}$)
reduces to:

\[
|t_{p}|^{2}=\left|\frac{4\bar{n}_{cav}g_{0}^{2}}{4\bar{n}_{cav}g_{0}^{2}+\Gamma_{m}\kappa-2i(\Omega-\Omega_{{\rm m}})\kappa}\right|^{2},
\]
where for simplicity we assumed critical coupling, i.e. $\eta=1/2$.
Evaluating the above expression for $\Omega=\Omega_{m}$ yields:

\[
\left|t_{p}\right|^{2}=\left(\frac{C}{C+1}\right)^{2}
\]
Here $C$ is the optomechanical cooperativity: $C=4g_{0}^{2}\bar{n}_{cav}/(\kappa\Gamma_{m})$.
Thus a cooperativity of unity is required to change the transmission
to 50\%. It is noted that the assumption underlying these theoretical
considerations, of two coherent drive fields, requires from an experimental
point of view two laser fields that are coherent for times longer
than the effective mechanical damping time.

The physical origin of the transparency window can be understood by
realizing that the beat of the probe field and the coupling laser
induces a time varying radiation pressure force. If the beat frequency
matches the mechanical oscillation frequency, then the mechanical
oscillator is driven resonantly. The driven oscillator in turn creates
sidebands on the intracavity field. Considering the strong coupling
laser only, in the resolved sideband limit, the lower motional sideband
is far off cavity resonance and can be neglected. In contrast, the
upper sideband of the coupling laser, created by the mechanical motion,
has precisely the same frequency as the probe field and is moreover
phase-coherent with the probe field. This leads to an interference
that yields a cancellation of the intracavity field on resonance,
giving rise to the transparency window. The phenomenon thereby results
from the destructive interference between reflection amplitudes for
photons scattered from the coupling laser and photons of the probe
field. 

The width of the (Lorentzian) transparency feature in the weak coupling
regime (where $\Gamma_{eff}\ll\kappa$) is given by the total effective
mechanical damping rate:

\[
\Gamma_{{\rm OMIT}}=\Gamma_{m}+4g_{0}^{2}\bar{n}_{cav}/\kappa=\Gamma_{{\rm eff}}\,.
\]
We note that at stronger drive this feature smoothly evolves into
the normal mode splitting discussed above (Sec.~\ref{sub:CoolingStrongCouplingNormalModeSplitting}).
The narrow transparency window is concomitant with a rapid variation
of the transmission phase of the probe beam. This implies that a pulse
with a bandwidth smaller than $\Gamma_{{\rm OMIT}}$ will experience
a group delay (without distortion of the pulse). The change of the
phase of the transmitted light is given by the expression (for $\Delta=-\Omega_{{\rm m}}$):

\[
\phi=\arctan\left(\frac{2(\Omega-\Omega_{{\rm m}})\kappa}{4\bar{n}_{cav}g_{0}^{2}+\Gamma_{m}\kappa}\right)
\]
Evaluating this expression for zero detuning ($\Omega=\Omega_{{\rm m}}$)
leads to a group delay of

\[
\tau_{g}=\frac{d\phi}{d\Omega}=\frac{1}{\Gamma_{m}}\frac{2}{C+1}=\frac{2}{\Gamma_{OMIT}}
\]
However, the bandwidth of the pulse needs to be smaller than the transparency
window for undistorted pulse propagation, which limits the delay-bandwidth
product to $\tau_{g}\Gamma_{{\rm OMIT}}\approx2$. Using an array
of optomechanical systems, the delay-bandwidth product can be increased
and may therefore serve as a way to store an optical waveform in long
lived phonons. Such cascaded optomechanical systems can be realized
using e.g. photonic crystals \cite{2011_ChangPainter_SlowingStoppingLightArray}.

\begin{figure}
\includegraphics[width=1\columnwidth]{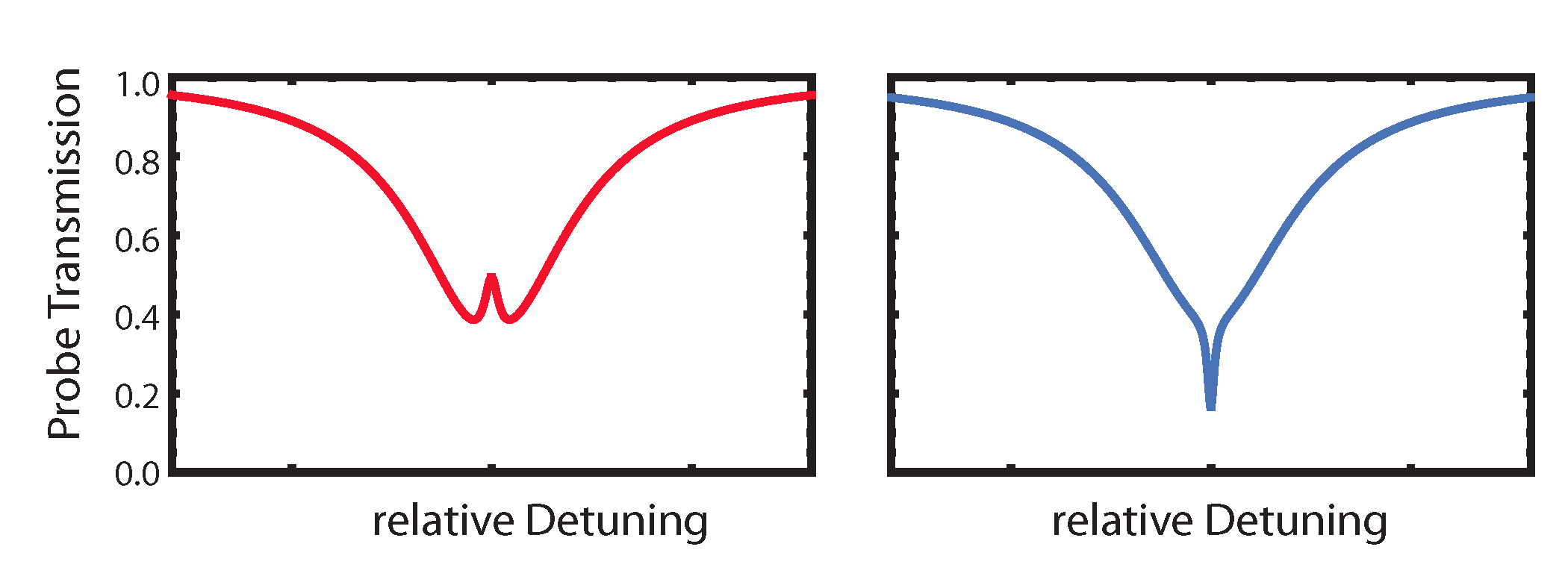}

\caption{Transmission of the probe beam in the presence of a strong control
field on the lower sideband (left panel, optomechanically induced
transparency, OMIT) and on the upper sideband (right panel, electromagnetically
induced absorption, EIA). }
\end{figure}

\subsubsection{Optomechanically induced amplification and electromagnetically induced
absorption }

If instead we consider the control laser being injected on the upper
sideband ($\Delta=\Omega_{m}$), an additional optical signal in the
probe beam is amplified \cite{Massel2011a,2011_Painter_EITOptomechanics}.
The analogous effect in atomic physics is referred to as electromagnetically
induced absorption \cite{Lezema1999} (EIA). The amplification process
can parametrically amplify a small signal (provided the resonance
condition is met), by virtue of the constructive interference of the
light scattered from the pump (control) to the signal (probe) frequency,
in direct analogy to the above phenomenon of optomechanically induced
transparency. Theoretically the phenomenon can be described by the
same equations as the effect of transparency, except for the fact
that now the mechanical damping is reducing with increasing power
on the upper sideband. The maximum gain is set by the maximum power
which can be injected onto the upper sideband ($\Delta=+\Omega_{m}$
), which is limited by the onset of the parametric oscillatory instability,
in which the coherent amplification of mechanical motion from the
noise occurs. The maximum average gain in this case is given by by
$G_{av}(\Delta=0)\approx4(4\Omega_{m}/\kappa)^{2}$\cite{Massel2011a}.
As for any non-degenerate parametric amplifier, the amplification
process has to add half a quantum of noise for fundamental reasons
\cite{Clerk2008a}, and the total added noise is given by $n_{add}=\bar{n}_{m}+1/2\approx k_{B}T/\hbar\Omega_{m}$
in the presence of thermal fluctuations. In the ideal case the relative
phase between pump and signal is not important for the EIA process.
However, any relative phase fluctuations between pump and signal need
to take place on a timescale long compared to the inverse effective
mechanical damping rate.

However, this optical amplification process does not lead to a \textit{stimulated}
optical amplification process, as in the case of optical Brillouin
scattering. The reason is that unlike the optical Brillouin scattering
case, the optical dissipation is \textit{larger} than the mechanical
dissipation ($\kappa\gg\Gamma_{m}$ ). This implies that the mechanical
mode can experience exponential buildup, while in a Brillouin laser
the opposite is the case%
\footnote{In the EIA scenario a stimulated optical amplification is expected
(and subsequent optomechanical Brillouin lasing) when the mechanical
oscillator is more strongly damped than the optical one, i.e when
the hierarchy $\Gamma_{m}\gg\kappa$ of dissipation is satisfied.
This ensures exponential amplification of the optical field above
the lasing threshold and prevents the mechanical mode from building
up significantly.%
}. 

\begin{figure}
\includegraphics[width=1\columnwidth]{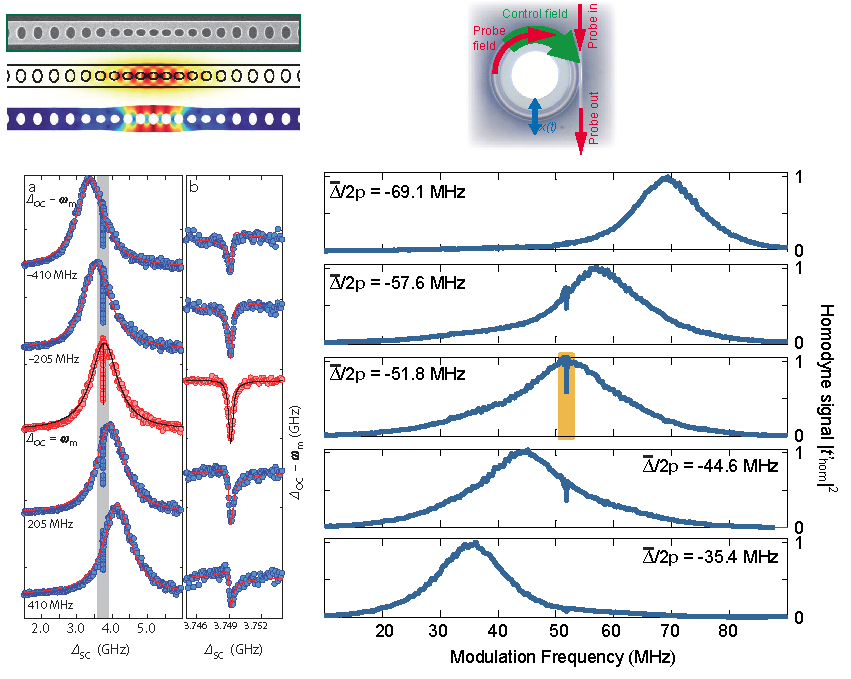}

\caption{\label{fig:OMIT}Optomechanically induced transparency observed in
the experiment. Data from a microtoroid setup \cite{2010_Weis_ScienceOMIT}
(left panel) and in photonic crystals \cite{2011_Painter_EITOptomechanics}
(right panel). Note that in the case of a photonic crystal the OMIT
signature is visible both in the reflected and transmitted signal. }
\end{figure}

\section{Classical Nonlinear Dynamics}

\label{sec:NonlinearDynamics}

\label{sec:VIII}

\begin{figure}
\includegraphics[width=1\columnwidth]{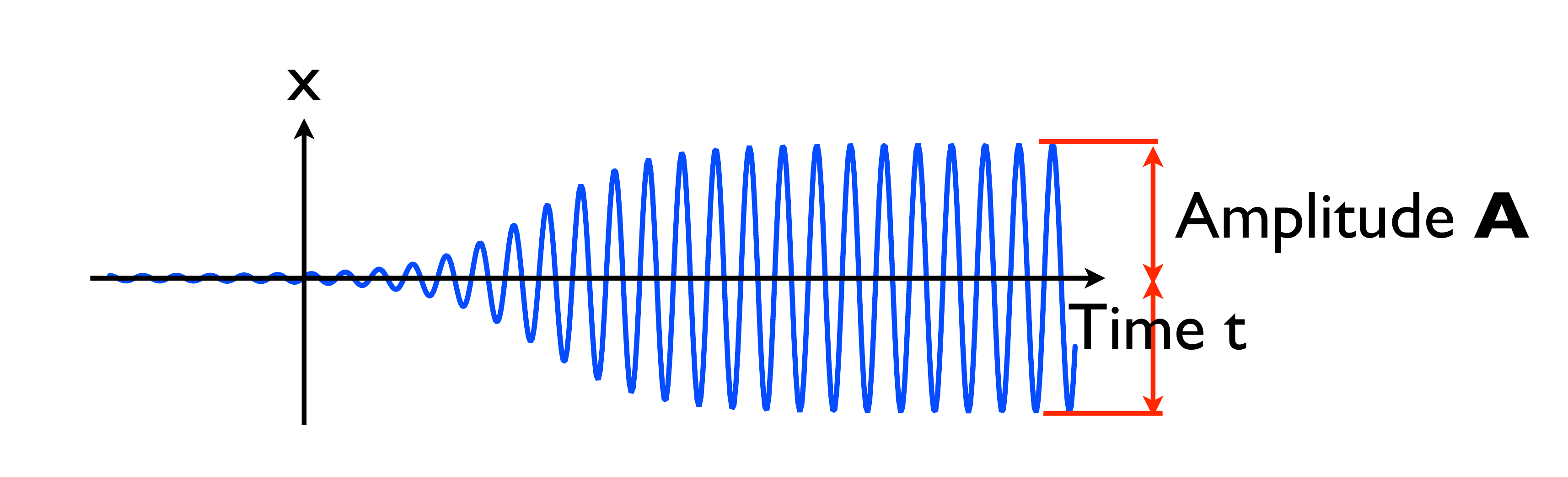}

\caption{\label{fig:InstabilityGrowth}Optomechanical parametric instability
towards {}``self-induced oscillations'' ({}``mechanical lasing''):
displacement $x$ vs. time $t$. In a system with a sufficiently strong
blue-detuned laser drive, the mechanical oscillations can display
anti-damping ($\Gamma_{{\rm eff}}<0$). This leads first to exponential
growth of any initial fluctuations, which then finally saturates due
to nonlinear effects, resulting in self-sustained mechanical oscillations
at a stable amplitude $A$.}
\end{figure}

Up to now we have mostly discussed effects that can be fully understood
within the linearized equations of motion, i.e. within the quadratic
approximation (\ref{eq:LinearizedInteraction}) to the optomechanical
Hamiltonian. However, the approximation itself can be used to predict
its breakdown: In the blue-detuned regime ($\Delta>0$), $\Gamma_{{\rm opt}}$
becomes negative, decreasing the overall damping rate. At first, this
leads to heating (instead of cooling), enhancing the oscillator's
effective temperature. Once the overall damping rate $\Gamma_{{\rm m}}+\Gamma_{{\rm opt}}$
becomes negative, an instability ensues. In that case, any tiny initial
(e.g. thermal) fluctuation will at first grow exponentially in time.
Later, nonlinear effects will saturate the growth of the mechanical
oscillation amplitude (Fig.~\ref{fig:InstabilityGrowth}). A steady-state
regime is reached, with oscillations proceeding at a fixed amplitude
$A$. These are called self-induced (back-action induced) optomechanical
oscillations. In fact, they are analogous to lasing action, but now
in a mechanical system and with the incoming laser radiation providing
the pump. Therefore, this optomechanical effect can also be understood
as one variant of {}``mechanical lasing'' or {}``phonon lasing''.
Just as for lasers/masers, the optomechanical system displays a threshold
and linewidth narrowing above threshold, described by a Schawlow-Townes
type limit. 

\subsection{Parametric instability and attractor diagram}

\label{sub:NonlinearDynamicsInstabilityAttractorDiagram}

In this section, we present the classical theory of the optomechanical
instability produced by radiation pressure backaction, following \cite{Marquardt2006,Ludwig2008a}.
Our main goal is to discuss the amplitude $A$ of the steady-state
mechanical oscillations, as a function of system parameters such as
laser power, detuning, and mechanical damping rate. We will find that
for a fixed set of parameters, $A$ can in general take on multiple
stable values, corresponding to several stable attractors of this
dynamical system. That effect is known as dynamical multistability,
and in experiments it may lead to hysteretic behaviour. Our discussion
will be directly applicable for the radiation-pressure induced parametric
instability. However, note that very similar physics and analogous
formulas apply for the case when this instability is induced by photothermal
forces. In fact, if the thermal decay time $\tau_{{\rm th}}$ is long
($\tau_{{\rm th}}^{-1}\ll\Omega_{{\rm m}}$), one can re-use most
parts of the discussion below, effectively only replacing $\kappa$
by $\tau_{{\rm th}}^{-1}$ (see \cite{Marquardt2006} for a discussion
of the differences).

The threshold of the instability can be obtained in a linear analysis
by demanding that

\begin{equation}
\Gamma_{{\rm m}}+\Gamma_{{\rm opt}}=0\,.\label{eq:ThresholdInstability}
\end{equation}
This will define the limits of an interval where $\Gamma_{{\rm m}}+\Gamma_{{\rm opt}}<0$,
i.e. where the system is unstable. This interval widens as the laser
power is increased.

A simple argument can now be used to obtain the amplitude $A$. We
start from the ansatz

\begin{equation}
x(t)=\bar{x}+A\cos(\Omega_{{\rm m}}t)\,\label{eq:AnsatzSelfInducedOscillations}
\end{equation}
for the self-induced oscillations. This is good for typical experimental
parameters, where $\Gamma_{{\rm m}},\Gamma_{{\rm opt}}\ll\Omega_{{\rm m}}$,
such that both damping and optomechanical effects only show up after
many oscillation periods. We will discuss the breakdown of this ansatz
in the following section on chaotic dynamics.

From Eq.~(\ref{eq:AnsatzSelfInducedOscillations}), one can obtain
the time-dependence of the radiation pressure force $F(t)$ (which
will depend on $A$ and $\bar{x}$). In steady-state, the time-averaged
power input due to this force, $\left\langle F\dot{x}\right\rangle $,
must equal the power lost due to friction, $m_{{\rm eff}}\Gamma_{{\rm m}}\left\langle \dot{x}^{2}\right\rangle $.
This can be recast into a condition resembling Eq.~(\ref{eq:ThresholdInstability}),
by defining an amplitude-dependent effective optomechanical damping
rate:

\begin{equation}
\Gamma_{{\rm opt}}(A)\equiv\frac{-\left\langle F\dot{x}\right\rangle }{m_{{\rm eff}}\left\langle \dot{x}^{2}\right\rangle }\,.
\end{equation}
In the low-amplitude limit $A\rightarrow0$, this reduces to the standard
definition of $\Gamma_{{\rm opt}}$ used up to now. Then the power
balance condition is simply

\begin{equation}
\Gamma_{{\rm m}}+\Gamma_{{\rm opt}}(A)=0\,,\label{eq:PowerBalanceDampingRates}
\end{equation}
which is an implicit equation for $A$. This strategy can be used
for arbitrary optomechanical systems, also containing more optical
modes or other types of radiation forces.

We need yet another condition, to fix the oscillation offset $\bar{x}$,
which is not identical with the unperturbed oscillator equilibrium
position. The time-averaged radiation pressure force deflects the
harmonic oscillator:

\begin{equation}
\left\langle F\right\rangle =m_{{\rm eff}}\Omega_{{\rm m}}^{2}\bar{x}\,.\label{eq:ForceBalance}
\end{equation}

In general, Eqs.~(\ref{eq:PowerBalanceDampingRates}) and (\ref{eq:ForceBalance})
need to be solved simultaneously for the unknowns $A$ and $\bar{x}$.
However, if $\Gamma_{{\rm m}}\ll\Omega_{{\rm m}}$, one can already
see the instability in a regime where the shift $\bar{x}$ is small
and can be neglected, such that only Eq.~(\ref{eq:PowerBalanceDampingRates})
is relevant.

We still have to obtain $F(t)$. This can be deduced by solving the
classical equation for the light field amplitude (Sec.~\ref{sub:OptomechanicalEquationsOfMotion}),

\begin{equation}
\dot{\alpha}=-\frac{\kappa}{2}(\alpha-\alpha_{{\rm max}})+i(\Delta+Gx(t))\alpha\,,
\end{equation}
where we defined $\alpha_{{\rm max}}$ to be the amplitude reached
inside the cavity right at resonance (in terms of Sec.~\ref{sub:OptomechanicalEquationsOfMotion},
we have $\alpha_{{\rm max}}=2\alpha_{{\rm in}}\sqrt{\kappa_{{\rm ex}}}/\kappa$).
After inserting the ansatz (\ref{eq:AnsatzSelfInducedOscillations}),
the solution \cite{Marquardt2006} can be written in a Fourier series
$\alpha(t)=e^{i\varphi(t)}\sum_{n}\alpha_{n}e^{in\Omega_{{\rm m}}t}$,
with coefficients

\begin{equation}
\alpha_{n}=\frac{\alpha_{{\rm max}}}{2}\frac{J_{n}(-\frac{GA}{\Omega_{{\rm m}}})}{in\frac{\Omega_{{\rm m}}}{\kappa}+\frac{1}{2}-i(G\bar{x}+\Delta)/\kappa}\,,
\end{equation}
where $J_{n}$ is the Bessel function of the first kind and the global
phase $\varphi(t)=(GA/\Omega_{{\rm m}})\sin(\Omega_{{\rm m}}t)$.
Now the force $F(t)=\hbar G\left|\alpha(t)\right|^{2}$ and the time
averages $\left\langle F\dot{x}\right\rangle $ and $\left\langle F\right\rangle $
can be calculated. One has $\left\langle \left|\alpha(t)\right|^{2}\right\rangle =\sum_{n}\left|\alpha_{n}\right|^{2}$
and $\left\langle \left|\alpha(t)\right|^{2}\dot{x}\right\rangle =A\Omega_{{\rm m}}{\rm Im}\sum_{n}\alpha_{n}^{*}\alpha_{n+1}$.
These series can be efficiently summed numerically, to obtain the
explicit dependence of Eqs.~(\ref{eq:PowerBalanceDampingRates})
and (\ref{eq:ForceBalance}) on $A,\,\bar{x}$ and the system parameters.

\begin{figure}[ptb]
 \centering\includegraphics[width=3in]{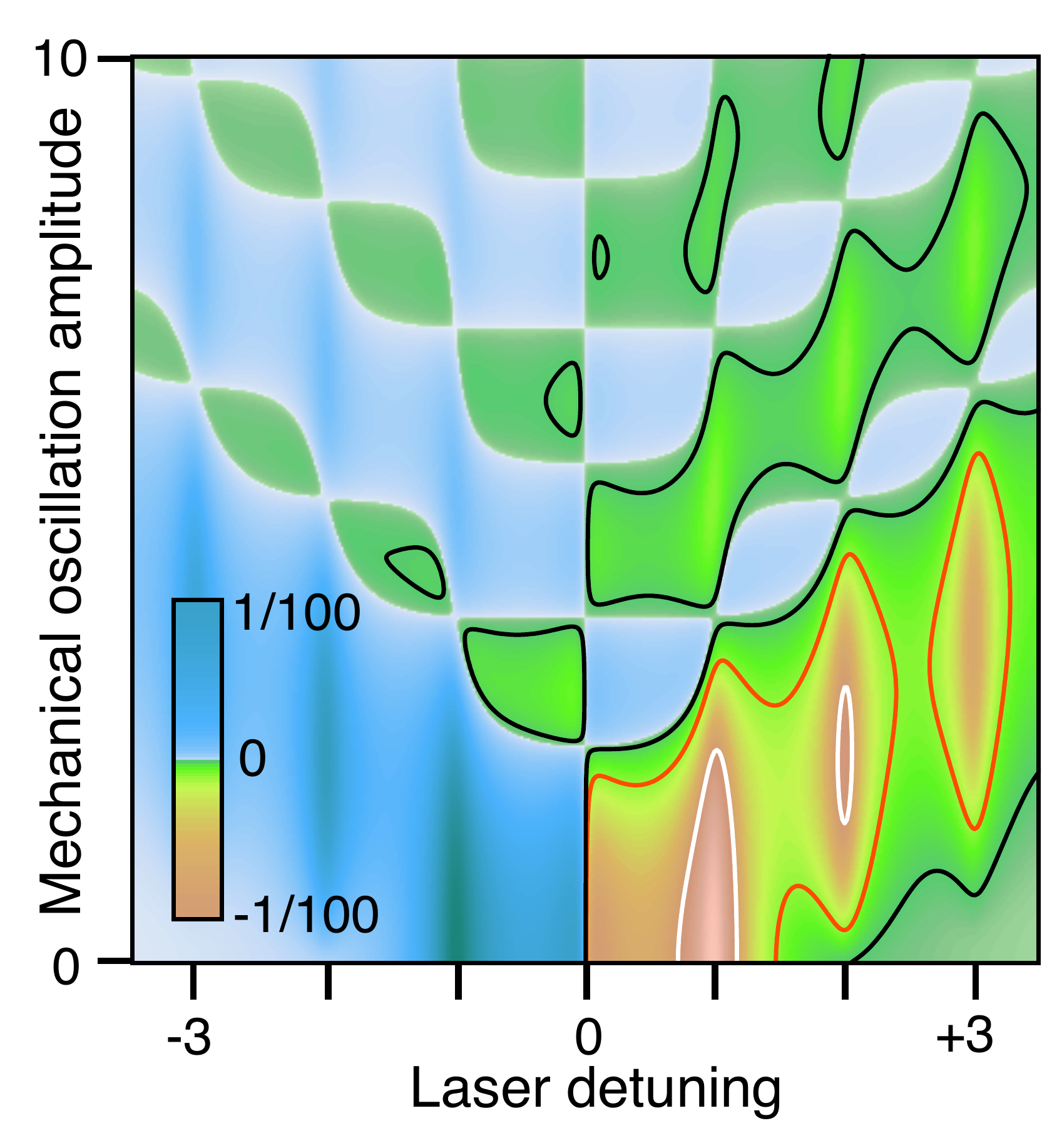}\caption{\label{fig:NonlinearAttractorDiagram}Attractor diagram for {}``mechanical
lasing'' in an optomechanical system. We show the damping rate $\Gamma_{{\rm opt}}$
as a function of both mechanical oscillation amplitude $A$ and laser
detuning $\Delta$, for a system in the resolved sideband regime ($\kappa/\Omega_{{\rm m}}=0.2$).
Contour lines of $\Gamma_{{\rm opt}}$ correspond to possible attractors.
For each line, only the upper part is stable, i.e. those points on
the curve where $\Gamma_{{\rm opt}}$ increases when going towards
higher amplitudes. The offset $\bar{x}$ has been assumed to be negligible.
(Plot shows $\Gamma_{{\rm opt}}m_{{\rm eff}}\Omega_{{\rm m}}^{2}/(2\hbar G^{2}\alpha_{{\rm max}}^{2})$
vs. $GA/\Omega_{{\rm m}}$ and $\Delta/\Omega_{{\rm m}}$; contours
shown at levels $-10^{-2}$/white, $-10^{-3}$/orange, $-10^{-4}$/black
)}
\end{figure}

The result is best discussed graphically. In Fig.~\ref{fig:NonlinearAttractorDiagram},
we show the attractor diagram, i.e. the possible amplitudes $A$ as
a function of any system parameter (in this case, the detuning $\Delta$).
This diagram can be generated by plotting the value of $\Gamma_{{\rm opt}}(A,\Delta)$
and then showing the contour lines $\Gamma_{{\rm opt}}=-\Gamma_{{\rm m}}$
that indicate possible attractors. Note that stable attractors are
only those where $|\Gamma_{{\rm opt}}|$ grows with $A$ (the upper
half of each line). In the case shown here ($\kappa/\Omega_{{\rm m}}=0.2$),
the structure of sidebands at $\Delta=n\Omega_{{\rm m}}$ both for
red and blue detuning shows clearly. Remarkably, one can have stable
self-induced oscillations even on the red-detuned side ($\Delta<0$),
but only for finite amplitude $A>0$. This is consistent with the
fact that the linearized theory there predicts cooling at $A\rightarrow0$.

Mathematically, the onset of small-amplitude oscillations, starting
from $A=0$, is an example of a Hopf bifurcation. In this regime,
$A\propto\sqrt{I-I_{{\rm th}}}$, where $I$ is any system parameter
(such as the detuning or the laser power), and $I_{{\rm th}}$ its
threshold value. 

An important feature is the dynamical multistability, i.e. the existence
of several stable solutions for a fixed set of external parameters.
This is observed for sufficiently good mechanical quality factors,
when higher-amplitude attractors become stable. It leads to hysteresis
in experiments, and might also be used for high-sensitivity {}``latching''
measurements. For more on this and the effects of noise and slow dynamics
of the amplitude, see \cite{Marquardt2006}. 

In experiments, an optomechanical instability due to retarded light
forces was first studied in a low-finesse setup with photothermal
forces \cite{HoehbergerProceedings2004}, where the retardation is
due to finite thermal conductivity and the theory described here applies
with appropriate modifications, see \cite{Marquardt2006,Ludwig2008}.
Subsequent studies of the photothermal setup observed parts of the
attractor diagram, confirmed dynamical bistability, and uncovered
a new regime where more than one mechanical mode gets involved in
the nonlinear dynamics \cite{Metzger2008}. 

The parametric instability driven by radiation pressure backaction,
as discussed here, was first demonstrated in a microtoroid setup \cite{Carmon2005,Kippenberg2005,Rokhsari2005a}.
The full attractor diagram still has to be observed in an experiment.

A recent experiment has demonstrated mechanical lasing (i.e. coherent
oscillations) in a setup where two optical modes are involved and
photon transitions between those modes provide the power to feed the
mechanical oscillations \cite{GrudininPhononLaser2010}. The attractor
diagram for the parametric instability in systems involving more than
one optical mode (including a 'membrane in the middle') will display
qualitatively new features due to the effects of optical Landau-Zener
dynamics \cite{HuaizhiWu2011}.

Just as in a laser, the phase of the self-induced oscillations is
arbitrary. Thus, external noise, including thermal Langevin forces
acting on the mechanical oscillator and radiation pressure shot noise,
will impart a slow phase diffusion. The effect of a force $\delta F$
on the phase scales inversely with the amplitude, $\delta\varphi\propto\delta F/A$,
as can be seen easily in a phase-space diagram. Thus, the diffusion
constant for the phase scales with $1/A^{2}$, diverging just above
threshold \cite{VahalaLinewidth2008}. This is the optomechanical
analogue of the Schawlow-Townes result for the linewidth of a laser
\cite{SchawlowTownes1958}. More precisely, the scenario is closer
to the case of a maser, since the thermal noise is not negligible
in the mechanical oscillator:

\begin{equation}
\Gamma_{{\rm m}}^{{\rm osc}}=\frac{\Gamma_{{\rm m}}}{\bar{n}}(2\bar{n}_{{\rm th}}+1)\,.
\end{equation}
A full discussion of the linewidth narrowing and phase diffusion in
optomechanical oscillations can be found in \cite{2010_Armour_PhaseDiffusion},
where the effects of photon shot noise are taken into account as well.

Up to now, we have discussed exclusively the classical dynamics. In
the quantum regime \cite{Ludwig2008a}, the parametric instability
threshold is broadened due to quantum fluctuations, with strong amplification
of fluctuations below threshold. The existence of attractors with
finite amplitude shows up, e.g., in phase-space plots of the Wigner
density, and it changes the phonon (as well as photon) statistics
\cite{2011_Qian_NegativeWignerDensity}.

Interesting new collective physics results if many such optomechanical
oscillators are coupled together. In particular, this can lead to
synchronization, which we will discuss in some more detail in the
section on multimode optomechanics (Sec.~\ref{sec:MultimodeOptomechanics}).

\begin{figure}[ptb]
\includegraphics[width=3in]{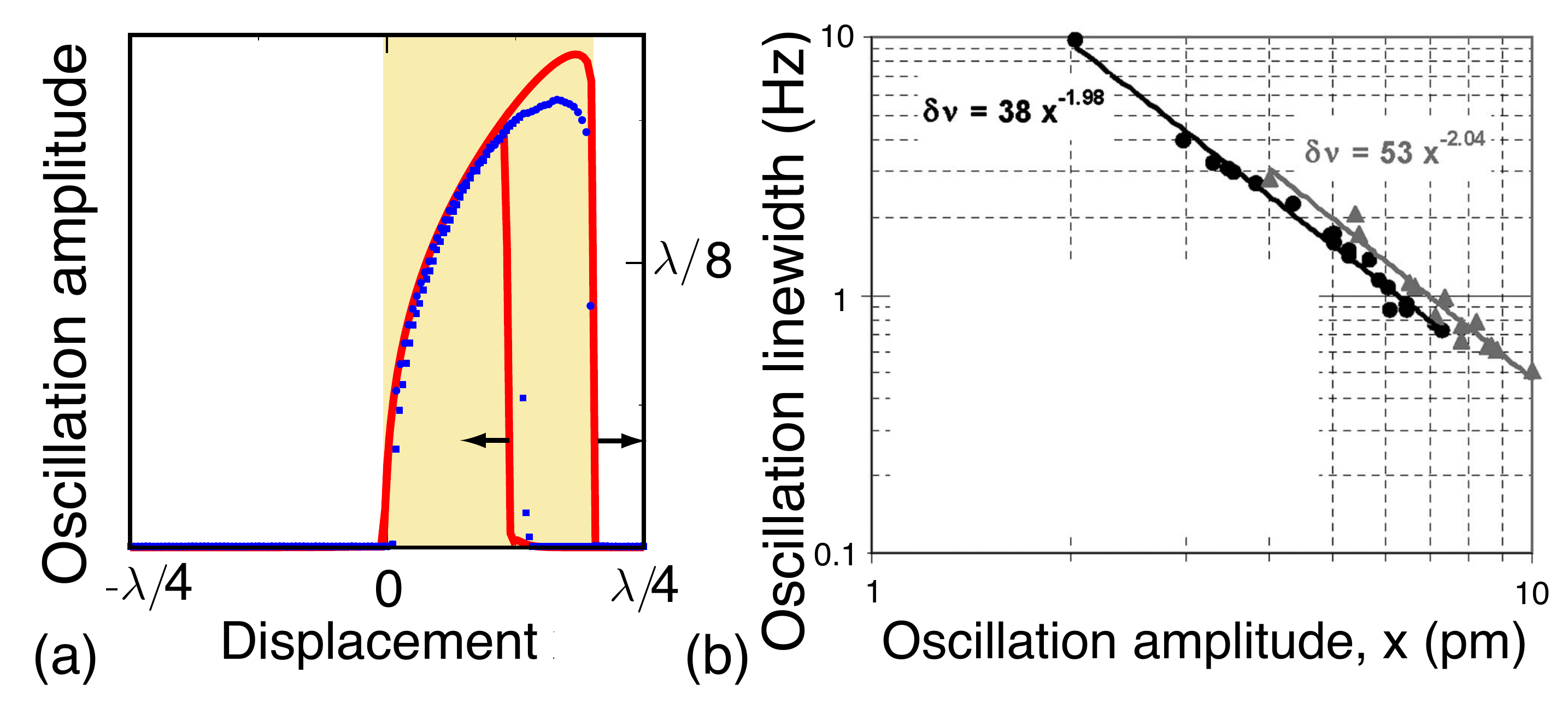}

\caption{\label{fig:NonlinearAttractorDiagramExperiment}(a) Experimental observation
of dynamical multistability for optomechanical oscillations. The mechanical
oscillation amplitude vs. static displacement (or equivalently, detuning)
displays hysteresis upon sweeping in different directions. $\lambda$
is the optical wavelength for this setup. Data from \cite{Ludwig2008},
where the radiation force was of photothermal origin. (b) Experimental
results for the linewidth narrowing above threshold, as a function
of oscillation amplitude, obtained for a microtoroidal setup in \cite{2006_Rokhsari_APL_PhaseDiffusionRate}.}
\end{figure}

\subsection{Chaotic dynamics}

\label{sub:NonlinearDynamicsChaos}

If the laser input power is increased sufficiently, the coupled motion
of the light field and the mechanical oscillator becomes chaotic.
In that regime, amplitude and phase fluctuate in a seemingly random
fashion that depends sensitively on initial conditions, even in the
absence of noise sources. Technically, the ansatz (\ref{eq:AnsatzSelfInducedOscillations})
of sinusoidal oscillations breaks down, and the full dissipative driven
dynamics of four degrees of freedom ($x,\, p$ and the complex light
amplitude $\alpha$) has to be taken into account. The chaotic regime
is characterized by positive Lyapunov exponents, where any tiny deviation
from the initial trajectory grows exponentially with time. 

Chaotic motion in optomechanical systems has been explored relatively
little so far, although it had been observed early on \cite{Carmon2005}
and has been studied more systematically in a subsequent experiment
\cite{Carmon2007a}.

\begin{figure}
\includegraphics[width=1\columnwidth]{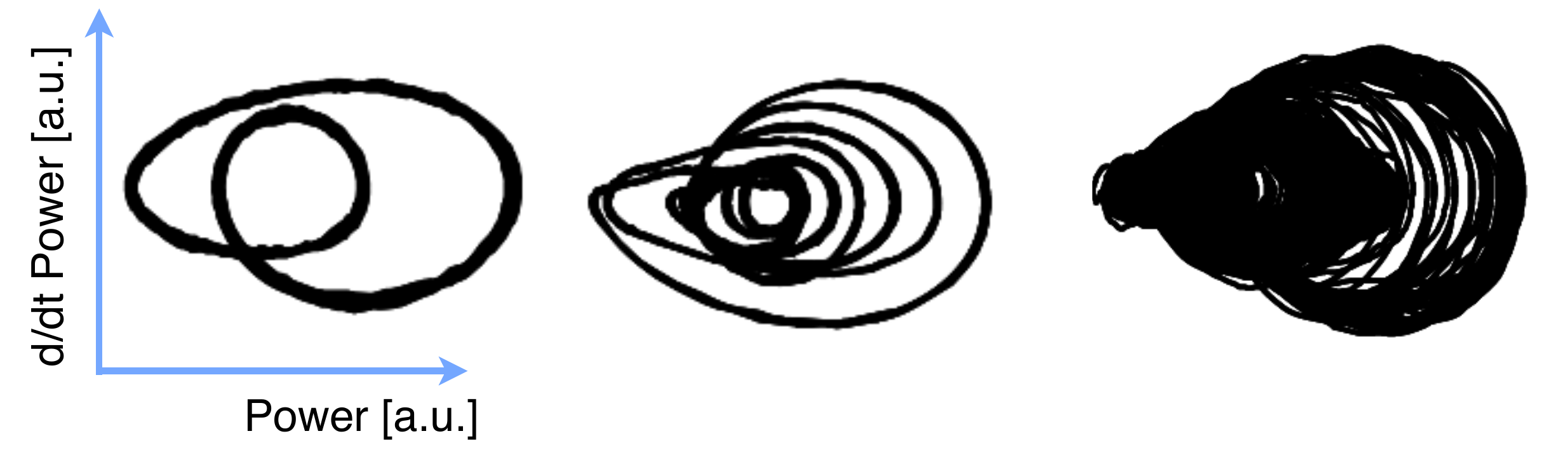}

\caption{The transition towards chaotic motion in an optomechanical system
has been observed in \cite{Carmon2007a} (courtesy Tal Carmon). }
\end{figure}

\section{Multimode optomechanics}

\label{sec:MultimodeOptomechanics}

Up to now we have almost exclusively considered one optical mode coupled
to one mechanical mode. This is the {}``minimal model'' of cavity
optomechanics, captured by the Hamiltonian (\ref{eq:StandardHamiltonianRotatingFrame}).
Of course, it is clear in principle that every mechanical resonator
has a multitude of normal modes, and every optical resonator likewise
has many different modes as well. It is relatively straightforward
to write down the appropriate extension of (\ref{eq:StandardHamiltonianRotatingFrame})
to the more general case:

\begin{align}
\hat{H} & =\sum_{k}\hbar\omega_{{\rm cav},k}\hat{a}_{k}^{\dagger}\hat{a}_{k}+\sum_{j}\hbar\Omega_{j}\hat{b}_{j}^{\dagger}\hat{b}_{j}\nonumber \\
 & -\hbar\sum_{j,k,l}[g_{0}]_{kl}^{j}\hat{a}_{k}^{\dagger}\hat{a}_{l}(\hat{b}_{j}+\hat{b}_{j}^{\dagger})+\ldots\label{eq:MultimodeHamiltonian}
\end{align}
Here the various optical ($\hat{a}_{k}$) and mechanical ($\hat{b}_{j}$)
modes interact according to the optomechanical coupling constant tensor
$[g_{0}]_{kl}^{j}=[g_{0}]_{lk}^{j*}$, whose entries depend on the
details of the optical and vibrational modes and their mutual interactions.
We left out the laser drive and the coupling to the radiation and
mechanical environments. 

Before we go to the general case, it should be noted that restricting
one's attention to the minimal model is often justified for many purposes.
The incoming monochromatic laser drive will select one optical resonance.
With regard to the mechanical motion, all the mechanical resonances
will show up in the RF spectrum obtained from a displacement measurement,
but we may choose to focus on one of those resonances, as long as
they are well separated. Likewise, cooling or heating in the resolved
sideband regime singles out a particular mechanical mode via the choice
of laser detuning. In the bad cavity limit, multiple modes can be
cooled (or amplified) simultaneously \cite{Ludwig2008,Bagheri2011}. 

\begin{figure}
\includegraphics[width=1\columnwidth]{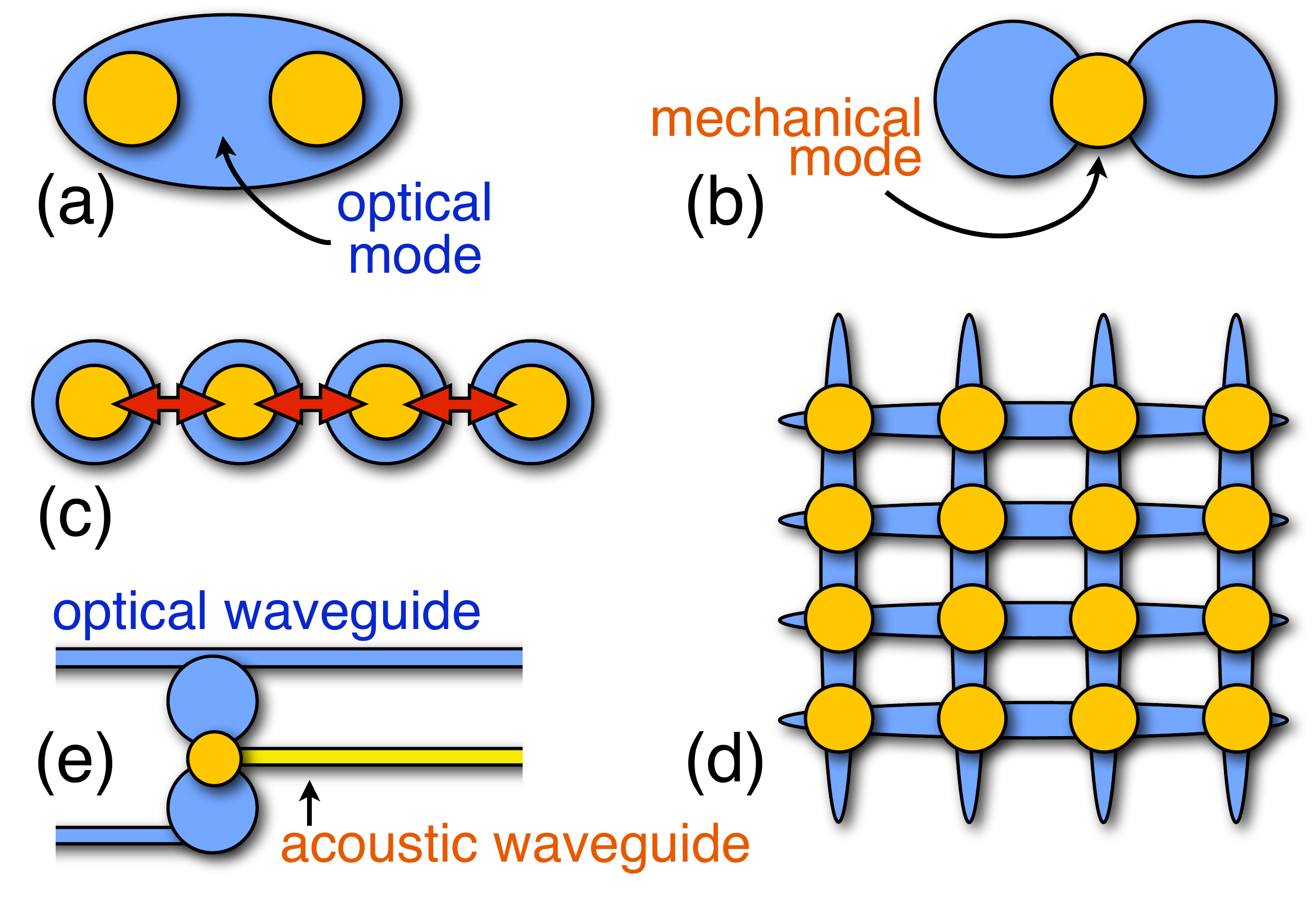}

\caption{\label{fig:MultimodeStructures}Schematic illustrating the possible
variety of multimode structures for optomechanical circuits. (a) Two
mechanical modes coupled to a common optical mode (for entanglement
etc.) (b) Two optical modes coupled to a mechanical mode (QND phonon
detection etc.) (c) 1D array, with mechanical couplings between the
cells. (d) 2D array, with coupling via common optical modes. (d) Optical
and acoustic waveguides feeding into localized optical and mechanical
modes (e.g. photon-phonon translator).}
\end{figure}

In the following, we want to discuss some scenarios and features where
it becomes crucial to go beyond the minimal model. It is clear that
going to structures with two or more mechanical or optical modes leads
to a wealth of different possible schemes (Fig.~\ref{fig:MultimodeStructures}),
only a few of which have been explored so far. Further examples can
be found in the next section, on quantum optomechanics (Sec.~\ref{sec:QuantumOptomechanics}). 

It was pointed out by Braginsky that scattering of photons between
two optical modes can lead to a parametric instability \cite{Braginsky2001}.
This analysis was intended primarily for interferometric gravitational
wave observatories, where the free spectral range may match relevant
mechanical frequencies. 

We have already mentioned a setup with a membrane in the middle of
a cavity (Sec.~\ref{sub:MeasurementsQNDFockState}), in the context
of QND phonon detection \cite{Thompson2008,Jayich2008,2009_Chen_QuantumLimit}.
This setup can be viewed as consisting of two optical modes, where
photons can tunnel between those modes via transmission through the
membrane. Having two optical modes is essential for the quadratic
dependence of optical frequency on displacement that arises from the
avoided crossing in the optical mode spectrum. 

Besides the QND scheme, setups with two relevant optical modes can
be realized in a large variety of implementations and have many additional
interesting features. Mechanical oscillations can take the system
through the avoided crossing, potentially resulting in Landau-Zener-Stueckelberg
physics and optical Rabi oscillations for the optical two-state system,
coherently shuffling photons between the two branches \cite{2010_Heinrich_LandauZener}.
Cooling and squeezing in the presence of quadratic optomechanical
coupling has been analyzed \cite{2010_Nunnenkamp_CoolingSqueezingQuadraticCoupling}.
Moreover, displacement sensing in a two-mode setup can be much more
efficient than the standard scheme \cite{Dobrindt2010}. The optomechanical
system of two coupled optical modes can also be viewed as a photonic
version of the Josephson effect and its classical dynamics can give
rise to chaos \cite{Larson2011}. 

If the mechanical resonance frequency matches the transition between
the optical branches, one can implement a version of optomechanical
phonon lasing \cite{GrudininPhononLaser2010} that is directly based
on a population inversion between the two optical levels, just like
in a real laser. The Landau-Zener physics mentioned above will significantly
modify the dynamics of these self-oscillations for large amplitudes
\cite{HuaizhiWu2011}. An interesting variant of this situation can
be implemented for whispering gallery optical and acoustical modes
in toroids or spheres in a stimulated Brillouin scattering scheme
\cite{2009_Carmon_StimulatedBrillouinScattering,2011_Carmon_StimulatedBrillouinScatteringNComm}.
In the same kind of Brillouin setup, the reverse process has been
demonstrated as well, i.e. cooling via a photon transition between
the lower and upper optical mode \cite{2012_Carmon_SpontaneousBrillouinCooling,2011_TomesCarmon_TheoryBrillouinCooling}.
Recently it has been pointed out theoretically that schemes with two
optical modes can be exploited to enhance the quantum nonlinearities
in optomechanical systems \cite{2012_Ludwig_TwoModeScheme,2012_Stannigel_TwoModeScheme},
see Sec.~\ref{sub:QuantumOptomechanicsNonlinear}. The photon-phonon
translator discussed in Sec.~\ref{sub:QuantumOptomechanicsQuantumProtocols}
is another example of a possible device that employs two optical modes.

Likewise, there are many schemes where more than only a single mechanical
mode is relevant. These could be the various normal modes of a given
mechanical structure, or several vibrating objects placed inside a
cavity or coupling to one optical mode (see Sec.~\ref{sub:QuantumOptomechanicsNonlinear}).
We mention only a few illustrative examples. An array of multiple
membranes inside a cavity was studied theoretically in \cite{Bhattacharya2008a,Hartmann2008,Xuereb2012}
with regard to mechanical normal modes, entanglement and collective
interactions. A recent experiment demonstrated tripartite optomechanical
mixing between one microwave mode and two mechanical modes \cite{Massel2012}.
Effects of multiple mechanical modes can also be studied for levitated
setups. It has been proposed that an array of levitated dumbbell-shape
dielectric objects can undergo an ordering transition.

The nonlinear dynamics of the self-oscillations ({}``mechanical lasing'';
Sec.~\ref{sec:NonlinearDynamics}) become particularly interesting
when more mechanical modes are involved. It has been shown experimentally
\cite{Metzger2008} that collective self-oscillations may arise when
several mechanical modes are excited simultaneously, using a strong
blue-detuned drive for a {}``bad cavity'' whose linewidth encompasses
several mechanical normal modes of the structure (a cantilever in
that case). More recently, it was pointed out that optomechanical
systems are very promising for observing synchronization phenomena
\cite{2011_Heinrich_SynchronizationPRL}. Assume an array of optomechanical
oscillators, each of which consists of a mechanical mode coupled to
an optical mode that is driven by a blue-detuned laser beam such as
to go into the {}``mechanical lasing'' regime (Sec.~\ref{sec:NonlinearDynamics}).
If these {}``clocks'' are coupled mechanically or optically, their
mechanical frequencies can lock to each other, even if they have been
distinct initially. Under appropriate conditions, a variant of the
Kuramoto-model can be derived for optomechanical systems \cite{2011_Heinrich_SynchronizationPRL},
which is a paradigmatic model of synchronization physics. For two
mechanically coupled cells, the equation turns out to be of the form

\[
\delta\dot{\varphi}=\delta\Omega-K\sin(2\delta\varphi)\,,
\]
where $\delta\Omega$ is the intrinsic frequency difference, $\delta\varphi$
the difference of mechanical oscillation phases, and $K$ an effective
coupling constant that can be related to microscopic parameters. If
$K$ is large enough, synchronization ensues ($\delta\dot{\varphi}=0$).
If, one the other hand, several oscillators couple to the same optical
mode, the behaviour can become of a form that is not described by
any Kuramoto-type model. This has been analyzed recently in detail
\cite{2011_Milburn_Synchronization}.

Synchronization may be important for metrological applications, where
several synchronized optomechanical {}``clocks'' of this type are
expected to be more stable against noise \cite{2010_Carmon_PhaseNoiseOptomOsc}.
Experimentally, some signs of synchronization in an optomechanical
device have been observed recently \cite{2011_Lipson_Sync} for two
optically coupled optomechanical oscillators, each of them implemented
as a double-disk SiN structure. Mechanical spectra showed the onset
first of self-oscillations and then of synchronization as a function
of laser detuning.

Two coupled mechanical modes of widely different damping rates can
give rise to Fano lineshapes in their excitation spectrum when they
hybridize. This has been demonstrated in an optomechanical system
\cite{2010_LinPainter_CoherentMixingMechanical}, and suggested for
information storage and retrieval in long-lived mechanical {}``dark''
states.

Optomechanical photonic crystal structures present an opportunity
to design more complex optomechanical circuits. These might be 1D
or 2D array structures, where many optical and mechanical modes are
arranged in a periodic layout and coupled to each other. Alternatively,
one can think of more intricate circuits, more similar to computer
chips, where different elements fulfill various functionalities (sensing,
amplification, general signal processing). 

Optomechanical arrays of this type have been studied theoretically
only in a few works so far, both with respect to their collective
classical nonlinear dynamics \cite{2011_Heinrich_SynchronizationPRL},
their quantum many-body dynamics \cite{Ludwig2012c} and for engineering
quantum dissipation \cite{Tomadin2012}, as well as with respect to
quantum applications: A suitably engineered array of optical and mechanical
modes coupled to a waveguide can slow down and store light \cite{2011_ChangPainter_SlowingStoppingLightArray}.
Photon-phonon entanglement in an optomechanical array of three cells
was studied in \cite{2011_Milburn_EntanglementOptomechanicalArray}.
Many nanomechanical modes in an array geometry can be entangled via
the light field, using a suitable parametric drive to select mode
pairs \cite{2012_Schmidt_NanomechanicalCircuits}. These studies pave
the way towards future architectures for (continuous variable) quantum
information processing with optomechanical circuits.

\section{Quantum Optomechanics}

\label{sec:QuantumOptomechanics}

Quantum mechanics has already figured at several places in our discussion
so far, notably in setting the limits for displacement sensing or
cooling. We will now turn to discuss potential future optomechanical
experiments where quantum behaviour will take center stage. We will
discuss ways to create interesting quantum states both in the optical
and mechanical system, and to create entanglement between the various
subsystems. We will then turn to nonlinear quantum effects whose understanding
requires us to go beyond the linearized optomechanical interaction,
i.e. beyond the quadratic Hamiltonian of Eq.~(\ref{eq:LinearizedInteraction}).
Finally, we will see how these effects are envisaged to form the ingredients
of future optomechanically aided protocols for quantum information
processing.

\subsection{Light-assisted coherent manipulation of mechanics}

\label{sub:QuantumOptomechanicsManipulationOfMechanics}

The light field can be employed in principle to generate arbitrary
quantum states of the mechanical oscillator. In this section, we will
restrict ourselves to the action of the linearized interaction Hamiltonian
of Eq.~(\ref{eq:LinearizedInteraction}),

\begin{equation}
\hat{H}_{{\rm int}}^{(lin)}=-\hbar g_{0}\sqrt{\bar{n}_{{\rm cav}}}(\delta\hat{a}^{\dagger}+\delta\hat{a})(\hat{b}+\hat{b}^{\dagger})\,.
\end{equation}
When injecting Gaussian optical states, as is the case for the usual
laser drive, this can produce arbitrary mechanical Gaussian states,
i.e. coherent and squeezed states out of the ground state (which has
to be reached first, either via optomechanical or bulk cooling techniques).
As long as the linearized Hamiltonian is valid, non-Gaussian mechanical
states can only be produced from non-Gaussian optical states. The
parameters that can be varied easily are the laser detuning $\Delta$
and the laser input power. Any pulse of light will generate a radiation
pressure force pulse that shifts the oscillator's wave function and
thereby permits to create a coherent state.

It is only slightly more difficult to create a squeezed state. As
we saw in Sec.~\ref{sub:DynamicalBackactionOpticalSpring}, a far-detuned
light beam creates an {}``optical spring effect'', i.e. a change
in the mechanical frequency by $\delta\Omega_{{\rm m}}=2\bar{n}_{{\rm cav}}g_{0}^{2}/\Delta$.
As is well known, a time-dependent modulation of $\Omega_{{\rm m}}$,
i.e. a parametric driving of the mechanical oscillator creates a squeezed
state: see \cite{Mari2009} for an analysis in the case of optomechanical
systems, where one can employ a modulation of the laser power. For
$\delta\Omega_{{\rm m}}(t)=\delta\Omega\cos(2\Omega_{{\rm m}}t)$,
the resulting effective mechanical Hamiltonian turns out to be (in
a frame rotating at $\Omega_{{\rm m}}$, and in rotating wave approximation)
the standard squeezing Hamiltonian

\begin{equation}
\hat{H}_{{\rm mech}}=\frac{\delta\Omega}{2}(\hat{b}^{2}+(\hat{b}^{\dagger})^{2})\,,
\end{equation}
such that $\hat{b}(t)=\cosh(\delta\Omega t)\hat{b}(0)-i\sinh(\delta\Omega t)\hat{b}^{\dagger}(0)$
in the absence of dissipation. Thus, in this time-dependent scheme,
squeezing grows exponentially with time. However, a realistic analysis
needs to take into account the initial thermal population, as well
as dissipation and decoherence, and distinguish between the steady-state
situation and the transient case. When several mechanical modes are
coupled to the same optical mode, a modulated laser drive will generate
two-mode squeezing or beam-splitter interactions between pairs of
modes that are selected according to their frequency. This can form
the basis for continuous variable quantum state processing in optomechanical
arrays \cite{2012_Schmidt_NanomechanicalCircuits}.

Measurements can also be used to generate interesting mechanical states
in a probabilistic manner, i.e. conditioned on the measurement result.
This includes squeezed states via single-quadrature detection (Sec.~\ref{sub:MeasurementsQNDSingleQuadrature})
or mechanical Fock states via phonon number readout (Sec.~\ref{sub:MeasurementsQNDFockState}).
Further below we will comment on other ways to generate more nonclassical
states (including non-Gaussian states), by nonlinear effects (Sec.~\ref{sub:QuantumOptomechanicsNonlinear})
or various state transfer protocols (Sec.~\ref{sub:QuantumOptomechanicsQuantumProtocols}).

\subsection{Mechanics-assisted readout and manipulation of light}

\label{sub:QuantumOptomechanicsReadoutManipulationOfLight}

The optomechanical interaction can be exploited to detect and manipulate
the quantum state of the light field.

An example that was suggested early on is the possibility of a QND
detection \cite{Braginsky1980} for the light intensity circulating
inside the cavity \cite{1977_Braginsky_QNDmeasurementOfLight,Jacobs1994,Pinard1995}.
The displacement of the end-mirror, induced by the radiation pressure
force, can serve as a noiseless meter for the light intensity. In
addition, as the radiation pressure force is proportional to the photon
number, it increases in discrete steps. If the photons are sufficiently
long-lived and the interaction strong enough, this may even enable
QND photon detection, by registering the resulting mechanical displacement.
In practice, however, that regime is extremely challenging to reach,
as it would require $g_{0}^{2}/(\kappa\Omega_{{\rm m}})\gg1$ (see
the section on quantum nonlinear dynamics below, Sec.~\ref{sub:QuantumOptomechanicsNonlinear}).

Regarding the manipulation of the light's quantum state, one of the
most straightforward applications of optomechanics consists in squeezing
the noise of the light beam \cite{Fabre1994b,Mancini1994}. In this
context, the change of the cavity length due to the intensity-dependent
radiation force can be compared to the effect of a Kerr medium inside
a rigid cavity. The resulting physical picture depends sensitively
on the detuning and the frequency at which the noise of the light
beam is analyzed. In the simplest case, one may imagine that a temporary
fluctuation in the incoming intensity of the light beam induces a
change in the cavity length via the radiation pressure force. This,
in turn, shifts the optical resonance and thereby affects the circulating
and outgoing intensities, potentially suppressing the noise. It should
be noted that for a single-sided optical cavity (without internal
losses) there can be no change of the amplitude noise at zero frequency,
since the incoming and outgoing intensities have to be equal. However,
there can be amplitude squeezing at finite frequencies (and phase
squeezing down to zero frequency). The squeezing effect diminishes
towards frequencies above the mechanical resonance. Finite temperatures
degrade squeezing, as the thermal motion of the mirror imprints extra
noise on the light beam. At low frequencies, and for detunings on
the order of $\kappa$, the light beam's noise is increased \cite{Fabre1994b}
by a factor $\bar{n}_{{\rm th}}/Q_{{\rm m}}$, where $\bar{n}_{{\rm th}}$
is the equilibrium phonon number. 

A recent review of the current efforts towards demonstrating radiation
pressure shot noise effects and squeezing in optomechanical experiments
can be found in \cite{Verlot2011}. First experiments in this direction
have simulated the quantum fluctuations by classical intensity noise
\cite{2010_Marino_SimulatingPonderomotiveSqueezingExperiment}. More
recently, the very strong coupling and low temperatures attainable
in realizations of cavity optomechanics with atomic clouds have allowed
to obtain first signatures of genuine optical squeezing \cite{2011_StamperKurn_SqueezingViaOptomechanics}
at the quantum level.

\subsection{Optomechanical entanglement}

\label{sub:QuantumOptomechanicsEntanglement}

The optomechanical interaction can be used to engineer entanglement
between the light beam and the mechanical motion, or between several
light beams or several mechanical modes. However, a prerequisite for
all of these approaches is to cool the mechanical oscillator to near
its ground state and to achieve sufficiently strong coupling.

In the following, we first describe entanglement in the single-photon
strong coupling regime (for more on this regime, see Sec.~\ref{sub:QuantumOptomechanicsNonlinear})
and then turn to continuous variable entanglement.

A particularly simple physical picture applies in an idealized situation
where we imagine starting with a superposition of photon states inside
the cavity. Then, the radiation pressure force assumes different,
discrete values for each photon number, displacing the mechanical
harmonic oscillator potential by $2x_{{\rm ZPF}}g_{0}/\Omega_{{\rm m}}$
per photon. If initially the mechanics was in its ground state, it
will evolve into a coherent state $\left|\alpha_{n}(t)\right\rangle _{{\rm vib}}$,
oscillating around the new, displaced origin that depends on the photon
number. Thus, we immediately arrive at a non-factorizable, i.e. entangled
state of the form:

\begin{equation}
\left|\Psi(t)\right\rangle =\sum_{n=0}^{\infty}c_{n}e^{i\varphi_{n}(t)}\left|n\right\rangle _{{\rm cav}}\otimes\left|\alpha_{n}(t)\right\rangle _{{\rm vib}}\label{eq:optomech_schroedingercat_state}
\end{equation}
Here $n$ is the photon number, $c_{n}$ are the arbitrary initial
amplitudes for the photon field (e.g. corresponding to a coherent
state), and $\varphi_{n}(t)$ a phase-shift that can be obtained by
solving the time-dependent Schrödinger equation for this problem.
The state (\ref{eq:optomech_schroedingercat_state}) can be interpreted
as a {}``Schrödinger cat'' type state, where a {}``microscopic''
degree of freedom (the optical cavity mode) is entangled with a {}``macroscopic''
(or mesoscopic) degree of freedom, the vibrating mirror. This picture
has been first analyzed in \cite{Bose1997,Mancini1997}. 

\begin{figure}
\includegraphics[width=1\columnwidth]{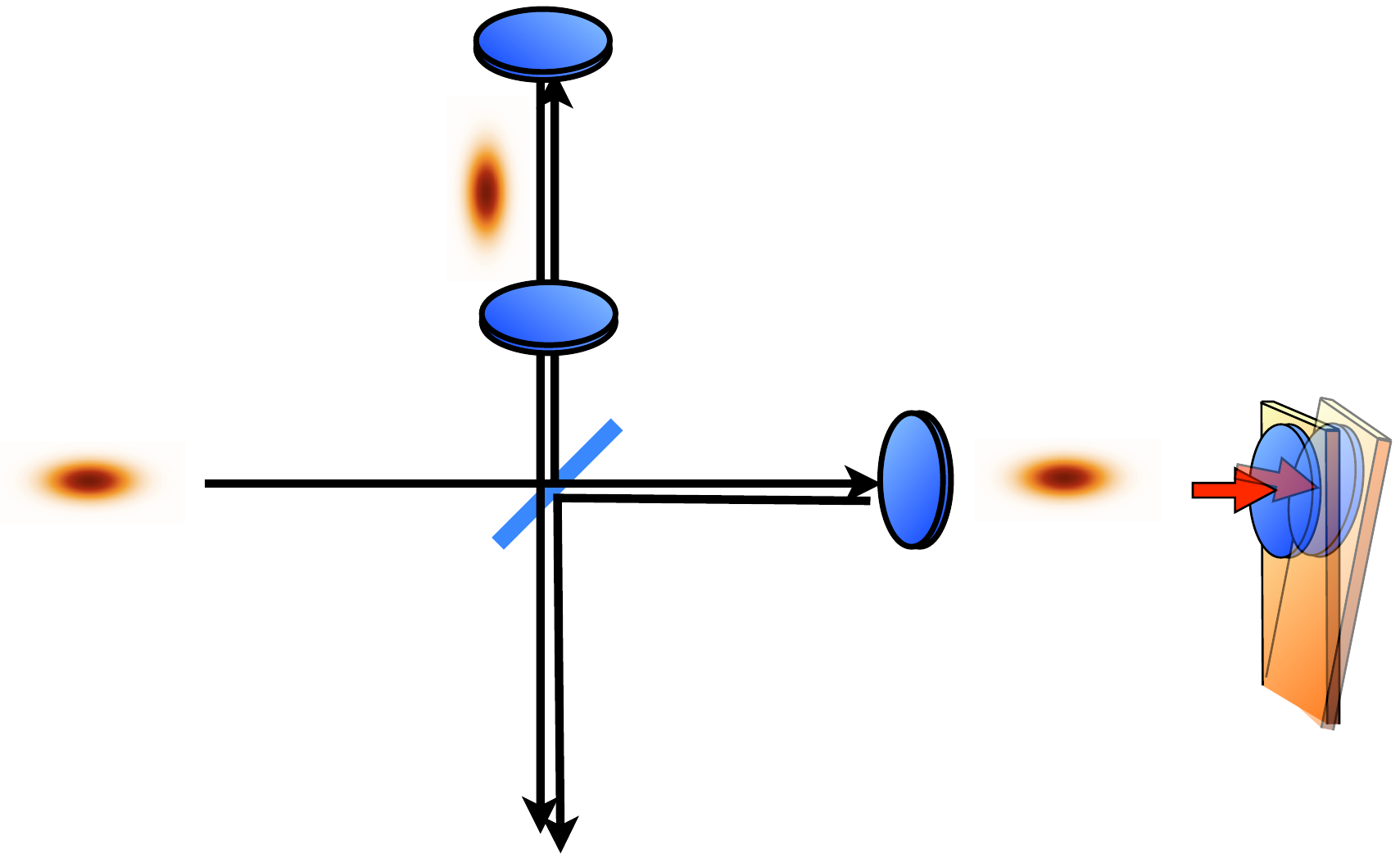}

\caption{\label{fig:PenroseProposal}An optomechanical which-path experiment
can be employed to test for the decoherence of mechanical superposition
states \cite{Marshall2003}. A photon would leave behind a vibration
in the mirror of an optomechanical cavity, which destroys the photon's
coherence with the other path. However, if the photon exits after
a full mirror oscillation period, it will be fully coherent again,
unless mechanical decoherence has occured in the meantime.}
\end{figure}

Several signatures of entanglement exist. In this example, where the
overall state is pure, we can simply trace out the mechanical vibrations,
arriving at the reduced density matrix of the optical field, which
will be found in a mixed state whenever there is entanglement. However,
at multiples of the mechanical period, light and mechanics completely
disentangle, since the coherent state $\left|\alpha_{n}(t)\right\rangle _{{\rm vib}}$
will have returned back to the origin, independent of photon number.
At these times, the photon state becomes pure again, even independent
of the mirror's effective temperature. The optical field's coherence
thus demonstrates decay and revivals. This can in principle be tested
in an interferometric optomechanical {}``which-way'' experiment,
where a photon can take either of two paths, one of which contains
an optomechanical cavity (see Fig.~\ref{fig:PenroseProposal}). The
revivals can be observed in the interference visibility, as a function
of the time the photon has spent inside the cavity. However, any mechanical
decoherence occuring in the meantime will spoil these perfect revivals
of the photon field's coherence. This can in principle be employed
for highly sensitive optical tests of sources of mechanical dissipation
and decoherence and fundamental quantum physics in general \cite{1998_Bouwmeester_IndiaProposal,Bose1999,Marshall2003,Folman2001}.
Experiments of this kind could quantify the decoherence of superpositions
of heavy objects (i.e. the mirror), and thus potentially shed new
light on the quantum-to-classical transition (see Sec.~\ref{sub:FuturePerspectivesFoundational}).
In particular, models of gravitationally induced decoherence might
be tested \cite{Bose1999,Marshall2003}, where the (admittedly tiny)
hypothetical extra decoherence rate becomes potentially observable
only for relatively massive objects. Such experiments will require
the challenging regime of $g_{0}/\Omega_{{\rm m}}>1$ (when the displacement
induced by a single photon may become on the order of the mechanical
zero-point motion). It has been recently suggested that nested interferometry
allows to reduce this requirement significantly \cite{Pepper2012}.

On the other hand, entanglement can also exist on the level of continuous
variables \cite{Braunstein2005}, of the type first proposed in the
famous Einstein-Podolsky-Rosen article \cite{EPR1935}. The analysis
of entanglement in this context has to be performed by taking into
account the dissipative nature of the systems involved, and typically
relies on solving the linearized quantum equations of motion (including,
if needed, input-output theory for treating the reflected or transmitted
light beam). In the regime where linearized equations of motion are
valid, Gaussian states of the mechanics and the light field will be
produced, and their entanglement (of the continuous-variable type)
can be characterized completely once the correlations between the
various mechanical and optical quadratures are known. A typical measure
of entanglement that is commonly applied here is the logarithmic negativity,
which can be calculated easily for Gaussian states \cite{2002_VidalWerner_EntanglementGaussianStates},
both pure and mixed. 

The entanglement between mechanical vibrations and the optical cavity
field (already described above in a picture appropriate for strong
coupling) has been analyzed in more detail for the continuous-variable
case in \cite{Vitali2007a,Paternostro2007}. Recently, it was pointed
out that a suitable time-dependent modulation of the drive can improve
the efficiency of photon-phonon entanglement \cite{2011_MariEisert_ModulationForEntanglement}.

It is also possible to create entanglement of two spatially separate
mirrors. When a strong pump beam runs through a nonlinear optical
$\chi^{(2)}$ medium acting as a nondegenerate optical parametric
amplifier, two-mode squeezing produces entanglement between the quadratures
of pairs of emanating light beams. This entanglement could then be
transferred via the radiation pressure force onto two spatially separated
mirrors that are part of optomechanical cavities on which these light
beams impinge \cite{Zhang2003}. In this way, optomechanics would
help to create mechanical EPR-type entanglement at a distance. The
verification of EPR entanglement between macroscopic test masses by
sensitive measurements has been studied in more detail for the context
of gravitational wave interferometer setups \cite{2009_Chen_MacroscopicEntanglement,2010_Chen_MacroscopicQuantumStates}.
In the context of optomechanics, we usually consider the system to
be driven by a coherent laser beam. However, it is natural to ask
whether special opportunities arise when the light that is injected
displays quantum features. For example, injecting squeezed light can
be beneficial for entangling nanomechanical resonators via the optomechanical
interaction \cite{2009_Agarwal_EntanglingViaSqueezedLight}.

In another approach, one could achieve the same goal without the optical
entanglement created by a nonlinear medium, and instead perform an
optical measurement after the interaction has taken place, in an entanglement
swapping scheme. If two independent light beams interact with separate
optomechanical cavities, then the beams can afterwards be brought
to interfere at a beamsplitter, and a suitable Bell state measurement
can then be used to generate entanglement between the distant mechanical
resonators \cite{Pirandola2006} (see \cite{2011_Borkje_Entanglement}
for a similar proposal). 

Alternatively, the driven optical field inside the cavity automatically
induces an effective interaction between several mechanical modes,
thus providing yet another way to generate mechanical entanglement,
without the need for any optical nonlinearities or entanglement swapping
schemes. In the case of two mechanical resonators (or two normal modes
of one resonator) coupling to the same driven cavity mode, this may
be understood as a consequence of the optical spring effect (Sec.~\ref{sub:DynamicalBackactionOpticalSpring}).
For the case of a single mechanical mode $\hat{b}$, eliminating the
driven cavity mode by second-order perturbation theory creates an
effective interaction term $\hbar(g_{0}^{2}\bar{n}_{{\rm cav}}/\Delta)(\hat{b}+\hat{b}^{\dagger})^{2}$.
Proceeding through the same argument for the case of two mechanical
modes gives rise to an effective mechanical interaction. It is of
the form

\begin{equation}
\hat{H}_{{\rm eff}}^{{\rm int}}=\hbar\frac{g_{0}^{2}\bar{n}_{{\rm cav}}}{\Delta}[(\hat{b}_{1}+\hat{b}_{1}^{\dagger})+(\hat{b}_{2}+\hat{b}_{2}^{\dagger})]^{2}\,,\label{eq:OpticalSpringMechanicalCoupling}
\end{equation}
if we assume for simplicity that both mechanical oscillators couple
equally strongly to the optical mode. 

In order to successfully entangle different mechanical modes, one
has to laser-cool those modes, since the mechanical vibrations would
be far from their ground state for typical bulk temperatures. The
optically induced steady-state entanglement between two movable mirrors
under simultaneous laser-cooling has been studied carefully in \cite{Pinard2005},
and later for somewhat different setups in \cite{2008_SchnabelChen_Entanglement,Hartmann2008}.
There is an interesting caveat for such studies: It is insufficient
to apply the Markov approximation to describe the dissipative dynamics
of the mechanical vibrations in this context, even though for many
other purposes in optomechanics that is a very reliable approach.
In fact, it can be shown that in a proper treatment \cite{Ludwig2010}
there is an optimum intermediate laser-cooling strength for which
entanglement is maximized (an effect entirely missed by the common
Markovian treatments).

The light-induced interaction given in Eq.~(\ref{eq:OpticalSpringMechanicalCoupling})
can form the basis of a general scheme for quantum state processing
with many nanomechanical modes. In order to selectively address pairs
of such modes for entanglement and state transfer, one simply has
to modulate the coupling strength (i.e. the laser intensity) at sum
and difference frequencies of those modes. Such a parametric scheme
only requires one appropriately modulated laser input to address whole
arrays of modes, if the proper layout is chosen \cite{2012_Schmidt_NanomechanicalCircuits}.

Another possibility consists in exploiting optomechanics to entangle
two light-beams. In these cases, the optomechanical interaction essentially
serves the purpose of a $\chi^{(2)}$ nonlinear medium. This has been
proposed e.g. for an optomechanical setup where two degenerate, orthogonally
polarized cavity modes are driven strongly and their interaction with
the moveable mirror creates EPR correlations \cite{2001_Mancini_EPR_TwoLightBeams}
between the quadrature variables of the beams emanating from these
modes. Another option is to entangle the two sidebands reflected from
a vibrating mirror, which works even in the absence of a cavity, for
a strong short incoming laser pulse \cite{Pirandola2003}. For this
situation, a more detailed analysis of the entanglement between mirror
vibrations and the full light field (infinitely many degrees of freedom)
was performed in \cite{2010_Chen_EntanglementWithContinuousFields}.
Optically trapped mirrors in a cavity optomechanics setup can als
be exploited to entangle light beams \cite{2008_Nergis_EntanglementOfMirrors}.

\subsection{Quantum hybrid systems}

\label{sub:QuantumHybridSystems}

Optomechanical systems already represent a quantum hybrid system,
i.e. a coupling between two quantum systems of a different physical
nature: light and mechanical vibrations. In general, hybrid approaches
may be useful for purposes such as quantum information processing,
in order to combine the advantages of different physical systems in
one architecture. Some systems may be strongly interacting (good for
computation), some are very coherent (good for long-term storage),
and yet others are easily transported over long distances (good for
communication). 

In principle, other components may be added easily to optomechanical
setups. This is because both the light field and the mechanical vibrations
are quite versatile in coupling to a variety of systems, such as cold
atoms, spins, superconducting and other electronic qubits, etc. Consequently,
there are already several proposals along these lines.

In a cloud of atoms, the total spin can sometimes be treated as a
harmonic oscillator, identifying its small fluctuations around a preferred
direction as position and momentum quadratures. This picture is useful
when discussing experiments where the state of the light field is
transferred to the atomic spin state and back again \cite{HammererRMP}.
More recently, it has been suggested that light might also be used
to couple the collective spin of an atom cloud to a nanomechanical
oscillator \cite{Hammerer2009a}. In such a setup, a light beam passing
first through an optomechanical cavity and then through an atom cloud
carries information in its quadratures about the sum and the difference
of position and momentum variables of the mechanics and the spin state.
A subsequent QND measurement then is able to prepare the two systems
in an EPR state, conditioned on the measurement result. As another
example, coupling to the internal transitions of atoms can enhance
cooling \cite{2009_VitaliRitsch_CoolingViaAtoms}. One can also consider
more involved internal level schemes for the atoms. For example, a
vibrating mirror could be coupled strongly to the collective spin
of a cloud of three-level atoms displaying electromagnetically-induced
transparency phenomena \cite{2011_Dantan_AtomCloudEITAndMirror}.
In addition, it has been suggested that coupling the collective spin
of an atom cloud parametrically to a resonator can lead to the phenomena
known from optomechanics \cite{2010_StamperKurn_SpinOptodynamicsOptomechanics}
(amplification and cooling of the spin, frequency shifts, and squeezing
of light), with the collective spin replacing the mirror motion.

After the pioneering experiments on optomechanics using cold atoms
(Sec.~\ref{sub:ExperimentsUltracoldAtoms}), a number of different
possibilities have been explored theoretically to couple the motion
of atoms to other systems. For example, the light field inside the
cavity may be used to couple the motion of a single atom to the vibrations
of an end-mirror or a membrane \cite{Hammerer2009,Wallquist2010},
where the strong coupling regime seems to be within reach. This could
be the basis for exploiting all the well-known tools for manipulating
and reading out the motion of a single trapped atom to gain access
to the membrane motion. It would also be an interesting system to
observe the entanglement between a microscopic and a macroscopic degree
of freedom. Another model system that has been studied \cite{Hammerer2010}
and is now being implemented \cite{2011_Treutlein_ColdAtomsAndMembrane}
is a cloud of atoms in a standing light wave reflected from a vibrating
mirror (without a cavity). This could allow long-distance coupling
between atoms and mechanical objects spaced apart by macroscopic distances.

A variety of other ideas exist for merging cold atom systems with
optomechanics. Both the optomechanics of Bose-Einstein condensates
\cite{2010_Meystre_BEC_Optomechanics,2011_Meystre_BackActionFromBEC,2011_Meystre_OptomechanicsBECAntiferromagnet}
and degenerate cold atom Fermi gases \cite{2010_Meystre_OptomechanicsFermiGas}
have been explored theoretically in some depth. Even the Mott-insulator
to superfluid transition of atoms in an optical lattice coupled to
a vibrating mirror has been analyzed, as an example of a strongly
interacting quantum system subject to the optomechanical interaction
\cite{2009_Meystre_MottInsulator,Larson2008}.

The idea of doing optomechanics on trapped atoms has found an interesting
counterpart in proposals for doing optomechanics on levitated dielectric
objects (Sec.~\ref{sub:ExperimentsNanoObjects}). The promise of
this approach lies in a drastically enhanced mechanical quality factor.
Some proposals considered trapping and cooling a mirror \cite{Bhattacharya2007,Bhattacharya2007a,Bhattacharya2008,Singh2010},
which has the advantage (over other objects) that scattering of the
light into unwanted directions is greatly reduced. Alternatively,
one can have dielectric spheres or other particles trapped in an optical
lattice or by other means \cite{Chang2010,Romero-Isart2010,Barker2010,2010_Barker_DopplerCoolingMicroSphere}.
A more detailed analysis of fundamental applications and protocols
can be found in \cite{2011_RomeroIsart_LevitationLong,2011_RomeroIsart_SuperpositionsPRL,2011_PenderMonteiroBarker_LevitatedSpheresCooling,Romero-Isart2011a}.
If the promise of very long mechanical coherence times is fulfilled,
then these platforms could offer the best means to test novel decoherence
mechanisms (see the discussion in Sec.~\ref{sub:QuantumOptomechanicsEntanglement}).
Some early experiments and studies on cavity optomechanics with sub-wavelength
nano-objects \cite{Favero2009,Favero2008} (and Sec.~\ref{sub:ExperimentsNanoObjects})
have already explored the optical coupling and some of the scattering
mechanisms that may become relevant in this domain, even though they
did not yet benefit from a suppression of mechanical dissipation. 

Connecting the world of superconducting or other solid-state qubits
to optomechanical systems represents an intriguing possibility in
the context of quantum information processing. This has become particularly
relevant since the pioneering experiment at UCSB \cite{O'Connell2010}
that demonstrated strong coupling between a superconducting phase
qubit and the ${\rm GHz}$ oscillations of a piezoelectric nanoresonator,
swapping a single excitation from the qubit into the resonator. First
experiments have demonstrated how to manipulate a mechanical nanoresonator
both via the optomechanical interaction and electrically \cite{Lee2010,Winger2011},
which is an important ingredient for a future hybrid platform of solid-state
qubits with electrical interactions coupled to mechanical nanoresonators
coupled to the light field. Theoretical proposals have already pointed
out how to use systems of this type to map solid-state quantum information
into photons and back again \cite{2010_Stannigel_OptomechanicalTransducersPRL,2011_Stannigel_OptomechanicalTransducersLongVersion,2011_SafaviPainter_PhotonPhononTranslator,2011_WangClerk_StateTransfer,2011_Tian_StateConversion,2011_RegalLehnert_HybridMWOptomechanics}.
More elementary, the nanomechanical structure can serve as an intermediary
to generate entanglement between microwave and optical fields \cite{2011_MilburnVitali_OpticalMicrowaveEntanglement}.

An equally promising avenue is to merge the fields of solid-state
quantum optics and cavity optomechanics. First experiments deliberately
introducing semiconductor materials in cavity optomechanical setups
now exist. These include $2\mu m$-diameter GaAs vibrating disk structures
with very high optomechanical coupling strength \cite{2010_Favero_GaAsOptomechanics,Ding2011}
and coupling of the light field to a semiconductor nanomembrane \cite{Usami2012}.
The excitonic transitions of quantum dots embedded in such materials
could couple to the mechanical vibrations either directly via deformation
potentials or indirectly via the light field \cite{Rundquist2011}.

\subsection{Quantum protocols}

\label{sub:QuantumOptomechanicsQuantumProtocols}

The previous sections introduced some basic quantum-physical features
in optomechanical systems, such as producing and reading out nonclassical
states of light and mechanics. As soon as this can be achieved reliably,
one may envisage building a toolbox for quantum manipulation in these
systems, and exploit it for purposes of quantum communication and
quantum information processing. This would follow the pioneering ideas
and efforts in the ion trap community, where it has been suggested
early on that one can exploit the motional degrees of freedom to facilitate
quantum gates between the internal states of ions \cite{Cirac1995,Leibfried2003}.
Micro- and nanomechanics offers the added value that they can be functionalized
and hence couple to many different physical degrees of freedom. Optomechanical
devices therefore offer a fruitful addition to the vast array of physical
systems that are being explored for quantum information processing
\cite{Zoller2005}.

One of the most prominent protocols is quantum state transfer. In
the context of optomechanics, this would allow to realiably convert
an optical pulse into a mechanical excitation (and vice versa). In
principle, this is straightforward, since the linearized optomechanical
interaction describes a coupling between two oscillators (mechanical
and driven cavity mode), which can be tuned via the laser intensity
(Sec.~\ref{sub:OptomechanicalCouplingHamiltonianFormulation}). For
the red-detuned case, $\Delta=-\Omega_{{\rm m}}$, in the resolved-sideband
limit, we found a beam-splitter type of interaction:

\begin{equation}
\hat{H}=-\hbar\Delta\delta\hat{a}^{\dagger}\delta\hat{a}+\hbar\Omega_{{\rm m}}\hat{b}^{\dagger}\hat{b}-\hbar g(\delta\hat{a}^{\dagger}\hat{b}+\delta\hat{a}\hat{b}^{\dagger})+\ldots\,.\label{eq:BeamSplitterInteractionAgain}
\end{equation}
Thus, the excitations $\delta\hat{a}$ on top of the strong coherent
laser drive can be swapped onto the mechanical resonator. Such an
operation would be performed in a pulsed scheme. The coupling $g(t)$
becomes time-dependent via the laser intensity, and it would be switched
on for just the right amount of time to perform a complete state swap
between the two oscillators ($\delta\hat{a}$ and $\hat{b}$). Two
laser pulses are needed for this scheme. A red-detuned control pulse
at $\omega_{L}^{{\rm control}}=\omega_{{\rm cav}}-\Omega_{{\rm m}}$
determines the time-dependent coupling $g(t)$. A second, {}``signal''
pulse serves to excite the $\delta\hat{a}$-oscillator into some target
state that then will be written onto the mechanics. Since $\delta\hat{a}$
oscillates at the frequency $-\Delta=\Omega_{{\rm m}}$ in the frame
rotating at $\omega_{L}^{{\rm control}}$ , the signal pulse has to
be injected at a frequency $\omega_{L}^{{\rm signal}}=\omega_{L}^{{\rm control}}+\Omega_{{\rm m}}=\omega_{{\rm cav}}$,
i.e. right at the cavity resonance. Ideally, the whole swapping pulse
sequence is shorter than the cavity decay time, which however requires
a two-mode setup with different decay rates for the modes (since the
control pulse cannot be shorter than $1/\kappa$ for a single mode).
A detailed analysis of swapping protocols can be found in \cite{2011_WangClerk_StateTransfer}.

\begin{figure}
\includegraphics[width=1\columnwidth]{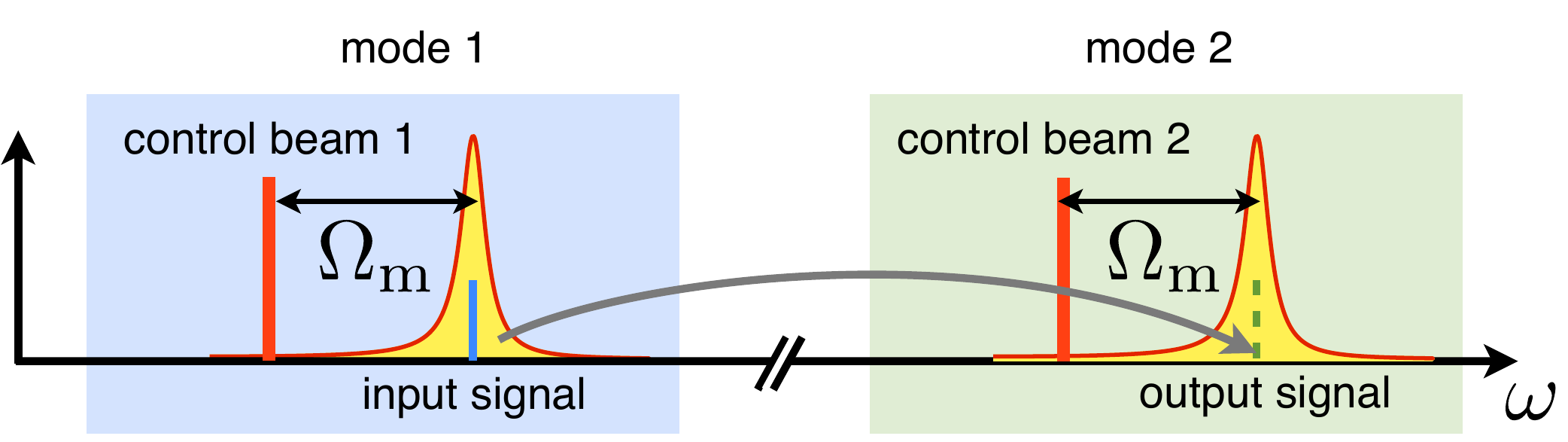}

\caption{\label{fig:Optical-wavelength-conversion}Optical wavelength conversion
in an optomechanical setup: Two modes at different frequencies are
illuminated by strong red-detuned control beams. When a signal is
injected at the resonance of mode 1, it is converted into a signal
emanating from mode 2.}
\end{figure}

There is also a simple classical picture for what happens during such
a pulse sequence: The superposition of signal and control beam leads
to a beat-note in the intensity at $\Omega_{{\rm m}}$. This translates
into a radiation pressure force that resonantly excites the mechanical
oscillator.

The optomechanical interaction can also be exploited for conversion
between different optical wavelengths. That scheme is illustrated
schematically in Fig.~\ref{fig:Optical-wavelength-conversion}. It
involves two optical resonances, each of which is driven by a strong
control beam, red-detuned by $\Omega_{{\rm m}}$. When an input signal
(e.g. a pulse) is injected at the resonance of mode 1, it will be
converted to an output signal emanating from the resonance of mode
2. This has been demonstrated recently in \cite{Hill2012,Dong2012}.

First experimental proof-of-principle demonstrations have shown these
concepts in the high-temperature, classical regime \cite{2011_FioreWang_PulsedOptomechanics,Verhagena}.
In this context it was pointed out \cite{2011_RomeroIsart_LevitationLong,Verhagena}
that the requirement for coherent state transfer of this type is $g>\kappa,\,\Gamma_{{\rm m}}\bar{n}_{{\rm th}}$.
The next challenge will be to drastically improve the fidelity and
demonstrate true quantum state transfer \cite{Parkins1999,Ritter2012},
e.g. by reconstructing the mechanical quantum state (Sec.~\ref{sub:MeasurementsQNDSingleQuadrature}). 

Recently, state transfer between a mechanical mode and an itinerant
microwave coherent state has been reported \cite{Lehnert2012}. In
this experiment, a suitably shaped microwave pulse was written onto
the motional state of a micromechanical membrane and later retrieved
via quickly switched control beams, hence realizing a coherent mechanical
memory for microwave pulses in the weak coupling regime. Another protocol
that does not require the strong coupling regime is based on quantum
state teleportation \cite{2011_RomeroIsart_LevitationLong,Hofer2011}.

If a single photon is sent into the setup and transferred to the mechanical
resonator, this will prepare the resonator in a Fock state or some
nonclassical state in general \cite{2010_Milburn_SinglePhoton,2010_KhaliliChen_NonGaussianQuantumStatesProtocol}.
That may be the most efficient route towards generating nonclassical
mechanical states in optomechanical systems as long as the single-photon
strong coupling regime (Sec.~\ref{sub:QuantumOptomechanicsNonlinear})
has not been reached.

\begin{figure}[ptb]
\includegraphics[width=3in]{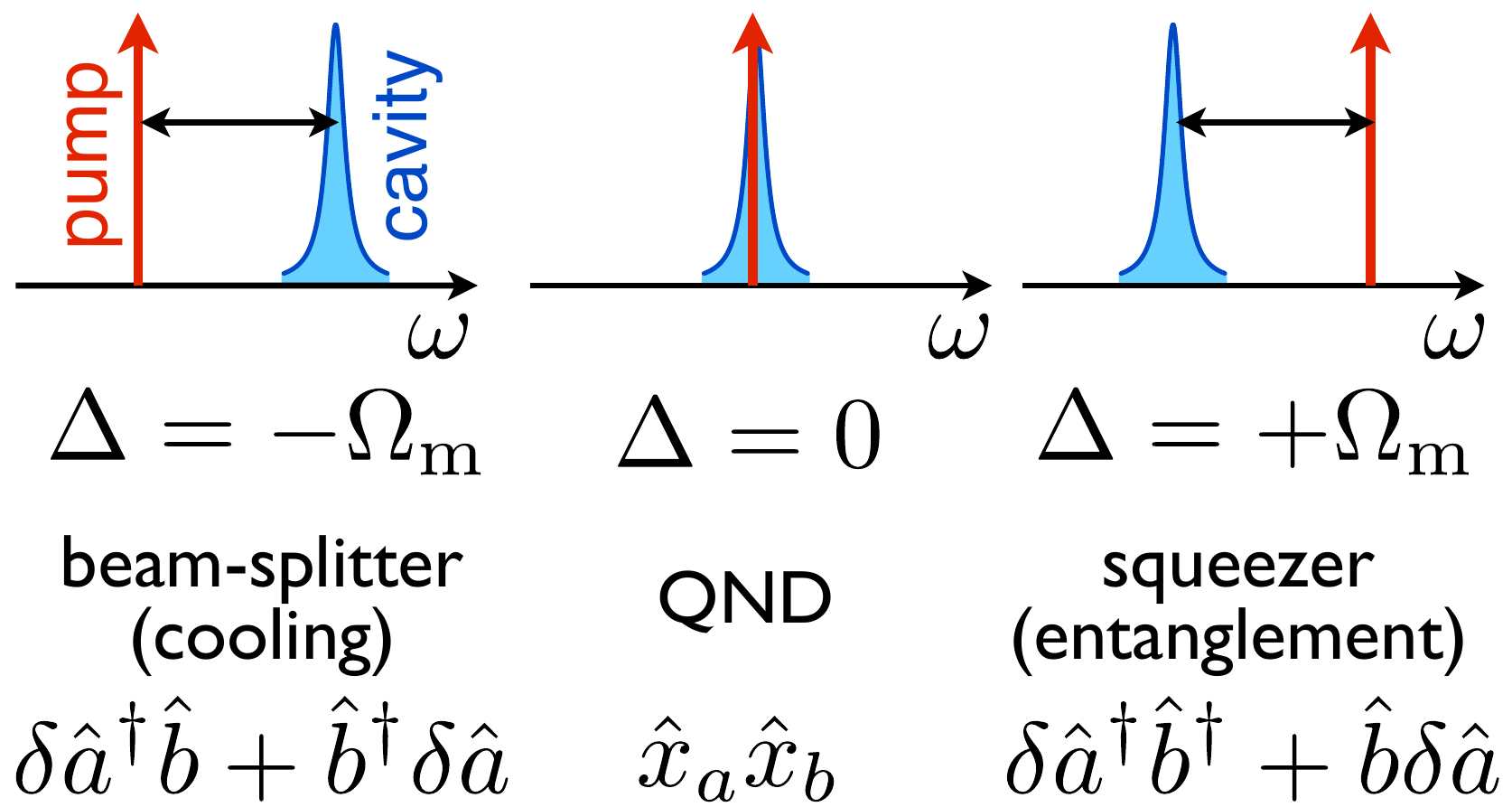}\caption{\label{fig:QuantumOptomechanicsHamiltonian}The linearized Hamiltonian
of cavity quantum optomechanics describes three different kinds of
interaction between the mechanical resonator and the driven cavity
mode, depending on which laser detuning is chosen. This is the basis
for elementary quantum protocols, such as storing optical information
into the mechanical degree of freedom.}
\end{figure}

Up to now, we have only described how single localized phonons (stored
inside the mechanical resonator) can be converted to photons and back
again. An equally interesting, or perhaps even more useful, scheme
would take traveling phonons and convert them to photons. Such a device
has been proposed recently and has been termed an optomechanical {}``traveling
wave phonon-photon translator'' \cite{2011_SafaviPainter_PhotonPhononTranslator}.
Although the frequency is shifted by many orders of magnitude, the
wave function of the outgoing single photon is designed to be a faithful
replica of the incoming phonon's wave function. 

The basic idea is the following: Phonons are traveling down a phononic
waveguide and enter a localized phononic mode, where they experience
the usual optomechanical interaction. Frequency upconversion is achieved
by having a high-intensity stream of incoming photons. In a Raman-type
scattering process, the phonon combines with one of those photons
to form a single photon at a slightly different frequency, so energy
conservation is obeyed. The strong pump beam and the weak stream of
outgoing converted photons can be efficiently separated by being coupled
to two different optical modes. Indeed, a suitably engineered optomechanical
structure \cite{2011_SafaviPainter_PhotonPhononTranslator} has two
optical modes coupling to the phonon displacement field in the following
way (which is conceptually identical to the membrane-in-the-middle
setup discussed in Sec.~\ref{sub:MeasurementsQNDFockState}):

\begin{equation}
\hat{H}=-\hbar g_{0}(\hat{b}+\hat{b}^{\dagger})(\hat{a}_{1}^{\dagger}\hat{a}_{2}+\hat{a}_{2}^{\dagger}\hat{a}_{1})+\ldots.
\end{equation}
If mode $\hat{a}_{1}$ is pumped strongly, we can replace the field
operator by the classical amplitude: $\hat{a}_{1}(t)\mapsto\bar{\alpha}e^{-i\omega_{L}t}$
(assume $\bar{\alpha}$ real-valued). If $\omega_{2}\approx\omega_{1}+\Omega_{{\rm m}}$,
then the resonant terms to be retained are:

\begin{equation}
\hat{H}=-\hbar g_{0}\bar{\alpha}(\hat{b}\hat{a}_{2}^{\dagger}e^{-i\omega_{L}t}+\hat{b}^{\dagger}\hat{a}_{2}e^{+i\omega_{L}t})+\ldots\label{eq:PhotonPhononEffectiveCoupling}
\end{equation}
This clearly displays the elementary process of converting a single
phonon to a single photon (and back). It also shows that the conversion
rate can be tuned via the pump strength $\bar{\alpha}$. All the remaining
challenge lies in ensuring that 100\% of the phonons arriving at the
device are indeed converted into photons. This is essentially an impedance-matching
problem, since the coupling of the phonon mode to the phonon waveguide
(ideally set by $\Gamma_{{\rm m}}$ if other losses can be neglected)
is usually much weaker than the coupling of the optical mode to the
photon waveguide (set by $\kappa_{2}$). Without extra fine-tuning,
most of the phonons would be reflected. This can be overcome by a
judicious choice of the coupling. Indeed, from the point of view of
the phonon mode, the coupling (\ref{eq:PhotonPhononEffectiveCoupling})
to the lifetime-broadened photon mode $\hat{a}_{2}$ gives rise to
a Fermi golden rule transition rate $4g_{0}^{2}\bar{\alpha}^{2}/\kappa_{2}$.
This is the rate at which a given localized phonon would be converted
and decay into the photonic waveguide, and it is identical to the
optomechanical cooling rate in the sideband-resolved regime. By matching
this to the coupling to the phononic waveguide, i.e. demanding $4g_{0}^{2}\bar{\alpha}^{2}/\kappa_{2}=\Gamma_{m}$,
one creates a situation that is equivalent to a two-sided cavity with
equal mirrors, where 100\% transmission can be achieved on resonance.
Consequently, in the present setup ideally 100\% of the phonons can
be converted if $2g_{0}\bar{\alpha}=\sqrt{\kappa_{2}\Gamma_{{\rm m}}}$.
A detailed analysis\cite{2011_SafaviPainter_PhotonPhononTranslator}
considers the full scattering matrix that describes scattering of
incoming phonons into the photon waveguide (or reflection back into
the phonon waveguide), and it includes the unwanted effects of extra
intrinsic losses and noise%
\footnote{In that reference, $\kappa$ and $\gamma$ refer to half the photonic
and phononic intensity decay rates $\kappa$ and $\Gamma_{{\rm m}}$
employed in our notation.%
}.

On a classical level, the device described above takes the slow amplitude
and phase modulations of the phonon field, i.e. of a soundwave traveling
down the waveguide, and transposes them into the optical domain by
shifting the carrier frequency from mechanical frequencies (e.g. ${\rm GHz}$)
up to optical frequencies. The fact that the bandwidth is set by the
smallest damping rate in the problem, which is $\Gamma_{{\rm m}}$,
can be an advantage if one uses the device as a narrow-bandwidth frequency
filter. For example, two photon-phonon translators in series (the
first one operated in reverse, i.e. going from photons to phonons)
implement a potentially very narrow filter in the optical domain.

Phonon-photon conversion has been analyzed for transferring solid-state
quantum information to the optical domain \cite{2010_Stannigel_OptomechanicalTransducersPRL,2011_Stannigel_OptomechanicalTransducersLongVersion,2011_WangClerk_StateTransfer}
and for transferring optical pulses between different wavelengths
and pulse shaping \cite{2010_TianWang_WavelengthConversion,2011_SafaviPainter_PhotonPhononTranslator,2011_Tian_StateConversion}. 

One example for the more advanced possibilities of optomechanical
quantum protocols is to perform continuous variable quantum teleportation.
Besides employing this for generating entanglement between distant
mechanical oscillators (as mentioned above), one may also teleport
an arbitrary input state of the light field onto the mechanics. The
generic idea is to start with the entanglement between the mechanical
motion and a light beam, and then to let the beam interfere at a beamsplitter
with another light beam, carrying an arbitrary input state. A subsequent
measurement in both output ports of the beamsplitter (Bell measurement)
then yields a classical measurement result on the basis of which one
manipulates the mechanical state, leaving it in a final quantum state
that equals the arbitrary input state. Such a scheme has been analyzed
for a strong short laser pulse impinging on a vibrating mirror in
a free-beam setup (without a cavity) \cite{Mancini2003,Pirandola2003},
where the reflected optical Stokes and anti-Stokes modes at $\omega_{{\rm L}}\pm\Omega_{{\rm m}}$
get entangled with the vibrations, and for a time-dependent drive
of an optomechanical cavity \cite{2011_RomeroIsart_LevitationLong,Hofer2011}.

\subsection{Nonlinear quantum optomechanics}

\label{sub:QuantumOptomechanicsNonlinear}

The optomechanical interaction

\[
-\hbar g_{0}\hat{a}^{\dagger}\hat{a}(\hat{b}+\hat{b}^{\dagger})
\]
is cubic in field operators, i.e. the corresponding Heisenberg equations
of motion are nonlinear. However, in experiments this nonlinearity
so far has only played a role in the classical regime of large amplitude
oscillations (both mechanical and with regard to the light field),
see Sec.~\ref{sub:NonlinearDynamicsInstabilityAttractorDiagram}.
In the quantum regime, we have thus far resorted to the {}``linearized''
description, with a quadratic interaction Hamiltonian of the type
$-\hbar g_{0}\sqrt{\bar{n}_{{\rm cav}}}(\delta\hat{a}^{\dagger}+\delta\hat{a})(\hat{b}+\hat{b}^{\dagger})$,
see Eq.~(\ref{eq:LinearizedInteraction}). This linearized approach
is good enough to understand many facets of cavity optomechanics:
displacement detection down to the SQL (Sec.~\ref{sub:MeasurementsDisplacementSensingAndSQL}),
the theory of optomechanical ground-state cooling (Sec.~\ref{sub:CoolingQuantumTheory}),
light/mechanics hybridization in the strong-coupling regime $g_{0}\sqrt{\bar{n}_{{\rm cav}}}>\kappa$
(Sec.~\ref{sub:CoolingStrongCouplingNormalModeSplitting}), optomechanically
induced transparency (Sec.~\ref{sub:CoolingOptomechanicallyInducedTransparency}),
optomechanical squeezing of light (Sec.~\ref{sub:QuantumOptomechanicsReadoutManipulationOfLight}),
and almost all of the various entanglement and state transfer schemes
presented in the previous sections. The experimental advantage of
the linearized interaction is that its strength $g=g_{0}\sqrt{\bar{n}_{{\rm cav}}}$
can be tuned at will by the incoming laser power. In this way, a small
value of $g_{0}$ (fixed by the setup) may be compensated for by a
stronger laser drive, until technical constraints become important.

The disadvantage of relying on the linearized interaction is that,
by itself, it will always turn Gaussian states (of the light field
and the mechanics) into Gaussian states. These may be squeezed or
entangled, but they will never have a negative Wigner density, which
may be required for certain quantum applications. It should be noted
that there are some ways around this restriction, by introducing a
nonlinearity at some other stage of the experiment: For example, one
may send in single-photon pulses and then transfer these Fock states
onto the mechanics using the linearized interaction. Another, sometimes
equivalent, option is to perform single-photon detection at the end,
thereby creating nonclassical states via post-selection of events.
These strategies are therefore related to the general schemes that
have been exploited already for linear optics quantum computation
by adding single-photon sources and photodetectors \cite{2001_KnillLaflammeMilburn_LinearOpticsQuantumComputation}.

Thus far, we have encountered only two ideas on true nonlinear quantum
optomechanics: The optical QND detection of the phonon number (Sec.~\ref{sub:MeasurementsQNDFockState})
is such an example, and it indeed would prepare (probabilistically)
Fock states of the mechanical oscillator. Another example, discussed
already very early in the literature, is the optomechanical {}``Schrödinger
cat'' type of entanglement, where a single photon should ideally
be able to displace the mechanical oscillator by about a mechanical
zero-point width (Sec.~\ref{sub:QuantumOptomechanicsEntanglement}).

These examples require a large value of $g_{0}$, which is a challenge.
We remind the reader that in a typical (Fabry-Perot type) setup the
value of $g_{0}$ can be estimated as

\begin{equation}
g_{0}=\omega_{{\rm opt}}\frac{x_{{\rm ZPF}}}{L}=\omega_{{\rm opt}}\frac{1}{L}\sqrt{\frac{\hbar}{2m_{{\rm eff}}\Omega_{{\rm m}}}}.\label{eq:g0_value}
\end{equation}
Here $L$ is the effective size of the cavity , and $m_{{\rm eff}}$
the effective mass. Both can be made small by miniaturizing the setup,
and consequently record values of $g_{0}$ are achieved in micrometer-size
devices, like photonic crystal nanobeams or very small disks and toroids.
This is even in spite of the fact that miniaturization also drives
up the mechanical frequency $\Omega_{{\rm m}}$. In such setups, $g_{0}$
currently takes values on the order of up to some ${\rm MHz}$ (see
Sec.~\ref{sub:OptomechanicalCouplingParameters}).

Next, we should discuss in which sense $g_{0}$ can be {}``large''.
The steady-state displacement produced by a single photon on average
($\bar{n}_{{\rm cav}}=1$) is

\begin{equation}
\frac{\delta x}{x_{{\rm ZPF}}}=2\frac{g_{0}}{\Omega_{{\rm m}}}.\label{eq:SinglePhotonDisplacement}
\end{equation}
Thus, to displace by more than the zero-point width (mechanical ground
state width), one needs $g_{0}>\Omega_{{\rm m}}$. However, if the
photon decay rate $\kappa$ is large, then one can see only the average
displacement produced by the photon number fluctuating around $\bar{n}_{{\rm cav}}=1$,
and one would not resolve the granularity of the photon stream. Such
a situation should still be well described within the linearized approximation.
To obtain truly nonlinear effects, one would rather like to make sure
the following picture applies. Take any single photon entering the
cavity. If its lifetime is large enough ($\Omega_{{\rm m}}\gg\kappa$,
the resolved-sideband regime), it will displace the oscillator by
the amount given in Eq.~(\ref{eq:SinglePhotonDisplacement}). This
then implies $g_{0}\gg\kappa$ as a necessary (but not sufficient)
condition.

These considerations directly lead us to consider the ratio

\begin{equation}
\frac{g_{0}}{\kappa}\,.
\end{equation}
If that ratio is larger than one, then the presence of a single phonon
would shift the optical frequency by more than a cavity linewidth.
More precisely, the mechanical displacement produced by a superposition
of 0 and 1 phonons would be measured so efficiently that the passage
of a single photon through the cavity already destroys the superposition.
We can also look at the light field's back-action: In the {}``bad
cavity'' limit ($\kappa\gg\Omega_{{\rm m}}$), a single passing photon
with a lifetime $\kappa^{-1}$ gives an average momentum kick $\delta p=2p_{{\rm ZPF}}g_{0}/\kappa$
to the mechanical oscillator, which would be larger than its zero-point
momentum uncertainty $p_{{\rm ZPF}}=m_{{\rm eff}}\Omega_{{\rm m}}x_{{\rm ZPF}}$
if $g_{0}>\kappa$. One can make a connection to the Lamb-Dicke parameter
used in ion-trap physics, by defining the ratio of momentum kick to
momentum zero-point fluctuations: $\eta_{{\rm Lambd-Dicke}}\equiv\delta px_{{\rm ZPF}}/\hbar=\delta p/(2p_{{\rm ZPF}})=g_{0}/\kappa$. 

The ratio $g_{0}/\kappa$ has been called the {}``granularity parameter''
\cite{Murch2008}, as $g_{0}/\kappa>1$ allows to reveal the granularity
of the photon stream (the discreteness of individual photons). It
has reached values on the order of and larger than one in experiments
with clouds of ultracold atoms (see Sec.~\ref{sub:ExperimentsUltracoldAtoms}).

There are other ways of interpreting $g_{0}/\kappa>1$ as well: The
(sideband-resolved) cooling rate $g_{0}^{2}\bar{n}_{{\rm cav}}/\kappa$
would be so large that one enters the strong-coupling regime (Sec.~\ref{sub:CoolingStrongCoupling})
already for $\bar{n}_{{\rm cav}}=1$. Alternatively, $g_{0}/\kappa$
can be written as the ratio between the mechanical zero-point fluctuations
and the width of the optical cavity resonance, expressed in terms
of a displacement (where the optomechanical coupling enters):

\[
\frac{g_{0}}{\kappa}=\frac{x_{{\rm ZPF}}}{\delta x_{{\rm cav}}}\,,
\]
where $\delta x_{{\rm cav}}=\kappa/G=\kappa x_{{\rm ZPF}}/g_{0}$.

There is an additional interesting aspect about $g_{0}/\kappa$. It
can serve as a {}``quantumness'' parameter \cite{Ludwig2008a},
with larger values denoting a gradual classical-to-quantum crossover.
All the parameters of any given standard optomechanical setup can
be boiled down to the following five dimensionless combinations:

\begin{equation}
\frac{\kappa}{\Omega_{{\rm m}}},\, Q_{{\rm m}}=\frac{\Omega_{{\rm m}}}{\Gamma_{{\rm m}}},\,\frac{\Delta}{\Omega_{{\rm m}}},\,\frac{g}{\kappa},\,\frac{g_{0}}{\kappa}
\end{equation}
Here the first four do not depend on the value of Planck's constant.
This is obvious for the first three (the sideband-resolution ratio,
the mechanical quality factor, and the laser detuning in units of
mechanical frequency). It is less obvious for $g/\kappa$. However,
this can be written as

\begin{equation}
\frac{g}{\kappa}=\frac{g_{0}}{\kappa}\sqrt{\bar{n}_{{\rm cav}}}=\sqrt{\frac{E_{{\rm cav}}}{2m_{{\rm eff}}L^{2}}\frac{\omega_{{\rm opt}}}{\Omega_{{\rm m}}}}.
\end{equation}
Here $E_{{\rm cav}}=\hbar\omega_{{\rm opt}}\bar{n}_{{\rm cav}}$ is
the light energy stored inside the cavity, which is connected to the
laser driving power. In this sense, $g/\kappa$ serves as a dimensionless
classical measure of laser power. Only the ratio $g_{0}/\kappa$ depends
on $\hbar$, as can be seen from Eq.~(\ref{eq:g0_value}): $g_{0}\propto\sqrt{\hbar}$.

Thus, one can imagine keeping all four classical ratios fixed, and
only increasing $g_{0}/\kappa$ by changing parameters in the setup.
This is then completely equivalent to increasing Planck's constant,
allowing one to resolve more and more quantum features as $g_{0}/\kappa$
grows. It should be noted that of course even for $g_{0}/\kappa\ll1$
one can observe quantum effects, but only in the linearized regime.
Proving the quantumness of these effects produced by linearized interactions
usually requires a quantitative comparison (e.g. with the light field's
or the oscillator's zero-point fluctuations). In contrast, some features
observed for larger values of $g_{0}/\kappa$ may even be qualitatively
distinct from classical predictions. An example noted above are the
quantum jumps of phonon number that could be observed for $g_{0}>\kappa_{{\rm abs}}$
(Sec.~\ref{sub:MeasurementsQNDFockState}).

The third important ratio involving the coupling is

\[
\frac{g_{0}^{2}}{\Omega_{{\rm m}}\kappa}.
\]
This is the ratio between the strength $g_{0}^{2}/\Omega_{{\rm m}}$
of the effective photon-photon interaction induced by the mechanics,
and the optical linewidth. When this starts to be larger than one
(and $\kappa\ll\Omega_{{\rm m}}$), then the presence of one photon
shifts the resonance sufficiently that a second photon cannot enter
the cavity. That leads to the photon-blockade phenomenon (see below).

In \cite{Ludwig2008a}, the regime of $g_{0}\sim\kappa\sim\Omega_{{\rm m}}$
was considered using both full numerical master equation simulations
and a quantum Langevin approach. There, it was found that for increasing
values of $g_{0}/\kappa$ quantum fluctuations start to have a pronounced
effect on the mechanical lasing instability that is observed at blue-detuned
laser driving. The strongly enhanced susceptibility of the system
just below the threshold amplifies the effects of these fluctuations,
and the threshold is smeared and shifted. In this quantum regime,
the co-existence of several attractors (known from the classical case,
see Sec.~\ref{sub:NonlinearDynamicsInstabilityAttractorDiagram})
results in non-Gaussian mechanical Wigner densities and mechanical
states with non-Poissonian phonon distributions and large Fano factors
\cite{Ludwig2008a}. It was found that in this regime for appropriate
parameters one can even generate true nonclassical mechanical states,
with partially negative Wigner densities \cite{2011_Qian_NegativeWignerDensity}.
These states are present in the steady state (under constant drive),
so the Wigner densities could then be read out according to the schemes
presented in Sec.~\ref{sec:QuantumOpticalMeasurementsOfMechanicalMotion}.

In \cite{2011_Nunnenkamp_SinglePhotonOptomechanics}, master equation
simulations were further extended to discuss the full range of detunings
and the excitation spectrum of the cavity. Multiple optical sidebands
are found, and the mechanical state of the oscillator is seen to develop
non-Gaussian states particularly at detunings which drive multi-photon
transitions. 

The nonlinear quantum optomechanical regime leads to very interesting
photon correlations. In particular, under appropriate conditions one
may observe optomechanically induced photon blockade \cite{2011_Rabl_PhotonBlockadeEffect}.
This shows up in the photon-photon correlations as strong anti-bunching,
$g^{(2)}(t=0)<1$, which has been calculated and discussed in \cite{2011_Rabl_PhotonBlockadeEffect}
for the case of weak laser driving. The regime of optomechanical photon
blockade requires sideband resolution ($\kappa<\Omega_{{\rm m}}$)
as well as strong single-photon coupling with $g_{0}>\kappa$ and
$g_{0}^{2}>\Omega_{{\rm m}}\kappa$. In a more recent work, the analysis
of photon correlations in this regime was extended to cover the full
temporal evolution of $g^{(2)}(t)$ and the Fano factor, as well as
higher moments of the photon counting statistics \cite{2012_Kronwald_FullPhotonStatistics}.

In \cite{2011_Chen_SinglePhotonDevices}, an exact solution was presented
for the regime where only a single photon is transmitted through such
a strongly coupled optomechanical setup, which is important for the
generation of entanglement (Sec.~\ref{sub:QuantumOptomechanicsEntanglement}).

The understanding of the strongly nonlinear quantum regime is aided
by the following picture, already discussed partially in Sec.~\ref{sub:QuantumOptomechanicsEntanglement},
and first employed in \cite{Bose1997,Mancini1997}: For any given
photon number $n_{{\rm cav}}$, the mechanical equilibrium position
is shifted by an amount $2x_{{\rm ZPF}}n_{{\rm cav}}g_{0}/\Omega_{{\rm m}}$.
One should thus consider the mechanical Fock states in this new displaced
parabolic potential. 

If an additional photon enters the cavity, the potential suddenly
shifts, but the mechanical wave function at first remains the same.
Thus, the overlaps of the given initial wave function and the displaced
new Fock states will determine the strength of possible transitions.
These overlap integrals are known as Franck-Condon factors from the
theory of molecules, where vibrations may be excited during electronic
transitions.

We conclude by briefly describing the formal treatment via the well-known
polaron transformation, as applied to an optomechanical system \cite{Mancini1997,Groblacher2009a,2011_Rabl_PhotonBlockadeEffect,2011_Nunnenkamp_SinglePhotonOptomechanics}.
We consider the standard optomechanical Hamiltonian (in the frame
rotating with the laser frequency)

\[
\hat{H}=-\hbar\Delta\hat{a}^{\dagger}\hat{a}+\hbar\Omega_{{\rm m}}\hat{b}^{\dagger}\hat{b}-\hbar g_{0}\hat{a}^{\dagger}\hat{a}(\hat{b}+\hat{b}^{\dagger})+\hbar\alpha_{L}(\hat{a}+\hat{a}^{\dagger})+\hat{H}_{{\rm bath}},
\]
where $\alpha_{L}$ is proportional to the laser amplitude. We complete
the square to obtain 

\begin{equation}
\hbar\Omega_{{\rm m}}\left(\hat{b}-\frac{g_{0}}{\Omega_{{\rm m}}}\hat{a}^{\dagger}\hat{a}\right)^{\dagger}\left(\hat{b}-\frac{g_{0}}{\Omega_{{\rm m}}}\hat{a}^{\dagger}\hat{a}\right)-\hbar(g_{0}^{2}/\Omega_{{\rm m}})(\hat{a}^{\dagger}\hat{a})^{2}
\end{equation}
for the first three terms in $\hat{H}$. This shows two things: First,
an effective photon-photon interaction is generated, viz. the $(\hat{a}^{\dagger}\hat{a})^{2}$
term, which is crucial for nonlinear effects and quantum gates.

\begin{figure}
\includegraphics[width=0.8\columnwidth]{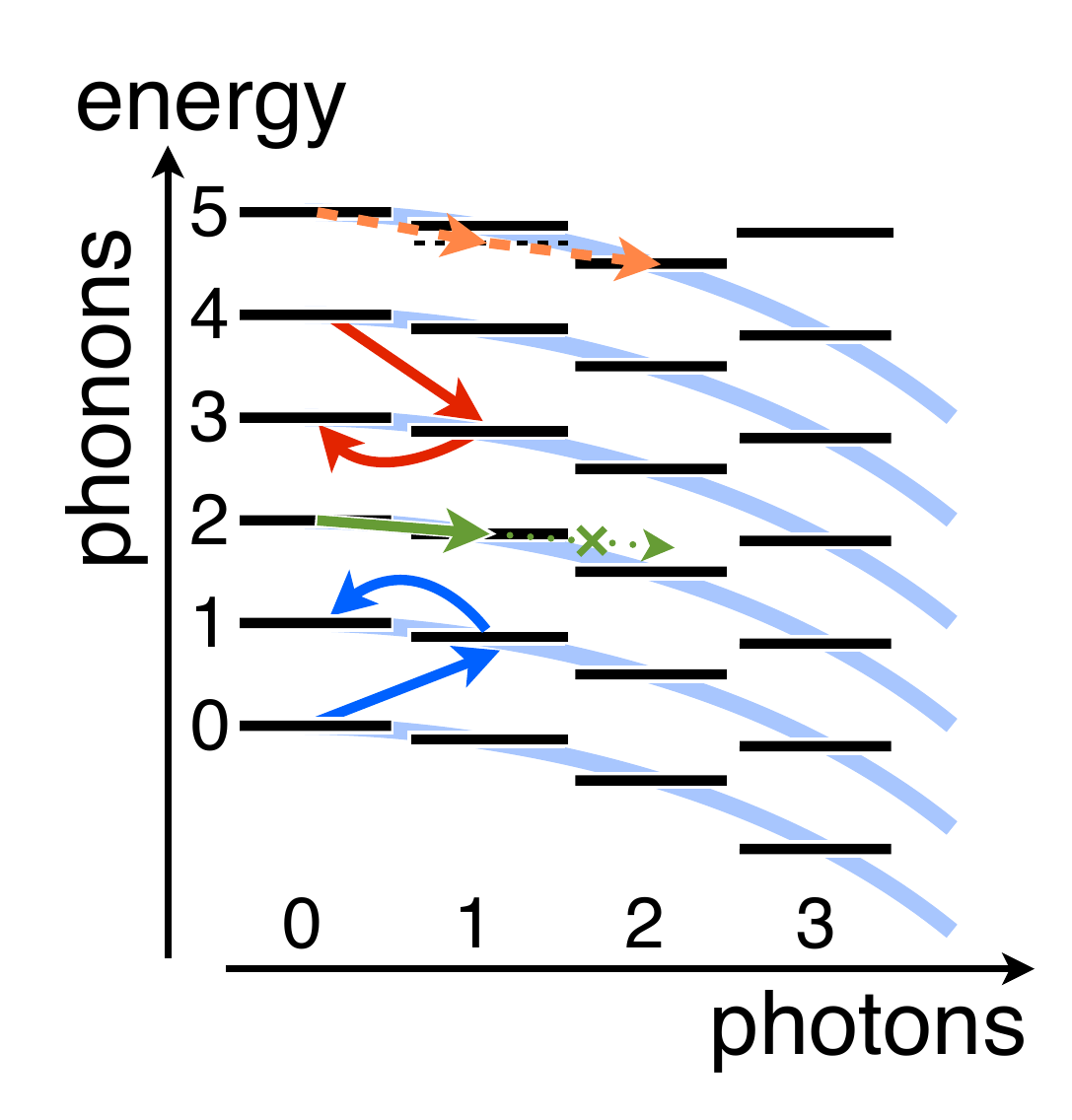}

\caption{\label{fig:LevelSpectrumNonlinearOptomechanics}The energy level spectrum
of an optomechanical system at zero detuning $\Delta=0$: $E(n_{{\rm phonon}},n_{{\rm cav}})/\hbar=-(g_{0}^{2}/\Omega_{{\rm m}})n_{{\rm cav}}^{2}+\Omega_{{\rm m}}n_{{\rm phonon}}$.
Some possible processes are indicated: Blue-detuned laser drive at
the first sideband ($\Delta\approx+\Omega_{{\rm m}}$) leading to
creation of a phonon and subsequent escape of the photon out of the
cavity (blue), a similar cycle at red-detuned drive ($\Delta\approx-\Omega_{{\rm m}}$)
leading to cooling by annihilation of a phonon (red), photon-blockade
prohibiting transitions $0\rightarrow1\rightarrow2$ towards two photons
(green), and a resonant two-photon transition $0\rightarrow2$ (orange)
via a virtual intermediate state (dashed). Note that the last two,
nonlinear effects require $\Omega_{{\rm m}},g_{0}^{2}/\Omega_{{\rm m}}>\kappa$.}
\end{figure}

Second, $n$ photons shift $\hat{b}$ by $(g_{0}/\Omega_{{\rm m}})n$,
i.e. they shift the equilibrium oscillator position by $2x_{{\rm ZPF}}(g_{0}/\Omega)n$
to the right. This can be accomplished by a unitary $\hat{U}=\exp[(\hat{b}^{\dagger}-\hat{b})(g_{0}/\Omega)\hat{a}^{\dagger}\hat{a}]$
acting on the wave functions. After applying that transformation to
the Hamiltonian, via $\hat{H}^{{\rm new}}=\hat{U}^{\dagger}\hat{H}\hat{U}$,
we obtain a Hamiltonian that would be diagonal in the absence of driving
and decay:

\begin{align}
\hat{H} & =-\hbar(\Delta+\frac{g_{0}^{2}}{\Omega}\hat{a}^{\dagger}\hat{a})\hat{a}^{\dagger}\hat{a}+\hbar\Omega\hat{b}^{\dagger}\hat{b}+\nonumber \\
 & \alpha_{L}(\hat{a}^{\dagger}e^{(\hat{b}-\hat{b}^{\dagger})(g_{0}/\Omega)}+h.c.)+\hat{U}^{\dagger}\hat{H}_{{\rm bath}}\hat{U}\,.
\end{align}
Note that in these new coordinates the addition of one photon shifts
the wave function to the left, whereas the center of the mechanical
oscillator potential is now forced to remain fixed (in contrast to
the actual physical situation, where the potential shifts and the
wave function is fixed). The exponential (polaron operator) in the
laser driving term generates the Franck-Condon overlap factors.

\begin{figure}
\includegraphics[width=1\columnwidth]{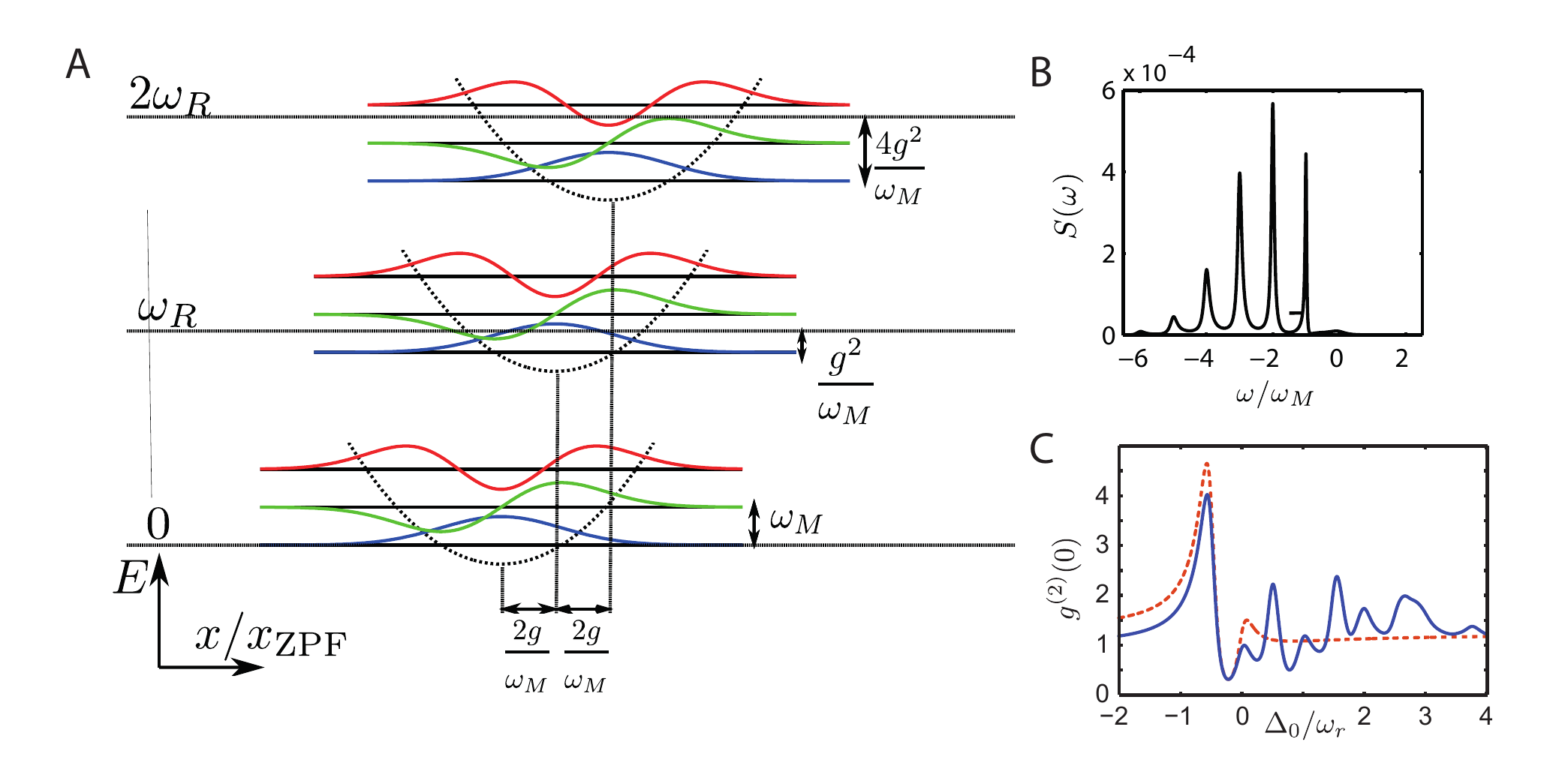}

\caption{Single-photon strong coupling regime. (a) Illustration of displacement
depending on photon number (leading to Franck-Condon physics). (b)
Transmission sidebands. (c) Photon-photon correlator as a function
of detuning (for weak drive). {[}(a),(b) courtesy of A. Nunnenkamp,
(c) courtesy of P. Rabl; $g$ in this figure refers to our $g_{0}${]}}
\end{figure}

The energy level scheme for the nonlinear quantum optomechanical regime
(Fig.~\ref{fig:LevelSpectrumNonlinearOptomechanics}) displays equally
spaced phonon ladders whose offset shifts depending on the photon
number: $E(n_{{\rm phonon}},n_{{\rm cav}})/\hbar=-(\Delta+g_{0}^{2}/\Omega_{{\rm m}}n_{{\rm cav}})n_{{\rm cav}}+\Omega_{{\rm m}}n_{{\rm phonon}}$.
In this scheme, the driving laser only induces energy-conserving transitions,
horizontal in the diagram. Different transitions are activated upon
changing the detuning. Multi-photon transitions become possible via
virtual non-resonant intermediate states (e.g. adding two photons
at once).

\section{Future Perspectives}

\label{sec:FuturePerspectives}The fast experimental and theoretical
advances in cavity optomechanics during the recent years are constantly
opening up new avenues with respect to applications and tests on the
foundations of physics. Here we remark briefly on the broader outlook.

\subsection{Foundational aspects}

\label{sub:FuturePerspectivesFoundational}

The ability to achieve coherent quantum control over the center of
mass motion of massive mechanical objects provides a fresh approach
to fundamental tests of quantum theory in a hitherto unachieved parameter
regime. Specifically, quantum optomechanics offers a universal scheme
for experiments in the quantum regime of massive mechanical objects
from clouds of $10^{5}$ atoms or nanometer-sized solid-state devices
of $10^{7}$ atoms and a mass of $10^{-20}$ kg, to micromechanical
structures of $10^{14}$ atoms and $10^{-11}$ kg, to macroscopic
centimeter-sized objects for gravitational wave detectors comprising
more than $10^{20}$ atoms and weighing up to several kg. In principle,
this offers a range of almost 20 orders of magnitude in mass and
6 orders of magnitude in size for macroscopic quantum experiments. 

A specific example where quantum optomechanics provides a new direction
for future experiments is the quantum measurement problem, which addresses
the question why quantum superpositions do not seem to occur at the
level of macroscopic objects \cite{Leggett2002a}. Various new theories
and phenomena beyond quantum theory have been suggested in order to
achieve an irreversible decay of superposition states, i.e. decoherence,
into well-defined classical states \cite{Karolyhazy1966,Diosi1984,Penrose1996,Bassi2003,Adler2009}.
Each of these approaches predicts a particular scaling of the decoherence
rates with particle number or mass, and with the actual distinctness
of the states involved in the superposition. For sufficiently macroscopic
systems and sufficiently distinct superposition states, these predictions
deviate significantly from the decoherence rates expected from standard
quantum theory \cite{Zurek1991,Zurek2003,Schlosshauer2008}. Current
matter-wave experiments with molecules may soon start to enter such
a regime \cite{Gerlich2011,Nimmrichter2011}. Systematic tests of
the validity of quantum theory necessarily also involve tests of such
scaling laws and the large  mass range offered by quantum optomechanical
systems will provide a unique opportunity. One way of producing quantum
states involving superpositions of mechanical states is via optomechanical
entanglement (see Sec. \ref{sub:QuantumOptomechanicsEntanglement}).
Probing the decoherence of such optomechanical superposition states,
for example via the interference visibility in a single-photon interferometer
\cite{1998_Bouwmeester_IndiaProposal,Bose1999,Marshall2003,Pepper2012},
may allow decisive tests of specific 'collapse' models \cite{Bassi2005,Kleckner2008}.
A particularly exciting perspective is to extend these experiments
to the large masses that are available in gravitational wave interferometers
(see \cite{Chen2013} for a recent review) and that were recently
cooled to 200 thermal quanta above the quantum ground state \cite{Abbott2009}.
Another route that has been suggested is to analyse the contrast of
matter-wave interference of levitated nano-objects, where superpositions
of macroscopically distinct position states are generated via optomechanics
\cite{Romero-Isart2011}. It has been shown that such experiments
would in principle allow to enter a regime in which all non-standard
decoherence theories can be systematically tested \cite{Romero-Isart2011a}.
The demanding experimental boundary conditions with respect to temperature
and background pressure (to minize the effects of standard decoherence)
might require the added benefit of a space enviroment. First studies
along this line are currently being performed \cite{Kaltenbaek2012}
.

Another fascinating long-term perspective is the possibility to make
use of the accessible large masses in quantum optomechanics experiments
to explore the scarcely studied interface between quantum physics
and gravity. Some of its aspects are already covered by the decoherence
tests discussed above, as the models of Karolyhazy \cite{Karolyhazy1966},
Diosi \cite{Diosi1984,Diosi2007} and Penrose \cite{Penrose1996,Penrose2000}
identify gravity as the dominant player of their state-vector collapse.
A completely different approach has been taken by a recent proposal
\cite{Pikovski2011} that suggests that quantum optomechancis experiments
could test directly predictions from quantum gravity. Specifically,
the availability of large masses in combination with quantum optical
state preparation and readout is shown to be sensitive to possible
deviations from the quantum commutation relation even at the Planck
scale. This would open the route to table-top quantum optics tests
of quantum gravity predictions. 

The possibility to interconnect optomechanical devices in large scale
arrays has already been discussed in the context of investigating
synchronization effects \cite{2011_Heinrich_SynchronizationPRL}.
The dynamics in such arrays may also enable the study of many-body
quantum effects, which could complement the current efforts in quantum
simulations, yet in a solid-state architecture \cite{Ludwig2012c,Tomadin2012}.
Another scarcely explored direction is to exploit the role of nonlinear
mechanical responses. For example, the double-well potential of bistable
mechanical resonators \cite{Bagheri2011} could be the starting point
for macroscopic tunnelling experiments. 

Finally, it is interesting to note that also in a broader context,
the topic of controlled photon-phonon interaction is receiving increasing
attention. For example, a recent experiment has demonstrated quantum
entanglement between optical phonon modes of two separate macroscopic
solids, specifically of milimiter-scale bulk diamond at room temperature
\cite{Lee2011}, which has been generated from photon-phonon entanglement.
Besides providing an interesting alternative to obtain quantum effects
involving macroscopic objects this is also of direct relevance for
applications, such as solid state quantum memories. Other examples
include new schemes to achieve coherent conversion of bosonic modes,
and even suggestions to exploit the measurement of optomechanical
recoil energies for mesoscopic mass standards\cite{Lan2013}.

\subsection{Applications}

\label{sub:FuturePerspectivesApplications}

Although the field of cavity optomechanics is only in its beginnings,
several domains of applications have become obvious already now. In
laser sciences, these include tunable optical filters, based on the
fact that optomechanical coupling can lead to extreme tuning of the
mechanical frequency up to several octaves, as well as optomechanical
implementations of laser stabilization \cite{Alegre2010}. In addition,
the compatibility of some optomechanical devices with silicon photonics
(see Sec. \ref{sec:ExperimentalRealizationAndParameters}) enables
on-chip optical architectures with added versatility. For example,
exploiting strong optical nonlinearities provided by optomechanical
cavities (see Sec.\ref{sec:BasicConsequencesOfOptomechanics}) adds
an important and long-sought feature to optical information processing.
Along the same lines, embedded optomechanical cavities have been shown
to serve as an all-optical memory element \cite{Bagheri2011,Cole2011b},
or have been proposed as a new technology for single-photon detection
\cite{2012_Ludwig_TwoModeScheme}. In the first case, the binary states
of a bistable nanomechanical resonator are controlled and monitored
by an optomechanical cavity; in the latter case, a single photon would
induce a measureable frequency shift to an optomechanical cavity,
thereby resulting in a detection that is in principle destruction-free
and photon-number resolving. In the long run, these features may provide
a new momentum to all-optical information processing \cite{Caulfield2010}.

For sensing applications, cavity optomechanics provides several new
aspects: for example, while damping of mechanical motion has been
used to increase the bandwidth of scanning microscopes since decades
\cite{Garbini1996,Bruland1996}, cavity optomechanical devices allow
both readout and damping of much higher mechanical frequencies, hence
providing faster sampling and scanning rates. At the same time, the
high sensitivity of the optical readout allows new integrated optomechanical
platforms for acceleration sensing. A recent demonstration using optomechanical
crystals achieved an on-chip acceleration resolution of 10 $\mu g/\sqrt{Hz}$with
a test mass of only a few ng\cite{Krause2012}. Optomechanical cooling
in combination with on-chip mechanical sensors has recently also been
suggested to provide a reduction in thermal noise for the optical
readout \cite{Winger2011}. In turn, the ability to coherently amplify
mechanical motion provides a route to radiation pressure driven coherent
oscillators with compact form factor and low power consumption. Finally,
the combination with optomechanical preparation of squeezed mechanical
states (see Sec.\ref{sub:QuantumOptomechanicsManipulationOfMechanics})
could lead to a new mechanical sensing technology with unprecedented
levels of sensitivity due to the reduced position variance of the
readout device.

From a quantum information processing perspective cavity optomechanics
offers a new architecture for coherent light-matter interfaces in
a solid-state implementation. Mechanical motion can serve as a universal
transducer to mediate long-range interactions between stationary quantum
systems (see Sec.\ref{sub:QuantumOptomechanicsQuantumProtocols}).
The specific trait of optomechanical systems is the interconversion
between stationary qubits and flying (photonic) qubits, which constitutes
one of the main elements of long-distance quantum communication and
a future quantum internet \cite{Kimble2008}. At the same time, strong
optomechanical coupling in the single-photon regime opens up the field
of non-Gaussian quantum optomechanics with a wealth of quantum operations
and protocols (see Sec.\ref{sub:QuantumOptomechanicsNonlinear}).
The phenomenon of optomechanically induced transparency enables slowing
of light pulses or even their storage, hence providing an interesting
solid-state implementation of a quantum memory (see Sec.\ref{sub:CoolingOptomechanicallyInducedTransparency}). 

Eventually, combining cavity optomechanics with other transduction
mechanisms will allow to exploit the full functionality of micro-
and nanomechanical devices. Such quantum hybrid systems utilize the
mechanical motion to achieve coupling between otherwise incompatible
or uncoupled quantum systems (see Sec. \ref{sub:QuantumHybridSystems}).
A particularly exciting perspective of opto-electromechanical hybrid
devices is their ability for coherent conversion between optical and
microwave frequencies. Cavity cooling in these hybrid structures could
also be applied to certain modes of a heat bath in integrated electronic
circuits, for example to suppress unwanted thermalization effects
in spintronic devices \cite{Usami2012}. Another interesting direction
is to couple individual qubits, for example single atoms or single
spins, to optomechanical devices. In combination with large mechanical
frequencies such structures could allow mechanically mediated qubit
interactions without additional laser cooling of the mechanical modes,
thereby significantly relaxing the experimental requirements for information
processing in qubit registers.

\section{Acknowledgements}

We would like to thank our many colleagues and co-workers in the field,
from whom we have learned a lot about this subject in numerous discussions.
We are grateful to Anton Zeilinger for enabling a stay at the {}``Internationale
Akademie Traunkirchen'' in the summer of 2010, during which this
review was started.

Among the many helpful remarks we received from colleagues, we would
like to thank in particular Sebastian Hofer, Max Ludwig, Oskar Painter,
Albert Schliesser and Dan Stamper-Kurn for their critical reading of
the manuscript and valuable feedback, and Uros Delic, David Grass
and Nikolai Kiesel for their assistance with figures and references.

Each of us acknowledges support by a Starting Grant of the European
Research Council (ERC StG). In addition, we like to acknowledge the
following funding sources: the DFG with the Emmy-Noether program (F.M.),
the DARPA ORCHID program (F.M., T.J.K.), the European FP7 STREP project
MINOS (M.A., T.J.K.) and the Marie-Curie ITN cQOM (M.A., T.J.K., F.M.).
T.J.K. acknowledges funding from the Swiss National Science Foundation
(SNF) and the NCCR of Quantum Engineering. M.A. acknowledges support
from the Austrian Science Fund FWF (projects FOQUS, START), from the
European Commission (Q-Essence) and from the European Space Agency
ESA.

\section{Appendix: Experimental Challenges}

\subsection{Influence of classical excess laser phase noise on laser cooling}

\label{Appendix:InfluenceOfLaserPhaseNoise}

We briefly consider the role of laser noise in optomechanical cooling.
Of particular interest is phase noise, described by the phase noise
spectral density $\bar{S}_{\phi\phi}(\Omega)$ (or alternatively described
by its equivalent frequency noise $\bar{S}_{\omega\omega}(\Omega$$)=\bar{S}_{\phi\phi}(\Omega)/\Omega^{2}$).
Excess phase (and amplitude) noise can be found in many laser systems
due to relaxation oscillations, which can even in the case of diode
lasers extend well into the GHz regime \cite{Kippenberg2012,Safavi2012,Wieman1991}.
Such excess phase noise has has been experimentally observed to heat
the mechanical oscillator \cite{Schliesser2008} and has been analyzed
theoretically \cite{Schliesser2008,Diosi2008,Rabl2009}. The spectral
density of force fluctuations caused by this noise when pumping on
the lower sideband in the resolved sideband regime, is given by (with
$\eta=\kappa_{{\rm ex}}/\kappa$):
\begin{equation}
\bar{S}_{FF}^{\mathrm{fn}}(\Omega_{\mathrm{m}})\approx\frac{4\eta^{2}G^{2}P^{2}}{\omega^{2}\Omega_{\mathrm{m}}^{4}}\bar{S}_{\omega\omega}(\Omega_{\mathrm{m}})
\end{equation}
 By comparing this force noise to an effective thermal Langevin force
of the laser ($\bar{S}_{FF}^{{\rm fn}}(\Omega_{m})=2m_{eff}\Gamma_{m}\bar{n}_{L}\hbar\Omega_{m}$)
near the mechanical resonance an equivalent laser noise occupation
$\bar{n}_{L}$ can be derived. The final occupancy of the mechanical
oscillator in the presence of optomechanical sideband cooling is subsequently
$n_{f}=\frac{\Gamma_{m}}{\Gamma_{m}+\Gamma_{opt}}(\bar{n}_{th}+\bar{n_{L}})$,
where $\bar{n}_{th}$ denotes the average occupancy of the thermal
bath. The excess contribution of the frequency noise is therefore:
\begin{equation}
n_{\min}^{excess}=\frac{\bar{n}_{{\rm cav}}}{\kappa}S_{\omega\omega}(\Omega_{m})\label{eq:}
\end{equation}
The lowest occupancy that can be attained in the presence of excess
phase noise is given by \cite{Rabl2009}

\begin{equation}
\bar{n}_{min}=\sqrt{\frac{\bar{n}_{th}\Gamma_{m}}{g_{0}^{2}}\bar{S}_{\omega\omega}(\Omega_{\mathrm{m}})}
\end{equation}
This expression can also be recast into a condition for the amount
of phase noise that would lead to unit occupancy, i.e. $P_{0}=0.5$.
Given the cavity photon number required to cool near to the ground
state ($\bar{n}_{f}=1$), we find that $n_{\min}^{excess}>1$ if the
frequency noise exceeds the level \cite{Rabl2009} 
\begin{equation}
\bar{S}_{\omega\omega}(\Omega_{m})>\frac{g_{0}^{2}}{k_{B}T/\hbar Q_{m}}
\end{equation}
If this level of noise is present, phase noise will preclude ground
state cooling. This formula also reveals that to mitigate the effect
of phase noise it is generally desirable to have a large vacuum optomechanical
coupling rate and a low mechanical damping rate, to increase the level
of tolerable phase noise.

\subsection{Influence of cavity frequency noise on laser cooling}

\label{Appendix:InfluenceOfCavityFrequencyNoise}

The cavity frequency fluctuations are driven by thermodynamical temperature
fluctuations. Considering the situation were light propagates inside
a dielectric cavity (microtoroid, microspheres, photonic crystals)
or penetrates a mirror surface, the temperature fluctuations of the
sampled volume $\mathcal{V}$ are determined both by the absolute
temperature and the heat capacity of that volume:
\[
\langle\delta T^{2}\rangle=\frac{T^{2}}{\rho C\mathcal{V}}
\]
Here $C$ is the specific heat capacity (per mass). In general, these
temperature fluctuations will exhibit a spectral density that depends
on the resonator geometry and boundary conditions for the thermal
transport.\ For some cases, such as silica microspheres, the spectral
densities $S_{TT}[\omega]$ are known analytically \cite{Gorodetsky2004}
and the corresponding frequency noise, $S_{\omega\omega}[\omega]=K_{{\rm th}}^{2}S_{TT}[\omega]$,
can be evaluated (where $K_{{\rm th}}=\frac{dn}{dT}\omega_{0}$ in
the case of thermorefractive noise and $K_{{\rm th}}=\frac{d\alpha}{dT}\omega_{0}$
in the case of thermoelastic noise). While it has been noted that
one can in principle compensate also thermorefractive noise \cite{kimble2008b},
the noise is of particular relevance to optomechanical cooling as
it provides a limit to the minimal occupancy. This can be understood
with the model referred to in the last section, ie. the fact that
cavity frequency noise will translate into radiation pressure force
fluctuations in the presence of a strong cooling laser. 

\textit{Finite thermal cavity occupancy. }A further source of cavity
noise can arise from the fact that the cooling field has thermal occupation.
This is in particular the case for microwave fields due to their low
frequency.\ If the cavity occupation is given by $\bar{n}_{{\rm cav}}^{{\rm th}}$,
then the final occupation is modified to (in the resolved sideband
limit):
\begin{equation}
\bar{n}_{f}=\bar{n}_{{\rm th}}\frac{\Gamma_{{\rm m}}}{\Gamma_{{\rm eff}}}+\bar{n}_{{\rm cav}}^{{\rm th}}+\frac{\kappa^{2}}{16\Omega_{m}^{2}}
\end{equation}
This implies that the final phonon number can never be below the effective
thermal occupation of the drive field \cite{Dobrindt2008}. It should
be noted that when the radiation field and the mechanical oscillator
initially have the same bath temperature (as will be the case in equilibrium,
without extra absorption), the equilibration of these two oscillators
of frequency $\omega_{{\rm cav}}$ and $\Omega_{m}$ will lead to
an effective cooling of the lower frequency mechanical oscillator,
as $\bar{n}_{{\rm th}}=\frac{k_{B}T_{bath}}{\hbar\Omega_{m}}\gg\bar{n}_{{\rm cav}}=\frac{k_{B}T_{bath}}{\hbar\omega_{{\rm cav}}}.$

\subsection{Influence of classical excess laser noise on sideband thermometry}

\label{Appendix:LaserNoiseSidebandThermometry}Excess noise of the
readout laser (characterized by an occupancy $n_{L}$ ) compromises
the self-calibration of the sideband thermometry method (Sec.~\ref{sub:CoolingQuantumTheory}),
as it can lead equally to an asymmetry, since in the presence of laser
noise $\dot{N}^{cav}(\Delta=\Omega_{m})=\frac{\kappa_{ex}}{\kappa}A^{+}(\bar{n}_{f}+1+n_{L})$
and $\dot{N}^{cav}(\Delta=-\Omega_{m})^{lower}=\frac{\kappa_{ex}}{\kappa}A^{-}(\bar{n}_{f}-n_{L})$.
This asymmetry can in this case be understood by arising from noise
squashing and anti-squashing of the classical and quantum noises alike.
This noise induced asymmetry on the upper and lower sideband can also
be viewed as originating from the effects of optomechanically induced
transparency (OMIT) and optomechanically induced amplification (or
EIA).

\bibliographystyle{apsrmp}
\bibliography{final_aspel_new_final,final_FMnew_final,final_RMP_final_final}

\end{document}